\newcommand*{\ATLASLATEXPATH}{}
\documentclass[PAPER,cernpreprint,atlasdraft=false,texlive=2016,UKenglish,titlesize=small]{\ATLASLATEXPATH atlasdoc}
 
\usepackage{\ATLASLATEXPATH atlaspackage}
\usepackage{\ATLASLATEXPATH atlasbiblatex}
 
\usepackage{\ATLASLATEXPATH atlasphysics}
 
\graphicspath{{logos/}{figures/}}
 
\usepackage{ANA-TOPQ-2018-17-PAPER-defs}

\AtlasTitle{Measurement of the $t\bar{t}$ production cross-section and
lepton differential distributions in $e\mu$ dilepton events from $pp$
collisions at $\sqrt{s}=13$\,\ensuremath{\text{Te\kern -0.1em V}} with the ATLAS detector}

\AtlasAbstract{
The inclusive top quark pair ($t\bar{t}$) production cross-section
$\sigma_{t\bar{t}}$ has been
measured in proton--proton collisions at $\sqrt{s}=13$\,\ensuremath{\text{Te\kern -0.1em V}}, using $36.1$\,fb$^{-1}$ of
data collected in 2015--16 by the ATLAS experiment at the LHC.
Using events with an opposite-charge $e\mu$ pair and $b$-tagged
jets, the cross-section is measured to be:
\begin{equation}\nonumber
\sigma_{t\bar{t}} = 826.4 \pm 3.6\,\mathrm{(stat)}\ \pm 11.5\,\mathrm{(syst)}\ \pm 15.7\,\mathrm{(lumi)}\ \pm 1.9\,\mathrm{(beam)}\,\mathrm{pb},
\end{equation}
where the uncertainties reflect the limited size of the data sample,
experimental and theoretical systematic
effects, the integrated luminosity, and the LHC beam energy,
giving a total uncertainty of 2.4\%. The result is consistent
with theoretical QCD calculations at next-to-next-to-leading order.
It is used to determine
the top quark pole mass via the dependence of the predicted cross-section on
$m_t^{\mathrm{pole}}$, giving $m_t^{\mathrm{pole}}=173.1^{+2.0}_{-2.1}$\,\ensuremath{\text{Ge\kern -0.1em V}}. It is also combined
with measurements at $\sqrt{s}=7$\,\ensuremath{\text{Te\kern -0.1em V}} and $\sqrt{s}=8$\,\ensuremath{\text{Te\kern -0.1em V}} to derive ratios and double
ratios of $t\bar{t}$ and $Z$ cross-sections at different energies. The
same event sample is used to measure absolute and normalised differential
cross-sections as functions of single-lepton and dilepton kinematic variables,
and the results are compared with predictions from various Monte Carlo event
generators.
}
 
\author{The ATLAS Collaboration}
 
\AtlasRefCode{TOPQ-2018-17}
\PreprintIdNumber{CERN-EP-2019-203}

\AtlasCoverSupportingNote{Analysis documentation}{https://cds.cern.ch/record/2634819}

\AtlasJournalRef{Eur.\ Phys.\ J.\ C 80 (2020) 528}
\AtlasDOI{10.1140/epjc/s10052-020-7907-9}
 
\AtlasCoverAnalysisTeam{Richard Hawkings}
 
\AtlasCoverEdBoardMember{Frederic Derue, Kevin Finelli, Miriam Watson (chair)}

\AtlasCoverEgroupEditors{atlas-TOPQ-2018-17-editors@cern.ch}
 
\AtlasCoverEgroupEdBoard{atlas-TOPQ-2018-17-editorial-board@cern.ch}
\hypersetup{pdftitle={ATLAS document},pdfauthor={The ATLAS Collaboration}}
 
\begin{document}
 
\maketitle
 
\tableofcontents

\section{Introduction}\label{s:intro}
 
The study of top quark--antiquark (\ttbar) production forms a cornerstone
of the physics programme of the ATLAS experiment at the CERN Large Hadron
Collider (LHC), allowing quantum chromodynamics (QCD) to be probed at some
of the highest accessible energy scales. The large mass of the top quark,
close to the scale of electroweak symmetry breaking, gives it a unique
role in the Standard Model of particle physics and potential extensions,
and \ttbar\ production also forms an important background in many
searches for physics beyond the Standard Model. Precise measurements
of absolute rates and differential distributions in \ttbar\ production
are therefore a vital tool in fully exploiting the discovery potential
of the LHC.
 
Predictions for the inclusive \ttbar\ production cross-section in
proton--proton ($pp$) collisions, \xtt,
are available at next-to-next-to-leading-order (NNLO)
accuracy in the strong coupling constant \alphas, including the resummation
of next-to-next-to-leading logarithmic (NNLL) soft gluon terms
\cite{topxtheo1,topxtheo2,topxtheo3,topxtheo4,topxtheo5,topxtheo6}, and are in
excellent agreement with  measurements from ATLAS and CMS
at $\sqrt{s}=7$, 8 and 13\,\TeV\
\cite{TOPQ-2013-04,TOPQ-2016-08,TOPQ-2015-09,CMS-TOP-13-004,CMS-TOP-12-006,CMS-TOP-16-006,CMS-TOP-17-001}.
At \sxyt, and assuming a fixed top quark mass of $\mtop=172.5$\,\GeV,
the NNLO+NNLL prediction is $832\pm 35^{+20}_{-29}$\,pb, as calculated using
the {\tt Top++ 2.0} program \cite{toppp}. The first uncertainty
corresponds to parton distribution function (PDF) and \alphas\ uncertainties,
and the second to QCD scale variations. The former were
calculated using the PDF4LHC prescription \cite{pdf4lhc} with the
MSTW2008 \cite{mstwnnlo1,mstwnnlo2}, CT10 NNLO \cite{cttenpdf,cttennnlo}
and NNPDF2.3 5f FFN \cite{nnpdfffn} PDF sets.\footnote{The NLO prescription
from Ref. \cite{pdf4lhc} was used, but applied to the specified NNLO
PDF sets. The PDF uncertainty envelope was defined to cover the positive
and negative 68\% confidence level uncertainties of each considered PDF
set, and the \xtt\ central value was defined as the midpoint of the envelope.
The recommended \alphas\ value was used for each PDF set ($0.1170\pm 0.0014$
for MSTW2008 and $0.1180\pm 0.0012$ for CT10 and NNPDF2.3) and the
\alphas\ variations were included in the envelope uncertainties.}
The latter was calculated from the envelope of predictions with the QCD
renormalisation and factorisation scales
varied independently up or down by a factor of two from their default values
of $\muf=\mur=\mtop$, whilst never letting them differ by more than a factor
of two \cite{qcdscale1,qcdscale2}. The total uncertainty corresponds
to a relative precision of $^{+4.8}_{-5.5}$\%.
 
The predicted  cross-section also depends strongly on \mtop,
decreasing by 2.7\% for a 1\,\GeV\
increase in the top mass. The top quark mass parameter used in the cross-section
prediction is actually the pole mass \mtpole, corresponding to the definition
of the mass of a free particle. This allows \xtt\ measurements to
be interpreted as measurements of \mtpole, free of the theoretical
ambiguities linked to the direct reconstruction of \mtop\
from the invariant mass of its decay products
\cite{buckleypole,mochpole,hoangpole,nasonpole}.
Ratios of \ttbar\ production cross-sections at different centre-of-mass
energies are also of interest, e.g. $\rttyw=\xtt(13\,\TeV)/\xtt(7\,\TeV)$.
Predictions for such ratios benefit from significant
cancellations in the QCD scale and top quark mass uncertainties,
but are still sensitive to the choice of PDF. The NNLO+NNLL predictions with the
same set of assumptions as given for \xtt\ above, and a 1\,\GeV\
uncertainty in \mtop, are
$\rttyw=4.69\pm 0.16$ and $\rttyv=3.28\pm 0.08$, i.e. relative uncertainties
of 3.3\% and 2.5\%. Double ratios of \ttbar\ to $Z$ production
cross-sections allow the experimental uncertainties to be further reduced,
by normalising the \ttbar\ cross-section at each energy to the corresponding
cross-section for $Z$ boson production \cite{STDM-2016-02}.
 
Within the Standard Model, the top quark decays 99.8\% of the time to
a $W$ boson and $b$-quark \cite{ckmpdg},
making the final-state topologies in \ttbar\
production dependent on the decay modes of the $W$ bosons. The channel
with an electron--muon pair with opposite electric charges, i.e.
$\ttbar\rightarrow W^{+}bW^{-}\bar{b}\rightarrow e^+\mu^-\nu\bar{\nu}\bbbar$,
is particularly clean.\footnote{Charge-conjugate decay modes are implied
unless otherwise stated.} It was exploited to make the most precise
ATLAS measurements of \xtt\ at $\sqrt{s}=7$, 8 and 13\,TeV
\cite{TOPQ-2013-04,TOPQ-2015-09}, based on events with an opposite-sign
$e\mu$ pair and one or two jets tagged as likely to contain $b$-hadrons ($b$-tagged). The \sxyt\ measurement in
Ref. \cite{TOPQ-2015-09} was based on the
3.2\,\ifb\ dataset recorded in 2015. This paper describes a new measurement
of \xtt\ at \sxyt\ using the same final state, but applied to the
combined 2015--16 ATLAS dataset of
\intlumi\,\ifb.
The cross-section measurement is further used to determine the top quark pole
mass via the dependence of the prediction on \mtpole, complementing the
analogous measurement with the $\sqrt{s}=7$ and 8\,\TeV\ data
\cite{TOPQ-2013-04}.
This paper also updates the \ttbar\ cross-section ratios
\rttyw\ and \rttyv, the \sxyt\ $\ttbar/Z$ ratio \rttzy, and the
double ratios of \ttbar\ to $Z$ cross-sections
\rttzyw\ and \rttzyv,
using the new \xtt\ result, superseding those derived from
the previous \sxyt\ \xtt\ measurement in Ref. \cite{STDM-2016-02}.
 
The $e\mu+b$-tagged jets sample also allows precise measurements of the
differential distributions of the leptons produced in \ttbar\ events
to be made. ATLAS has published \cite{TOPQ-2015-02}
measurements at \sxvt\ of the
absolute and normalised differential cross-sections as functions of the
transverse momentum \ptl\ and absolute pseudorapidity \etal\ of the single
leptons\footnote{
ATLAS uses a right-handed coordinate system with its origin at
the nominal interaction point in the centre of the detector, and the $z$ axis
along the beam line. Pseudorapidity is defined in terms of the polar angle
$\theta$ as $\eta=-\ln\tan{\theta/2}$, and transverse momentum and energy
are defined relative to the beam line as $\pt=p\sin\theta$ and
$\et=E\sin\theta$. The azimuthal angle around the beam line is denoted by
$\phi$, and distances in $(\eta,\phi)$ space by
$\Delta R=\sqrt{(\Delta\eta)^2+(\Delta\phi)^2}$.
The rapidity is defined as $y=\frac{1}{2}\ln\left(\frac{E+p_z}{E-p_z}\right)$,
where $p_z$ is the $z$-component of the momentum and  $E$ is the energy
of the relevant object or system. The distance in $(y,\phi)$ space is given by
$\Delta R_y=\sqrt{(\Delta y)^2+(\Delta\phi)^2}$.}
(combined for electrons and muons), the \pt, invariant mass and
absolute rapidity of the $e\mu$ system (\ptll, \mll\ and \rapll), the absolute
azimuthal angle $|\Delta\phi|$  between the two leptons in the transverse plane
(\dphill), and the scalar sums of the transverse momenta (\ptsum) and energies
(\esum) of the two leptons. These distributions were found to be generally
well described by  predictions from a variety of leading-order (LO) multileg
and next-to-leading-order (NLO) \ttbar\ matrix-element event generators
interfaced to parton showers, and by NLO fixed-order QCD calculations.
The sensitivity of the data to the gluon PDF and to the top quark pole mass
was also demonstrated. This paper measures the same
distributions at \sxyt, using \ttbar\ samples which are about six times
the size of those available at \sxvt. Two-dimensional distributions of
\etal, \rapll\ and \dphill\ as functions of \mll\ are also reported.
The data are again compared with
the predictions of various NLO \ttbar\ matrix-element event generators,
but the interpretations in terms of PDF constraints and \mtpole\ are left
for future work.
 
The event selection, measurement methodology and uncertainty evaluations for
both the inclusive \ttbar\ cross-section and the differential distributions
are similar to those used at $\sqrt{s}=7$ and 8~\,\TeV\
\cite{TOPQ-2013-04,TOPQ-2015-02}, with the exception that the
minimum lepton transverse momentum requirement has been lowered from
25\,\GeV\ to 20\,\GeV, whilst still requiring at least one lepton
to be above the lepton trigger threshold of 21--27\,\GeV.
This increases the fraction of
$\ttbar\rightarrow e\mu\nu\bar{\nu}\bbbar$ events that are selected
by 16\%, thus reducing the extrapolation uncertainties in the
modelling of \ttbar\ production and decay.
The data and Monte Carlo simulation samples used in the analyses are described
in Section~\ref{s:datmc}, followed by the event reconstruction and selection
in Section~\ref{s:objevt}. The measurement methodology for both the inclusive
and differential cross-sections is described in Section~\ref{s:meas},
and the evaluation of systematic uncertainties in Section~\ref{s:syst}.
The inclusive cross-section results are given in Section~\ref{s:xsres},
together with the derivation of the top quark pole mass from \xtt,
and the corresponding \ttbar\ and $\ttbar/Z$ cross-section ratios.
The differential cross-section
results are discussed in Section~\ref{s:diffres}, and compared with the
predictions of several \ttbar\ event generators. Finally, conclusions
are discussed in Section~\ref{s:conc}.

\section{Data and simulated event samples} \label{s:datmc}
 
The ATLAS detector \cite{PERF-2007-01,atlasibl,IBLTDR} at the LHC covers
nearly the entire solid angle around the collision point.
It consists of an inner tracking
detector surrounded by a thin superconducting solenoid producing
a 2T axial magnetic field, electromagnetic and hadronic calorimeters,
and an external muon spectrometer incorporating three large toroidal
magnet assemblies.
The analysis was performed on samples of proton--proton collision data
collected at \sxyt\ in 2015 and 2016, corresponding to total integrated
luminosities of 3.2\,\ifb\ in 2015 and 32.9\,\ifb\ in 2016 after
data quality requirements. Events were required to pass a single-electron
or single-muon trigger \cite{TRIG-2016-01,TRIG-2018-05},
with transverse momentum thresholds that were
progressively raised during the data-taking as the instantaneous
luminosity increased. The electron trigger was fully efficient for
electrons with reconstructed $\pt>25$\,\GeV\ in 2015 and the first 6\,\ifb\
of 2016 data, and for $\pt>27$\,\GeV\ for the remainder. The corresponding
muon trigger thresholds were $\pt>21$\,\GeV\ for 2015 data,
$\pt>25$\,\GeV\ for the first 6\,\ifb\ of 2016 data  and $\pt>27$\,\GeV\ for
the rest. Each triggered event also includes the signals from on average
14 (25) additional inelastic $pp$ collisions in 2015 (2016) data, referred
to as pileup.
 
Monte Carlo simulated event samples were used to develop the analysis
procedures, to evaluate signal and background contributions, and to compare
with data. Samples were processed using either the full ATLAS
detector simulation \cite{SOFT-2010-01} based on GEANT4 \cite{geant4}, or
with a faster simulation making use of parameterised showers in the
calorimeters \cite{ATL-PHYS-PUB-2010-013}. The effects of pileup were
simulated by generating additional inelastic $pp$ collisions with
{\sc Pythia8} (v8.186) \cite{pythia8} using the A2 set of parameter values
(tune) \cite{ATL-PHYS-PUB-2011-014}
and overlaying them on the primary simulated events. These combined
events were then processed using the same reconstruction and analysis
chain as the data. Small corrections were applied to lepton trigger and
reconstruction efficiencies to improve agreement with the response
observed in data.
 
The baseline simulated \ttbar\ sample was produced using the NLO
matrix-element event generator {\sc Powheg-Box} v2 (referred to hereafter as
{\sc Powheg}) \cite{powheg,powheg2,powheg3,powheghvq}
with the NNPDF3.0 NLO PDF set \cite{nnpdf3}, interfaced to {\sc Pythia8}
(v8.210) with the NNPDF2.3 LO PDF set and the A14 tune
\cite{ATL-PHYS-PUB-2014-021}
for the parton shower, hadronisation and underlying-event modelling.
In the {\sc Powheg} configuration,
the \hdamp\ parameter, which gives a cut-off scale for the first gluon
emission, was set to $\frac{3}{2}\mtop$, and the factorisation and
renormalisation scales were set to
$\muf=\mur=\sqrt{(m_{t}^2+(p_{\mathrm{T},t})^2)}$,
where the top quark \pt\ is evaluated before radiation
\cite{ATL-PHYS-PUB-2016-020}.
 
Alternative \ttbar\ simulation samples used to assess systematic uncertainties
were generated with {\sc Powheg}
interfaced to {\sc Herwig7} (v7.0.4) \cite{herwig7} with the H7UE tune, and with
the {\sc MadGraph5\_aMC@NLO} (v2.3.3.p1) generator (referred to hereafter as
{\sc aMC@NLO}) \cite{amcnlo} with the NNPDF3.0 NLO PDF set, interfaced to
{\sc Pythia8} with the A14 tune. In the {\sc aMC@NLO} sample, the
renormalisation and factorisation scales were set in the same way as for
{\sc Powheg}, and the MC@NLO prescription \cite{mcatnlopre} was used
for matching the NLO matrix element to the parton shower. Uncertainties
related to the amount of initial- and final-state radiation were
explored using two alternative {\sc Powheg\,+\,Pythia8} samples: one with
\hdamp\ set to $3\mtop$, \muf\ and \mur\ reduced by a factor of two from
their default values, and
the A14v3cUp tune variation, giving more parton-shower radiation;
and a second sample with $\hdamp=\frac{3}{2}\mtop$, \muf\ and \mur\
increased by a factor of two and the A14v3cDo tune variation, giving less
parton-shower radiation. These parameter
variations were chosen in order to reproduce differential cross-section
and jet multiplicity distributions measured in \ttbar\ events,
as discussed in Ref. \cite{ATL-PHYS-PUB-2016-020}.
The top quark mass was set to 172.5\,\GeV\ in all these
samples, consistent with measurements from ATLAS \cite{TOPQ-2017-03}
and CMS \cite{CMS-TOP-14-022}. The $W\rightarrow\ell\nu$ branching
ratio was set to the Standard Model prediction of 0.1082 per lepton flavour
\cite{wpdg},
and the {\sc EvtGen} program \cite{evtgen} was used to handle the decays
of $b$- and $c$-flavoured hadrons.  All the samples were normalised using the
NNLO+NNLL inclusive cross-section prediction discussed in Section~\ref{s:intro}
when comparing simulation with data. Additional \ttbar\ samples with
different event generator configurations were used in comparisons with the
measured normalised differential cross-sections as discussed in
Section~\ref{ss:gencomp}.
 
Backgrounds in these measurements are classified into two types: those with
two real prompt leptons (electrons or muons) from $W$ or $Z$ boson decays
(including those produced by leptonic decays of $\tau$-leptons), and those
where at least one of the reconstructed leptons is misidentified, i.e.
a non-prompt lepton from the decay of a bottom or charm hadron, an electron
from a photon conversion, a hadronic jet misidentified as an electron,
or a muon produced from the decay in flight of a pion or kaon.
The background with two real prompt leptons is dominated by the
associated production of a $W$ boson and single top quark, $Wt$.
This process was simulated using {\sc Powheg} v1 \cite{wtinter1}
with the CT10 NLO PDF set \cite{cttenpdf}, interfaced
to {\sc Pythia6} (v6.428) \cite{pythia6} with the P2012 tune \cite{perugia}.
The `diagram removal' scheme \cite{wtdr} was used to handle the interference
between the \ttbar\ and $Wt$ final states that occurs at NLO. The sample
was normalised to a cross-section of $71.7\pm 3.8$\,pb, based on the
approximate NNLO calculation \cite{wttheo,wttheo2} using the
MSTW2008 NNLO PDF set \cite{mstwnnlo1,mstwnnlo2},
and taking into account PDF and QCD scale uncertainties.
Smaller backgrounds result from
$Z\rightarrow\tau\tau (\rightarrow e\mu)$+jets, and from diboson production
($WW$, $WZ$ and $ZZ$) in association with jets. These backgrounds were modelled
using {\sc Sherpa} 2.2.1 \cite{sherpa21} ($Z$+jets) and {\sc Sherpa 2.1.1}
(dibosons), as discussed in Ref.~\cite{TOPQ-2016-11}.
Production of \ttbar\ in association with a leptonically decaying $W$, $Z$
or Higgs boson gives a negligible contribution to the opposite-sign
$e\mu$ samples compared to inclusive \ttbar\ production, but
is significant in the same-sign control samples used to assess the
background from misidentified leptons. These processes were simulated
using {\sc aMC@NLO\,+\,Pythia8} ($\ttbar+W/Z$) or {\sc Powheg\,+\,Pythia8}
($\ttbar+H$) \cite{TOPQ-2016-11}.
 
Backgrounds with one real and one misidentified lepton arise from \ttbar\
events with one leptonically decaying and one hadronically decaying $W$,
including $W\rightarrow\tau\nu$ with a hadronic $\tau$ decay. These processes
were simulated with {\sc Powheg\,+\,Pythia8} in the same way
as for dileptonic \ttbar.
Similar backgrounds also arise from $W$+jets production, modelled with
{\sc Sherpa 2.2.1} as for $Z$+jets;
and $t$-channel single top quark production, modelled with
{\sc Powheg\,+\,Pythia6} \cite{powhegtchan}
with the CT10 PDF set and P2012 tune. The contributions
of these backgrounds to the opposite-sign samples were determined with
the help of the same-sign control samples in data. Other backgrounds,
including processes with two misidentified leptons, are negligible after the
event selections used in the analysis.

\section{Event reconstruction and selection}\label{s:objevt}
 
The analysis makes use of reconstructed electrons, muons and $b$-tagged jets.
Electron candidates were reconstructed from a localised cluster of energy
deposits in the
electromagnetic calorimeter matched to a track in the inner detector, passing
the `Tight' likelihood-based requirement of Ref. \cite{PERF-2017-01}.
They were required to have transverse energy $E_\mathrm{T}>20$\,\GeV\
and pseudorapidity
$|\eta|<2.47$, excluding the transition region between the barrel and endcap
electromagnetic calorimeters, $1.37<|\eta|<1.52$. To ensure they originated
from the event primary vertex, electrons were required to satisfy
requirements on the transverse impact parameter significance calculated
relative to the beam line of $|d_0|/\sigma_{d_0}<5$, and on the
longitudinal impact parameter calculated relative to the event primary
vertex of $|\Delta z_0 \sin\theta|<0.5$\,mm, where $\theta$ is the polar
angle of the track. The event primary vertex was
defined as the reconstructed vertex with the highest sum of $\pt^2$ of
associated tracks. To reduce background from non-prompt electrons, candidates
were further required to be isolated, using
\pt- and $|\eta|$-dependent requirements on the summed calorimeter energy
within a cone of size $\Delta R=0.2$ around the electron cluster, and on the sum
of track \pt\ within a cone of variable size
$\Delta R=\mathrm{min}(0.2,10\,\GeV/\pt(e))$ around the electron
track direction.
The selections were tuned to give a 90\% efficiency for electrons of
$\pt=25$\,\GeV\ in simulated $Z\rightarrow ee$ events, rising to 99\% at
60\,\GeV.
 
Muon candidates were reconstructed by combining matching tracks reconstructed
in the inner detector and muon spectrometer, and were required to
have $\pt>20$\,\GeV, $|\eta|<2.5$ and to satisfy the `Medium' requirements
of Ref. \cite{PERF-2015-10}. Muons were also required to be isolated using
calorimeter and track information in the same way as it was used
for electrons, except
that the track-based isolation was calculated with a cone of size
$\Delta R=\mathrm{min}(0.3,10\,\GeV/\pt(\mu))$. The selections were
again tuned to give efficiencies
varying from 90\% at $\pt=25$\,\GeV\ to 99\% at 60\,\GeV\ on
simulated $Z\rightarrow\mu\mu$ events. No requirements were made on the muon
impact parameters relative to the primary vertex, as they do not provide
any useful additional background rejection in this event topology.
 
Jets were reconstructed using the anti-$k_t$ algorithm \cite{antikt,fastjet}
with radius parameter $R=0.4$, starting from topological clusters in the
calorimeters \cite{PERF-2014-07}. After calibration using information from
both simulation and data \cite{PERF-2016-04}, jets were required to have
$\pt>25$\,\GeV\ and $|\eta|<2.5$, and jets with $\pt<60$\,GeV\
and $|\eta|<2.4$ were subject to additional pileup rejection criteria
using the multivariate jet-vertex tagger (JVT) \cite{PERF-2014-03}.
To prevent double counting of electron energy deposits as jets, the closest
jet to an electron candidate was removed if it was within $\Delta R_y=0.2$
of the electron. Furthermore, to reduce the contribution of leptons
from heavy-flavour hadron decays inside jets, leptons within $\Delta R_y=0.4$
of selected jets were discarded, unless the lepton was a muon and the jet
had fewer than three associated tracks, in which case the jet was discarded
(thus avoiding an efficiency loss for high-energy muons undergoing
significant energy loss in the calorimeters).
 
Jets likely to contain $b$-hadrons were $b$-tagged using the MV2c10 algorithm
\cite{PERF-2016-05}, a multivariate discriminant making use of track
impact parameters and reconstructed secondary vertices. A tagging working
point corresponding to 70\% efficiency for tagging $b$-quark jets from
top quark decays in simulated \ttbar\ events was used, corresponding to
rejection factors (i.e. the inverse of the mistag rates)
of about 400 against light-quark and gluon jets and
12 against jets originating from charm quarks.
 
Selected events were required to have exactly one electron and exactly
one muon passing the requirements detailed above, with at least
one of the leptons matched to a corresponding electron or muon trigger.
Events where  the electron and muon were separated in angle by
$|\Delta\theta|<0.15$ and $|\Delta\phi|<0.15$, or where at least one jet
with $\pt>20$\,\GeV\ failed quality requirements \cite{PERF-2012-01},
were rejected. Events with an opposite-sign $e\mu$ pair formed the main
analysis sample, whilst events with a same-sign $e\mu$ pair were used
in the estimation of background from misidentified leptons.
Table~\ref{t:evtselreq} summarises the main selection requirements.
 
\begin{table}
\centering
 
\begin{tabular}{l|ll}\hline
Object & Identification & Selection \\
\hline
Electrons & Tight likelihood & $E_\mathrm{T}>20$\,\GeV, $|\eta|<1.37$ or $1.52<|\eta|<2.47$, isolation \\
Muons & Medium & $\pt>20$\,\GeV, $|\eta|<2.5$, isolation \\
Jets & Anti-$k_t$ $R=0.4$ & $\pt>25$\,\GeV, $|\eta|<2.5$, $b$-tagging with MV2c10 at 70\% efficiency \\
\hline
Event  & & 1~electron+1~muon with opposite sign, 1 or 2 $b$-tagged jets \\
\end{tabular}
\caption{\label{t:evtselreq}Summary of the main object and event selection
requirements.}
\end{table}

\section{Cross-section measurement}\label{s:meas}
 
The same technique, employing the subsets of the opposite-sign $e\mu$ sample
with exactly one and exactly two $b$-tagged jets, was used to measure
both the inclusive \ttbar\ cross-section and the differential distributions.
The measurements are introduced in the following two subsections, followed
by a discussion of the background estimate in Section~\ref{ss:bkgd}
and the validation of the differential measurements using studies based on
simulation in Section~\ref{ss:diffval}.
 
\subsection{Inclusive cross-sections}\label{ss:incmeas}
 
The inclusive
\ttbar\ cross-section \xtt\ was determined by counting the numbers of
opposite-sign $e\mu$ events with exactly one ($N_1$) and exactly two ($N_2$)
$b$-tagged jets. The two event counts satisfy the tagging equations:
\begin{equation}
\begin{array}{lll}
N_1 & = & L \xtt\ \epsem 2\epsb (1-\cb\epsb) + \nib , \\*[2mm]
N_2 & = & L \xtt\ \epsem \cb\epsb^2 + \niib
\end{array}\label{e:tags}
\end{equation}
where $L$ is the integrated luminosity of the sample, \epsem\ the efficiency
for a \ttbar\ event to pass the opposite-sign $e\mu$
selection, and \cb\ is a tagging correlation coefficient close to unity.
The combined probability for a jet from the quark $q$ in the $t\rightarrow Wq$
decay to fall within the acceptance of the detector, be reconstructed as a jet
with transverse momentum above the selection threshold, and be tagged as a
$b$-jet, is denoted by \epsb. If the decays of the two top quarks
and the reconstruction of the two associated $b$-tagged jets are
completely independent, the probability \epsbb\ to reconstruct and tag
both $b$-jets is given by $\epsbb=\epsb^2$. In practice, small correlations
are present, due to kinematic correlations between the $b$-jets from the two
top quarks, or the production of extra \bbbar\ or \ccbar\ pairs in the
\ttbar\ events.
These effects are taken into account via the correlation coefficient
$\cb=\epsbb/\epsb^2$, or equivalently
$\cb=4\nttem N^{\ttbar}_2/(N^{\ttbar}_1+2N^{\ttbar}_2)^2$, where \nttem\ is the
number of selected $e\mu$ \ttbar\ events and $N^{\ttbar}_1$ and $N^{\ttbar}_2$
are the numbers of such events with one and two $b$-tagged jets. In the baseline
\ttbar\ simulation sample, $\epsem\approx 0.9$\%, including the branching
ratio for a \ttbar\ pair to produce an $e\mu$ final state. The corresponding
value of
\cb\ is $1.007\pm 0.001$ (the uncertainty coming from the limited size
of the simulation sample), indicating a small positive correlation between the
reconstruction and $b$-tagging of the two quarks produced in the top quark
decays.
Background from sources other than \ttbar\ events with two prompt leptons also
contributes to $N_1$ and $N_2$ and is given by the terms \nib\ and \niib,
evaluated using a combination of simulation and data control samples
as  discussed in Section~\ref{ss:bkgd} below. The values of \epsem\ and \cb\
were taken from \ttbar\ event simulation, allowing the tagging
equations~(\ref{e:tags}) to be solved to determine \xtt\ and \epsb.
 
The selection efficiency \epsem\ can be written as the product of two
terms: $\epsem=\aem\gem$. The acceptance $\aem\approx 1.7$\%
represents the fraction
of \ttbar\ events which have a true opposite-sign $e\mu$ pair from
$t\rightarrow W\rightarrow e/\mu$ decays, with each lepton having
$\pt>20$\,\GeV\ and $|\eta|<2.5$. The contributions via leptonic
$\tau$ decays ($t\rightarrow W\rightarrow\tau\rightarrow e/\mu$) are included.
The lepton four-momenta were taken after final-state radiation, and `dressed'
by including the four-momenta of any photons within a cone of size
$\Delta R=0.1$ around the lepton direction, excluding photons produced
from hadron decays or interactions with the detector material.
The reconstruction efficiency \gem\ represents the probability that the
two leptons are reconstructed and pass all the identification and
isolation requirements. A fiducial cross-section \xfid, for the production of
\ttbar\ events with an electron and a muon satisfying the requirements on
\pt\ and $\eta$, can then be defined as $\xfid=\aem\xtt$, and measured
by replacing $\xtt\epsem$ with $\xfid\gem$ in Eqs.~(\ref{e:tags}).
The fiducial cross-section definition makes no requirements on the
presence of jets, as the tagging formalism of Eqs.~(\ref{e:tags})
allows the number of \ttbar\ events with no reconstructed and $b$-tagged jets
to be inferred from the event counts $N_1$ and $N_2$.
Measurement
of the fiducial cross-section avoids the systematic uncertainties associated
with the evaluation of the acceptance, in particular estimation of
the fraction of
$\ttbar\rightarrow e\mu\nu\bar{\nu}\bbbar$ events where at least one lepton
has $\pt<20$\,\GeV\ or $|\eta|>2.5$.
 
A total of 40\,680 data events passed the opposite-sign $e\mu$ selection
in the 2015 data sample, and 358\,664 events in the 2016 data sample.
They were subdivided according to the number of $b$-tagged jets,
irrespective of the number of untagged jets. The numbers of events
with one and two $b$-tagged jets in each sample are shown in
Table~\ref{t:evtcount}, together with the expected
non-\ttbar\ contributions from $Wt$ and dibosons evaluated from simulation,
and $Z(\rightarrow\tau\tau\rightarrow e\mu)$+jets and misidentified leptons
evaluated using both data and simulation. The one $b$-tag sample
is expected to be about 88\% pure and the two $b$-tag sample 96\% pure
in \ttbar\ events, with the largest backgrounds in both samples
coming from $Wt$ production.
The distribution of the number of $b$-tagged jets is shown for the
2015 and 2016 data samples together in
Figure~\ref{f:dmcjlept}(a), and compared with the expectations from simulation,
broken down into contributions from \ttbar\ events (modelled using the
baseline {\sc Powheg\,+\,Pythia8} sample), and various background processes.
The predictions using alternative \ttbar\ generator configurations
({\sc Powheg\,+\,Pythia8}
with more or less parton-shower radiation, denoted by `RadUp' and `RadDn',
and {\sc aMC@NLO\,+\,Pythia8})
are also shown. All expected contributions are normalised to the
integrated luminosity of the data sample using the
cross-sections discussed in Sections~\ref{s:intro} and~\ref{s:datmc}.
The excess of data events over the prediction in the
zero $b$-tagged jets sample (which is not used in the measurement) was
also observed previously \cite{TOPQ-2013-04,TOPQ-2015-09} and is compatible
with the expected uncertainties in modelling diboson and $Z$+jets production.
 
\begin{table}
\centering
 
\begin{tabular}{l|
r@{$\,\pm\,$}l
r@{$\,\pm\,$}l|
r@{$\,\pm\,$}l
r@{$\,\pm\,$}l
}
Sample & \multicolumn{4}{c|}{2015} & \multicolumn{4}{c}{2016} \\
Event counts & \multicolumn{2}{c}{$N_1$} & \multicolumn{2}{c|}{$N_2$} & \multicolumn{2}{c}{$N_1$} & \multicolumn{2}{c}{$N_2$} \\
\hline
Data &  \multicolumn{2}{l}{14239} & \multicolumn{2}{l|}{8351} & \multicolumn{2}{l}{133977} &  \multicolumn{2}{l}{75853} \\
\hline
$Wt$ single top &   1329 & 92  &    261 & 86 & 12490 & 870 &   2430 & 810 \\
$Z(\rightarrow\tau\tau\rightarrow e\mu)$+jets &    123 & 15 &      7 & 2 &    910 &   110 &     37 & 9 \\
Diboson &    42 &     5 & 1 & 0 &   481 & 58 & 21 &  7 \\
Misidentified leptons &  164 & 54 & 58 & 37 &  1720 & 520 & 670 & 390 \\
\hline
Total background &   \phantom{9}1660 & 110 &  \phantom{9}327 & 94 &  \phantom{9}15600 & 1000 & \phantom{9}3160 & 890 \\
\end{tabular}
\caption{\label{t:evtcount}Observed numbers of opposite-sign $e\mu$ events
with one ($N_1$) and two ($N_2$) $b$-tagged jets in 2015 and 2016 data,
together with the estimates of backgrounds and associated uncertainties
described in Section~\ref{s:syst}. Uncertainties shown as zero are less than
0.5 events.}
\end{table}
 
\begin{figure}[htp]
\vspace{-7mm}
\parbox{83mm}{\includegraphics[width=76mm]{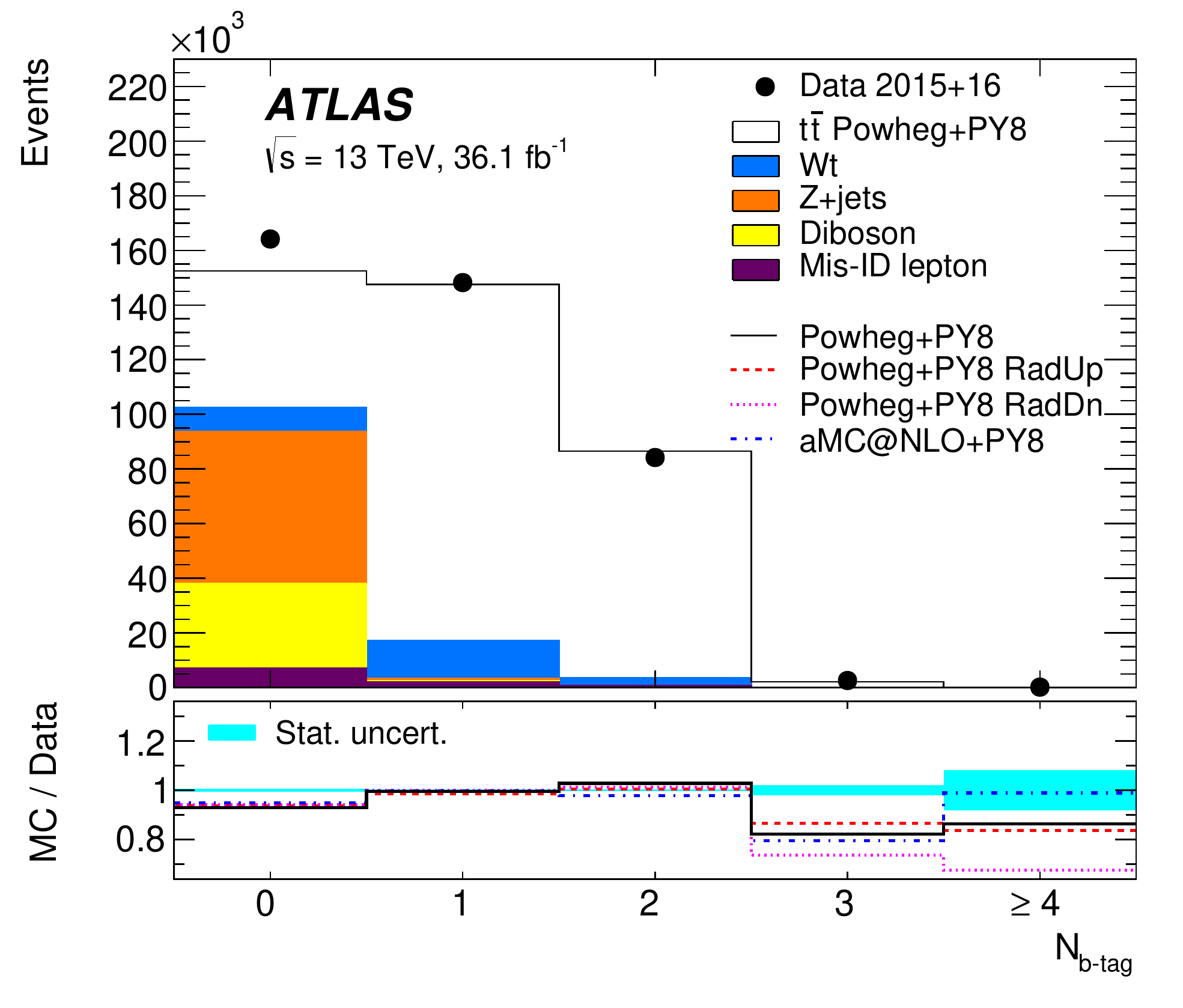}\vspace{-7mm}\center{(a)}}
\parbox{83mm}{\includegraphics[width=76mm]{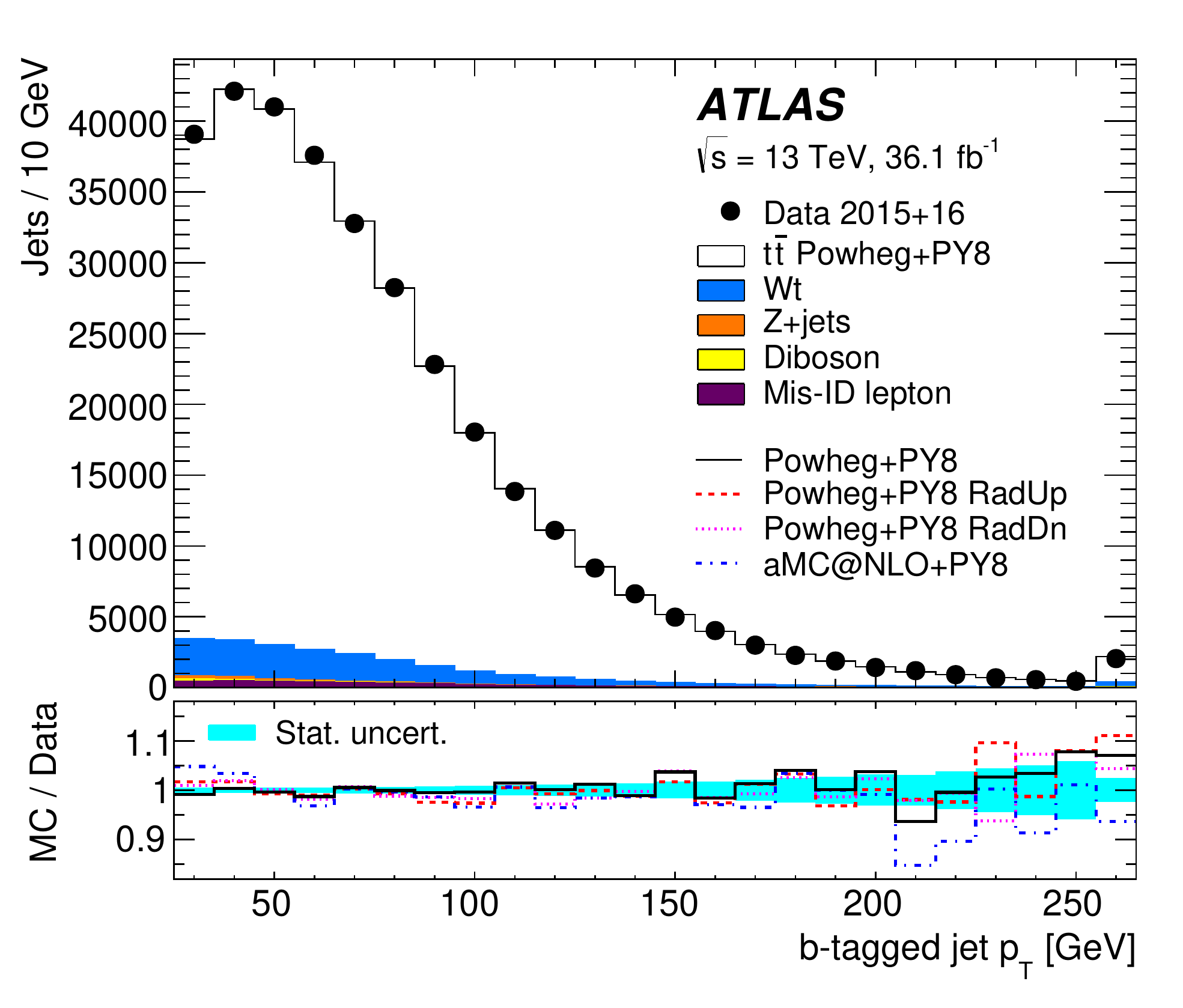}\vspace{-7mm}\center{(b)}}
\parbox{83mm}{\includegraphics[width=76mm]{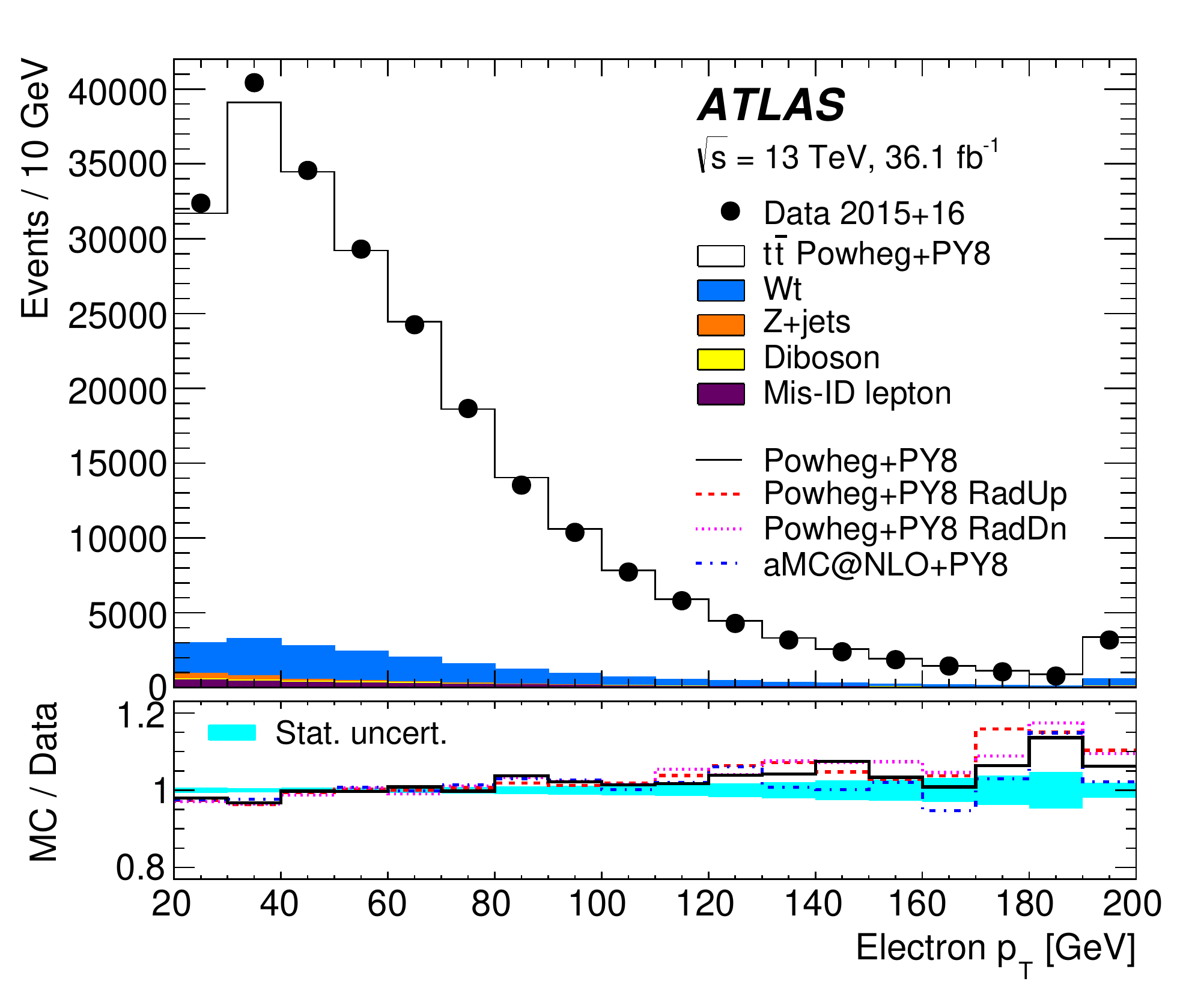}\vspace{-7mm}\center{(c)}}
\parbox{83mm}{\includegraphics[width=76mm]{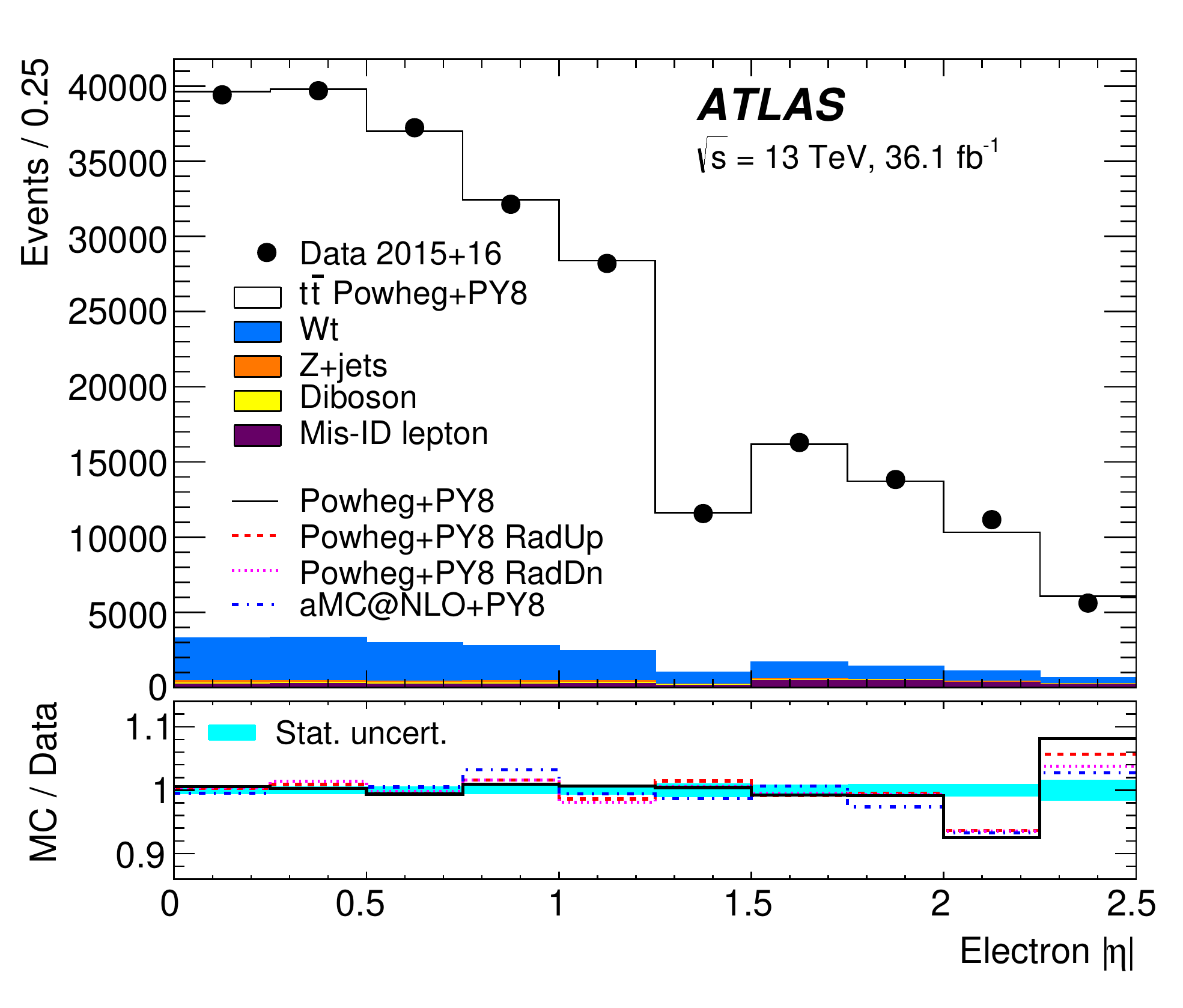}\vspace{-7mm}\center{(d)}}
\parbox{83mm}{\includegraphics[width=76mm]{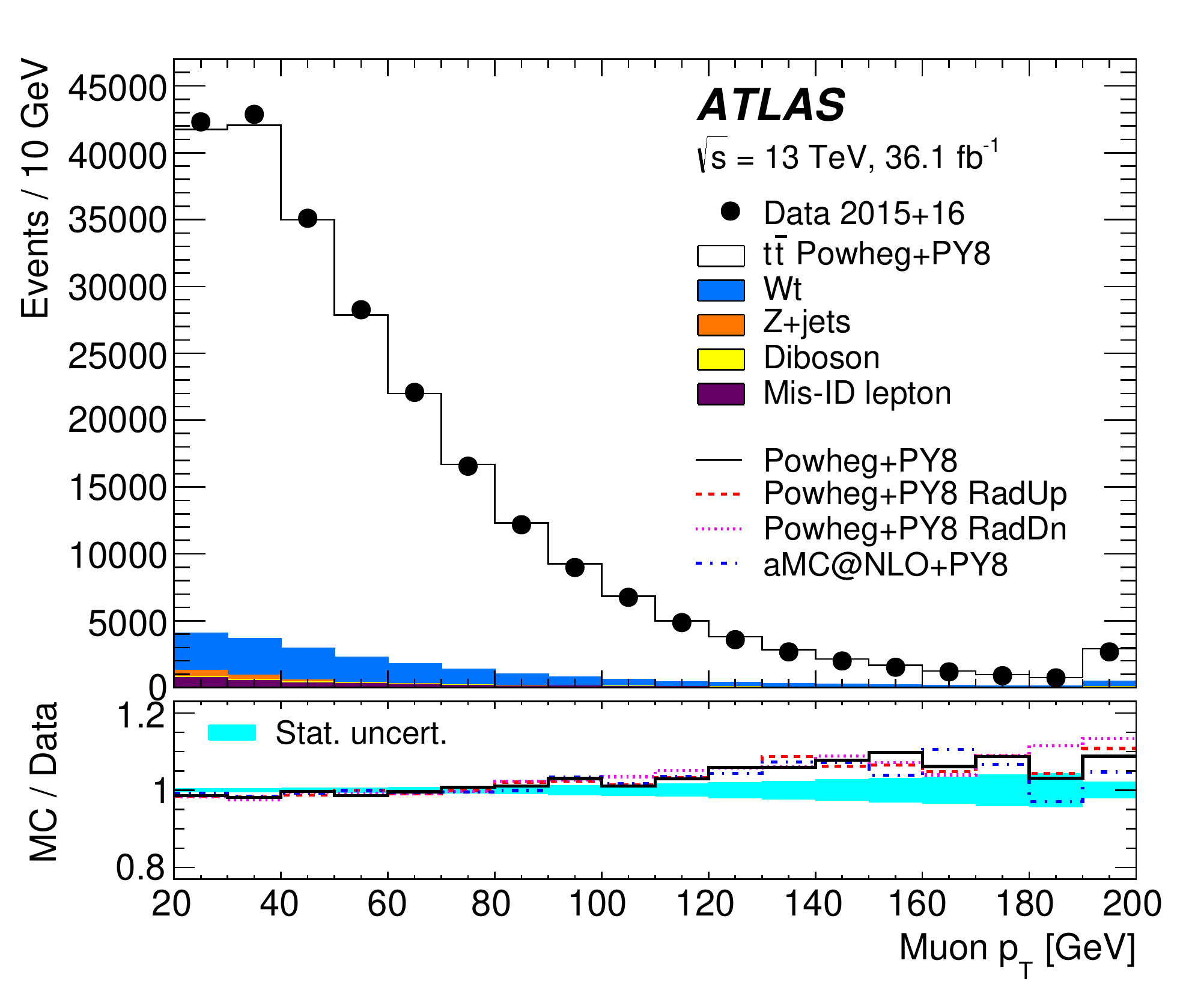}\vspace{-7mm}\center{(e)}}
\parbox{83mm}{\includegraphics[width=76mm]{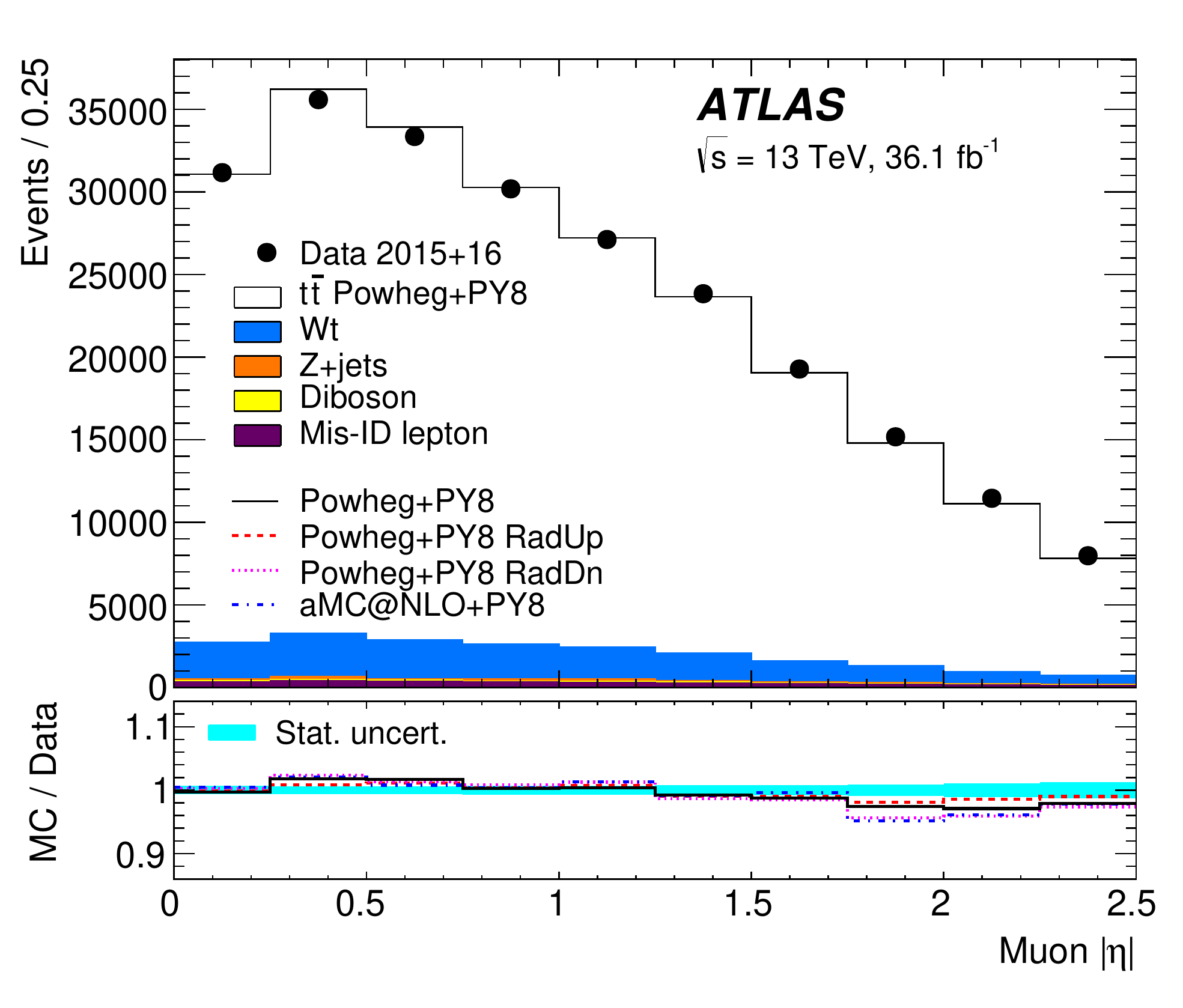}\vspace{-7mm}\center{(f)}}
\caption{\label{f:dmcjlept}Distributions of
(a) the number of $b$-tagged jets in selected opposite-sign $e\mu$ events;
and (b) the \pt\ of $b$-tagged jets,
(c) the \pt\ of the electron, (d) the $|\eta|$ of the electron,
(e) the \pt\ of the muon and
(f) the $|\eta|$ of the muon, in events with an opposite-sign $e\mu$ pair and
at least one $b$-tagged jet. The reconstruction-level
data are compared with the expectation from
simulation, broken down into contributions from
\ttbar\ \,({\sc Powheg\,+\,Pythia8}), $Wt$, $Z$+jets,
dibosons, and events with misidentified electrons or muons. The simulation
prediction is normalised to the same integrated luminosity as the data in
(a) and to the same number of entries as the data in (b--f).
The lower parts of the figure show the ratios of simulation to data,
using various \ttbar\ signal samples and with the cyan shaded band indicating
the statistical uncertainty.
The last bin includes the overflow in panels (b), (c) and (e).}
\end{figure}
 
Figures~\ref{f:dmcjlept}(b)--\ref{f:dmcjlept}(f) show distributions of the
\pt\ of the $b$-tagged jets, and the \pt\ and $|\eta|$ of the electron and
muon, in opposite-sign $e\mu$ events with at least one $b$-tagged jet, a sample
which is dominated by \ttbar\ events. The total simulation prediction is
normalised to the same number of events as the data to facilitate shape
comparisons. The $|\eta|$ distributions for electrons and muons
reflect the differences in acceptance and efficiency, in particular
the reduction in electron acceptance across the calorimeter transition region,
and the reduced acceptance for muons around $|\eta|\approx 0$.
In general, the simulation predictions give a good description of the data,
although the baseline {\sc Powheg\,+\,Pythia8} simulation predicts a
significantly harder lepton \pt\ distribution than seen in data.
 
The inclusive cross-section was determined separately from the 2015 and 2016
datasets, and the results were combined, taking into account correlations in
the systematic uncertainties. As the systematic uncertainties are much larger
than the statistical uncertainties, and not fully correlated between the two
samples
(true in particular for the uncertainty in the integrated luminosity),
this procedure gives a smaller
overall uncertainty than treating the 2015--16 data as a single sample.
The selection efficiency \epsem\ is about 10\% lower in the 2016
data compared to the 2015 data, due to the harsher pileup conditions
and higher-\pt\ trigger thresholds.

\subsection{Differential cross-sections}\label{ss:diffmeas}
 
The differential cross-sections as functions of the lepton and dilepton
variables defined in Section~\ref{s:intro} were measured
using an extension of Eqs.~(\ref{e:tags}), by counting the number of leptons or
events with one (\nxi) or two (\nyi) $b$-tagged jets where the lepton(s)
falls in bin $i$ of a differential distribution at reconstruction level.
For the single-lepton distributions \ptl\ and \etal, there are two counts
per event, in the two bins corresponding to the electron and muon. For the
dilepton distributions, each event contributes a single count corresponding
to the bin in which the appropriate dilepton variable falls. For each bin
of each differential distribution, these counts satisfy the tagging equations:
\begin{equation}
\begin{array}{lll}
\nxi & = &  L \xtti\ \gemi 2\epsbi (1-\cbi\epsbi) + \nibi , \\*[2mm]
\nyi & = &  L \xtti\ \gemi \cbi(\epsbi)^2 + \niibi  ,
\end{array}\label{e:fidtags}
\end{equation}
where \xtti\ is the absolute fiducial differential cross-section in bin $i$.
The reconstruction efficiency \gemi\ represents the ratio of the number
of reconstructed $e\mu$ events (or leptons for \ptl\ and \etal) in bin $i$
defined using the reconstructed lepton(s), to the number of true $e\mu$ events
(or leptons) in the same bin $i$ at particle level,
evaluated using \ttbar\
simulation. The true electron and muon were required to have $\pt>20$\,\GeV\
and $|\eta|<2.5$, but no requirements were made on reconstructed or
particle-level jets. The efficiency \gemi\ corrects for both the lepton
reconstruction efficiency and the effects of event migration, where events
in bin $j$ at
particle level appear in a different bin $i\neq j$ at reconstruction
level. The integral of any dilepton differential cross-section is equal to the
fiducial cross-section \xfid\ defined in Section~\ref{ss:incmeas}, and the
integrals of the single-lepton \ptl\ and \etal\ distributions are equal to
$2\xfid$.
The correlation coefficient $\cbi$ depends on the event counts in bin $i$
analogously to the inclusive \cb\ appearing in Eqs.~(\ref{e:tags}). The
values of \gemi\ were taken from \ttbar\ simulation, and are generally
around 0.5--0.6. The corresponding values of \cbi\ are always within 1--2\%
of unity, even at the edges of the differential distribution. The background
term \nibi\ varies from 11\% to 23\% of the total event count \nxi\ in
each bin, and \niibi\ varies from 3\% to 14\% of \nyi.
They were determined from simulation and data control
samples, allowing the tagging equations~(\ref{e:fidtags}) to be solved to
give the absolute fiducial differential cross-sections \xtti\ and
associated \epsbi\ values for each bin $i$ of each differential distribution.
 
The bin ranges for each differential distribution were based on those used
at \sxvt\ \cite{TOPQ-2015-02}, adding an additional bin for 20--25\,\GeV\
in the \ptl\ distribution and extending the lowest bin down to 40\,\GeV\ for
\ptsum\ and \esum\ to accommodate the reduced minimum lepton \pt\ requirement
of 20\,\GeV. The number and sizes of bins were chosen according to the
experimental resolution in order to keep the bin purities (i.e the fractions
of events reconstructed in bin $i$ that originate from bin $i$ at particle
level) above about 0.9, and to keep a maximum of around ten bins for
the angular distributions (\etal, \rapll\ and \dphill).  The variations
in the angular distributions  predicted by  different \ttbar\ models do not
motivate a finer binning, even though the experimental resolution would allow
it.
The chosen bin ranges can be seen in Tables~\ref{t:insXSec1}--\ref{t:insXSec4}
in the Appendix. The last bin of the \ptl, \ptll, \mll,
\ptsum\ and \esum\ distributions includes overflow events falling above
the last bin boundary.
 
The normalised fiducial differential cross-sections \xntti\
were calculated from the absolute cross-sections \xtti\ as follows:
\begin{eqnarray}\label{e:normx}
\xntti = \frac{\xtti}{\Sigma_j\ \xttj} = \frac{\xtti}{\xfid},
\end{eqnarray}
where \xfid\ is the cross-section summed over all bins of the
fiducial region, equal to the fiducial cross-section defined in
Section~\ref{ss:incmeas}, or twice that in the case of the single-lepton
distributions.  The \xntti\ values were then divided by the
bin widths $W_i$, to produce the
cross-sections differential in the variable $x$ ($x=\ptl$, \etal, etc.):
\begin{eqnarray}\label{e:binw}
\frac{1}{\sigma}\left( \frac{{\mathrm d}\sigma}{{\mathrm d}x}\right)_i =
\frac{\xntti}{W_i}\ . \label{e:diffxsec}
\end{eqnarray}
The normalised differential cross-sections are correlated between bins
because of the normalisation condition in Eq.~(\ref{e:normx}). The absolute
dilepton differential cross-sections are not statistically correlated between
bins, but kinematic correlations between the electron and muon within one
event introduce small correlations within the absolute
single-lepton \ptl\ and \etal\ distributions.
 
The larger number of selected \ttbar\ events compared to the \sxvt\
analysis allows double-differential cross-sections to be
measured, i.e. distributions that are functions of two variables.
Three such distributions were measured, with \etal, \rapll\ or \dphill\
as the first variable, and \mll\ as the second variable, effectively
measuring the \etal, \rapll\ and \dphill\ distributions in four bins of
\mll, chosen to be $\mll<80$\,\GeV, $80<\mll<120$\,\GeV,
$120<\mll<200$\,\GeV\ and $\mll>200$\,\GeV. The excellent resolution in
\etal, \rapll\ and \dphill\ results in migration effects being significant
only between \mll\ bins. The formalism of Eqs.~(\ref{e:fidtags}) was used,
with the index $i$ running over the two-dimensional grid of bins in both
variables. The normalised double-differential cross-sections were calculated
with the sum in the denominator of Eq.~(\ref{e:normx}) running over all bins,
making the integral of the normalised double-differential cross-section
equal to unity over the entire fiducial region, rather than normalising
e.g. the \etal\ distribution to unity in each \mll\ bin separately.
 
The measured differential cross-sections include contributions where one
or both leptons are produced via leptonic decays of $\tau$-leptons
($t\rightarrow W\rightarrow\tau\rightarrow e/\mu$). To enable comparisons
with theoretical predictions which only include direct
$t\rightarrow W\rightarrow e/\mu$ decays, a second set of cross-section results
was derived with a bin-by-bin multiplicative correction \fntaui\ to remove
the $\tau$ contributions:
\begin{eqnarray}\label{e:notau}
\xtti\,(\mbox{no-$\tau$}) = \fntaui\xtti\ ,
\end{eqnarray}
and similarly for the normalised cross-sections $\xntti\,(\mbox{no-$\tau$})$.
The corrections \fntaui\ were evaluated from the baseline
{\sc Powheg\,+\,Pythia8} \ttbar\ simulation as the fractions of leptons or
events in each particle-level bin which do not involve $\tau$-lepton decays.
They are typically in the range
0.8--0.9, the smaller values occurring in bins with a large contribution of
low-\pt\ leptons where the $\tau$ contributions are largest.
 
Since the uncertainties in most of the differential cross-section bins
are dominated by the data statistical uncertainties,
and the luminosity uncertainty
largely cancels out in the normalised differential cross-sections, the
2015--16 data were treated as a single sample in the differential analysis.
The varying lepton trigger thresholds and offline identification
efficiencies were taken into account
by calculating \gemi\ from an appropriately weighted mixture of simulated
events. Figure~\ref{f:dmcdilept} shows the reconstructed
dilepton distributions for events with at least one $b$-tagged jet,
comparing data with predictions using various \ttbar\ generator configurations.
As in Figures~\ref{f:dmcjlept}(b)--\ref{f:dmcjlept}(f),
the predictions generally describe the data well, although in some regions
there are significant differences between the data and all
predictions, which are discussed further in Section~\ref{ss:gencomp} below.
 
\begin{figure}
\vspace{-7mm}
\parbox{83mm}{\includegraphics[width=76mm]{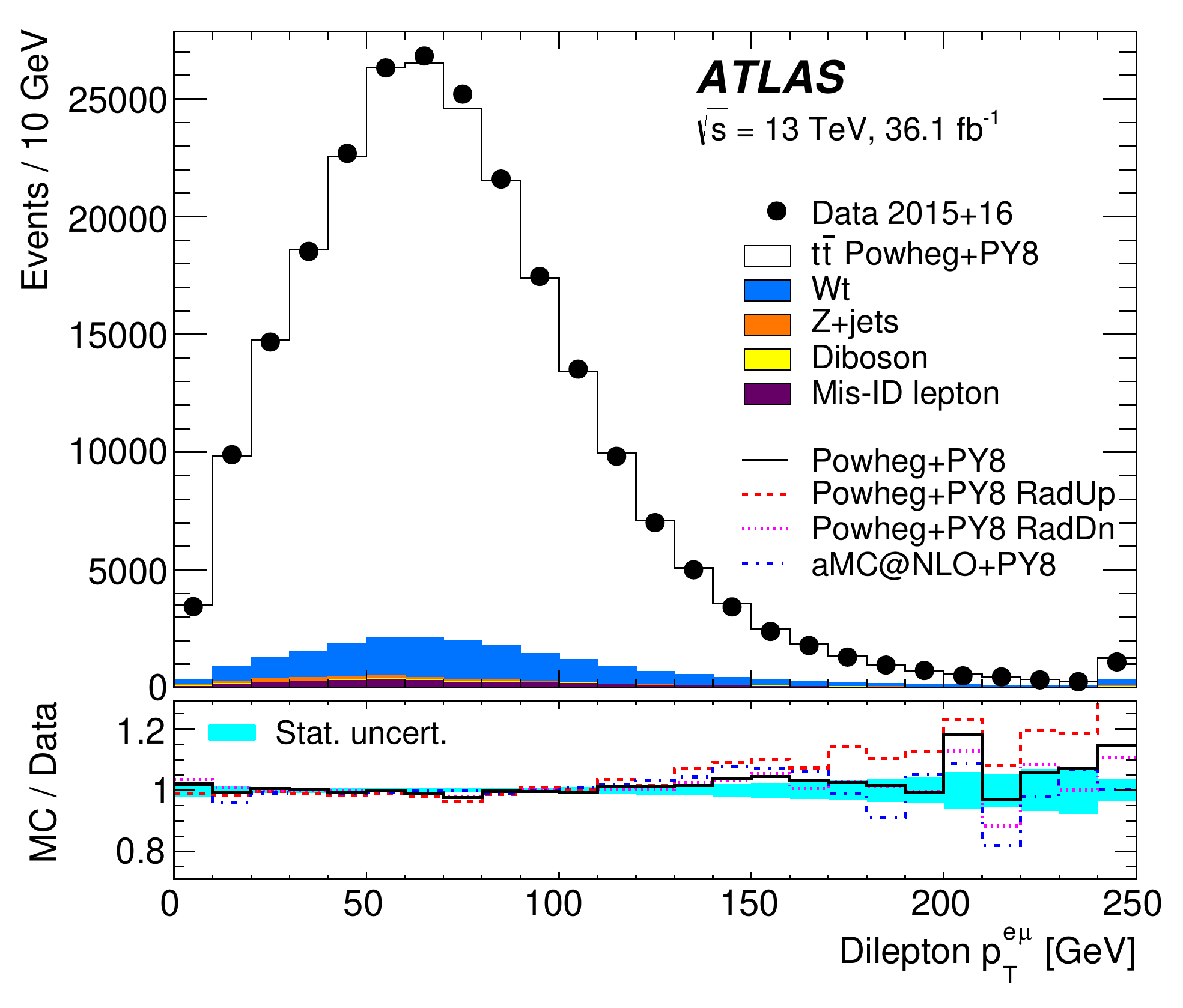}\vspace{-7mm}\center{(a)}}
\parbox{83mm}{\includegraphics[width=76mm]{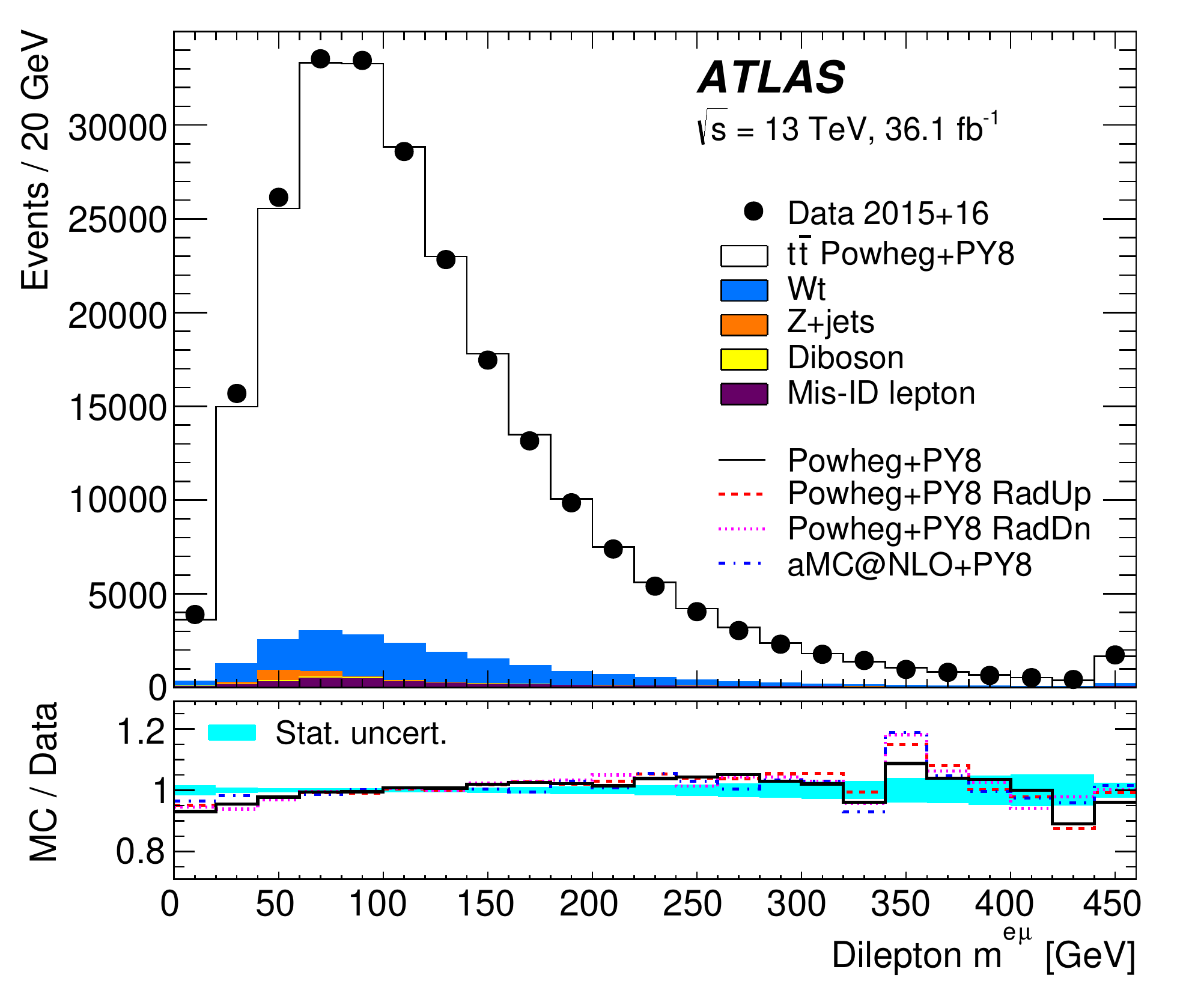}\vspace{-7mm}\center{(b)}}
\parbox{83mm}{\includegraphics[width=76mm]{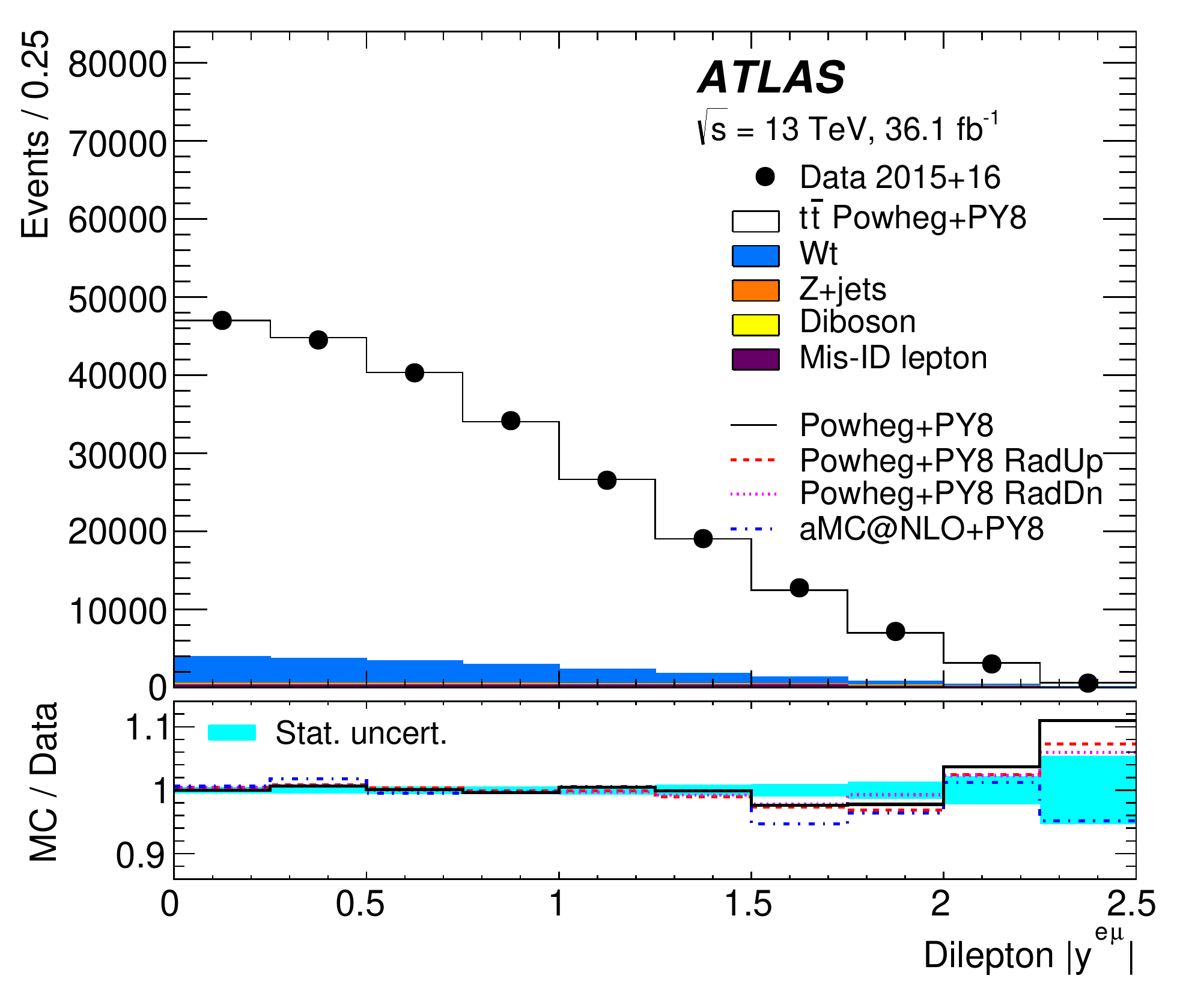}\vspace{-7mm}\center{(c)}}
\parbox{83mm}{\includegraphics[width=76mm]{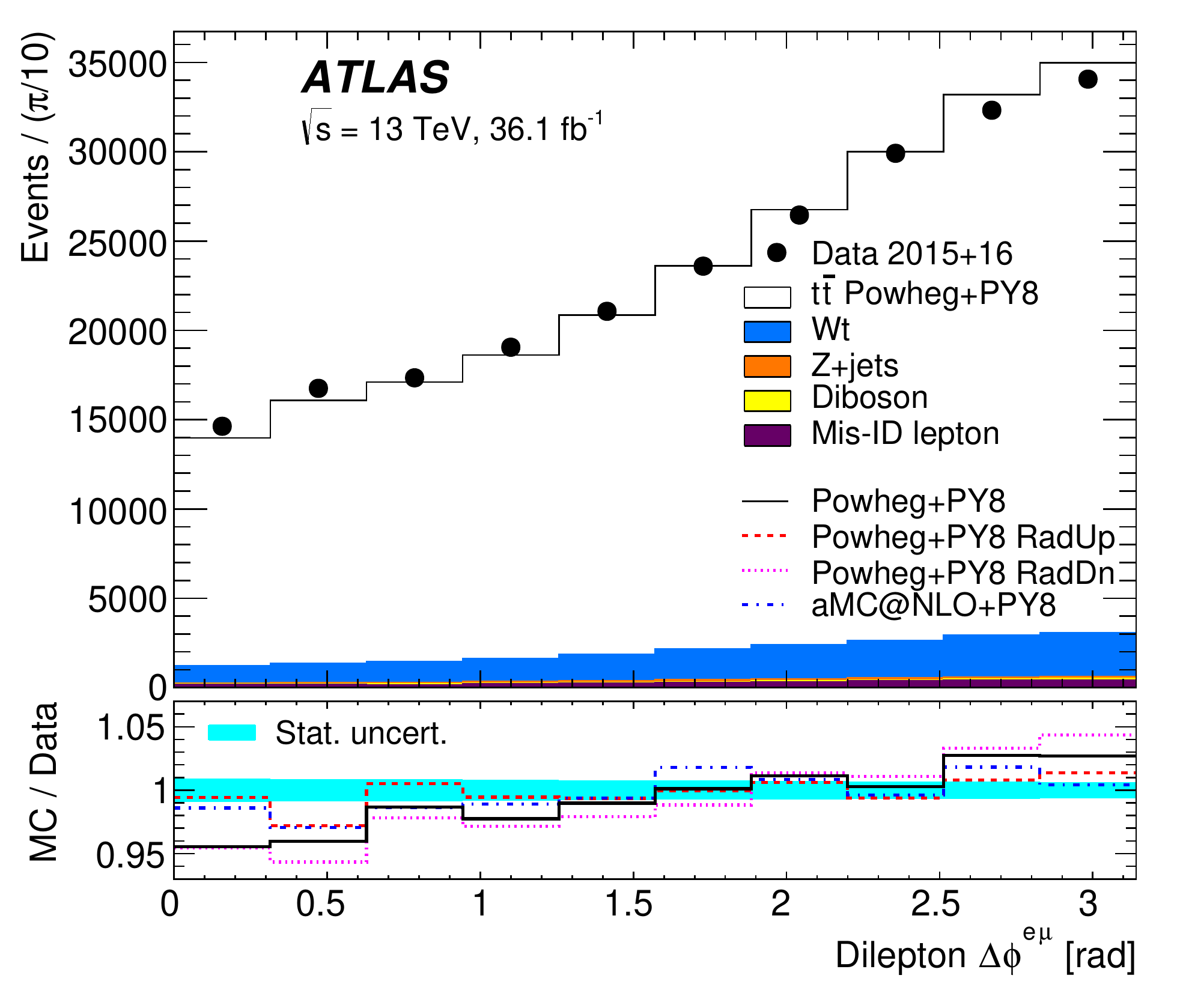}\vspace{-7mm}\center{(d)}}
\parbox{83mm}{\includegraphics[width=76mm]{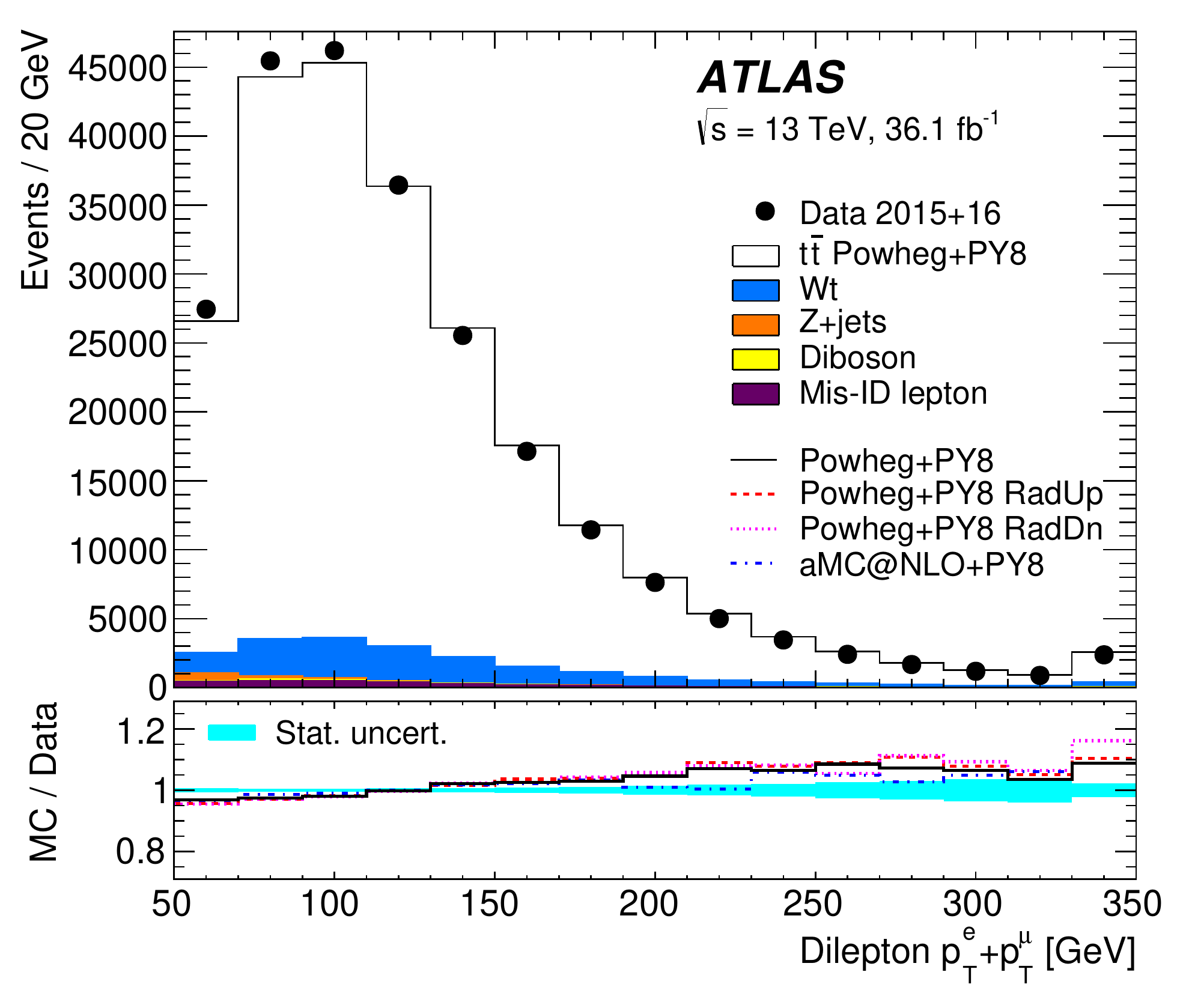}\vspace{-7mm}\center{(e)}}
\parbox{83mm}{\includegraphics[width=76mm]{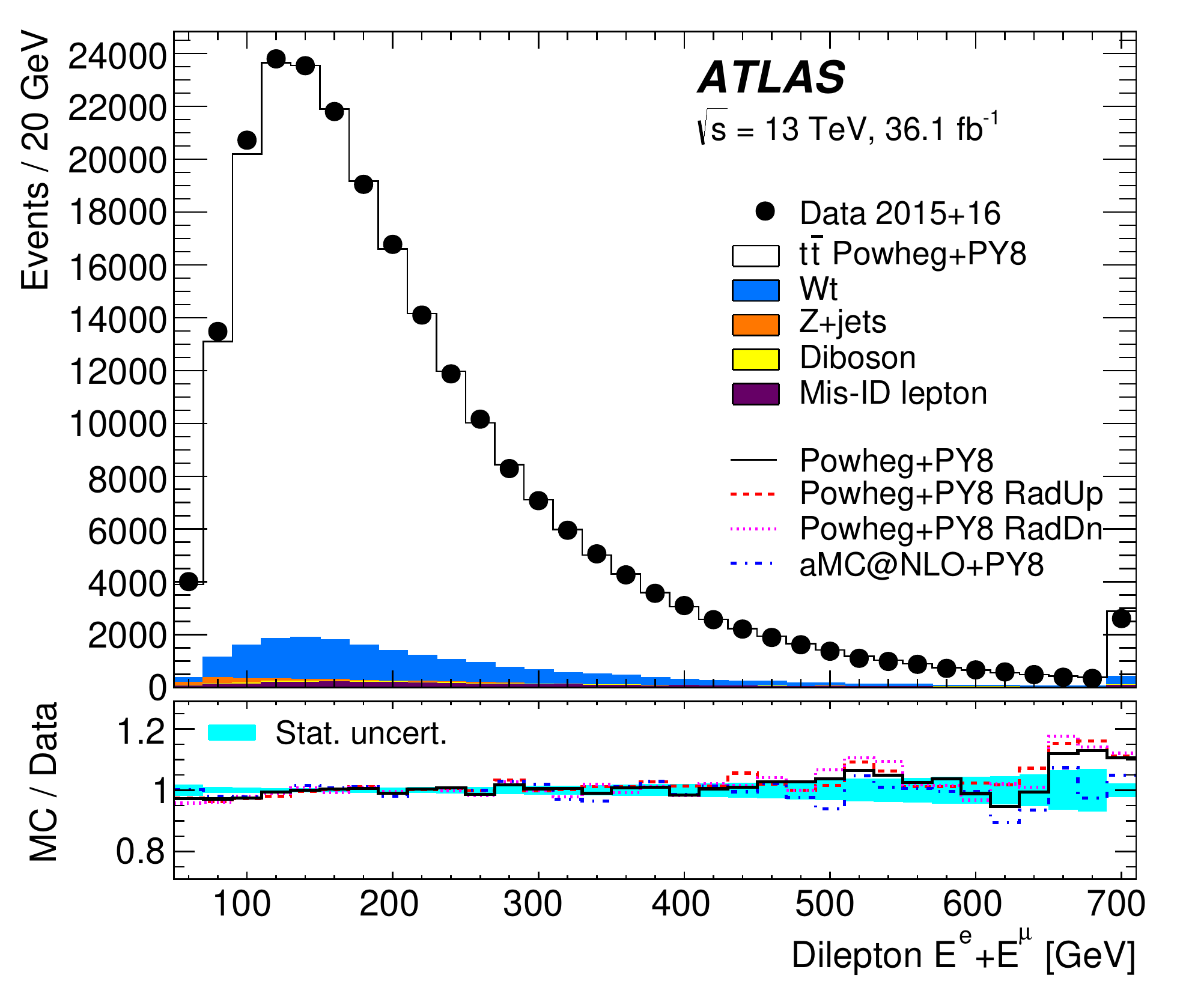}\vspace{-7mm}\center{(f)}}
\caption{\label{f:dmcdilept}Distributions of
(a) the dilepton \ptll, (b) invariant mass \mll, (c) rapidity \rapll,
(d) azimuthal angle difference \dphill, (e) lepton \pt\ sum \ptsum\ and
(f) lepton energy sum \esum, in events with an opposite-sign $e\mu$ pair and
at least one $b$-tagged jet. The reconstruction-level data are
compared with the expectation from simulation, broken down into contributions
from \ttbar\ \,({\sc Powheg\,+\,Pythia8}), $Wt$, $Z$+jets,
dibosons, and events with misidentified electrons or muons, normalised to the
same number of entries as the data.
The lower parts of the figure show the ratios of simulation to data,
using various \ttbar\ signal samples and with the cyan shaded band indicating
the statistical uncertainty. The last bin includes the overflow in panels
(a), (b), (e) and (f).}
\end{figure}

\subsection{Background estimates}\label{ss:bkgd}
 
The dominant background from $Wt$ production, and the smaller contribution
from diboson events (dominated by $WW$ production) were evaluated from
simulation, using the samples detailed in Section~\ref{s:datmc}.
The production of a $Z$ boson accompanied by heavy-flavour jets
is subject to large theoretical uncertainties,
so the background contributions in the one and two $b$-tag samples
predicted by {\sc Sherpa}  (normalised to the inclusive
$Z$ cross-section predictions from FEWZ \cite{fewz}) were further scaled by
factors of $1.10\pm 0.12$ (one $b$-tag) and $1.20\pm 0.12$ (two $b$-tags)
obtained from data. These
scale factors were derived from the ratio of data to simulation event
yields for $Z\rightarrow ee/\mu\mu$ accompanied by one or by two $b$-tagged
jets. The $Z\rightarrow ee/\mu\mu$ yields were obtained by requiring
two opposite-sign electrons or muons passing the selections detailed
in Section~\ref{s:objevt}, and performing a template fit to the
dilepton invariant mass distribution
in the range $30<m_{\ell\ell}<150$\,\GeV\ in order to subtract the contributions
from \ttbar\ events and misidentified leptons.
The uncertainties are dominated by variations in the scale factors
as functions of $Z$ boson \pt.
Further uncertainties of 5\% in the one $b$-tag sample
and 23\% in the two $b$-tag sample were assigned from the change in the
final background prediction when replacing the {\sc Sherpa} sample with
one generated using {\sc MadGraph} \cite{madgraph} interfaced to {\sc Pythia8},
including re-evaluation of the scale factors. Similar procedures were used
to evaluate the uncertainty in the $Z$+jets background prediction in every
bin of the differential distributions, including a comparison of the per-bin
predictions from {\sc Sherpa} and {\sc Madgraph} after normalising each
sample to data in the inclusive $Z\rightarrow ee/\mu\mu$ control regions.
 
The background from events with one real and one misidentified lepton
was evaluated with the help of the same-sign $e\mu$ control sample.
For the inclusive cross-section analysis, the contributions \njfake\ to the
total numbers $N_j$ of opposite-sign $e\mu$ events with $j=1,2$ $b$-tagged jets
are given by:
\begin{equation}
\begin{array}{rll}
\njfake & = & R_j ( \njdss-\njphss\,)\ , \\*[1mm]
R_j & = & \frac{\njfakeos}{\njfakess}\ ,
\end{array}\label{e:fakeest}
\end{equation}
where \njdss\ is the number of observed same-sign events,
\njphss\ is the number of same-sign events with two prompt leptons
estimated from simulation, and $R_j$ is the ratio in simulation of
the number of opposite-sign (\njfakeos) to same-sign (\njfakess)
events with misidentified leptons, all
with $j$ $b$-tagged jets. This formalism relies on simulation to predict
the ratio of opposite- to same-sign misidentified-lepton events, and the prompt
same-sign contribution, but not the absolute number of misidentified-lepton
events \njfake, which is calculated using the same-sign event counts in data.
The same formalism in bins $i$ of lepton differential variables was used to
estimate the misidentified-lepton
background contributions to \nibi\ and \niibi\ in each bin of the
differential cross-section analysis.
 
Table~\ref{t:fakelept} shows the estimates from simulation of
misidentified-lepton contributions to the opposite- and same-sign event counts
in the inclusive cross-section analysis, separately for the 2015 and 2016
selections.
The prompt contributions (corresponding to $\njphss$ in Eqs.~(\ref{e:fakeest}))
are about 25\% of the one $b$-tag and 35\% of the two $b$-tag
same-sign samples. They include `wrong-sign' contributions,
dominated by dilepton \ttbar\ events where the electron charge sign
has been misidentified, and `right-sign' contributions, with two genuine
same-sign prompt leptons, from $\ttbar+V$ events ($V=W$, $Z$ or $H$),
$WZ$, $ZZ$ or same-sign $WW$ production. The misidentified-lepton contributions
are dominated by electrons from photon conversions, shown separately
for events where the photon
was radiated from a prompt electron in a \ttbar\ dilepton event, or
came from some other background source. These contributions
are followed by electrons or muons from
the semileptonic decays of heavy-flavour hadrons (e.g $b$-hadrons produced from
the top quark decays, or charm hadrons produced from hadronic $W$ decays in
single-lepton \ttbar\ events), and other sources, such
as misidentified hadrons or decays in flight of pions and kaons.
Within each category and $b$-jet multiplicity, the numbers of opposite-
and same-sign events are comparable, but with up to a factor two more
opposite- than same-sign events in the major categories, and larger
variations for the small contributions labelled `Other'. The reasons for
this behaviour are complex, depending e.g. on details of the electron
reconstruction, or on charge correlations between the decay products of the
two top quarks.
 
The composition of the same-sign samples is also illustrated in
Figure~\ref{f:sslept}, which shows electron and muon \pt\ and $|\eta|$
distributions in same-sign data events with at least one $b$-tagged jet,
and the corresponding simulation predictions, broken down into prompt leptons
(combining the right- and wrong-sign categories of Table~\ref{t:fakelept})
and various misidentified-lepton categories (again combining `other' electrons
and muons into a single category).
Table~\ref{t:fakelept} shows that the simulation reproduces the observed
numbers of same-sign events well, and the distributions shown in
Figure~\ref{f:sslept} demonstrate that it also reproduces the general
features of the lepton kinematic distributions, the largest differences
in individual bins being around 20\%. These studies validate the
overall modelling of misidentified leptons by the simulation, even though
the background estimates determined via Eqs.~(\ref{e:fakeest}) do not
rely on the simulation providing an accurate estimate of the absolute rates
of such events.
Additional studies were performed using same-sign control
samples with relaxed electron or muon isolation criteria (increasing the
relative contribution of heavy-flavour decays), and changing the lepton
selection to $\pt>40$\,\GeV\ (enhancing the fraction of photon conversions),
and a similar level of agreement was seen both in rates and distribution
shapes.
 
\begin{table}
\small
 
\hspace{-5mm}
\begin{tabular}{l|
r@{$\,\pm\,$}l
r@{$\,\pm\,$}l
r@{$\,\pm\,$}l
r@{$\,\pm\,$}l|
r@{$\,\pm\,$}l
r@{$\,\pm\,$}l
r@{$\,\pm\,$}l
r@{$\,\pm\,$}l
}\hline
& \multicolumn{8}{c|}{2015} & \multicolumn{8}{c}{2016} \\
Component &
\multicolumn{2}{c}{OS $1b$} & \multicolumn{2}{c}{SS $1b$} & \multicolumn{2}{c}{OS $2b$} & \multicolumn{2}{c|}{SS $2b$} &
\multicolumn{2}{c}{OS $1b$} & \multicolumn{2}{c}{SS $1b$} & \multicolumn{2}{c}{OS $2b$} & \multicolumn{2}{c}{SS $2b$} \\
\hline
$t\rightarrow e\rightarrow\gamma$ conversion $e$ &   59 &   5 &   41 &   4 &   33 &   3 &   21 &   3 &  594 &  15 &  360 &  11 &  336 &  11 &  191 &   9 \\
Background conversion $e$ &   53 &   6 &   35 &   4 &   19 &   3 &   15 &   2 &  424 &  15 &  227 &  36 &  185 &   8 &  116 &   6 \\
Heavy-flavour $e$ &   27 &   3 &   26 &   3 &    3 &   1 &    2 &   1 &  208 &   8 &  188 &   8 &   20 &   3 &   11 &   2 \\
Other $e$ &    2 &   2 &    0 &   0 &    1 &   1 &    0 &   0 &   48 &   9 &    5 &   1 &   19 &   3 &    2 &   1 \\
Heavy-flavour $\mu$ &   50 &   5 &   46 &   5 &    8 &   2 &    2 &   1 &  434 &  14 &  335 &  12 &   79 &   6 &   27 &   4 \\
Other $\mu$ &   11 &   2 &    2 &   1 &    4 &   1 &    0 &   0 &   54 &  29 &  151 & 126 &   46 &   4 &   11 &   2 \\
\hline
Total misidentified &   201 &   10 &   149 &    8 &    69 &    5 &    40 &    4 &  1761 &   41 &  1266 &  132 &   684 &   16 &   358 &   12 \\\hline
Wrong-sign prompt & \multicolumn{2}{c}{-} &   24 &   3 & \multicolumn{2}{c}{-} &   12 &   2 & \multicolumn{2}{c}{-} &  224 &   9 & \multicolumn{2}{c}{-} &  113 &   6 \\
Right-sign prompt & \multicolumn{2}{c}{-} &   21 &   1 & \multicolumn{2}{c}{-} &    9 &   0 & \multicolumn{2}{c}{-} &  195 &   4 & \multicolumn{2}{c}{-} &   88 &   1 \\
\hline
Total & \multicolumn{2}{c}{-} &   194 &    9 & \multicolumn{2}{c}{-} &    61 &    4 & \multicolumn{2}{c}{-} &  1685 &  132 & \multicolumn{2}{c}{-} &   560 &   13 \\\hline
Data & \multicolumn{2}{c}{-} & \multicolumn{2}{l}{167} & \multicolumn{2}{c}{-} & \multicolumn{2}{l|}{55} & \multicolumn{2}{c}{-} & \multicolumn{2}{l}{1655} & \multicolumn{2}{c}{-} & \multicolumn{2}{l}{551} \\
\end{tabular}
 
\caption{\label{t:fakelept}Breakdown of estimated misidentified-lepton
contributions in simulation to the one ($1b$) and two ($2b$) $b$-tag opposite-
and same-sign (OS and SS) $e\mu$ event samples from 2015 and 2016 separately.
The various
misidentified-lepton categories are described in Section~\ref{ss:bkgd},
and the contributions labelled `Other' include all sources other than
photon conversions and heavy-flavour decays. For the same-sign
samples, the estimated contributions of wrong-sign (where the electron charge
sign is misidentified) and right-sign prompt lepton events are also shown,
and the total expectations are compared with the data. The uncertainties
are due to the limited size of the simulated samples, and values or
uncertainties shown as zero are less than 0.5 events.}
\end{table}
 
The ratios $R_j$ in Eqs.~(\ref{e:fakeest}) were evaluated to be $R_1=1.4\pm 0.3$
and $R_2=1.7\pm 0.9$ for the 2015 data sample, and $R_1=1.4\pm 0.4$ and
$R_2=1.9\pm 1.0$ for the 2016 sample. The uncertainties
encompass the range of $R_j$ values seen for the major sources of
misidentified-lepton events; as can be seen from the entries in
Table~\ref{t:fakelept},
the opposite- to same-sign event count ratios are different for the main
categories, and the uncertainty allows for their relative contributions
to be different from that predicted by the baseline simulation. The $R_j$
values seen in the control samples with loosened isolation, and the
predictions from alternative \ttbar\ simulation samples using {\sc Pythia6}
or {\sc Herwig7} instead of {\sc Pythia8} hadronisation were also considered.
A conservative 50\% uncertainty in the prompt lepton same-sign contribution
was also taken into account, covering the mismodelling of electron charge
misidentification in simulation and the uncertainties in the predicted
cross-sections
for $\ttbar+V$ and diboson processes. The final misidentified-lepton background
estimates for the 2015 and 2016 opposite-sign data samples in the inclusive
cross-section analysis are shown in Table~\ref{t:evtcount}.
 
\begin{figure}
\vspace{-7mm}
\parbox{83mm}{\includegraphics[width=76mm]{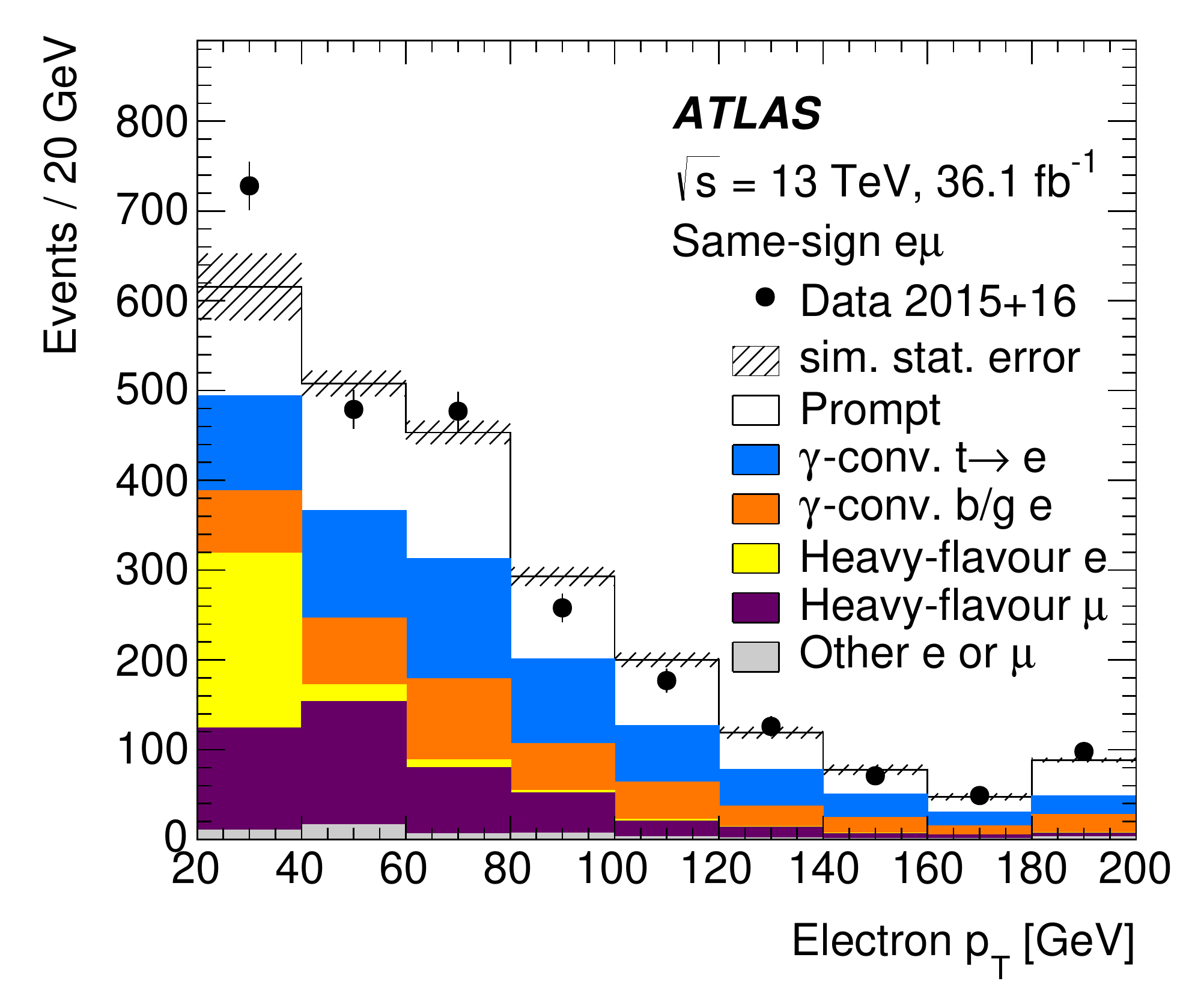}\vspace{-7mm}\center{(a)}}
\parbox{83mm}{\includegraphics[width=76mm]{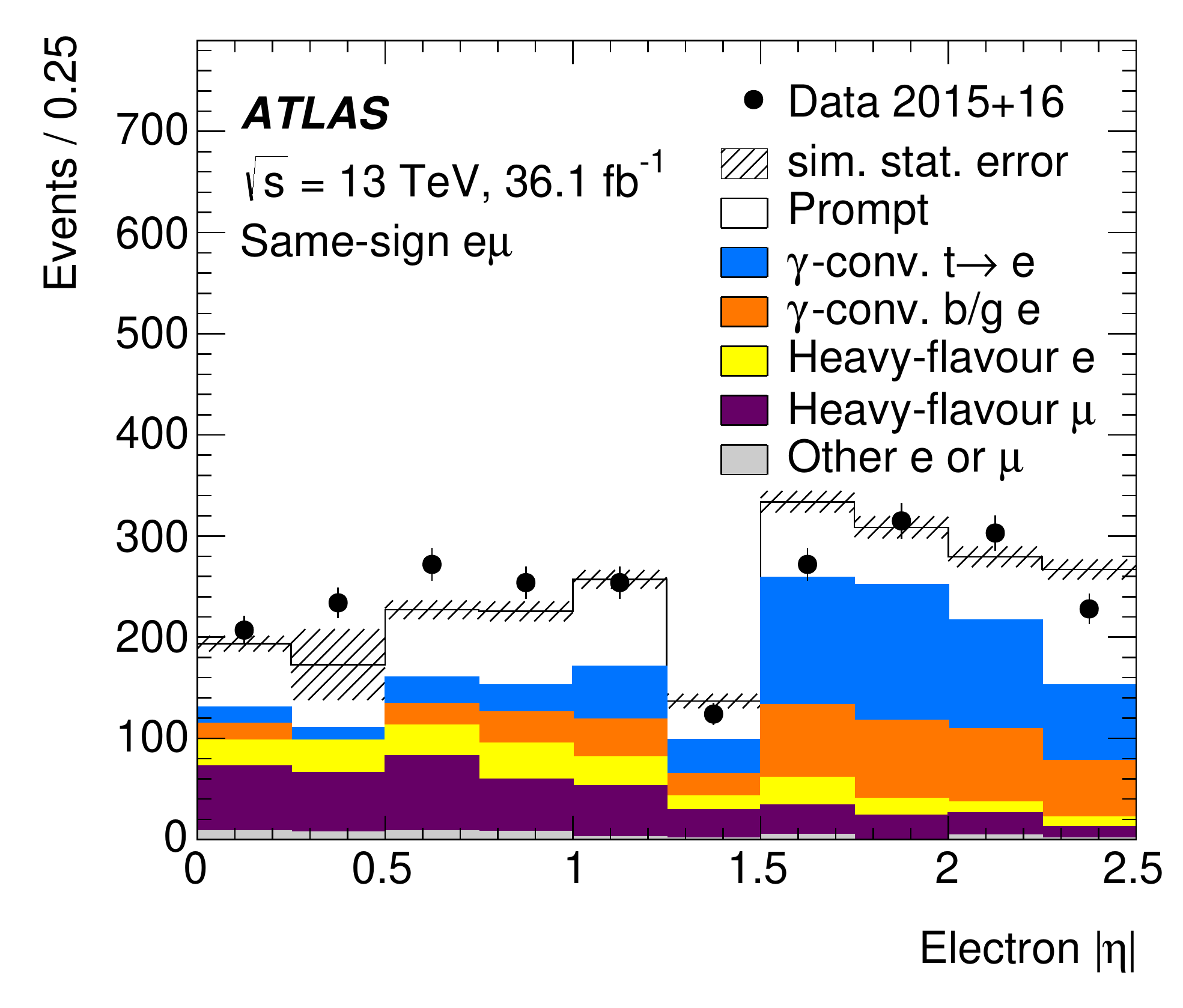}\vspace{-7mm}\center{(b)}}
\parbox{83mm}{\includegraphics[width=76mm]{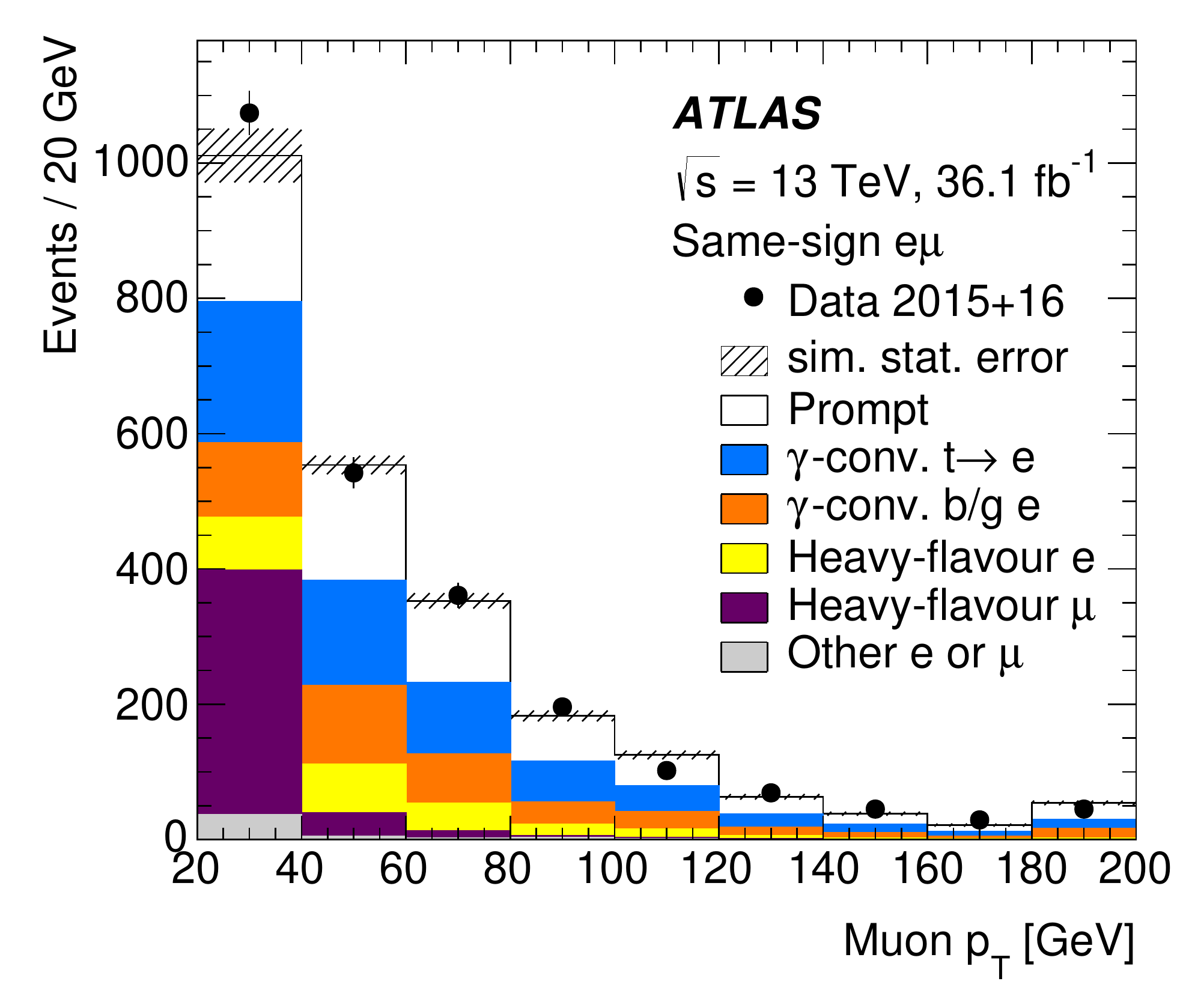}\vspace{-7mm}\center{(c)}}
\parbox{83mm}{\includegraphics[width=76mm]{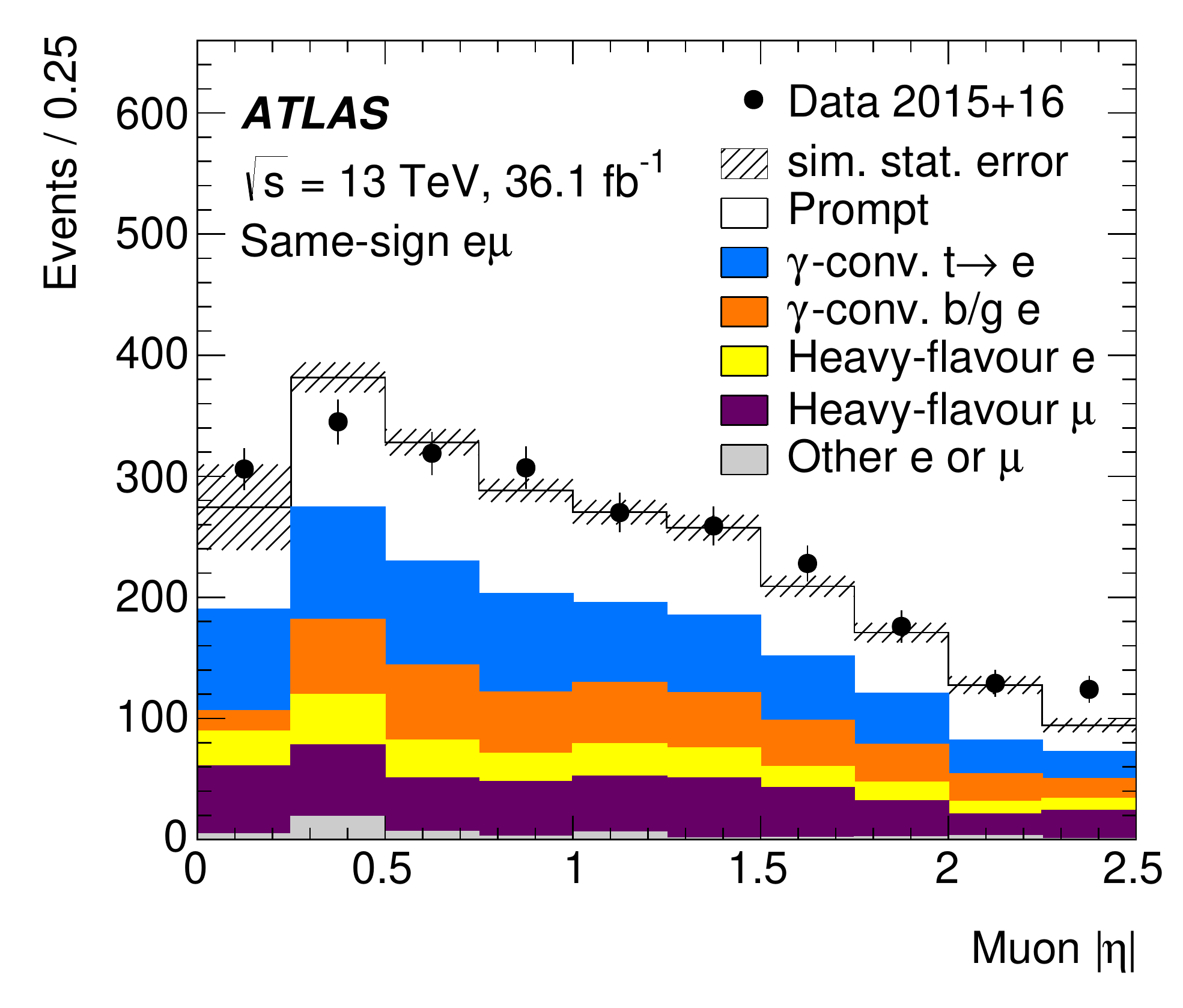}\vspace{-7mm}\center{(d)}}
\caption{\label{f:sslept}Distributions of
(a) the electron \pt, (b) the electron $|\eta|$, (c) the muon \pt\ and (d)
the muon $|\eta|$, in events with a same-sign $e\mu$ pair and
at least one $b$-tagged jet. The simulation prediction is normalised to the
same integrated luminosity as the data, and broken down into contributions
where both leptons are prompt, or one is a misidentified lepton from a photon
conversion originating from a top quark decay or from background, from
heavy-flavour decay or from other sources. The statistical uncertainty in the
total simulation prediction is significant in some bins,
and is shown by the hatching.
In the \pt\ distributions, the last bin includes the overflows.}
\end{figure}
 
Figure~\ref{f:ssdilept} shows the corresponding same-sign event distributions
for the dilepton variables, showing a similar quality of modelling of these
kinematic distributions by the simulation as seen for the single-lepton
variables in Figure~\ref{f:sslept}. The $R^i_1$ and $R^i_2$ values in the binned
version of Eqs.~(\ref{e:fakeest}) do not vary in simulation beyond
the uncertainties assigned above to the inclusive $R_1$ and $R_2$, so the
same relative uncertainties in $R_1$ and $R_2$
were also used for the differential analysis, and taken to be correlated
across all bins.
 
In the opposite-sign sample, the total non-\ttbar\ background fraction
from all sources varies
significantly as a function of some of the differential variables, but remains
dominated by $Wt$ events in all bins. It reaches around 20\% in the one
$b$-tag sample and 10\% of the two $b$-tag sample at the high ends of
the single-lepton \ptl\ and dilepton \ptll\ distributions, but varies little
with lepton \etal.
 
\begin{figure}
\vspace{-7mm}
\parbox{83mm}{\includegraphics[width=76mm]{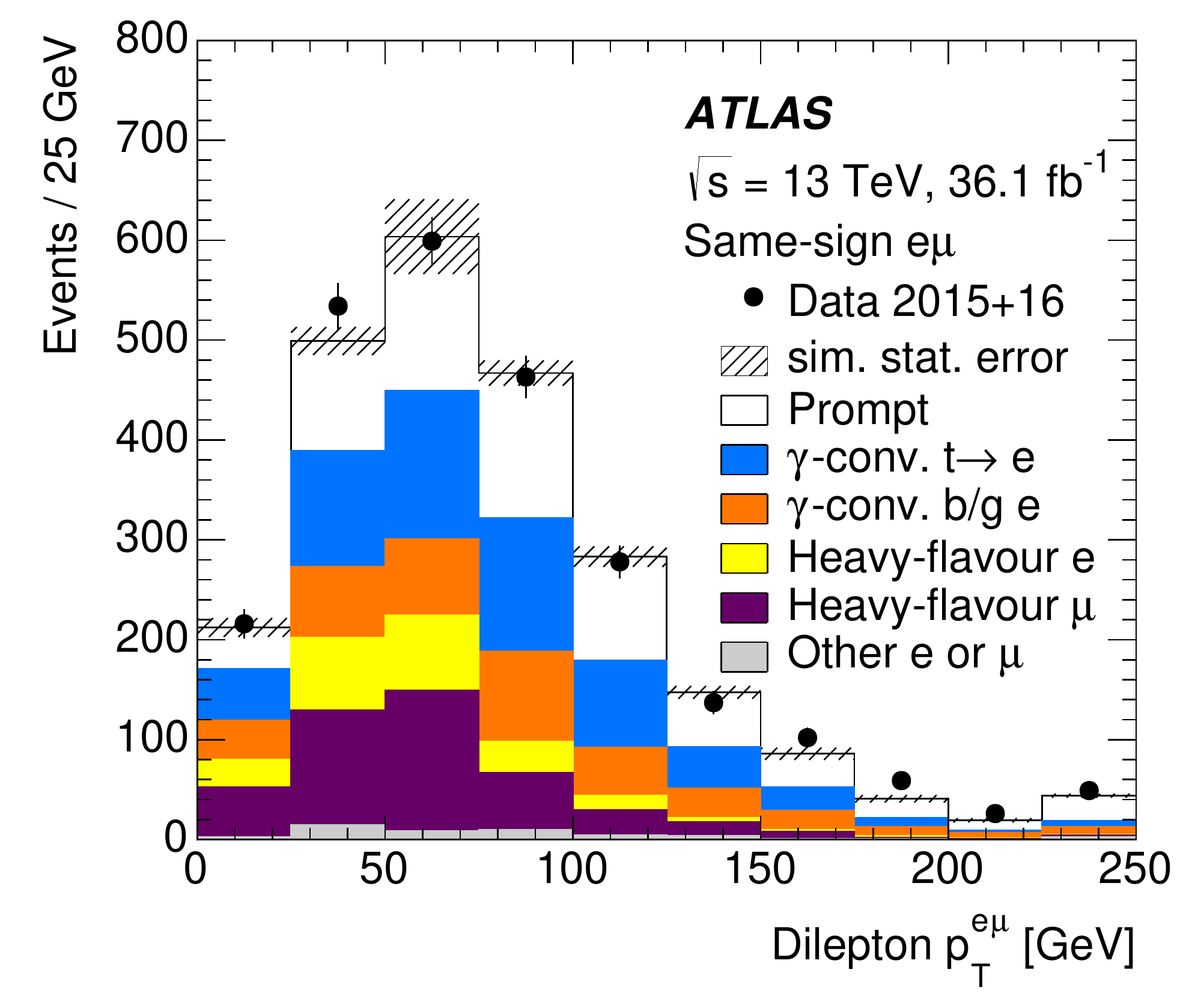}\vspace{-7mm}\center{(a)}}
\parbox{83mm}{\includegraphics[width=76mm]{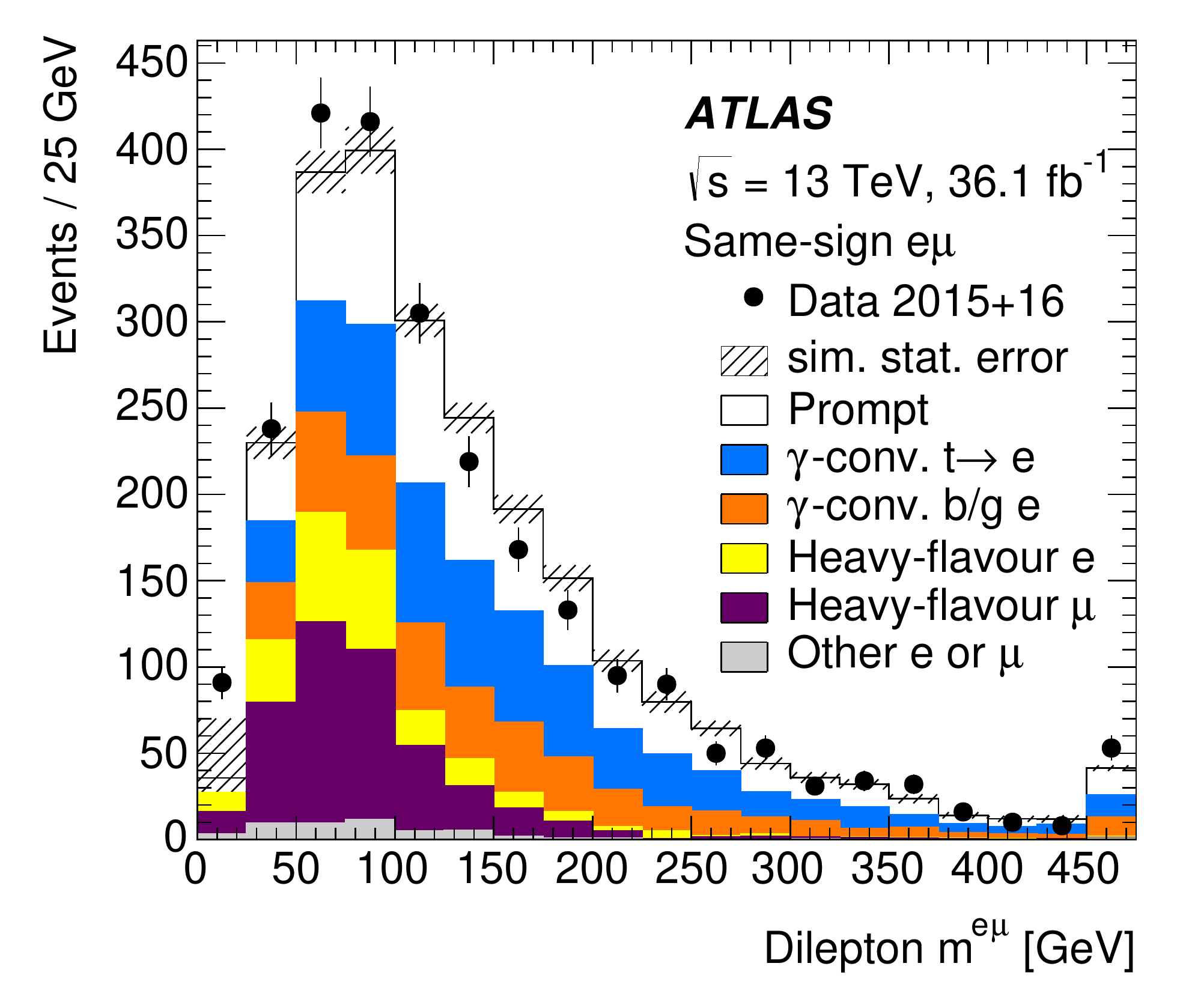}\vspace{-7mm}\center{(b)}}
\parbox{83mm}{\includegraphics[width=76mm]{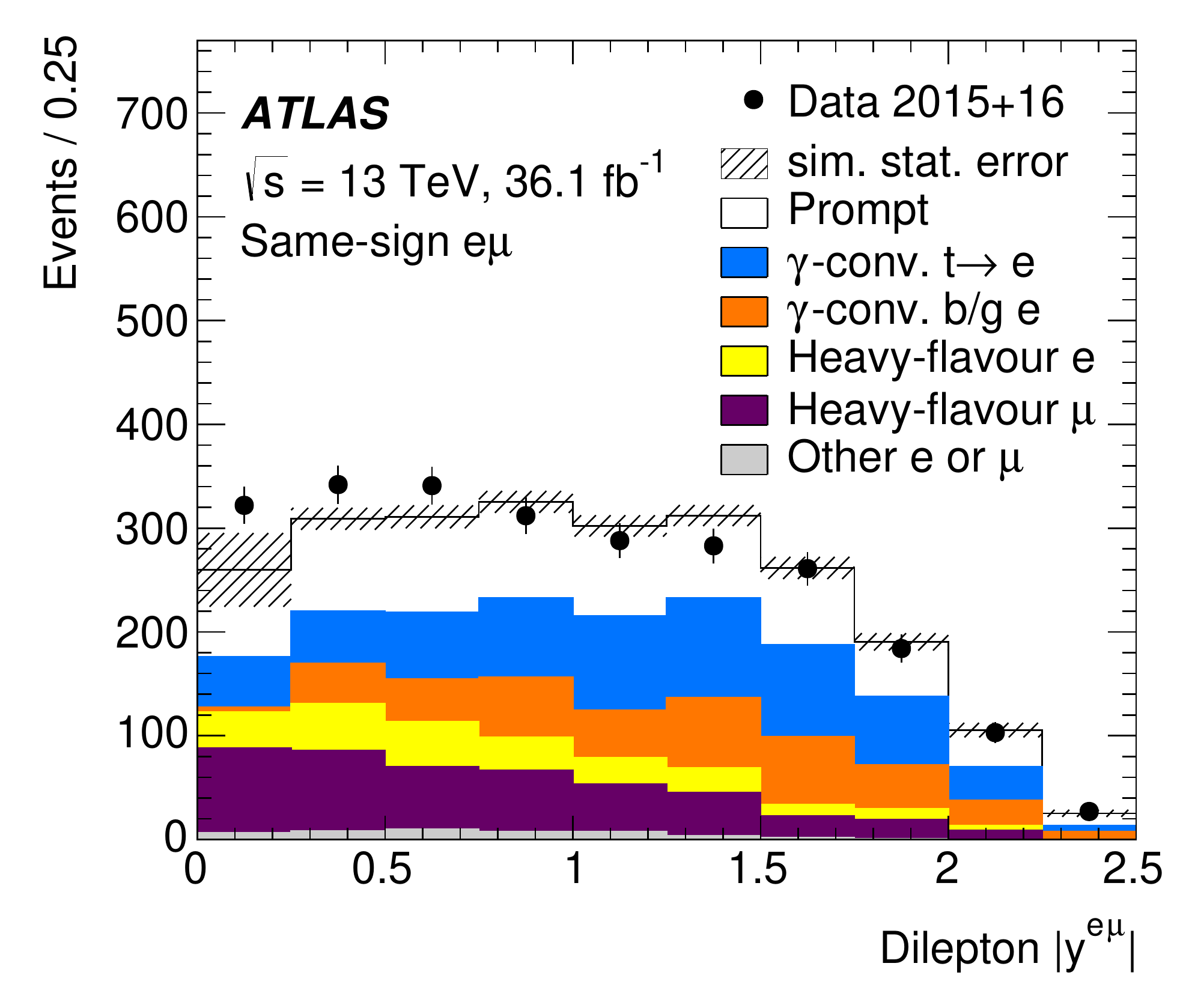}\vspace{-7mm}\center{(c)}}
\parbox{83mm}{\includegraphics[width=76mm]{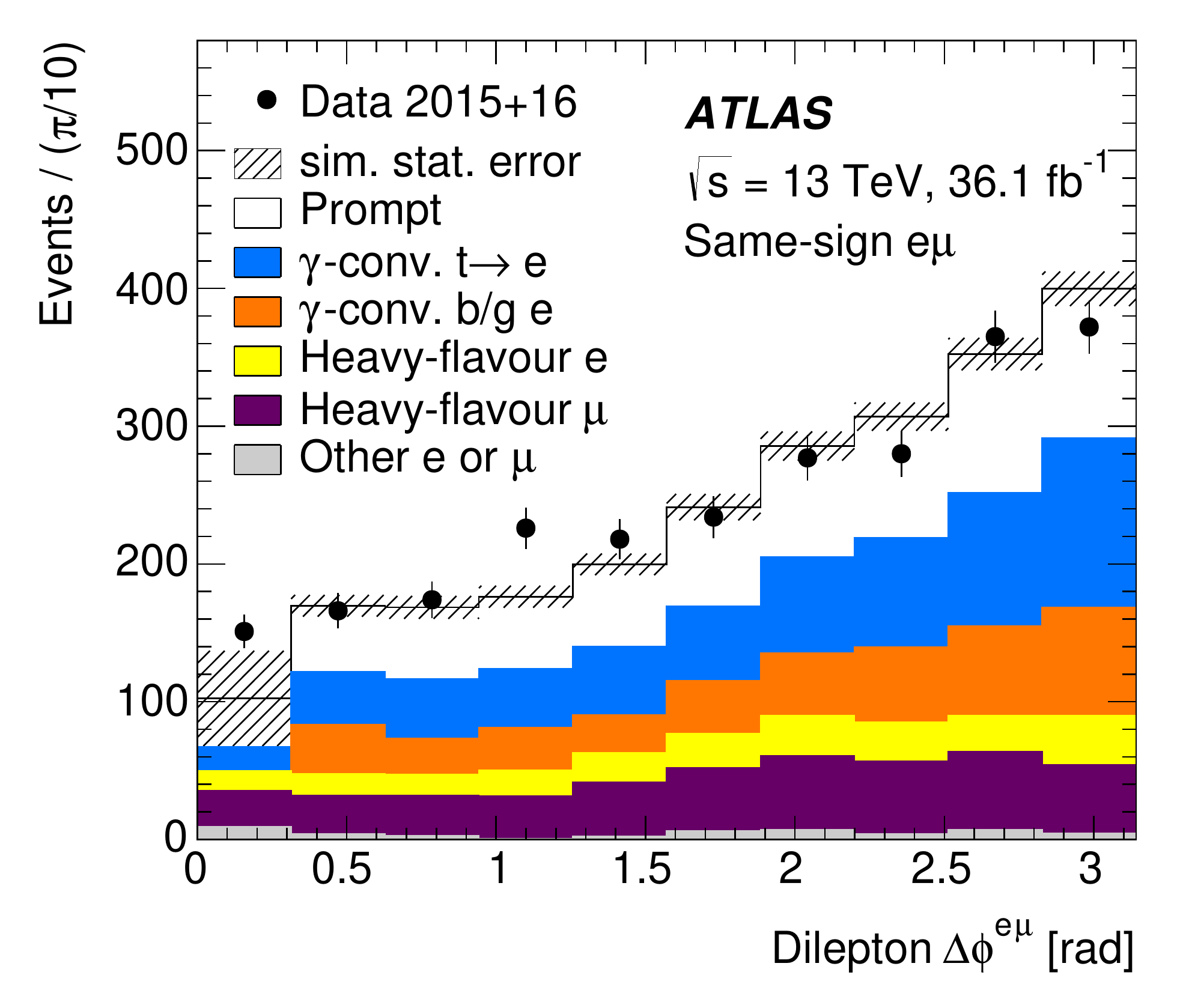}\vspace{-7mm}\center{(d)}}
\parbox{83mm}{\includegraphics[width=76mm]{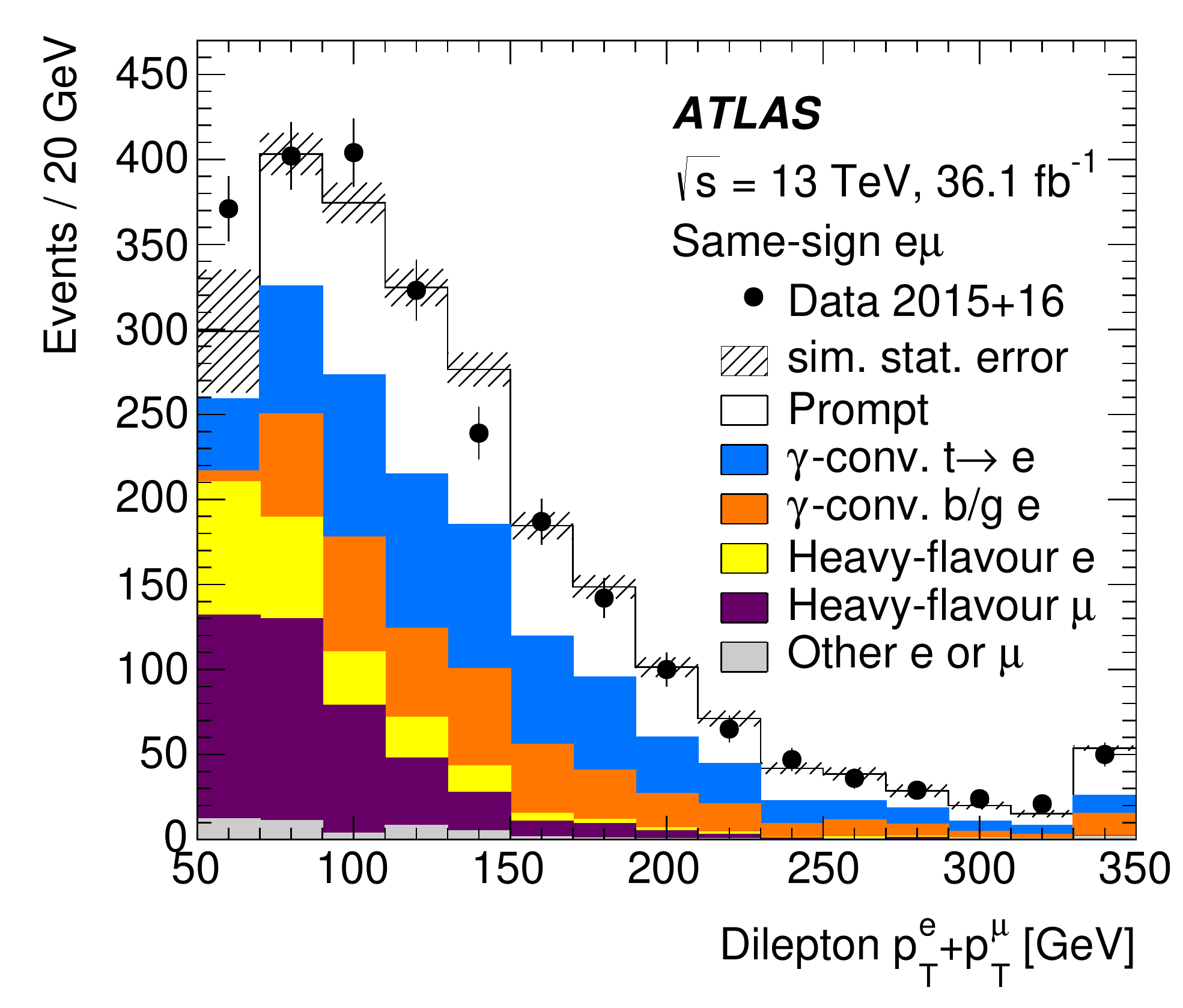}\vspace{-7mm}\center{(e)}}
\parbox{83mm}{\includegraphics[width=76mm]{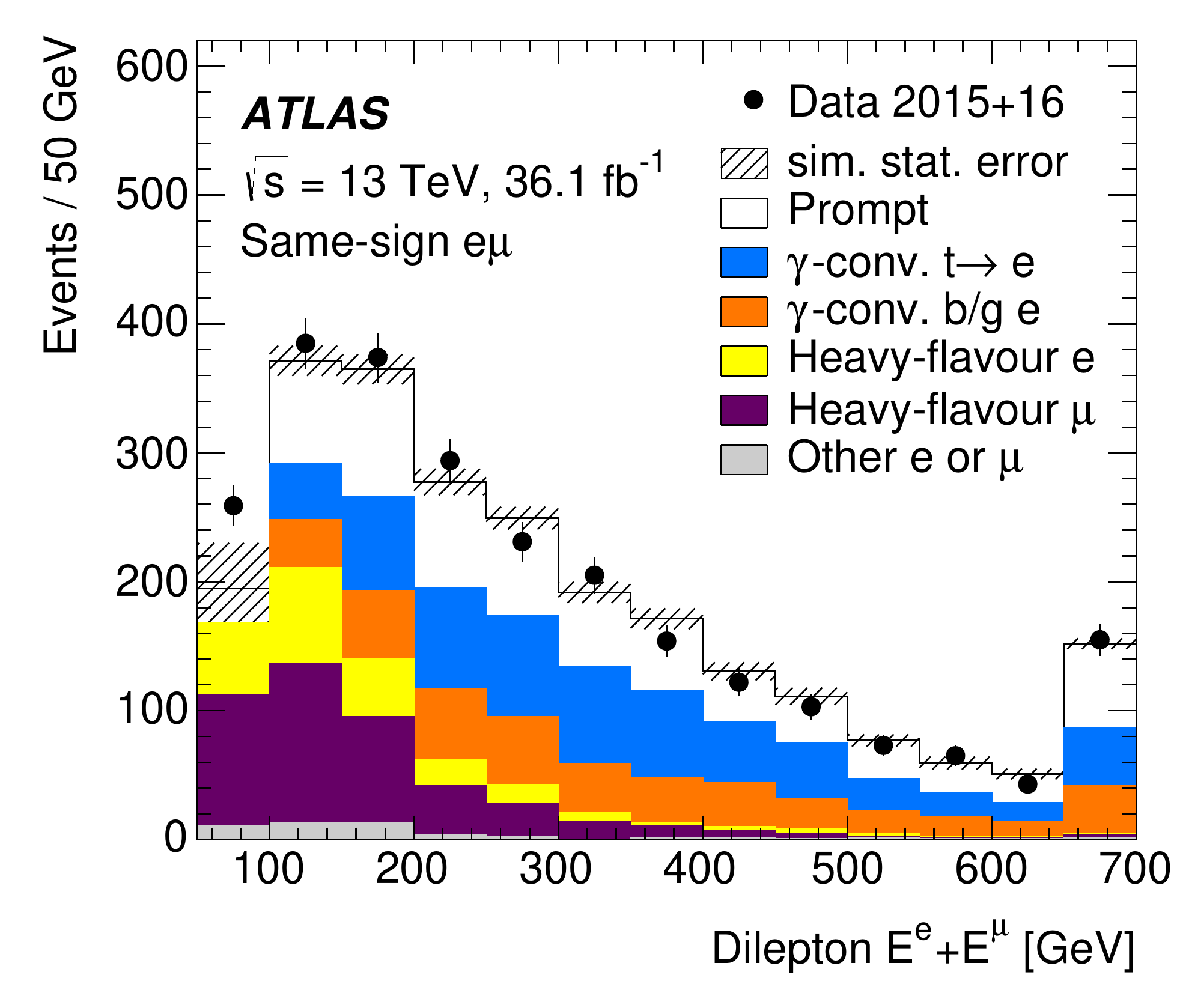}\vspace{-7mm}\center{(f)}}
\caption{\label{f:ssdilept}Distributions of
(a) the dilepton \ptll, (b) invariant mass \mll, (c) rapidity \rapll,
(d) azimuthal angle difference \dphill, (e) lepton \pt\ sum \ptsum\ and
(f) lepton energy sum \esum, in events with a same-sign $e\mu$ pair and
at least one $b$-tagged jet. The simulation prediction is normalised to the
same integrated luminosity as the data, and broken down into contributions
where both leptons are prompt, or one is a misidentified lepton from a photon
conversion originating from a top quark decay or from background, from
heavy-flavour decay or from other sources. The statistical uncertainty in the
total simulation prediction is significant in some bins,
and is shown by the hatching. In the \ptll, \mll, \ptsum\
and \esum\ distributions, the last bin includes the overflows.}
\end{figure}
 
\subsection{Validation of the differential measurements}\label{ss:diffval}
 
A set of tests using pseudo-experiment datasets generated from
simulation were used to validate the analysis procedures for the differential
measurements, as documented in detail for the \sxvt\
analysis \cite{TOPQ-2015-02}.
These tests demonstrated that the method is unbiased and correctly
estimates the statistical uncertainties in each bin of each distribution.
Figure~\ref{f:biastest} shows examples for the \ptl, \ptll, \etal\ and
\rapll\ distributions. The filled black points show the relative differences
between the mean normalised differential cross-sections obtained from
1000 pseudo-experiments and the true cross-sections in each bin, divided
by the true cross-sections to give fractional differences. The
pseudo-experiments were generated from a reference \ttbar\ sample, and the
reference sample was also used to determine the values of \gemi\ and \cbi\
in each
bin $i$ of the distributions. The compatibility of the filled black points with
zero within the statistical uncertainty of the reference sample confirms
that the method is unbiased for this sample. The open red points and dotted
lines show the mean pseudo-experiment results and true values for an
alternative sample with different underlying distributions,
again expressed as fractional deviations from
the true cross-sections in the reference sample, and obtained
using \gemi\ and \cbi\ values from the reference sample. The alternative samples
were chosen in order to produce a large variation in the distribution
under test. An independent
\ttbar\ simulation sample with $\mtop=175$\,\GeV\ was used for the \ptl\
and \ptll\ distributions, and the baseline \ttbar\ sample generated
with NNPDF3.0 was reweighted to the predictions of the CT14 PDF set \cite{ct14}
for \etal\ and \rapll. In all cases, the results are
consistent with the true values within the statistical uncertainties
of the alternative samples, demonstrating that the simple bin-by-bin
correction procedure correctly recovers the alternative distributions,
without the need for iteration or a matrix-based unfolding technique.
Similar results were obtained for the analogous validation tests
performed on the double-differential cross-section measurements.
The various distributions shown in Figure~\ref{f:biastest}
also illustrate the sensitivity of the
normalised differential cross-sections to \mtop\ and different PDF sets.
 
\begin{figure}
\parbox{83mm}{\includegraphics[width=76mm]{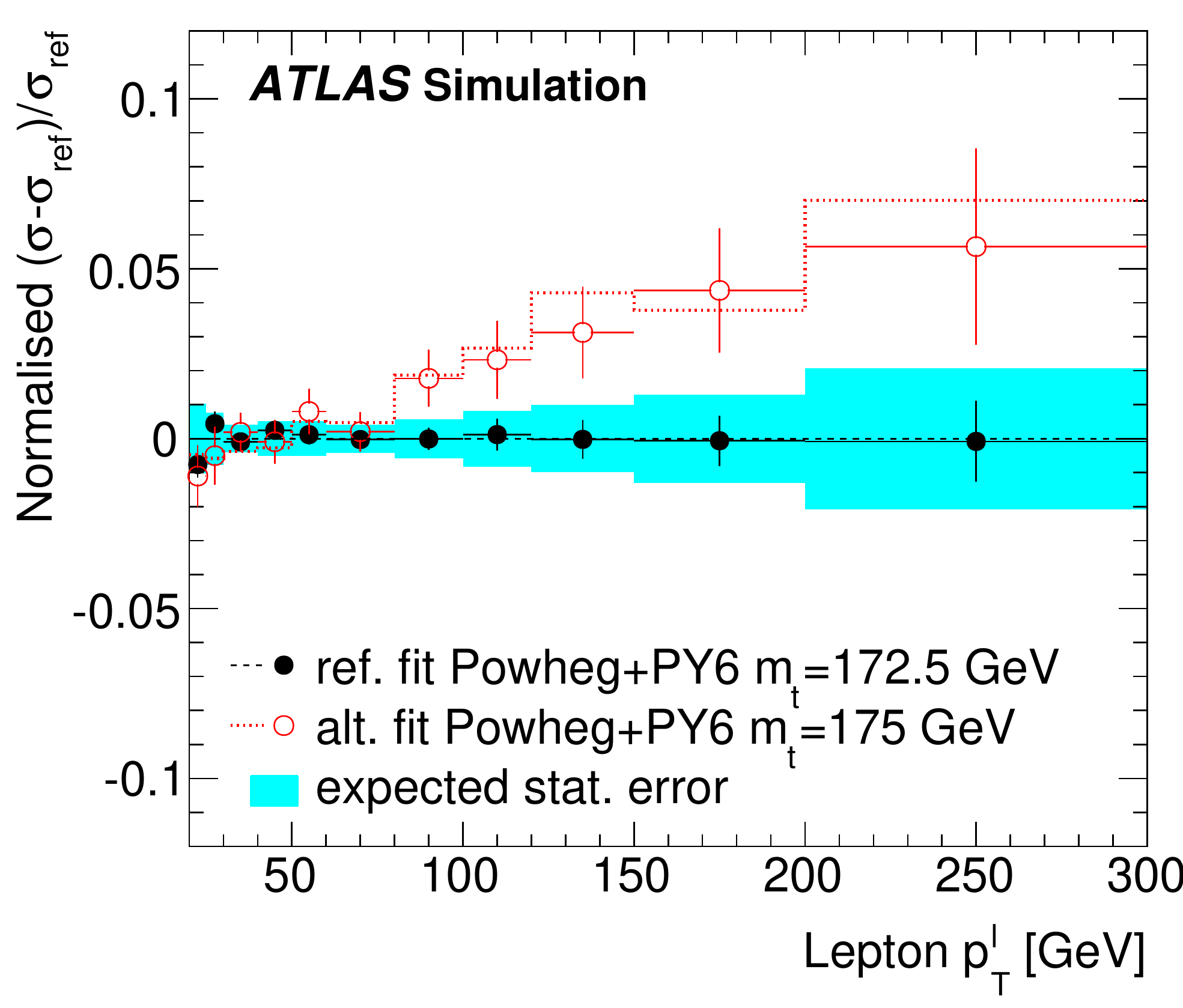}\vspace{-7mm}\center{(a)}}
\parbox{83mm}{\includegraphics[width=76mm]{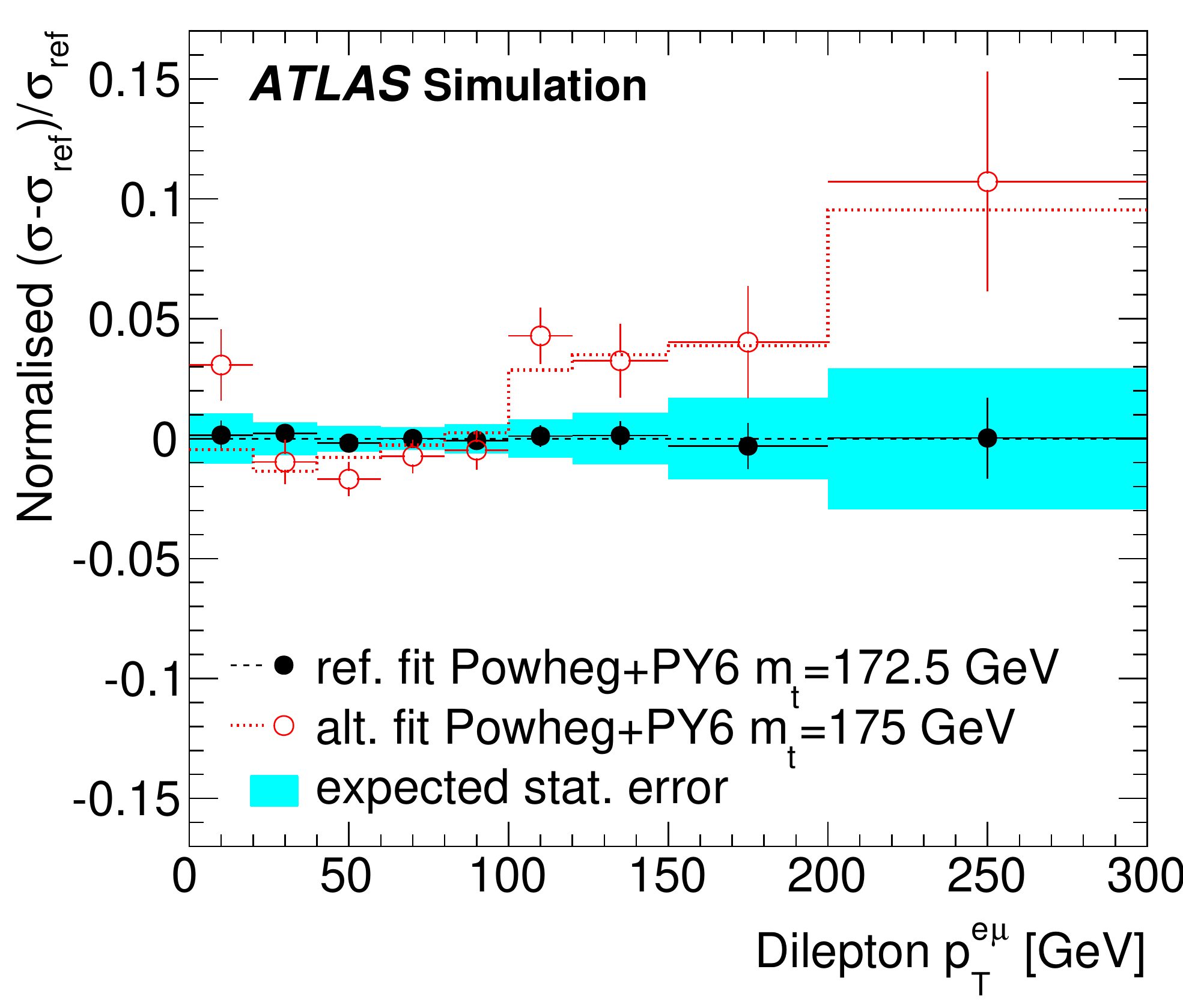}\vspace{-7mm}\center{(b)}}
\parbox{83mm}{\includegraphics[width=76mm]{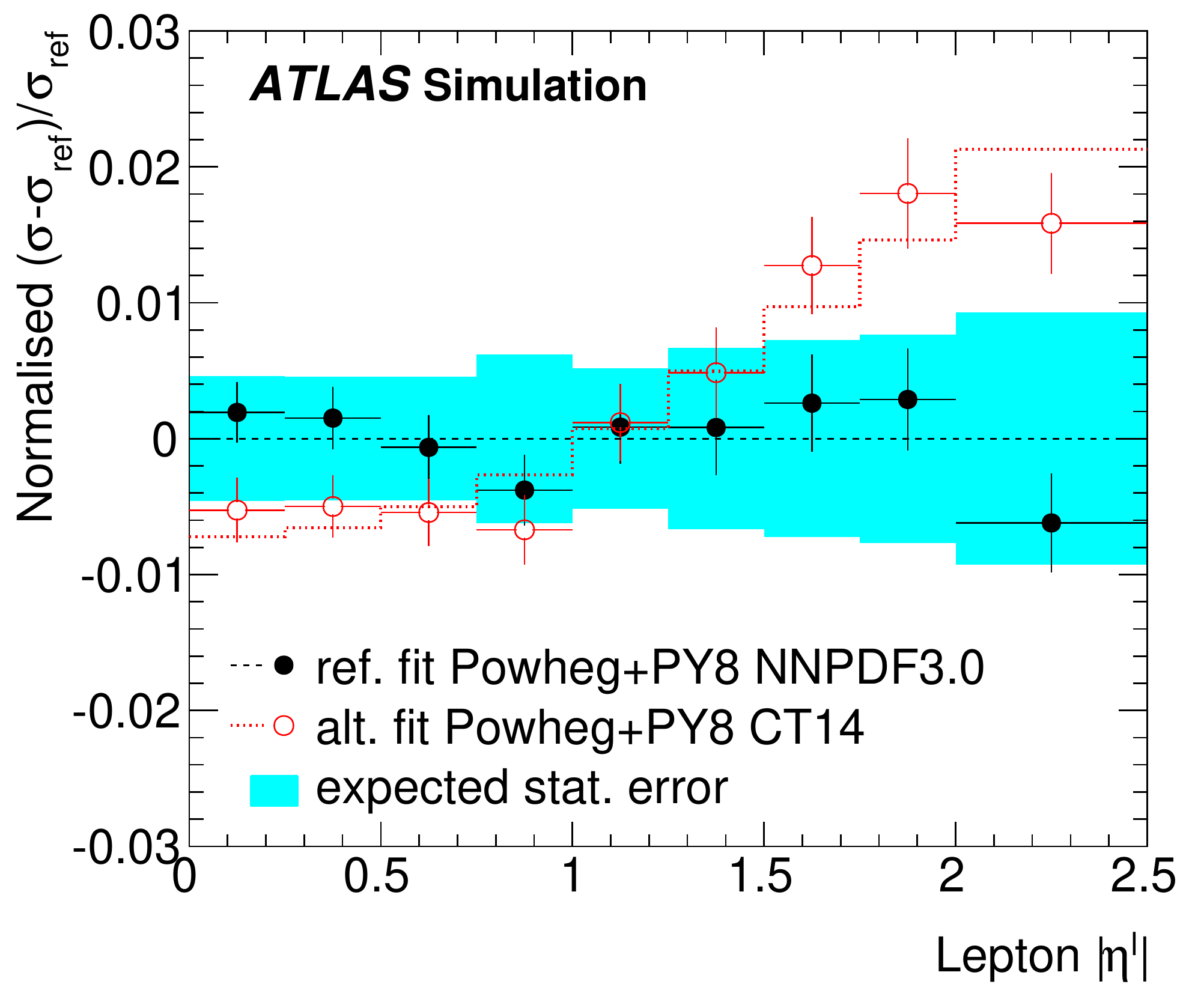}\vspace{-7mm}\center{(c)}}
\parbox{83mm}{\includegraphics[width=76mm]{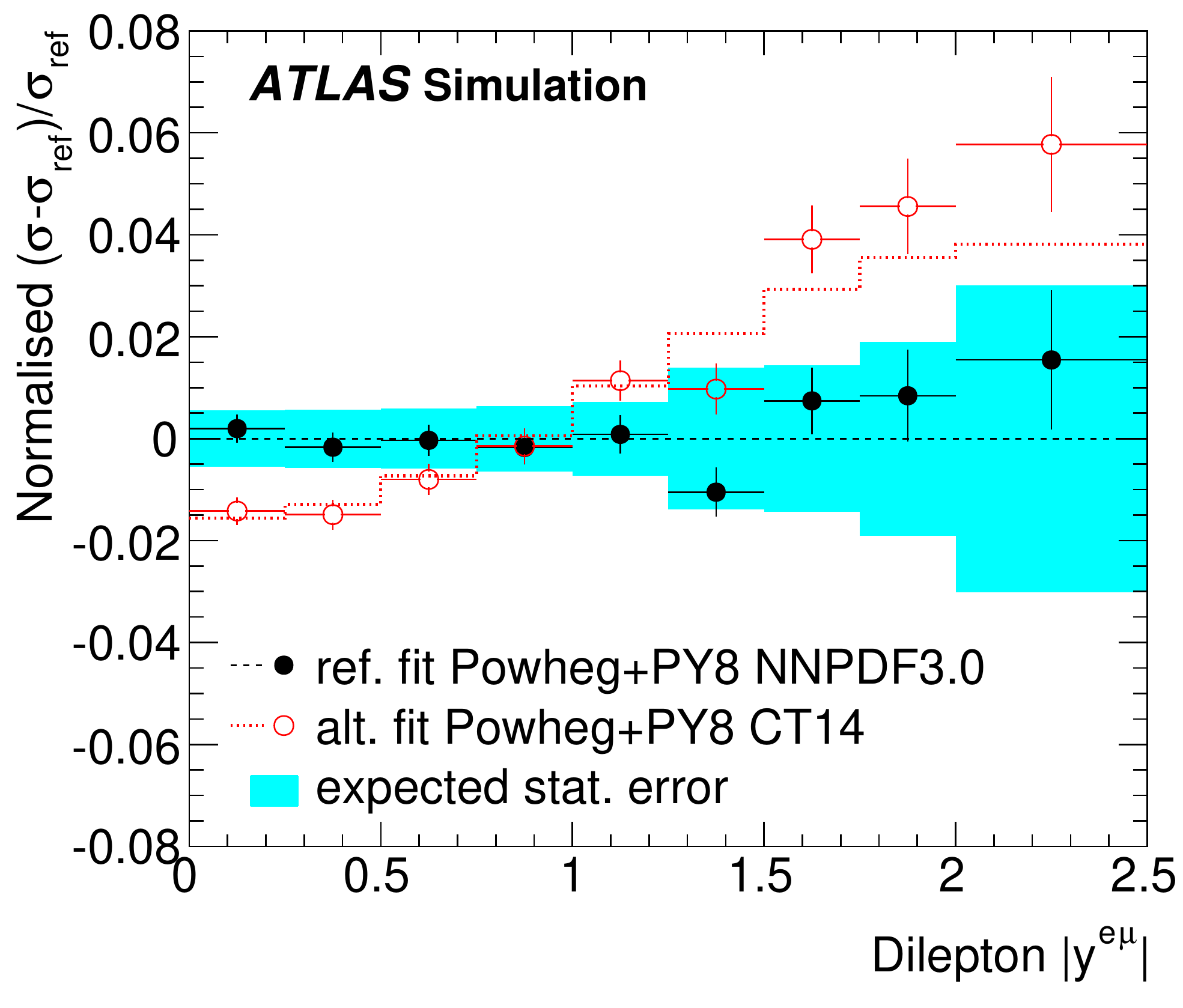}\vspace{-7mm}\center{(d)}}
\caption{\label{f:biastest}Results of pseudo-experiment studies on simulated
events for the extraction of the normalised differential cross-section
distributions for (a) \ptl, (b) \ptll, (c) \etal\ and (d) \rapll, shown as
relative deviations $(\sigma-\sigma_{\mathrm{ref}})/\sigma_{\mathrm{ref}} $
from the reference cross-section values in the
{\sc Powheg\,+\,Pythia6} CT10 (a,b) or {\sc Powheg\,+\,Pythia8} NNPDF3.0
(c,d) samples with $\mtop=172.5$\,\GeV.
The black filled points show the mean deviations from the reference values
of the results from
pseudo-data samples generated with the reference simulation sample, with
error bars indicating the uncertainties due to the limited number of simulated
events. The cyan shaded bands indicate the expected statistical uncertainties
for a single
sample corresponding to the data integrated luminosity. The open red points
show the mean deviations from the reference values obtained from
pseudo-experiments
generated from an alternative simulation sample with $\mtop=175$\,\GeV\ (a, b)
or by reweighting the baseline sample to the CT14 PDF (c,d).
The red error bars represent the uncertainty due to the limited size
of these alternative samples, and the red dotted lines show the
true deviations from the reference in the alternative samples.}
\end{figure}
 
\section{Systematic uncertainties}\label{s:syst}
 
Systematic uncertainties in the measured inclusive cross-section
arise from uncertainties in the input quantities \epsem,
\cb, \nib,  \niib\ and $L$ appearing in Eqs.~(\ref{e:tags}), and the
corresponding quantities in Eqs.~(\ref{e:fidtags}) for the differential
cross-sections. Each source of systematic uncertainty was evaluated by
changing all relevant input quantities coherently and re-solving the tagging
equations, thus taking into account systematic correlations between the
different inputs (and between different bins in the differential analysis).
The sources of systematic uncertainty are divided into the five groups
discussed below, and are shown in detail for the inclusive and fiducial \ttbar\
cross-sections in Table~\ref{t:incsyst}. The uncertainties are shown
in groups for each bin of the single- and double-differential cross-sections in
Tables~\ref{t:insXSec1}--\ref{t:insXSec10}, and the uncertainties for the
normalised single-differential cross-sections are also shown in
Figure~\ref{f:fracsyst}.
 
\begin{table}
\centering
 
\begin{tabular}{ll|ccc|cc}\hline
& Uncertainty source & $\Delta\epsem/\epsem$ & $\Delta\gem/\gem$ &
$\Delta\cb/\cb$ & $\Delta\xtt/\xtt$ & $\Delta\xfid/\xfid$ \\
&  & (\%) & (\%) & (\%) & (\%) & (\%) \\
\hline
&                        Data statistics &      &      &      & 0.44 & 0.44 \\
\hline
\ttbar\ mod. &                     \ttbar\ generator & 0.38 & 0.05 & 0.05 & 0.43 & 0.10 \\
&                 \ttbar\ hadronisation & 0.24 & 0.42 & 0.25 & 0.49 & 0.67 \\
&         Initial/final-state radiation & 0.30 & 0.26 & 0.16 & 0.45 & 0.41 \\
&      \ttbar\ heavy-flavour production & 0.01 & 0.01 & 0.26 & 0.26 & 0.26 \\
&         Parton distribution functions & 0.44 & 0.05 &  -   & 0.45 & 0.07 \\
&                 Simulation statistics & 0.22 & 0.15 & 0.17 & 0.22 & 0.18 \\
Lept. &                 Electron energy scale & 0.06 & 0.06 &  -   & 0.06 & 0.06 \\
&            Electron energy resolution & 0.01 & 0.01 &  -   & 0.01 & 0.01 \\
&               Electron identification & 0.34 & 0.34 &  -   & 0.37 & 0.37 \\
&                Electron charge mis-id & 0.09 & 0.09 &  -   & 0.10 & 0.10 \\
&                    Electron isolation & 0.22 & 0.22 &  -   & 0.24 & 0.24 \\
&                   Muon momentum scale & 0.03 & 0.03 &  -   & 0.03 & 0.03 \\
&              Muon momentum resolution & 0.01 & 0.01 &  -   & 0.01 & 0.01 \\
&                   Muon identification & 0.28 & 0.28 &  -   & 0.30 & 0.30 \\
&                        Muon isolation & 0.16 & 0.16 &  -   & 0.18 & 0.18 \\
&                        Lepton trigger & 0.13 & 0.13 &  -   & 0.14 & 0.14 \\
Jet/$b$ &                      Jet energy scale & 0.02 & 0.02 & 0.06 & 0.03 & 0.03 \\
&                 Jet energy resolution & 0.01 & 0.01 & 0.04 & 0.01 & 0.01 \\
&                       Pileup jet veto &  -   &  -   &  -   & 0.02 & 0.02 \\
&                $b$-tagging efficiency &  -   &  -   & 0.04 & 0.20 & 0.20 \\
&                    $b$-tag mistagging &  -   &  -   & 0.06 & 0.06 & 0.06 \\
Bkg. &              Single-top cross-section &  -   &  -   &  -   & 0.52 & 0.52 \\
&       Single-top/\ttbar\ interference &  -   &  -   &  -   & 0.15 & 0.15 \\
&                  Single-top modelling &  -   &  -   &  -   & 0.34 & 0.34 \\
&                $Z$+jets extrapolation &  -   &  -   &  -   & 0.09 & 0.09 \\
&                Diboson cross-sections &  -   &  -   &  -   & 0.02 & 0.02 \\
&                     Diboson modelling &  -   &  -   &  -   & 0.03 & 0.03 \\
&                 Misidentified leptons &  -   &  -   &  -   & 0.43 & 0.43 \\
\hline
&                  Analysis systematics & 0.91 & 0.75 & 0.44 & 1.39 & 1.31 \\
\hline
$L/E_\mathrm{b}$ &                 Integrated luminosity &  -   &  -   &  -   & 1.90 & 1.90 \\
&                           Beam energy &  -   &  -   &  -   & 0.23 & 0.23 \\
\hline
&                     Total uncertainty & 0.91 & 0.75 & 0.44 & 2.40 & 2.36 \\
\hline
\end{tabular}
\caption{\label{t:incsyst}Breakdown of the relative systematic uncertainties
in \epsem, \gem\ and \cb, and the statistical, systematic (excluding
luminosity and beam energy) and total uncertainties in the inclusive
and fiducial \ttbar\ cross-section measurements. The five groups
of systematic uncertainties corresponding to the discussion in
Sections~\ref{ss:sytt} to~\ref{ss:sylumieb} are indicated in the leftmost
column.}
\end{table}
 
\begin{figure}[tp]
\parbox{83mm}{\includegraphics[width=76mm]{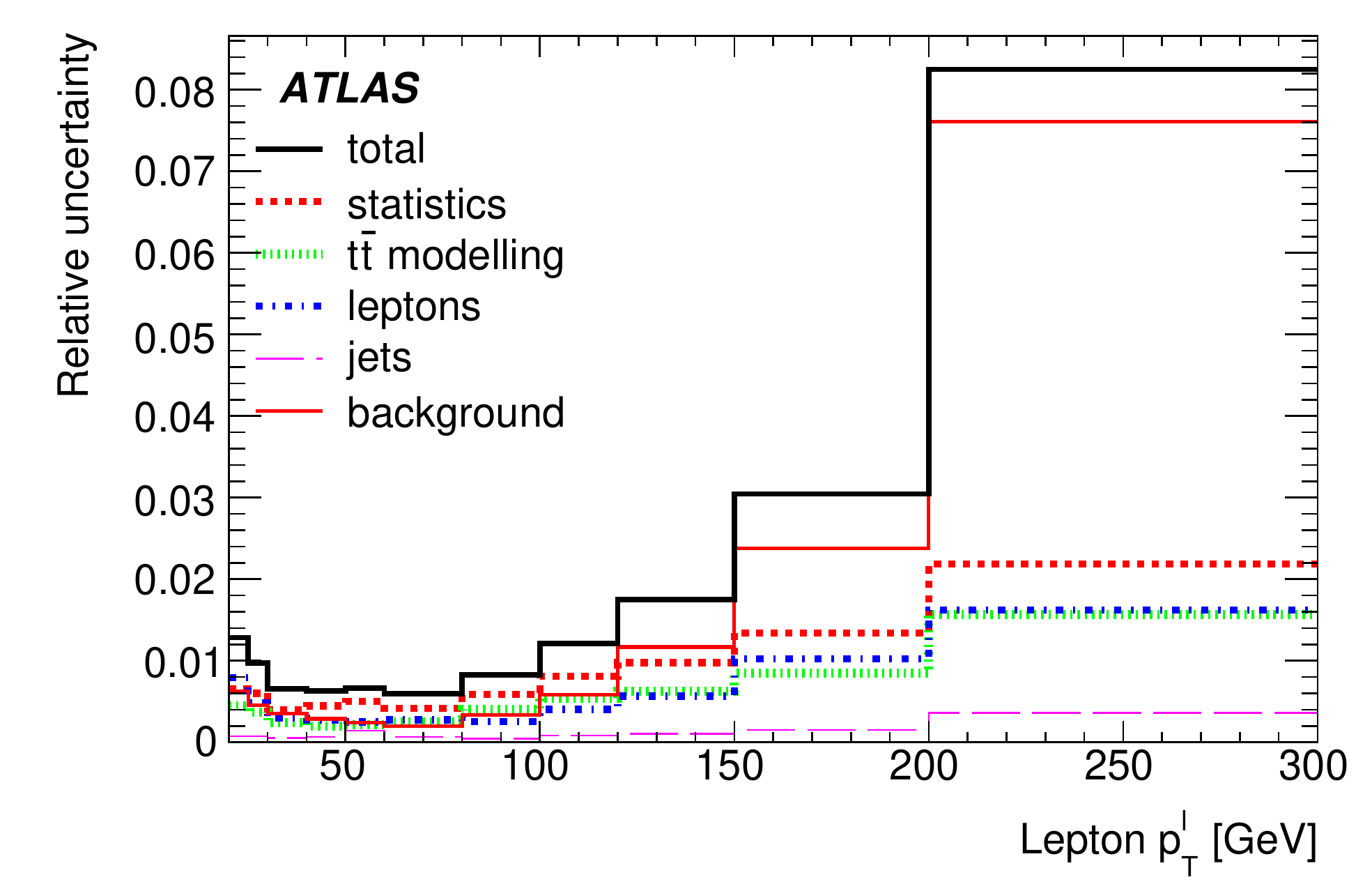}\vspace{-7mm}\center{(a)}}
\parbox{83mm}{\includegraphics[width=76mm]{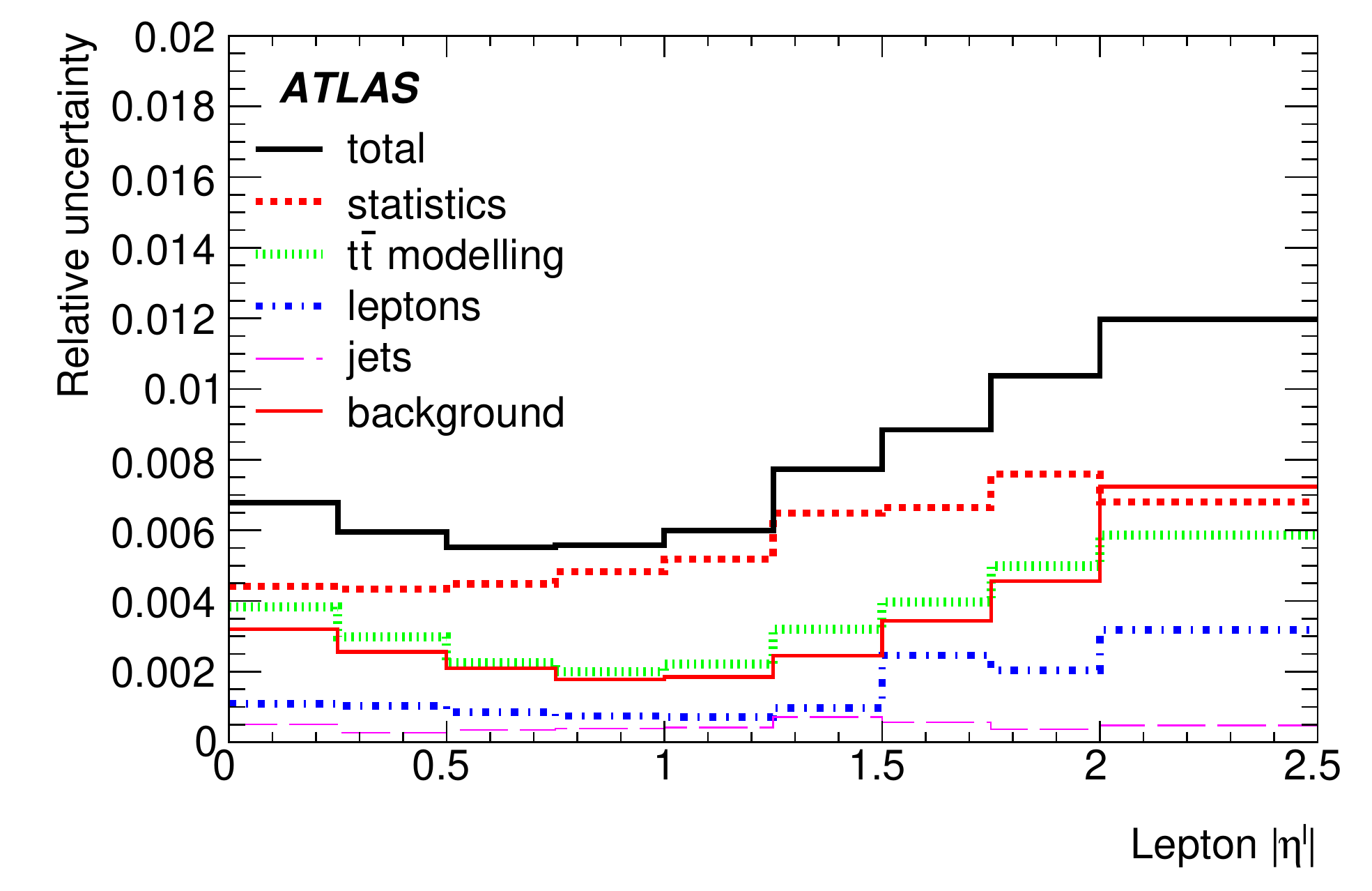}\vspace{-7mm}\center{(b)}}
\parbox{83mm}{\includegraphics[width=76mm]{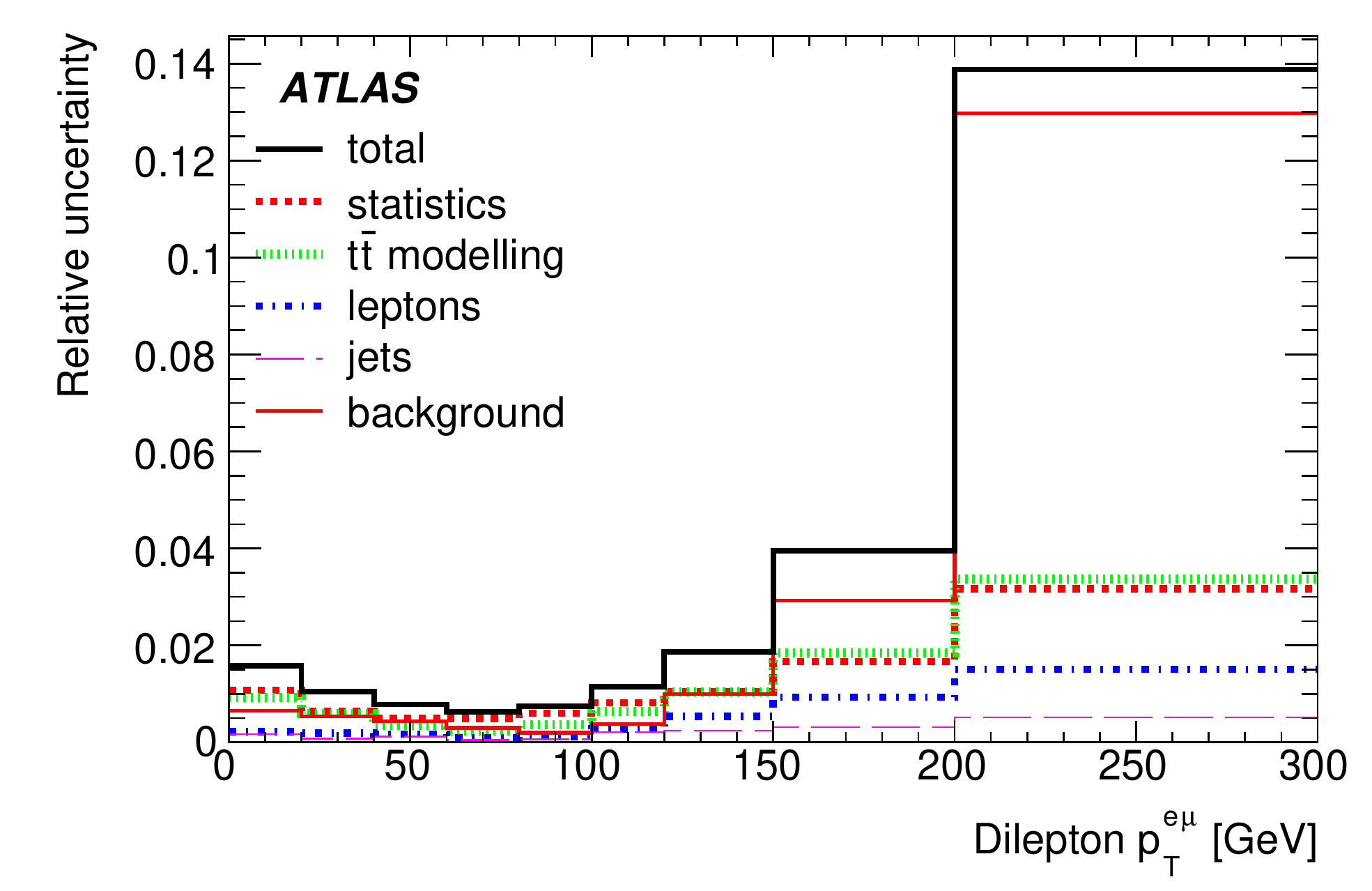}\vspace{-7mm}\center{(c)}}
\parbox{83mm}{\includegraphics[width=76mm]{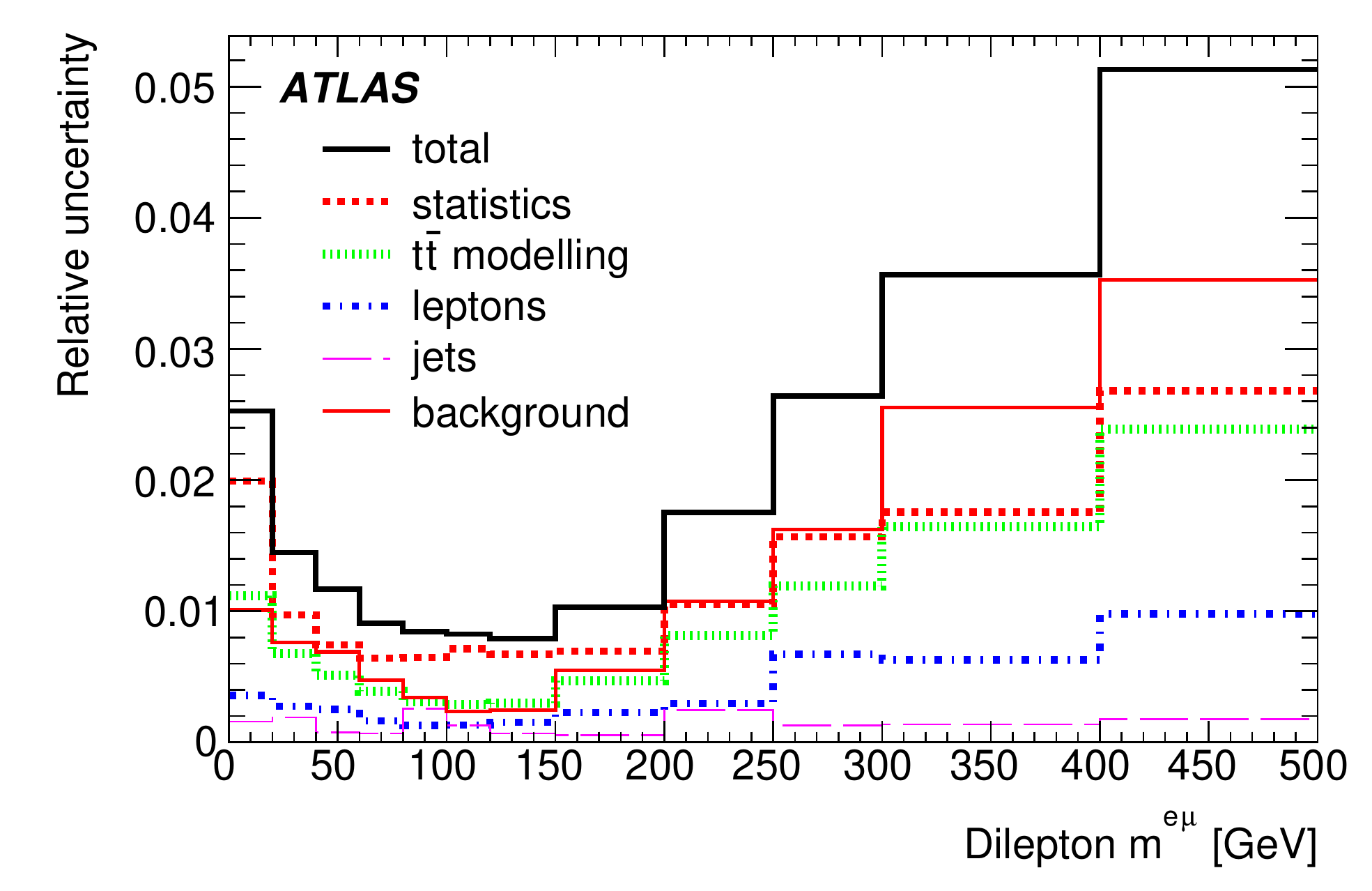}\vspace{-7mm}\center{(d)}}
\parbox{83mm}{\includegraphics[width=76mm]{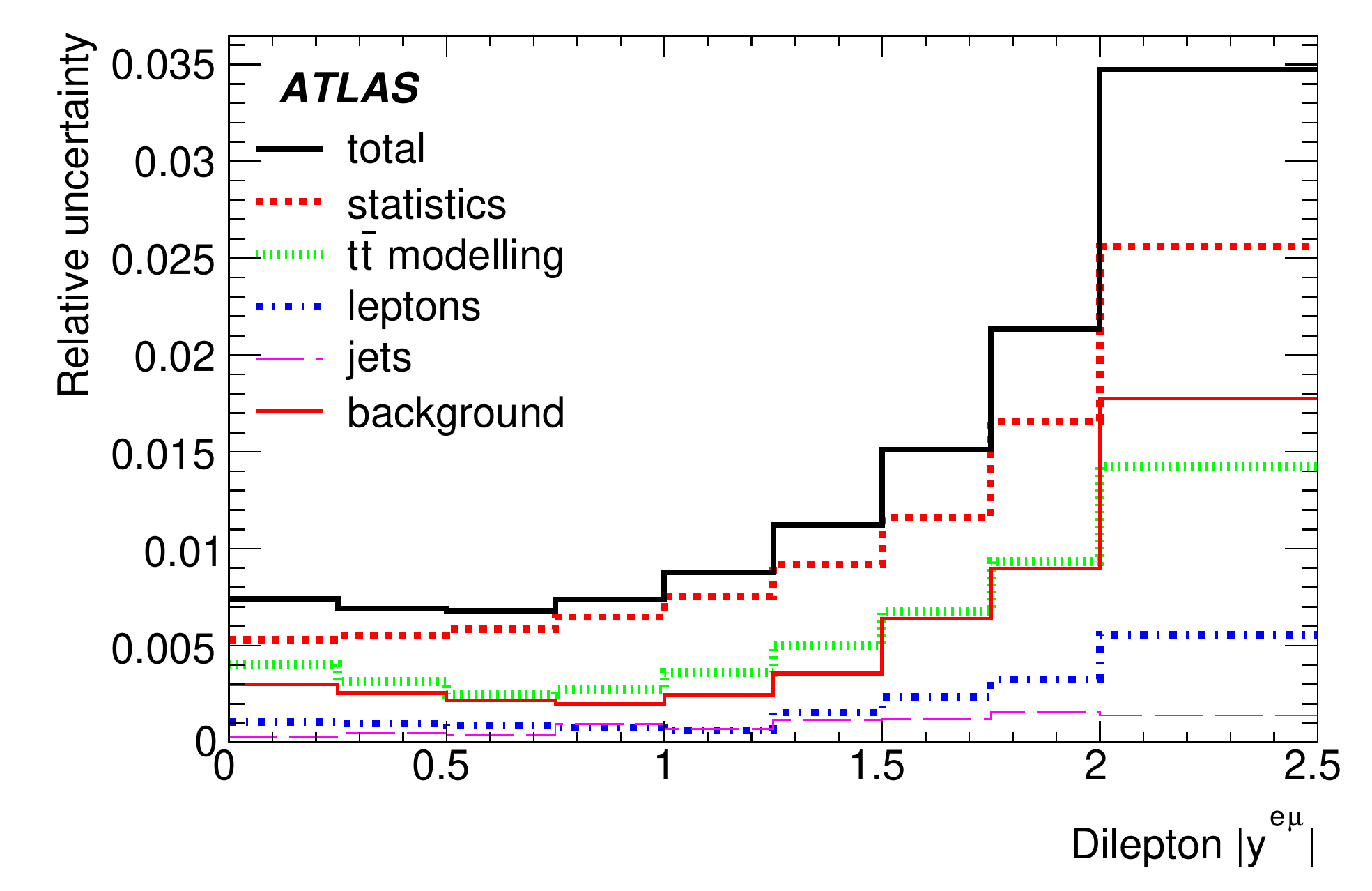}\vspace{-7mm}\center{(e)}}
\parbox{83mm}{\includegraphics[width=76mm]{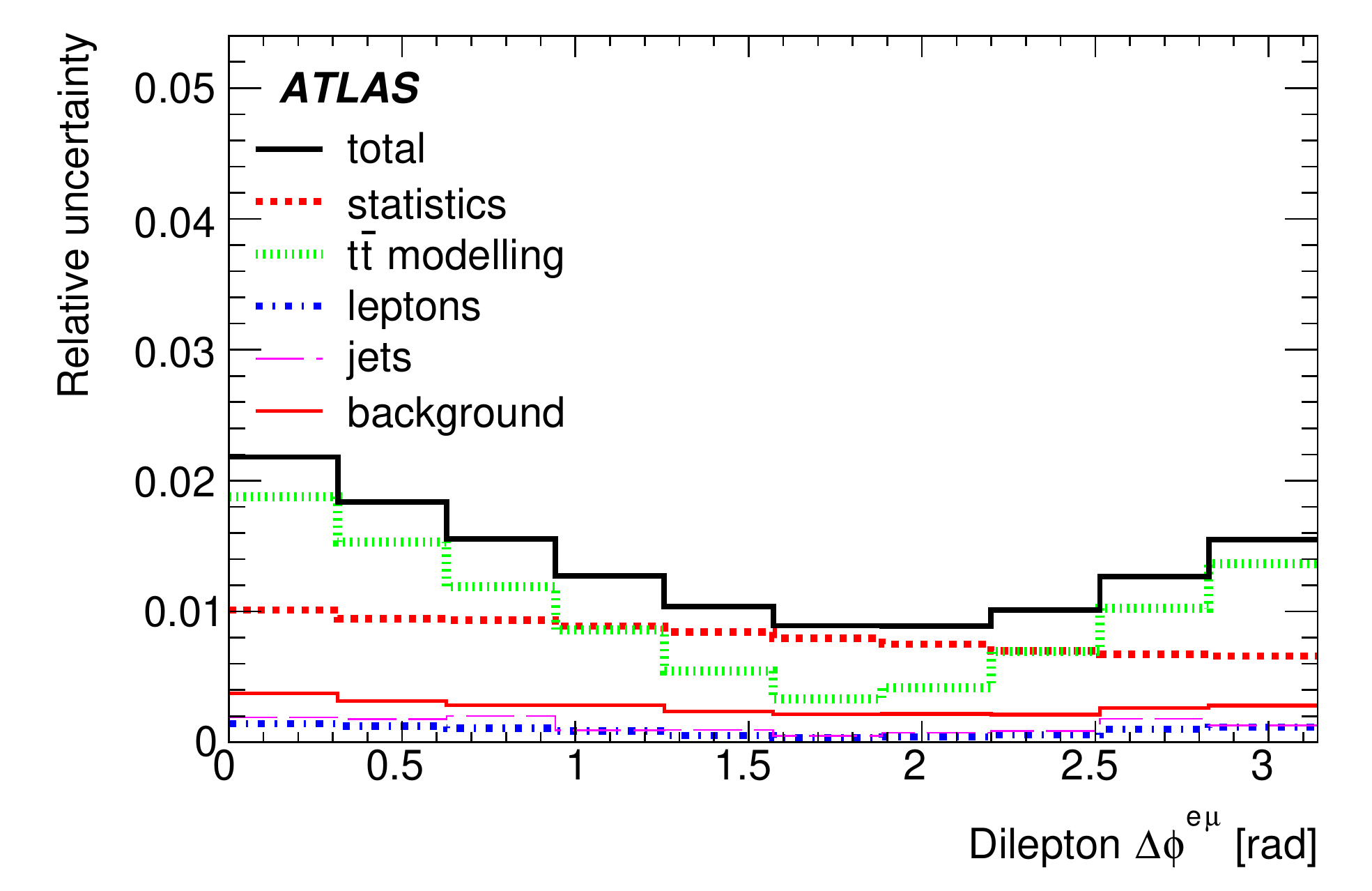}\vspace{-7mm}\center{(f)}}
\parbox{83mm}{\includegraphics[width=76mm]{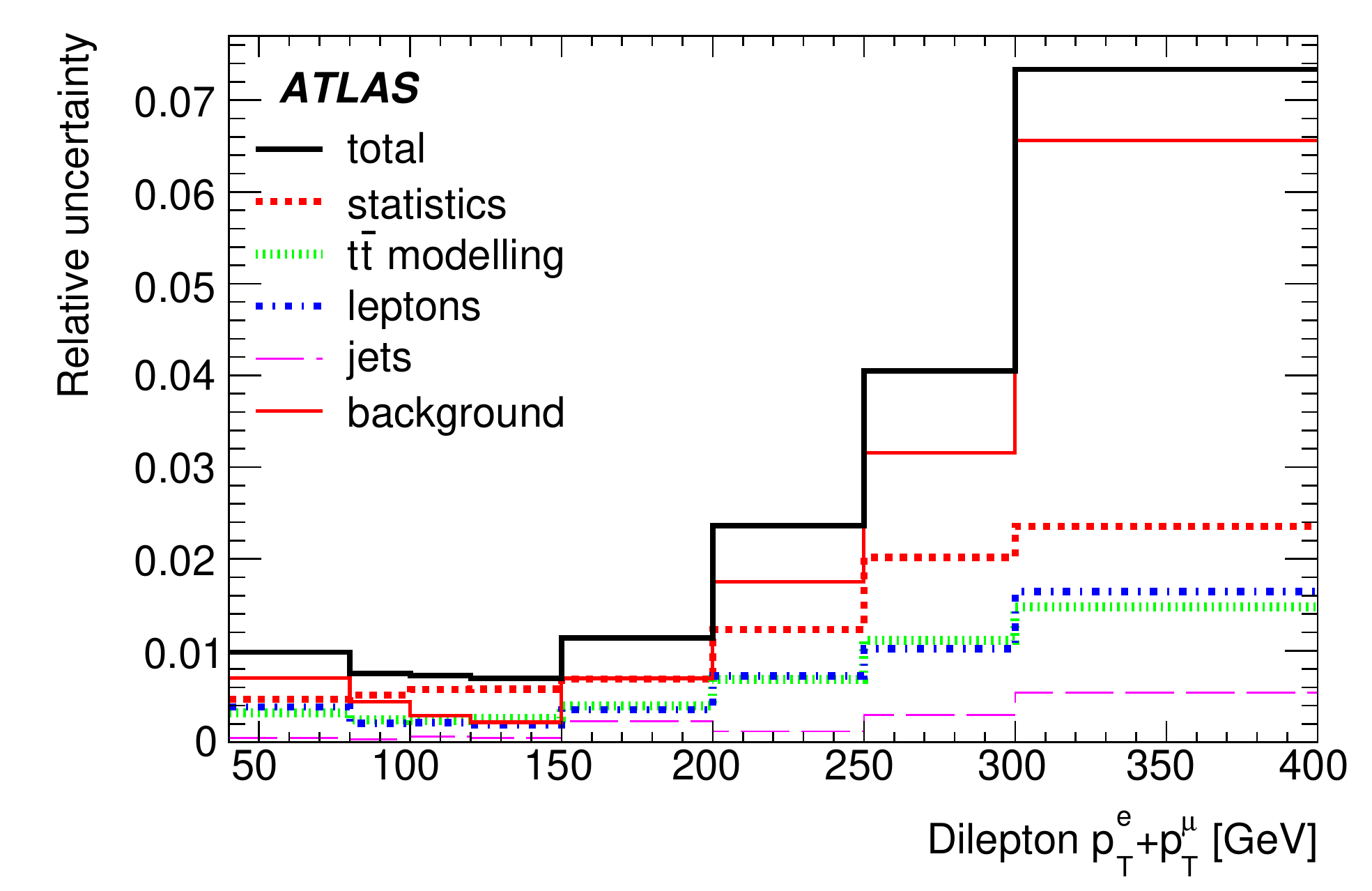}\vspace{-7mm}\center{(g)}}
\parbox{83mm}{\includegraphics[width=76mm]{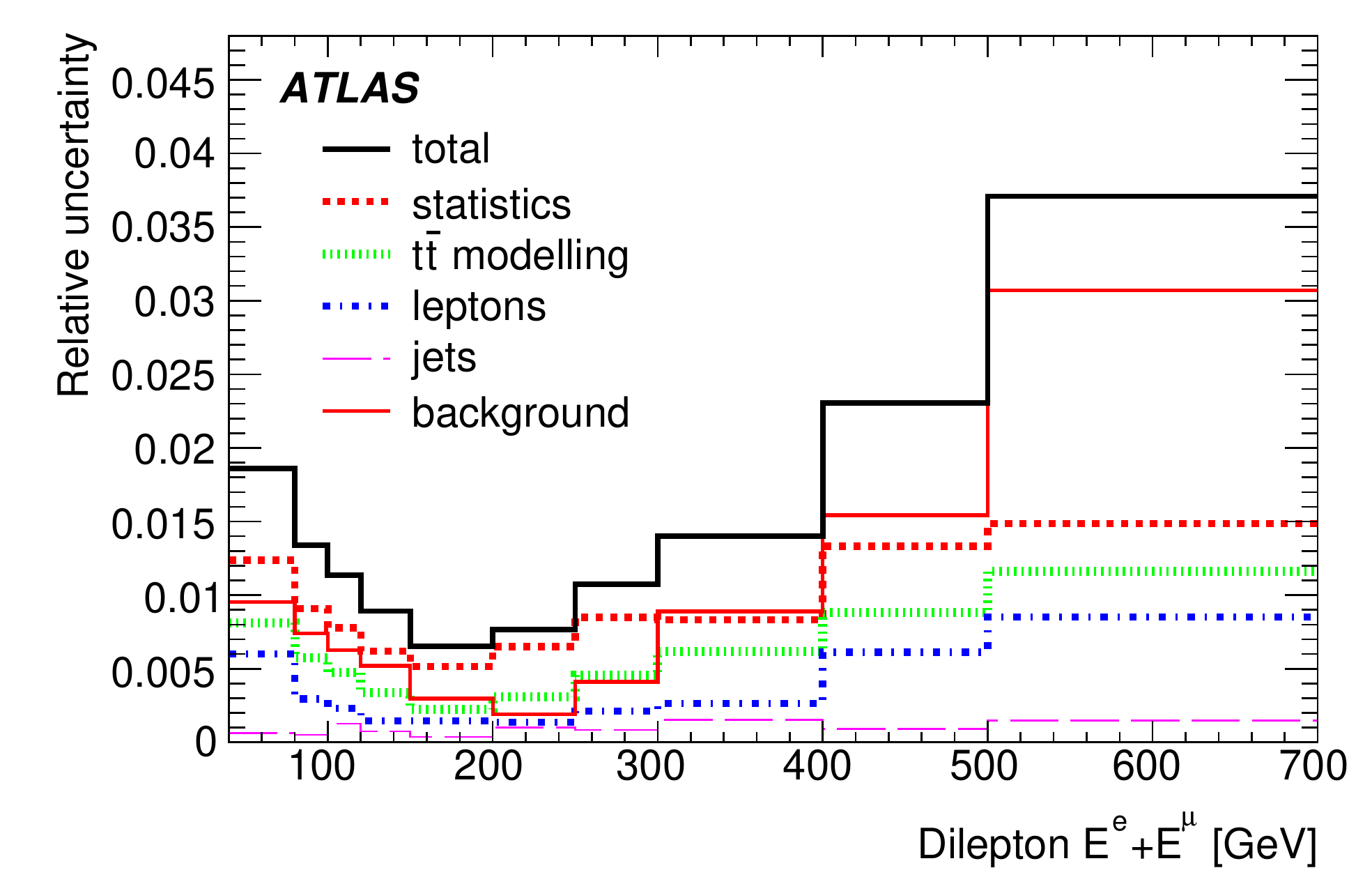}\vspace{-7mm}\center{(h)}}
\caption{\label{f:fracsyst}Relative uncertainties in the measured normalised
differential cross-sections coming from data statistics,
\ttbar\ modelling, leptons, jets and background,  as a function of each lepton
or dilepton
differential variable. The total uncertainty is shown by the thick black lines,
and also includes small contributions from the integrated luminosity and
LHC beam energy uncertainties.}
\end{figure}
 
\subsection{\ttbar\ modelling}\label{ss:sytt}
 
The uncertainties in \epsem, \gem, \gemi, \cb\ and \cbi\ (and \fntaui\ for
the $\tau$-corrected cross-sections) were evaluated using the alternative
\ttbar\ samples described in Section~\ref{s:datmc}. The \ttbar\
generator uncertainty was determined by comparing the baseline
{\sc Powheg\,+\,Pythia8} sample with {\sc aMC@NLO\,+\,Pythia8}.
The parton shower, hadronisation and underlying event uncertainty (referred to
as `hadronisation' below) was evaluated by comparing the baseline with
{\sc Powheg\,+\,Herwig7}. The initial/final-state radiation uncertainty
was evaluated as half the difference
between the {\sc Powheg\,+\,Pythia8} model variations with more or less
parton-shower radiation; as discussed in Section~\ref{s:datmc},
these samples also include variations of \muf\ and \mur.
As shown in Table~\ref{t:incsyst}, the \ttbar\ generator
uncertainty is larger for \epsem\ than for \gem, as the {\sc Powheg\,+\,Pythia8}
and {\sc aMC@NLO\,+\,Pythia8} samples predict different particle-level
acceptances \aem. In contrast, the differences in \aem\ and
\gem\ for the \ttbar\ hadronisation uncertainty have opposite signs, leading
to a smaller shift in \epsem\ than in \gem.
In the differential analyses, the bin-by-bin shifts
in \gemi\ and \cbi\ were fitted with polynomial functions to reduce
statistical fluctuations. All these comparisons were carried out without
applying the lepton isolation requirements, as the isolation efficiencies were
measured in situ in data as discussed in Section~\ref{ss:sylept}, and the
simulation was only used to predict the lepton reconstruction, identification
and overlap removal uncertainties. This procedure also reduces the sensitivity
to the modelling of hadronisation, the underlying event and colour reconnection.
 
The values of \cb\ and \cbi\ are sensitive to the fraction of \ttbar\
events with extra \bbbar\ or \ccbar\ pairs. Such \ttbar\ plus heavy-flavour
production gives rise to events with three or more $b$-tagged jets;
as can be seen from Figure~\ref{f:dmcjlept}(a) and also measured in
a dedicated analysis \cite{TOPQ-2017-12},  this rate is underestimated
by the available \ttbar\ models that only produce extra \bbbar\ or \ccbar\
pairs through the parton shower. The potential effect on \cb\ was studied
by reweighting the baseline {\sc Powheg\,+\,Pythia8} \ttbar\ sample so as to
increase the fraction of events with at least three $b$-jets at generator
level by 40\%, an enhancement which reproduces both the rate of events with
three $b$-tags and the \pt\ and $\eta$ distributions of the third highest-\pt\
$b$-tagged jet in these events.
The resulting shifts in \cb\ and \cbi\ were assigned as
additional systematic uncertainties due to the modelling of heavy-flavour
production in \ttbar\ events.
 
Parton distribution function uncertainties were evaluated by reweighting
the baseline {\sc Powheg\,+\,Pythia8} \ttbar\ sample using generator
weights associated with each of the 100 variations (replicas) provided
by the NNPDF3.0 authors \cite{nnpdf3}, and calculating the RMS of the
changes induced in \epsem, \gem\ and \gemi. The resulting uncertainties are
0.45\% in \xtt, but less than 0.1\% in \xfid, as variations of the
PDF mainly affect the acceptance rather than the reconstruction efficiency.
Similar uncertainties were found for the PDF4LHC15\_NLO\_30 meta-PDF
\cite{pdf4lhc2}, which
is based on a Monte Carlo combination of the NNPDF3.0, CT14 \cite{ct14}
and MMHT14 \cite{mmht} PDF sets. The central values from all these PDF
sets lie within the uncertainty band obtained from NNPDF3.0.
 
The prediction for \epsem\ is also sensitive to the assumed value of
the top quark mass, as a heavier top quark increases the average lepton \pt\
and makes their $|\eta|$ slightly more central. This effect was evaluated
using \ttbar\ simulation samples with \mtop\ variations from 170 to
177.5\,\GeV, giving a relative change in \epsem\ of 0.3\% for a 1\,\GeV\
change in \mtop. The effect is partially counterbalanced by changes in
the $Wt$ background prediction, which decreases with increasing \mtop.
By convention, the inclusive \ttbar\ cross-section \xtt\ is quoted at
a fixed top quark mass value, but a $\pm 1$\,\GeV\ variation in \mtop\ is
included in the uncertainties for the lepton differential distributions.
 
The total \ttbar\ modelling uncertainties also include the small contributions
due to the limited size of the baseline \ttbar\ simulation sample, and are shown
for the differential distributions by the green dotted lines in
Figure~\ref{f:fracsyst}.
 
\subsection{Lepton identification and measurement}\label{ss:sylept}
 
The modelling of the electron and muon identification efficiencies
was studied using $Z\rightarrow ee/\mu\mu$ events,
as described in Refs. \cite{PERF-2017-01,PERF-2015-10}. Small
corrections were applied to the simulation, and  the correlations in the
associated
systematic uncertainties as a function of lepton \pt\ and $\eta$ were taken
into account and propagated to all differential distributions.
Similar procedures were used to measure the electron and muon trigger
efficiencies with $Z\rightarrow ee/\mu\mu$ decays. Since only one lepton was
required to pass the trigger requirements in order to accept the event, the
trigger efficiencies for events passing the offline selection are high, around
97\% for 2015 data and 94\% for 2016 data. Most of the efficiency
loss comes from events where one lepton has a transverse momentum
below the trigger threshold and the other lepton is above the threshold
but fails the trigger selection. The electron
charge misidentification probability was measured as a function of
\pt\ and $|\eta|$ using the ratio of same- to opposite-sign reconstructed
$Z\rightarrow ee$ events, and the full difference between data and simulation,
which is only significant for forward electrons with $|\eta|>1.5$,
was assigned as an uncertainty. The electron and muon energy/momentum
scales and resolutions were determined using
$Z\rightarrow ee/\mu\mu$, $Z\rightarrow\ell\ell\gamma$,
$J/\psi\rightarrow ee/\mu\mu$ and $\Upsilon\rightarrow\mu\mu$ decays
\cite{PERF-2017-03,PERF-2015-10}, and the residual uncertainties are typically
much smaller than those associated with the lepton efficiency measurements.
 
The lepton isolation efficiencies were measured directly in the
\ttbar-dominated $e\mu$ plus $b$-tagged jet samples, by determining the
fractions of events where either the electron or muon fails the isolation
cut, as functions of lepton \pt\ and separately for the barrel ($|\eta|<1.5$)
and forward regions. The  samples of leptons failing isolation cuts
have significant contamination
from misidentified leptons, reaching up to 10\% in the $e\mu$ plus two
$b$-tagged jet sample at low lepton \pt, and up to 50\% in the one $b$-tagged
jet sample, but in both cases decreasing strongly with increasing lepton \pt.
The results were corrected for this contamination,
estimated from data with the aid of leptons with large
impact parameter significance ($|d_{0}|/\sigma_{d_{0}}>5$), that provide
a control sample enriched in misidentified leptons. Templates for the
impact parameter significance distributions of misidentified leptons were
obtained from the same-sign $e\mu$ samples, subtracting estimated
prompt lepton contributions using simulation.
The total uncertainties on the measured isolation
efficiencies are up to 0.9\% for electrons and 0.6\% for muons
at low lepton \pt, dominated by the dependence on the choice of impact parameter
significance cut, and reduce to 0.1\% at high \pt, where the isolation
efficiency is around 98\% and the misidentified lepton contributions
are very small. The method was validated by using the
various alternative \ttbar\ simulation samples (which predict different
isolation efficiencies) as pseudo-data, and by explicitly changing the
lepton isolation efficiencies in simulation and verifying that the measurement
procedure recovered the changes.
 
The isolation efficiencies measured on data are shown in Figure~\ref{f:isolefi},
together with the prediction from simulation.
The baseline \ttbar\ simulation sample gives a good modelling of the muon
isolation efficiency, but underestimates the electron isolation efficiency
in data
by up to 1\% at low lepton \pt, leading to a total correction of about
0.4\% for \epsem. The residual uncertainties on the \pt-integrated
corrections are around 0.2\% for both
electrons and muons, dominated by the subtraction of misidentified-lepton
background at low \pt. For comparison, the differing lepton isolation
efficiency predictions from the various \ttbar\ simulation samples would lead to
differences in \epsem\ of up to about 0.4\%.
The corresponding corrections as a function of lepton
\pt\ and $|\eta|$ were propagated to the values of \gemi\ in each
bin of the differential distributions, and also applied to the
estimates for the dominant $Wt$ background.
The total lepton-related uncertainties are shown by the blue
dot-dashed lines in Figure~\ref{f:fracsyst}.
 
\begin{figure}[tp]
\parbox{83mm}{\includegraphics[width=76mm]{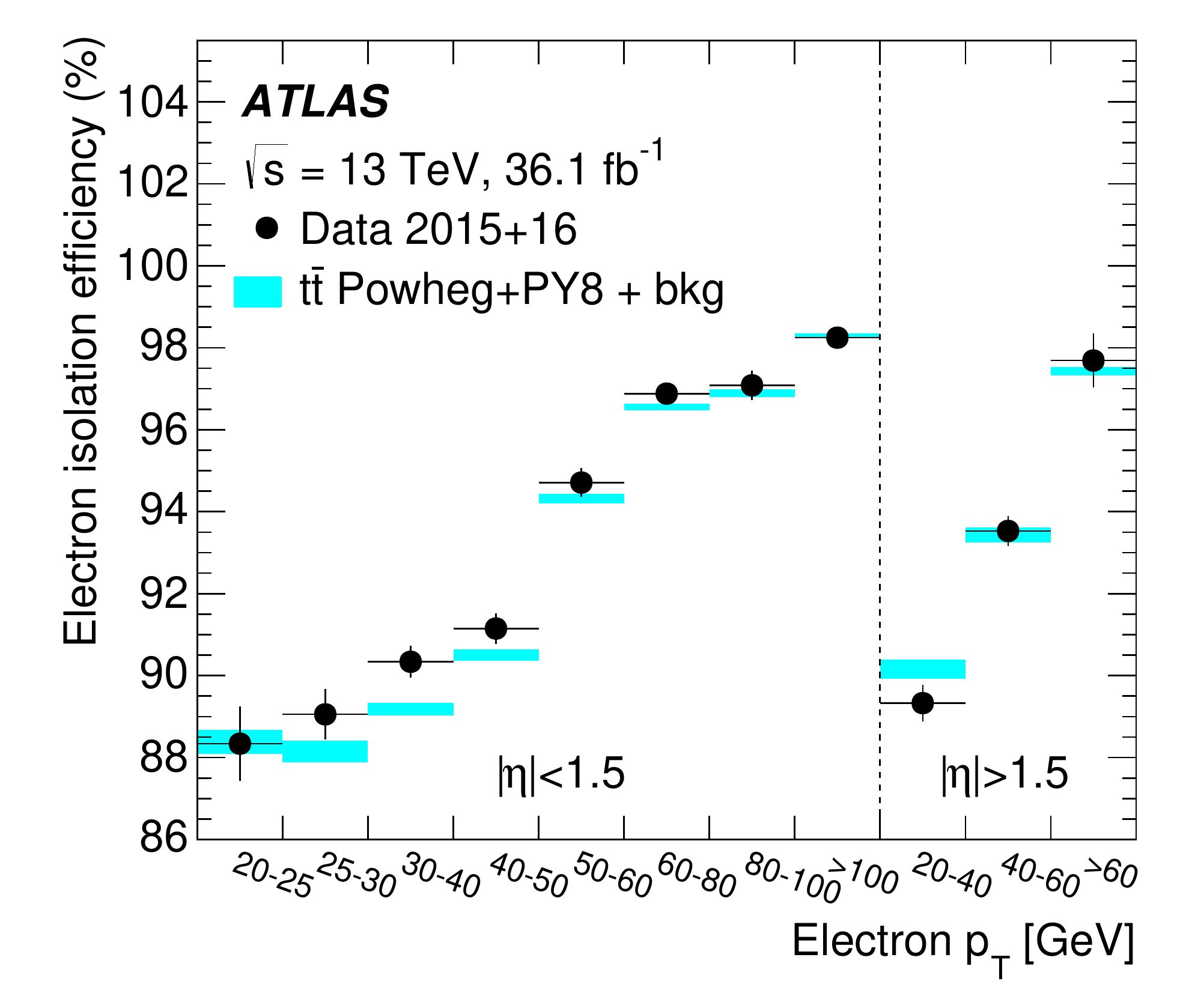}\vspace{-7mm}\center{(a)}}
\parbox{83mm}{\includegraphics[width=76mm]{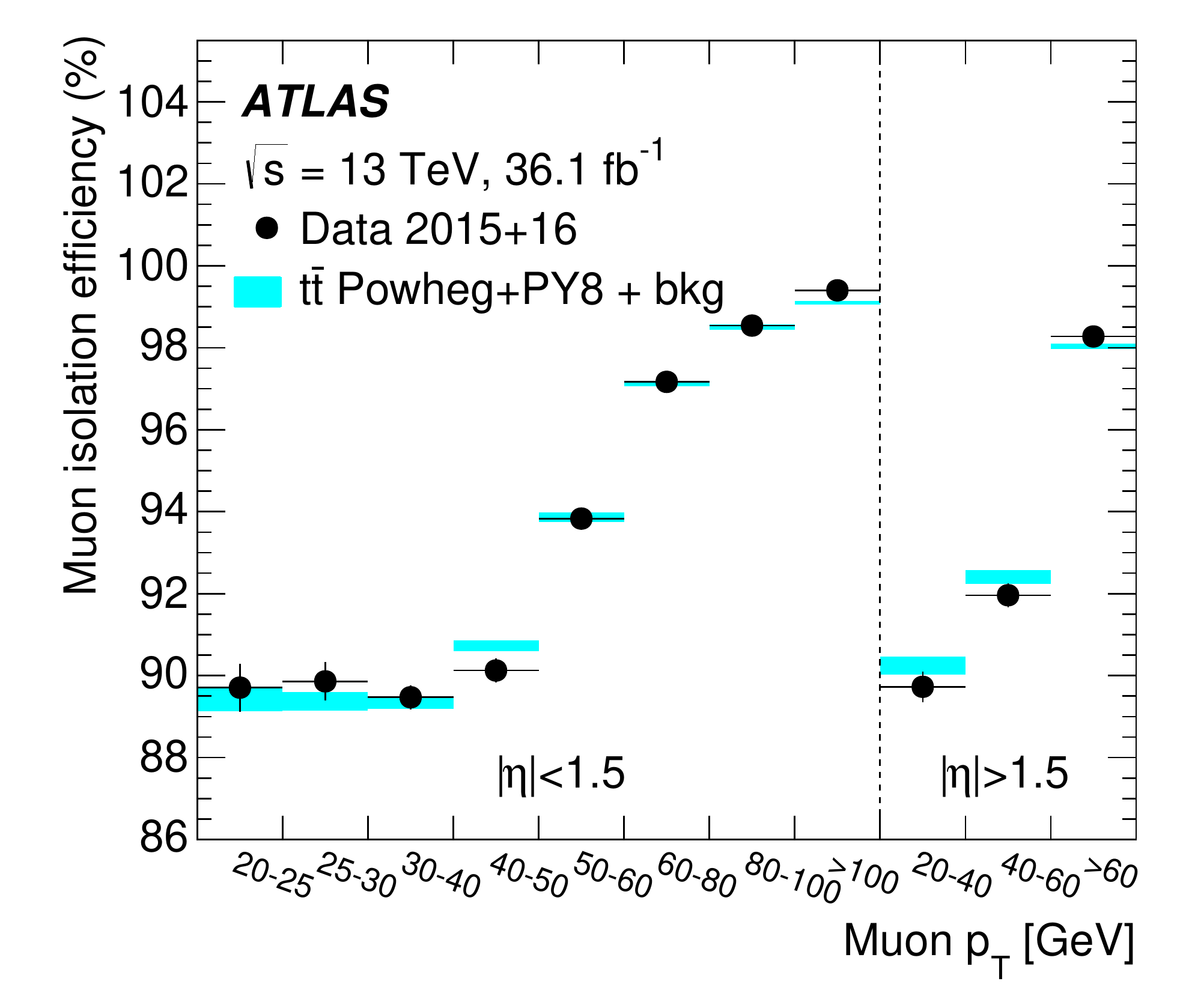}\vspace{-7mm}\center{(b)}}
\caption{\label{f:isolefi}Lepton isolation efficiencies for (a) electrons
and (b) muons measured in the \ttbar-dominated $e\mu$ plus $b$-tagged
jets sample. The data measurements are shown by the black points with
error bars indicating the total uncertainty, and the predictions of the
baseline simulation with {\sc Powheg\,+\,Pythia8} \ttbar\ events plus
background shown
by the cyan bands with width indicating the statistical uncertainty. The
measurements are shown for eight bins of lepton \pt\ in the barrel
region ($|\eta|<1.5$) and three bins in the forward region ($|\eta|>1.5$).}
\end{figure}

\subsection{Jet measurement and $b$-tagging}\label{ss:syjetb}
 
Uncertainties in jet reconstruction and calibration affect the estimates
of the background contributions from $Wt$ and diboson events, and the values
of \cb\ and \cbi. They also have a very small effect on \epsem, \gem\
and \gemi\ due to the removal of leptons within $\Delta R_y=0.4$  of selected
jets. The jet energy scale was determined using a combination of simulation,
test beam and in situ measurements \cite{PERF-2016-04} and the corresponding
uncertainties were evaluated using a model with 20~independent uncertainty
components. The jet energy resolution was measured using Run~1 data
\cite{PERF-2011-04} and the resulting uncertainties were extrapolated to the
\sxyt\ data samples. The modelling of the JVT requirement used to
reject jets coming from pileup was evaluated using jets in
$Z\rightarrow\mu\mu$ events \cite{PERF-2014-03}.
 
The efficiency for $b$-tagging jets in \ttbar\ events was extracted from the
data via Eqs.~(\ref{e:tags}), but simulation was used to predict the numbers
of $b$-tagged jets in $Wt$ and diboson background events. The values of
\cb\ and \cbi\ also depend weakly on the efficiencies for tagging both heavy-
and light-flavoured jets. The modelling of the $b$-tagging performance in
simulation was corrected using scale factors determined using
dileptonic \ttbar\ events for $b$-jets \cite{PERF-2016-05},
single-lepton \ttbar\ events for charm jets
\cite{ATLAS-CONF-2018-001}, and dijet events for light-quark and
gluon jets \cite{ATLAS-CONF-2018-006}.
The corresponding uncertainties were propagated to the background and
correlation coefficient estimates. The uncertainties related to jets and
$b$-tagging are shown by the purple dashed lines in Figure~\ref{f:fracsyst},
and are dominated by the effects of $b$-tagging uncertainties on the
background estimates.
 
\subsection{Background modelling}\label{ss:sybg}
 
The normalisation of the $Wt$ background was varied by 5.3\%,
corresponding to the PDF and QCD scale uncertainties on the
approximate NNLO cross-section
prediction discussed in Section~\ref{s:datmc}. The potential effects of
interference between the \ttbar\ and $Wt$ final states were assessed by
comparing the predictions of {\sc Powheg\,+\,Pythia6} samples with the diagram
removal and diagram subtraction approaches to handling this interference
\cite{wtdr,wtinter1,wtinter2}. The corresponding uncertainty in the inclusive
cross-section result is small, but the diagram subtraction method predicts
up to 30\% less $Wt$ background in the one $b$-tag sample and 60\% less in the
two $b$-tag sample at the high ends of the lepton \pt\ and dilepton
\ptll, \mll, \ptsum\ and \esum\ distributions, where interference
effects become large and dominate the total uncertainty
(see Figure~\ref{f:fracsyst}). However, a
dedicated study of events with two leptons and two $b$-tagged jets
\cite{TOPQ-2017-05} suggests that the data lie between the predictions
of the models with diagram removal and diagram subtraction in the region
where interference effects are important. Further modelling uncertainties
were assessed by comparing the predictions from the baseline $Wt$ sample
with those of {\sc aMC@NLO} interfaced to {\sc Herwig++} \cite{herwigpp},
with {\sc Powheg\,+\,Pythia6} samples with more or less parton-shower radiation,
and with {\sc Powheg\,+\,Herwig7}, in all cases normalising the total
production cross-section to the approximate NNLO prediction. The small
background acceptance uncertainties due to variations of the PDFs were
evaluated using NNPDF3.0 replicas in the same way as for the \ttbar\
signal. They were taken to be uncorrelated with the signal PDF uncertainties,
but are included in the `Parton distribution functions' entry
in Table~\ref{t:incsyst}.
 
Uncertainties in the diboson background
were assessed by varying the cross-sections by 6\% based on calculations
with MCFM \cite{mcfm} using the CT10 PDF set \cite{cttenpdf},
and changing the QCD factorisation, renormalisation, resummation and
CKKW matching scales by factors of two up and down within the {\sc Sherpa}
generator. The combined uncertainties amount to 12\% of the diboson
contribution to the one $b$-tag sample and 33\% for the two $b$-tag sample.
 
The backgrounds from $Z$+jets and events with misidentified leptons were
estimated using data control samples, and the corresponding uncertainties
were evaluated as discussed in Section~\ref{ss:bkgd}. The total
background-related uncertainties in the normalised differential cross-sections
are shown by the red solid lines in Figure~\ref{f:fracsyst}, and are dominated
by those in the $Wt$ background.
 
\subsection{Luminosity and beam energy}\label{ss:sylumieb}
 
The uncertainties in the integrated luminosity are 2.0\% for the 2015
and 2.1\% for the 2016 datasets, evaluated as discussed in
Ref. \cite{ATLAS-CONF-2019-021} using a calibration of the LUCID-2
detector \cite{lucid} obtained from $x$--$y$ beam-separation scans in
each year. For the inclusive cross-section analysis, the total luminosity
uncertainties were broken down into individual components which were
each considered correlated or uncorrelated between years, as appropriate, in
the combination of the cross-section results from the two datasets
\cite{ATLAS-CONF-2019-021}. A single
luminosity uncertainty of 2.1\% in the combined 2015--16 sample was
used for the differential cross-section analysis. In both cases, the
luminosity-induced uncertainties in the measured cross-sections are around
10\% larger than the uncertainty in the integrated luminosity itself,
as the integrated luminosity is needed both for the conversion of the
\ttbar\ event yields to \xtt, and in order to normalise the simulation-based
estimates of the $Wt$ and diboson backgrounds.
 
The LHC beam energy is known to be within 0.1\% of the nominal value
of exactly 6.5\,\TeV\ per beam for \sxyt\ collisions,
based on the LHC magnetic model and comparisons of the revolution
frequencies of proton and lead-ion beams \cite{ebeam2}. A 0.1\% variation
in $\sqrt{s}$ corresponds to a 0.23\% variation in \xtt, according
to the NNLO+NNLL predictions of {\tt Top++} \cite{toppp}. Following
the approach of previous analyses \cite{TOPQ-2013-04,TOPQ-2015-09},
this uncertainty is included in the experimental uncertainty of \xtt,
allowing the measurement to be compared with theoretical predictions for
\xtt\ at exactly \sxyt. The beam energy uncertainty also affects the
predictions for both the absolute and normalised  differential distributions,
as e.g.\ the lepton \pt\ distributions become slightly harder and the
\etal\ distributions slightly more forward as $\sqrt{s}$ increases. These shifts
were evaluated by reweighting the {\sc aMC@NLO\,+\,Pythia8} \ttbar\ sample
using PDF weights calculated using LHAPDF \cite{lhapdf} so as to vary
the effective $\sqrt{s}$ by $\pm 0.1$\%, and the resulting uncertainties
were included in the differential cross-section results. The combined
effects of the luminosity and beam energy uncertainties on the normalised
differential cross-sections are listed in
Tables~\ref{t:insXSec1}--\ref{t:insXSec4}, and are at most 0.3\%, always
small compared with the other systematic and statistical uncertainties
of the measurements.
 
\section{Inclusive cross-section results and interpretation}\label{s:xsres}
 
The results of the inclusive \ttbar\ cross-section analysis are given in
Section~\ref{ss:incres}, followed by the extraction of the top quark mass in
Section~\ref{ss:mass} and the determination of ratios of cross-sections at
different $\sqrt{s}$ values in Section~\ref{ss:ratio}.
The analyses were initially
performed `blind' by multiplying the \xtt\ values by an unknown, randomly chosen
scale factor which was only removed after verifying that consistent results were
obtained from the 2015 and 2016 datasets, and after finalising all
systematic uncertainties and stability studies. As a validation of the
analysis procedures, the yields of
$Z\rightarrow ee$ and $Z\rightarrow\mu\mu$ selections relative to the
expectations from {\sc Powheg\,+\,Pythia8}-based $Z\rightarrow\ell\ell$
simulation were also compared across all data-taking periods
and trigger selections, and found to be compatible within
the assigned systematic and very small statistical uncertainties.
 
\subsection{Total and fiducial cross-section results}\label{ss:incres}
 
Table~\ref{t:incres} shows the  results for \xtt\ and \xfid\ from the entire
2015--16 dataset treated as a single sample, the 2015 and 2016 datasets
separately, and the combination of 2015 and 2016
results. The latter was performed using the best linear unbiased estimator
technique \cite{blue,valassiblue},
taking into account correlations in the systematic uncertainties.
The combination gives the smallest total
uncertainty, 9\% smaller than that from all data treated as one sample, and
gives the final results:
\begin{eqnarray}
\xtt & = & \ttxval \pm \ttxstat \pm \ttxsyst \pm \ttxlumi \pm \ttxebeam\,\mathrm{pb},\ \mathrm{and} \nonumber \\
\xfid & = & \ttxfidval \pm \ttxfidstat \pm \ttxfidsyst \pm \ttxfidlumi \pm \ttxfidebeam\,\mathrm{pb},\nonumber
\end{eqnarray}
where the four uncertainties are due to data statistics, experimental and
theoretical  systematic effects internal to the analysis, the knowledge of
the integrated luminosity, and the knowledge of the LHC beam energy.
The  total relative uncertainties are \ttxrel\ for both \xtt\ and \xfid.
The 2015 and 2016 datasets
have relative weights of 0.49 and 0.51. The uncertainties due to the luminosity
are only partially correlated and are similar in magnitude in both datasets,
leading
to approximately equal weights despite the much larger data sample from 2016.
Other uncertainties are largely correlated between the two datasets, except
for the statistical components of uncertainties estimated from data,
such as the electron and muon
identification efficiencies, and the misidentified-lepton background estimate.
The $\chi^2$ for the combination of 2015 and 2016 data is 0.23 for one degree
of freedom, demonstrating good compatibility of the results. The
values of \epsb\ obtained in 2015 data and simulation are very similar,
and 1.6\% lower in 2016 data than simulation, well within the expected
uncertainties in the modelling of $b$-tagging performance \cite{PERF-2016-05}.
The result for \xtt\ is  reported for a fixed top quark mass of
$\mtop=172.5$\,\GeV, and depends on the assumed value according to
$(1/\xtt)\, \mathrm{d}\xtt/\mathrm{d}\mtop=-0.20$\%/\GeV.
The \mtop\ dependence of \xfid\ is negligible.
The fiducial cross-section was also corrected to
remove the contribution of events with leptons from leptonic $\tau$ decays
as discussed in Section~\ref{ss:diffmeas}, giving a result of
$\xfidnotau=\ttxntfidval\pm\ttxntfidstat\pm\ttxntfidsyst\pm\ttxntfidlumi\pm\ttxntfidebeam$\,pb.
 
\begin{table}[tp]
\centering
 
\begin{tabular}{l|c|c}\hline
Dataset & $\xtt$ [pb] & $\xfid$ [pb] \\
\hline
All data & $ 830.7 \pm  2.2\pm 11.6\pm 18.4\pm  1.9$ (22.0) &
$14.14 \pm0.04\pm0.19\pm0.31\pm0.03$ (0.37) \\
2015 data & $ 820.9 \pm  6.9\pm 11.9\pm 18.4\pm  1.9$ (23.1) &
$13.98 \pm0.12\pm0.19\pm0.31\pm0.03$ (0.39) \\
2016 data & $ 831.8 \pm  2.3\pm 11.6\pm 19.5\pm  1.9$ (22.9) &
$14.16 \pm0.04\pm0.19\pm0.33\pm0.03$ (0.39) \\
\hline
Combination & $ 826.4 \pm  3.6\pm 11.5\pm 15.7\pm  1.9$ (19.9) &
$14.07 \pm0.06\pm0.18\pm0.27\pm0.03$ (0.33) \\
\end{tabular}
\caption{\label{t:incres}Measurements of the inclusive total (\xtt) and
fiducial (\xfid) \ttbar\ production cross-sections at \sxyt\
using the full dataset,
the 2015 and 2016 datasets separately, and the combination of the 2015 and 2016
measurements. The fiducial cross-section requires an opposite-sign
$e\mu$ pair, with both leptons having $\pt>20$\,\GeV\ and $|\eta|<2.5$,
as discussed in Section~\ref{ss:incmeas}.
The four uncertainties for each measurement correspond to
the statistical, experimental and theoretical systematic,
integrated luminosity, and beam energy
uncertainties. The total uncertainty is given in parentheses after each
result.}
\end{table}
 
The breakdown of statistical and systematic uncertainties in the measurements
is given in Table~\ref{t:incsyst}, which also shows the average uncertainty
contributions to \epsem\ and \cb, weighted as in the combination.
The largest uncertainties come from the calibration of the integrated
luminosity, followed by \ttbar\ modelling (generator, hadronisation,
radiation and PDFs),
background modelling ($Wt$ single-top cross-section and misidentified leptons),
and lepton identification efficiencies.
The uncertainties due to \ttbar\ generator choices and PDFs are smaller
for \xfid\ than for \xtt, but are offset by
a larger uncertainty due to \ttbar\ hadronisation, such that the total
uncertainties in the two measurements are very similar.
 
The results are stable within the statistical uncertainties
when increasing the minimum jet \pt\ requirement from the nominal value
of 25\,\GeV\ up to 75\,\GeV, where the tagging correlations become much
stronger ($\cb=1.16$).
The results are also stable when tightening the jet selection
to $|\eta|<1.0$ and changing the $b$-tagging selection to use the 60\%
or 77\% efficiency working points. However, a significant trend was
found when tightening the lepton \pt\ requirement from the nominal
$\pt>20$\,\GeV\ in several steps up to $\pt>55$\,\GeV, where \epsem\ is reduced
by a factor 4.4 and \xtt\ changes by $-3.9\pm 0.7$\%, the uncertainty
corresponding to the uncorrelated statistical component only.
This is caused by the lepton \pt\ spectrum
in data being significantly softer than that in the baseline
{\sc Powheg\,+\,Pythia8} simulation (see Figures~\ref{f:dmcjlept}(c) and~\ref{f:dmcjlept}(e),
and Figure~\ref{f:distresa}(a) below). As discussed in Section~\ref{ss:gencomp}
and shown in Figure~\ref{f:rdistresa}, the \ptl\ distribution is better
described by the alternative {\sc aMC@NLO\,+\,Pythia8} \ttbar\ sample, or by
reweighting the baseline {\sc Powheg\,+\,Pythia8} sample to better
describe the measured top quark \pt\ spectrum \cite{TOPQ-2016-01}. Using either
of these \ttbar\ samples to calculate \epsem\ increases the measured \xtt\
with a lepton $\pt>20$\,\GeV\ requirement by about 0.5\%, and greatly improves
the stability of the result against changes in the lepton \pt\ requirement.
Since this change is similar to the already assigned \ttbar\ modelling
uncertainties (in particular from the {\sc aMC@NLO} vs {\sc Powheg} comparison),
no additional uncertainty was included.
 
The inclusive cross-section result, together with previous measurements
at \sxwt\ and \sxvt\ \cite{TOPQ-2013-04}, is compared in Figure~\ref{f:xsvse}
with the NNLO+NNLL QCD prediction described in Section~\ref{s:intro}.
The measurement agrees with the predictions using the CT10, MSTW2008
and NNPDF2.3 PDF sets combined with the PDF4LHC prescription.
It is significantly more precise than this prediction, demonstrating the
power of the measurement to constrain the gluon PDF at high Bjorken-$x$.
The lower ratio panel compares the measurements to predictions using the
CT14 \cite{ct14} and NNPDF3.1\_notop \cite{nnpdf31} NNLO PDF sets,
two recent PDF sets which do not use any \ttbar\ data in their fits.
The NNPDF3.1\_notop PDF set does not include any variations of \alphas\
from the nominal value of 0.118, so the \alphas\ uncertainty
obtained from an \alphas\ variation of $\pm 0.0012$ with the standard
NNPDF3.1 PDF set was added in quadrature to the PDF-alone uncertainty
calculated with NNPDF3.1\_notop. The \sxyt\ measurement is also in
good agreement with the predictions from these PDF sets.
 
The result is also consistent with, and supersedes, the previous ATLAS
measurement using the same technique applied to 2015 data alone, which
had an uncertainty of 4.4\% \cite{TOPQ-2015-09}. The smaller uncertainty
of \ttxrel\ in the updated analysis results from improvements in the modelling
of \ttbar\ production (including tuning to \ttbar\ data at \sxyt), more
precise calibration of the integrated luminosity and of the LHC beam energy,
and better understanding of the lepton identification efficiencies and
energy scales, as well as the larger data sample.
The new result is also consistent with results from CMS in the dilepton
\cite{CMS-TOP-17-001} and lepton+jets \cite{CMS-TOP-16-006} final states,
but again has higher precision.
 
\begin{figure}
\centering
 
\includegraphics[width=150mm]{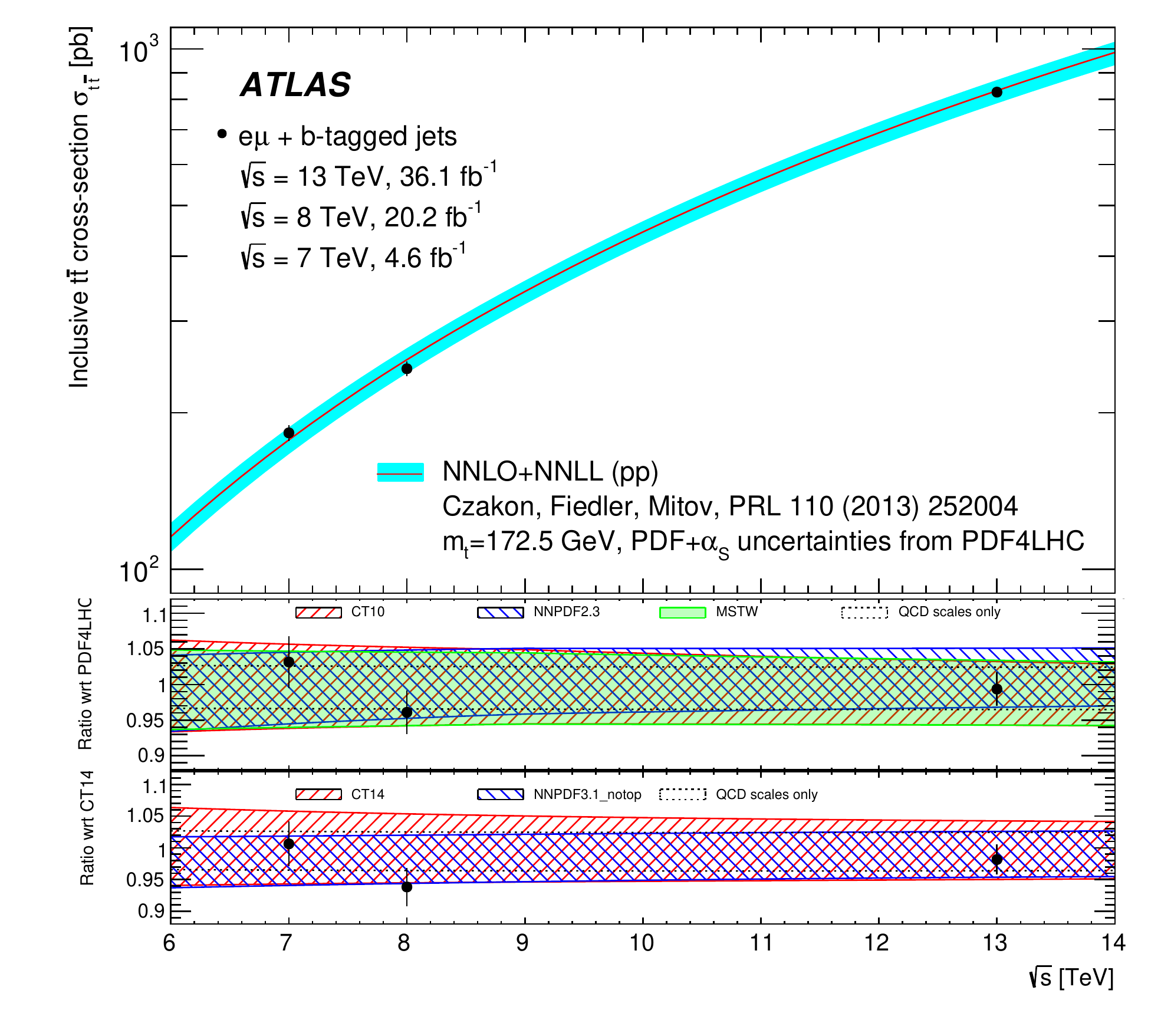}
\caption{\label{f:xsvse}The upper plot shows the inclusive \ttbar\
cross-section \xtt\ as a function
of centre-of-mass energy $\sqrt{s}$, comparing ATLAS results from the
$e\mu$ plus $b$-tagged jets final state at
$\sqrt{s}=7$, 8 and 13\,\TeV\ with NNLO+NNLL theoretical predictions
\cite{topxtheo5} calculated using {\tt Top++} \cite{toppp} using the PDF4LHC
prescription for PDF and \alphas\ uncertainties \cite{pdf4lhc},
and $\mtop=172.5$\,\GeV.
The middle plot shows the ratios of the measurements and predictions
to the central value of the prediction using PDF4LHC.
The total uncertainties when using the  individual NNPDF2.3, MSTW and CT10
PDFs are shown as overlapping hatched or coloured bands, and the dotted lines
show the QCD scale uncertainties alone.
The lower plot shows the ratios of the measurements and predictions from
the CT14 and NNPDF3.1\_notop PDFs to the central value from CT14.
The $\sqrt{s}=7$ and 8\,\TeV\ results are taken from
Ref. \cite{TOPQ-2013-04}, with the LHC beam energy uncertainties
reduced according to Ref. \cite{ebeam2}.}
\end{figure}

\subsection{Extraction of the top quark pole mass}\label{ss:mass}
 
The strong dependence of the inclusive \ttbar\ cross-section prediction
on the top quark pole mass \mtpole\ can be exploited to interpret precise
measurements of \xtt\ as measurements of \mtpole, as discussed in
Section~\ref{s:intro}. The ATLAS $\sqrt{s}=7$ and 8\,\TeV\ measurements
in the $e\mu$ channel were interpreted in this way, giving a combined value of
$\mtpole=172.9^{+2.5}_{-2.6}$\,\GeV\ \cite{TOPQ-2013-04}, and similar
measurements have been performed by CMS at $\sqrt{s}=7$, 8 and 13\,\TeV\
\cite{CMS-TOP-16-006,CMS-TOP-17-001,CMS-TOP-13-004},
as well as by D0 at the Tevatron $\bar{p}p$ collider \cite{d0pole}.
 
The NNLO+NNLL prediction for \xtt\ as a function of \mtpole\ at \sxyt\
was calculated using  {\tt Top++} \cite{toppp} and the
CT14 NNLO PDF set \cite{ct14} with uncertainties scaled to 68\% confidence
levels and $\alphas=0.1180\pm 0.0012$. CT14 was chosen as a recent
PDF set which does not use any \ttbar\ cross-section data as input.
The resulting dependence
was parameterised using the functional form proposed in Ref. \cite{topxtheo5}:
\begin{equation}\nonumber
\xtttheo(\mtpole)=\sigma(\mtref)\left(\frac{\mtref}{\mtpole}\right)^4
(1+a_1x+a_2x^2) \ .
\end{equation}
Here, $x=(\mtpole-\mtref)/\mtref$, the constant $\mtref=172.5$\,\GeV,
and $\sigma(\mtref)$, $a_1$ and $a_2$ are free parameters. The resulting
function is shown in Figure~\ref{f:mtpole}.
The measurement of \xtt\ given in Section~\ref{ss:incres} is also shown,
with its small dependence on \mtop\ due to variations of the experimental
acceptance and $Wt$ background discussed in Section~\ref{ss:sytt}. These
variations were studied using \ttbar\ and $Wt$ simulation samples with
several values of \mtop, and the corresponding dependencies of \epsem,
\nib\ and \niib\ on \mtop\ were parameterised with second-order polynomials.
The mass parameter used to characterise the dependence of the measured
\xtt\ on \mtop\ represents the top quark mass used in the Monte
Carlo event generators rather than \mtpole, but since the dependence of
the measured \xtt\ on \mtop\ is small, this approximation causes negligible
bias if \mtop\ and \mtpole\ differ by only a few \GeV. Under these
conditions, the intersection of the theoretical and experimental curves
shown in Figure~\ref{f:mtpole} gives an unambiguous extraction
of the top quark pole mass.
 
\begin{figure}
\centering
 
\includegraphics[width=120mm]{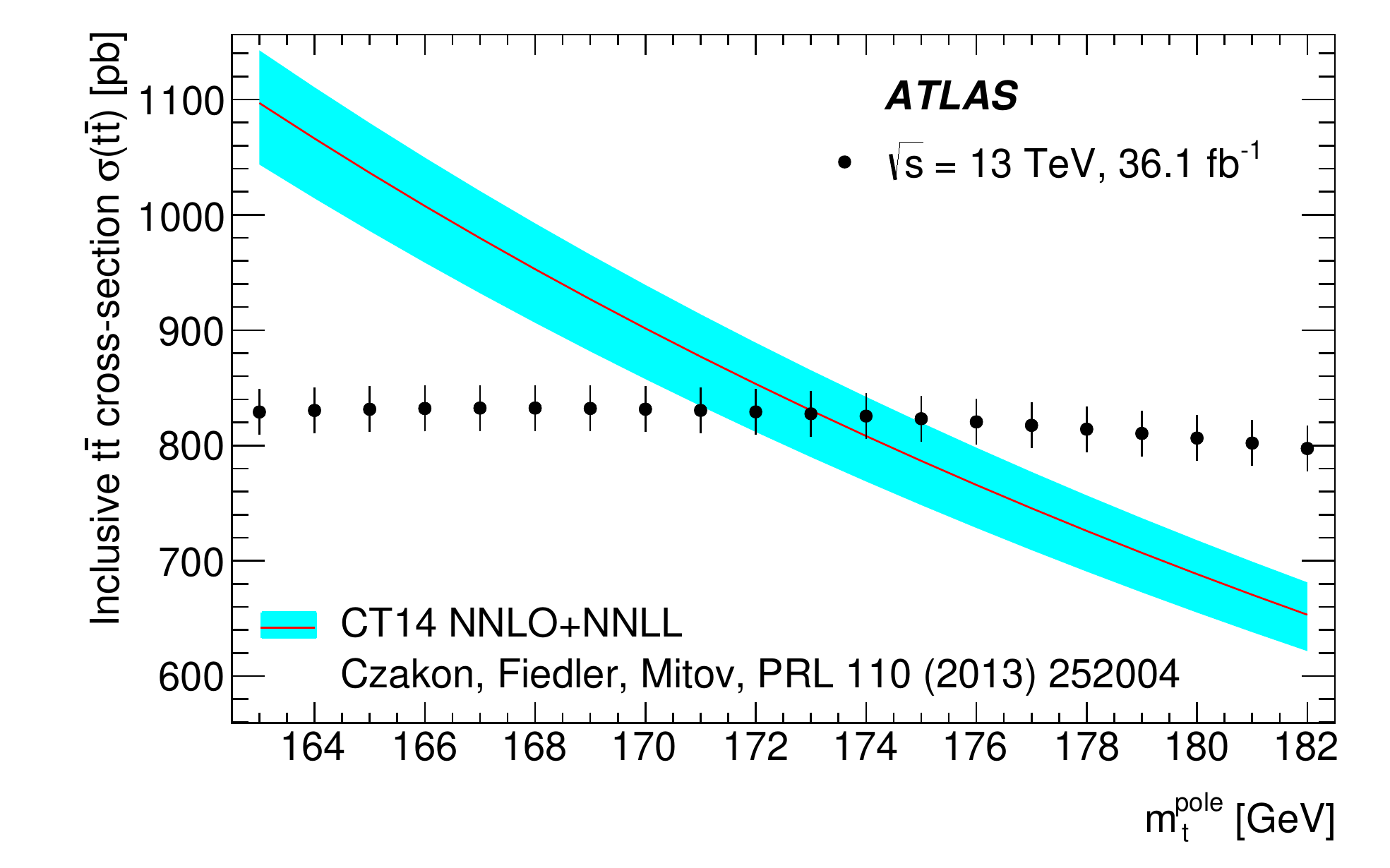}
\caption{\label{f:mtpole}Predicted inclusive \ttbar\ cross-section at \sxyt\
as a function of the top quark pole mass \mtpole, for the CT14 PDF set. The
cyan band indicates the total uncertainty in the prediction from PDF+\alphas\
and QCD scale uncertainties. The experimental measurement with its uncertainty
and dependence on the assumed value of \mtop\ through acceptance and background
corrections is shown by the black points with error bars.}
\end{figure}
 
The mass extraction was performed by maximising
the following Bayesian likelihood as a function of \mtpole:
\begin{equation}\label{e:mlike}
{\cal L}(\mtpole) = \int \gauss{\xttp}{\xtt(\mtpole)}{\xerrexp} \cdot
\gauss{\xttp}{\xtttheo(\mtpole)}{\xerrtheo\,}\ \mathrm{d}\xttp\ ,
\end{equation}
where $\gauss{x}{\mu}{s}$ represents a Gaussian probability density in the
variable $x$ with mean $\mu$ and standard deviation $s$. The first Gaussian
term in the integral represents the experimental measurement \xtt\ with its
dependence on \mtpole\ and uncertainty \xerrexp, and the second term
represents the theoretical prediction \xtttheo\ with its asymmetric
uncertainty \xerrtheo\ obtained from the quadrature sum of the combined
PDF plus \alphas\ uncertainty, and the QCD scale uncertainty,
each evaluated as described in Section~\ref{s:intro}.
The likelihood in Eq.~(\ref{e:mlike}) was maximised to obtain \mtpole\
when using the CT14 NNLO PDF set
to calculate \xtttheo, and also when using the NNPDF3.1\_notop NNLO PDF set,
with \alphas\ uncertainties inferred from NNPDF3.1 as discussed in
Section~\ref{ss:incres}. Results were also obtained using the individual
CT10, MSTW and NNPDF2.3 NNLO PDF sets to calculate \xtttheo, for
comparison with the $\sqrt{s}=7$ and 8\,\TeV\ results. The MMHT and
NNPDF3.0 PDF sets were not considered, as they include \ttbar\ cross-section
data in order to constrain the gluon PDF, and hence cannot also be used
to determine \mtpole\ without introducing a circular dependence
\cite{alphasxtt}.
In each case, the value of \xtt\ was recalculated using the corresponding NLO
PDF set to calculate the value of \epsem.
The results from each PDF set are shown in
Table~\ref{t:mtopres}, together with the result using the PDF4LHC
prescription to combine the CT10, MSTW and NNPDF2.3 results, keeping the
CT10 central value but enlarging the uncertainty to cover the envelope of the
positive and negative uncertainties of each individual PDF set.
The NNPDF3.1\_notop PDF set gives the smallest uncertainty
of $\pm 1.7$\,\GeV, demonstrating the power of recent improvements
in the gluon PDF determination to reduce the uncertainty on \mtpole. However,
given the approximate procedure used to evaluate the \alphas\ uncertainty
for this PDF set, the CT14 PDF set was chosen for the baseline result.
 
\begin{table}[tp]
\centering
 
\begin{tabular}{l|c}\hline
PDF set & \mtpole\ [GeV] \\
\hline
CT14 & $173.1^{+2.0}_{-2.1}$ \rule[-2mm]{0mm}{7mm}\\
\hline
NNPDF3.1\_notop & $172.9^{+1.7}_{-1.7}$ \rule[-2mm]{0mm}{7mm}\\
\hline
CT10 & $172.1^{+2.0}_{-2.0}$ \rule[-2mm]{0mm}{7mm}\\
MSTW & $172.3^{+2.0}_{-2.1}$ \rule[-2mm]{0mm}{7mm}\\
NNPDF2.3 & $173.4^{+1.9}_{-1.9}$ \rule[-2mm]{0mm}{7mm}\\
\hline
PDF4LHC & $172.1^{+3.1}_{-2.0}$ \rule[-2mm]{0mm}{7mm}\\
\end{tabular}
\caption{\label{t:mtopres}Top quark pole mass results for various NNLO
PDF sets, derived from the \ttbar\ cross-section measurement at \sxyt.
The uncertainties include PDF+\alphas, QCD scale and experimental sources.
The PDF4LHC result spans the uncertainties of the CT10, MSTW and NNPDF2.3
PDF sets.}
\end{table}
 
Table~\ref{t:mtopsyst} shows the breakdown of uncertainties in \mtpole\
calculated using the CT14 PDF set,
which are dominated by uncertainties in \xtttheo\ through
PDF+\alphas\ and QCD scale variations. Improving the experimental measurement
of \xtt\ further would therefore have little effect on the determination
of \mtpole\ via this method.
The result
is compatible with other measurements of \mtpole\ via lepton differential
distributions \cite{TOPQ-2015-02}, and via the reconstruction of top quark
differential distributions in inclusive \ttbar\ \cite{CMS-TOP-2018-004}
and \ttbar+jet \cite{TOPQ-2014-06,TOPQ-2017-09} events,
as well as previous measurements using the total \ttbar\ cross-section
\cite{TOPQ-2013-04,CMS-TOP-16-006,CMS-TOP-17-001,CMS-TOP-13-004,d0pole}.
It is also consistent with the Particle Data Group average of
$\mtpole=173.1\pm 0.9$\,\GeV\ \cite{pdg19} from a subset of these measurements.
The result using the CT14 PDF improves upon the previous ATLAS result
from $\sqrt{s}=7$ and 8\,\TeV\ data using the CT10, MSTW and NNPDF2.3 PDFs
combined with the PDF4LHC prescription \cite{TOPQ-2013-04}.
However, using the PDF4LHC prescription with the \sxyt\ data gives a larger
uncertainty
of $^{+3.1}_{-2.0}$\,\GeV, as the prediction of \xtt\ from NNPDF2.3 starts to diverge from that of CT10 and MSTW at higher $\sqrt{s}$ (see Figure~\ref{f:xsvse}),
leading to a larger spread in the \mtop\ values from the different PDF sets.
 
\begin{table}[tp]
\centering
 
\begin{tabular}{l|c}\hline
Uncertainty source & $\Delta\mtpole$ [GeV] \\
\hline
Data statistics & 0.2 \\
Analysis systematics & 0.6 \\
Integrated luminosity & 0.8 \\
Beam energy & 0.1 \\
\hline
PDF+$\alphas$ & $^{+1.5}_{-1.4}$ \rule[-2mm]{0mm}{7mm}\\
QCD scales & $^{+1.0}_{-1.5}$ \rule[-2mm]{0mm}{7mm}\\
\hline
Total uncertainty & $^{+2.0}_{-2.1}$ \rule[-2mm]{0mm}{7mm}\\
\end{tabular}
\caption{\label{t:mtopsyst}Uncertainties in the top quark pole mass extracted
from the \ttbar\ production cross-section measurement at \sxyt, using the
CT14 PDF set.}
\end{table}
 
\subsection{\ttbar\ and $\ttbar/Z$ cross-section ratios at different energies}\label{ss:ratio}
 
The ratios \rttyw\ and \rttyv\ were calculated using the \sxyt\ \xtt\
result discussed above and the $\sqrt{s}=7$ and 8\,\TeV\ results from
Refs. \cite{TOPQ-2013-04}, corrected to reduce the LHC beam energy
uncertainty to 0.1\% \cite{ebeam2}. The \xtt\
values and uncertainties are summarised in Table~\ref{t:xsecratinput};
the largest systematic uncertainties come in all cases from \ttbar\ modelling
and the knowledge of the integrated luminosity.
As the nominal \ttbar\ simulation sample used at
$\sqrt{s}=7$ and 8\,\TeV\ was {\sc Powheg\,+\,Pythia6} with the CT10 PDFs,
the \sxyt\ result was rederived using a similar \ttbar\ sample to calculate
\epsem\ and \cb, increasing the 13\,\TeV\ \xtt\ value by 0.46\%.
PDF uncertainties were evaluated for each of the error sets or replicas of the
CT10, MSTW and NNPDF2.3 PDF sets, considering the effect of each individual
variation to
be correlated between the numerator and denominator of the \xtt\ ratio.
Significant cancellations occur, leading to PDF uncertainties of about 0.5\%
in each ratio, significantly smaller than the 1\% uncertainties for the
$\sqrt{s}=7$ and 8\,\TeV\ \xtt\ measurements. The parton-shower radiation
uncertainties were similarly evaluated using {\sc Powheg\,+\,Pythia6} samples
with more and less parton-shower radiation in all datasets, giving
residual uncertainties of around 0.4\% in the ratios. Other \ttbar\
modelling uncertainties due to the choice of NLO generator and hadronisation
model were conservatively taken to be uncorrelated, due to the different
alternative generators used in the measurements. The uncertainties
due to the $Wt$ background cross-section and $\ttbar/Wt$ interference were
assessed in the same way at all $\sqrt{s}$ values and considered correlated.
Lepton, jet and $b$-tagging uncertainties were mainly considered uncorrelated,
due to the changes
in detector configuration and lepton identification algorithms between
measurements. The integrated luminosity measurements were based on different
primary  detectors at 7--8\,\TeV\ and 13\,\TeV, and the luminosity scale was
calibrated using individual beam-separation scans in each dataset
\cite{DAPR-2011-01,DAPR-2013-01,ATLAS-CONF-2019-021} with only a fraction
of the uncertainties being correlated. The total luminosity
uncertainties were therefore conservatively taken to be uncorrelated in the
\xtt\ ratio measurements. The beam energy uncertainties are correlated
between $\sqrt{s}$ values, but the varying dependence of \xtt\ on
$\sqrt{s}$ (see Figure~\ref{f:xsvse}) leads to a small ($<0.1$\%) residual
uncertainty on the ratios.
 
\begin{table}[tp]
\centering
 
\begin{tabular}{c|c|cc}\hline
$\sqrt{s}$ [\TeV] & \xtt [pb] & \xzfid{e} [pb] & \xzfid{\mu} [pb] \\
\hline
7 & $182.9\pm 3.1 \pm 4.2 \pm 3.6$  & $ 451.2\pm  0.5\pm  1.7 \pm  8.1$ & $ 450.0 \pm  0.4\pm  2.0 \pm  8.1$ \\
8 & $242.9 \pm 1.7 \pm 5.5 \pm 5.1$ & $ 507.0\pm  0.2\pm  4.3 \pm  9.6$ & $ 504.7\pm  0.2\pm  3.6 \pm  9.6$ \\
13 & $824.7\pm 6.9 \pm 12.1 \pm 18.4^{(a)}$ & $ 778.3\pm  0.7\pm  4.1 \pm 15.9$ & $ 774.4\pm  0.7\pm  6.3 \pm 15.8$ \\
& $830.2\pm 3.6 \pm 11.7 \pm 15.7^{(b)}$ & & \\
\hline
\end{tabular}
\caption{\label{t:xsecratinput}Input inclusive \ttbar\ and fiducial $Z\rightarrow ee/\mu\mu$ cross-sections used in the calculations of the \ttbar\ and
$\ttbar/Z$ cross-section ratios and double ratios shown in
Tables~\ref{t:xsecratres} and~\ref{t:ttzratio}.  The three uncertainties in
each cross-section are due to data statistics, experimental and theoretical
systematic effects (including the LHC beam energy uncertainties) and knowledge
of the integrated luminosities of the data samples. For \sxyt, the \ttbar\
cross-section labelled $(a)$ uses 2015 data only and is used for the
$\ttbar/Z$ ratio \rttzy\ and the double ratios \rttzyw\ and \rttzyv, whilst
the cross-section labelled $(b)$ uses the combination of 2015 and 2016 data,
and is used for the \ttbar\ cross-section ratios \rttyw\ and \rttyv. Both
\ttbar\ cross-sections have been calculated using a {\sc Powheg\,+\,Pythia6}
sample to derive the efficiencies (see text).}
\end{table}

The resulting cross-section ratios are shown in Table~\ref{t:xsecratres},
together with the NNLO+NNLL predictions calculated using {\tt Top++} as
described in Section~\ref{s:intro}, with the uncertainties
from the CT10, MSTW and NNPDF2.3 PDFs combined according to the PDF4LHC
prescription. The total uncertainties in the measurements are \rttywerr\
for \rttyw\ and \rttyverr\ for \rttyv, improving on the
uncertainties of 4.9\% and 4.7\% obtained using the 2015 \sxyt\ dataset
alone in Ref. \cite{STDM-2016-02}.
Figure~\ref{f:xratio} compares the measurements with the predictions using
the CT10, MSTW and NNPDF2.3 PDF sets, as well as the more
recent CT14 \cite{ct14}, ABM12LHC \cite{abm12pdf}, ABMP16 \cite{abmp16pdf},
ATLAS-epWZ12 \cite{STDM-2011-43}, HERAPDF2.0 \cite{herapdf20},
MMHT14 \cite{mmht} and NNPDF3.0 \cite{nnpdf3}  PDF sets,
some of which include some LHC data (including \ttbar\ cross-section
measurements in the cases of ABM12LHC, ABMP16, MMHT and NNPDF3.0). The
ratio \rttyw\ is lower than all the predictions,  and the ratio
\rttyv\ higher than all the predictions except ABM12LHC. However, both
ratios are compatible with all the predictions except ABM12LHC
within two standard deviations. Some of these results
are also reflected in Figure~\ref{f:xsvse}.
The behaviour of ABM12LHC is attributed to the lower gluon density at high
Bjorken-$x$ compared to the other considered PDF sets, which leads
to a larger relative increase in the \ttbar\ cross-section as a function of
$\sqrt{s}$. This behaviour is less apparent in the more recent
ABMP16 PDF set, which includes more precise constraints from
LHC top quark measurements. The current experimental
uncertainties, dominated by the luminosity uncertainties which do not cancel
in the ratios, do not allow the predictions using the other PDFs to be
distinguished.
 
\begin{table}[tp]
\centering
 
\begin{tabular}{c|cc}\hline
$\sqrt{s}$ values [\TeV] & Measured cross-section ratio & NNLO+NNLL prediction \\
\hline
13/7 & $ 4.54\pm  0.08 \pm  0.10 \pm  0.12$ (0.18) & $ 4.69 \pm  0.16$ \\
13/8 & $ 3.42\pm  0.03 \pm  0.07 \pm  0.10$ (0.12) & $ 3.28 \pm  0.08$ \\
8/7  & $ 1.33\pm 0.02 \pm 0.02 \pm 0.04$ (0.05) & $1.43\pm 0.01$ \\
\hline
\end{tabular}
\caption{\label{t:xsecratres}Ratios of inclusive \ttbar\ production
cross-sections measured at $\sqrt{s}=13$, 7 and 8\,\TeV,
together with the corresponding NNLO+NNLL predictions using
{\tt Top++} \cite{toppp} with the PDF4LHC prescription for PDF and
\alphas\ uncertainties \cite{pdf4lhc}.
The three uncertainties in the measured ratios are due to data statistics,
experimental and theoretical
systematic effects (including the LHC beam energy uncertainties)
and knowledge of the integrated luminosities of the data samples. The ratio of
$\sqrt{s}=8$ and 7\,\TeV\ results is taken from Ref. \cite{TOPQ-2013-04}.
The total uncertainty is given in parentheses after each result.}
\end{table}
 
\begin{figure}[tp]
\parbox{83mm}{\includegraphics[width=76mm]{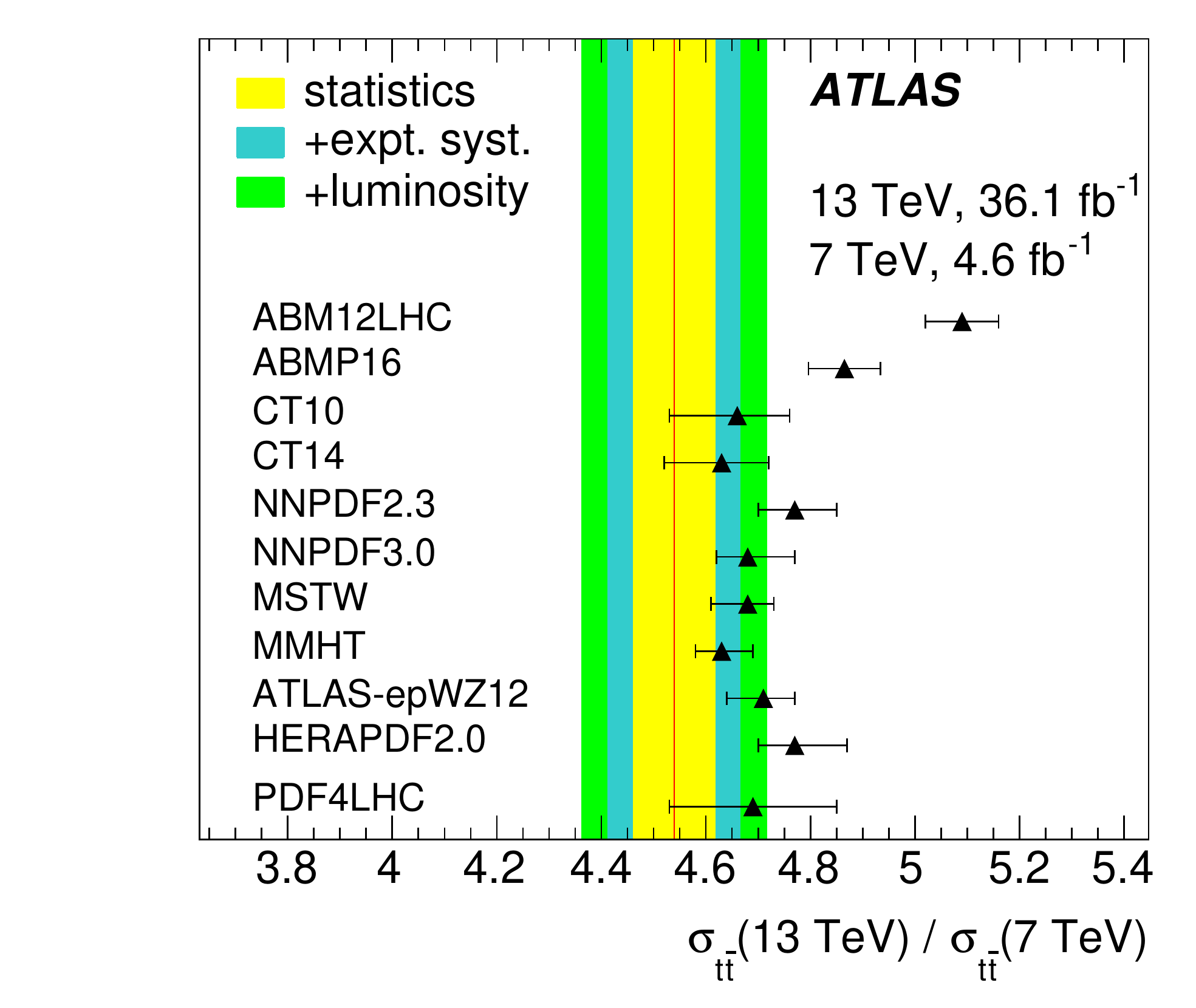}\vspace{-7mm}\center{(a)}}
\parbox{83mm}{\includegraphics[width=76mm]{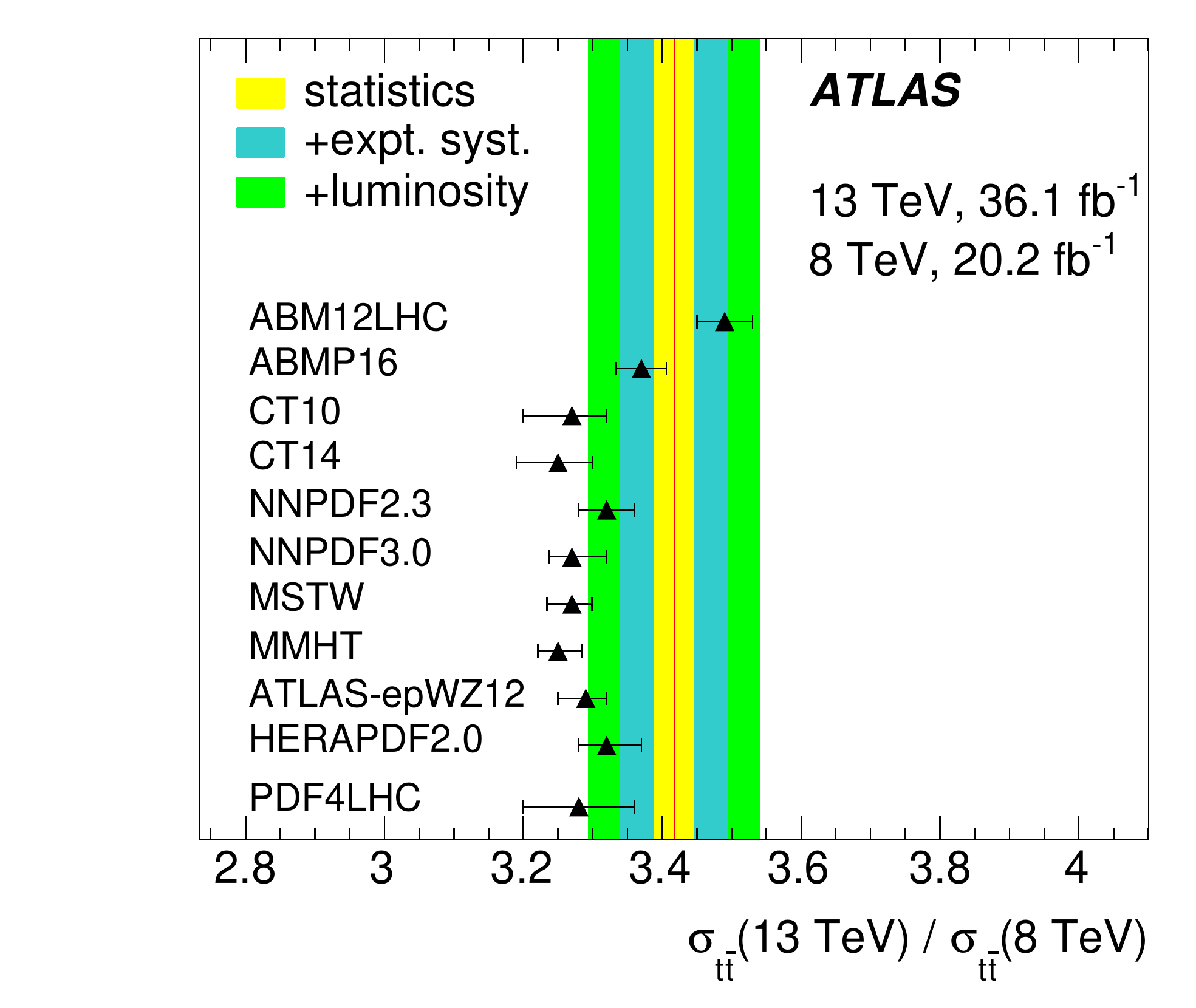}\vspace{-7mm}\center{(b)}}
\caption{\label{f:xratio}Ratios of \ttbar\ production cross-sections
at different energies: (a) \rttyw, (b) \rttyv.
The bands show the experimental measurements with
the statistical (inner yellow bands), statistical plus experimental and
theoretical systematic (middle cyan bands)
and total including luminosity (outer green bands) uncertainties. The black
triangles with error bars show the predictions and uncertainties from
various PDF sets. The last entry shows the prediction using the
PDF4LHC recipe, encompassing the predictions from the CT10, MSTW
and NNPDF2.3 PDF sets.}
\end{figure}
 
As discussed in Ref. \cite{STDM-2016-02}, double ratios of
\ttbar\ to $Z$ cross-sections at different energies can be used to reduce
the luminosity uncertainty, potentially enhancing the sensitivity to PDF
differences. The \ttbar\ cross-section at a given energy can be normalised to
the corresponding $Z\rightarrow\ell\ell$ fiducial cross-section \xzfid{\ell}
at the same energy by defining the ratio \rttz\ as:
\begin{equation}\label{e:rttz}
\rttz=\frac{\xtt}{0.5 \left( \xzfid{e} + \xzfid{\mu} \right)} \ ,
\end{equation}
where the use of the unweighted average of $Z\rightarrow ee$ and
$Z\rightarrow\mu\mu$ cross-sections maximises the potential cancellation
of electron- and muon-related systematic uncertainties when the \ttbar\
cross-section is measured using events with one electron and one muon.
Provided that the \ttbar\ and $Z$ cross-sections are measured using the same
data sample,
the integrated luminosity uncertainty cancels almost completely in the
ratio \rttz. Double ratios \rttzab{i}{j}\ of \rttz\ at two different energies
$i$ and $j$ can then be defined:
\begin{equation}\nonumber
\rttzab{i}{j} = \frac{\rttz (i)}{\rttz (j)}\ ,
\end{equation}
which benefit from cancellations of uncertainties between beam energies and
production processes.
In Ref. \cite{STDM-2016-02}, the previous measurement of \xtt\ at \sxyt\ from
Ref. \cite{TOPQ-2015-09} was used together with the $\sqrt{s}=7$ and 8\,\TeV\
\xtt\ measurements from Refs. \cite{TOPQ-2013-04} and corresponding
measurements of \xzfid{\ell} at each energy to derive double ratios
\rttzyw, \rttzyv\ and \rttzvw, which were compared  with the predictions from
various PDF sets. The $Z\rightarrow\ell\ell$
cross-sections were measured in a fiducial volume corresponding to the lepton
acceptance ($\pt>25$\,\GeV\ and $|\eta|<2.5$) with the dilepton invariant mass
in the range $66<m_{\ell\ell}<116$\,\GeV.
 
The precision of the $\ttbar/Z$ ratio and double ratios involving \sxyt\ data
were limited by the 4.4\% uncertainty in the corresponding \xtt\ measurement. These ratios have therefore
been updated using the more precise \sxyt\ \xtt\ measurement shown
in Table~\ref{t:incres}. The result from 2015 data alone (with an uncertainty
of 2.8\%) was used in order to maintain
the cancellation of luminosity uncertainties in Eq.~(\ref{e:rttz}), as the
corresponding $\xzfid{\ell}$ measurements only used the 2015 data sample.
Since the \xtt\ result was
derived using reoptimised lepton identification and updated calibrations,
the lepton
uncertainties were conservatively treated as uncorrelated between the
\sxyt\ \ttbar\ and $Z$ measurements. The largest uncertainties in
the double ratio are associated with \ttbar\ modelling, and these were treated
in the same way as for the updated \ttbar\ cross-section ratios discussed above,
including the 0.46\% increase of the \sxyt\ \xtt\ value corresponding to the
use of a {\sc Powheg\,+\,Pythia6} CT10 nominal \ttbar\ simulation sample.
All other uncertainties were treated according to the correlation model
described in Ref. \cite{STDM-2016-02}, with the LHC beam energy uncertainties
updated according to Ref. \cite{ebeam2}. The input cross-sections are
summarised in Table~\ref{t:xsecratinput}.
 
The resulting single and double ratios are shown in Table~\ref{t:ttzratio},
together with the predictions using the CT14 PDF set, calculated as described
in Ref. \cite{STDM-2016-02}.
The total uncertainties are \rttzyerr\ for \rttzy,
\rttzywerr\ for \rttzyw\ and \rttzyverr\ for \rttzyv, which are
significant improvements on the corresponding uncertainties of 3.5\%,
3.8\% and 3.6\% in Ref. \cite{STDM-2016-02}.
The largest uncertainties come from the
\xtt\ measurements, in particular the \ttbar\ modelling uncertainties, which
are mainly treated as uncorrelated between beam energies. Excluding PDF
uncertainties, where the correlations between beam energies are fully
accounted for, \ttbar\ modelling uncertainties contribute
1.7\% and 1.4\% to the uncertainties on the ratios \rttzyw\ and \rttzyv.
In principle, these uncertainties could be reduced by using a fully
coherent set of \ttbar\ simulation samples and uncertainty model at all
beam energies, but that has not been attempted here.
 
The results are compared with the predictions of the ABM12LHC, CT14,
NNPDF3.0, MMHT, ATLAS-epWZ12  and HERAPDF2.0 PDF sets
(the same sets as in Ref. \cite{STDM-2016-02}) in Figure~\ref{f:ttzratio}.
The measurement of the $\ttbar/Z$ cross-section ratio at \sxyt\ is
compatible with all the predictions within two standard deviations.
Although the experimental uncertainty is only \rttzyerr, the
predictions have common uncertainties of $^{+4.0}_{-4.6}$\% from
QCD scale and top quark mass variations, limiting the sensitivity to
PDF variations. The pattern for the double ratios
is similar to that seen for the \ttbar-only ratios in
Figure~\ref{f:xratio}; the normalisation to $Z\rightarrow\ell\ell$
cross-sections serves mainly to reduce the luminosity-related uncertainties.
The double ratio \rttzyw\ lies below all the predictions, and
\rttzyv\ lies above all the predictions except that of ABM12LHC.
However, the measurements are consistent with all the predictions within
two standard deviations, with the exception of ABM12LHC for \rttzyw. Similar
trends were seen in Ref. \cite{STDM-2016-02}, although with less separation
between PDFs due to the larger uncertainties in the double ratios.
 
\begin{table}
\centering
 
\begin{tabular}{c|cc}\hline
$\sqrt{s}$ value [\TeV] & $\ttbar/Z$ cross-section ratio & CT14 prediction \\
\hline
13 & $ 1.062\pm  0.009 \pm  0.016 \pm  0.002$ (0.018) & $1.132^{+0.078}_{-0.075}$
\rule[-2mm]{0mm}{6mm} \\
\hline\hline
$\sqrt{s}$ values [\TeV] & $\ttbar/Z$ cross-section double ratio & \\ 
\hline
13/7 & $ 2.617\pm  0.049 \pm  0.060 \pm  0.007$ (0.078) & $2.691^{+0.045}_{-0.058}$ \rule[-2mm]{0mm}{6mm} \\
13/8 & $ 2.212\pm  0.024 \pm  0.049 \pm  0.006$ (0.055) & $2.124^{+0.026}_{-0.035}$ \rule[-2mm]{0mm}{6mm} \\
\hline
\end{tabular}
\caption{\label{t:ttzratio}Measurements of the ratio of $\ttbar/Z$
cross-sections at \sxyt, and double ratios of $\ttbar/Z$
cross-sections at \sxyt\ and \sxwt\ or \sxvt, compared with predictions
using the CT14 PDF set. The three uncertainties in the measurements are due
to data statistics, experimental and theoretical systematic effects
(including the small uncertainty
due to the LHC beam energy uncertainties) and knowledge of the integrated
luminosities of the data samples. The total uncertainty is given in parentheses
after each result.}
\end{table}
 
\begin{figure}[tp]
\hspace{42mm}\parbox{83mm}{\includegraphics[width=76mm]{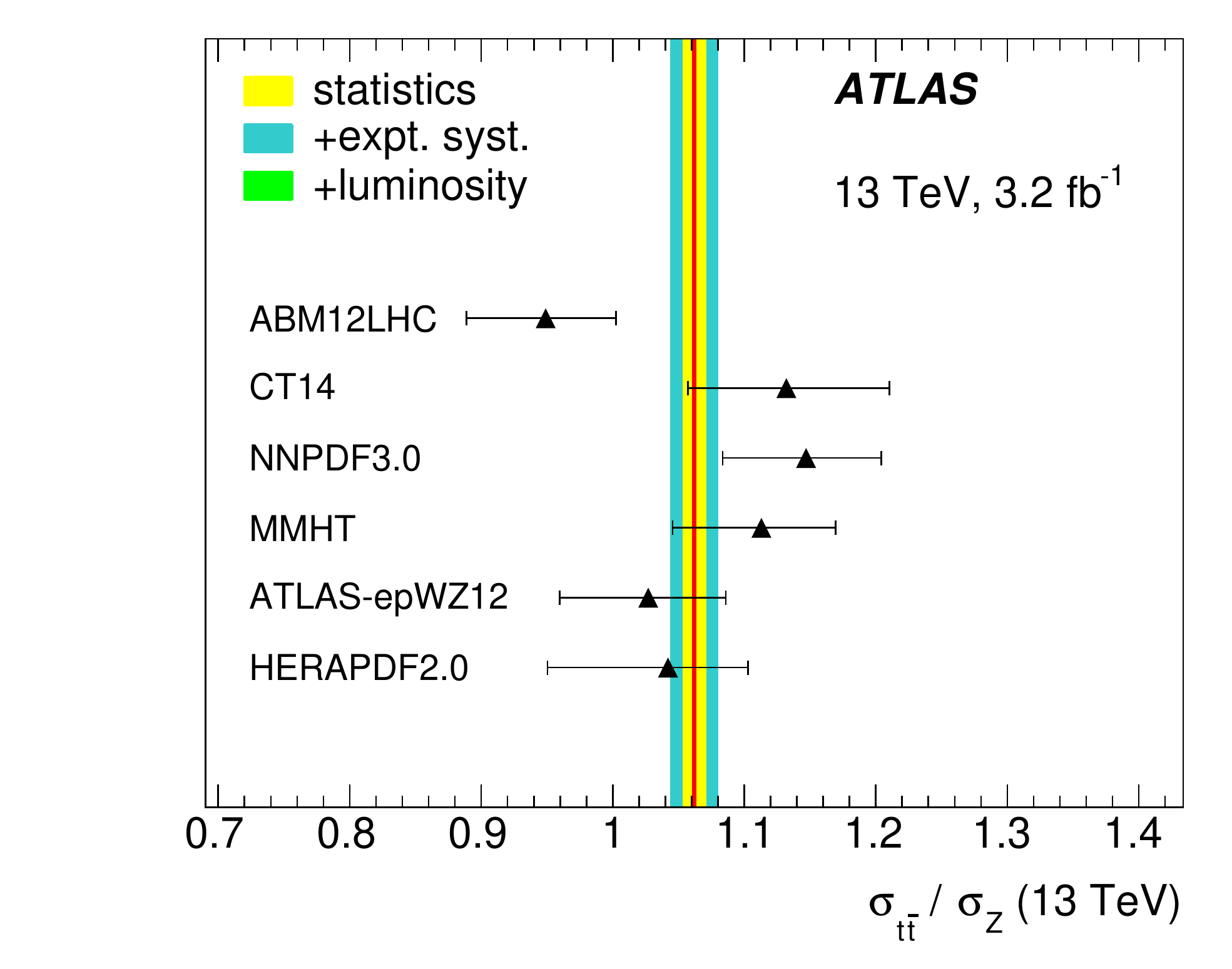}\vspace{-7mm}\center{(a)}}
 
\parbox{83mm}{\includegraphics[width=76mm]{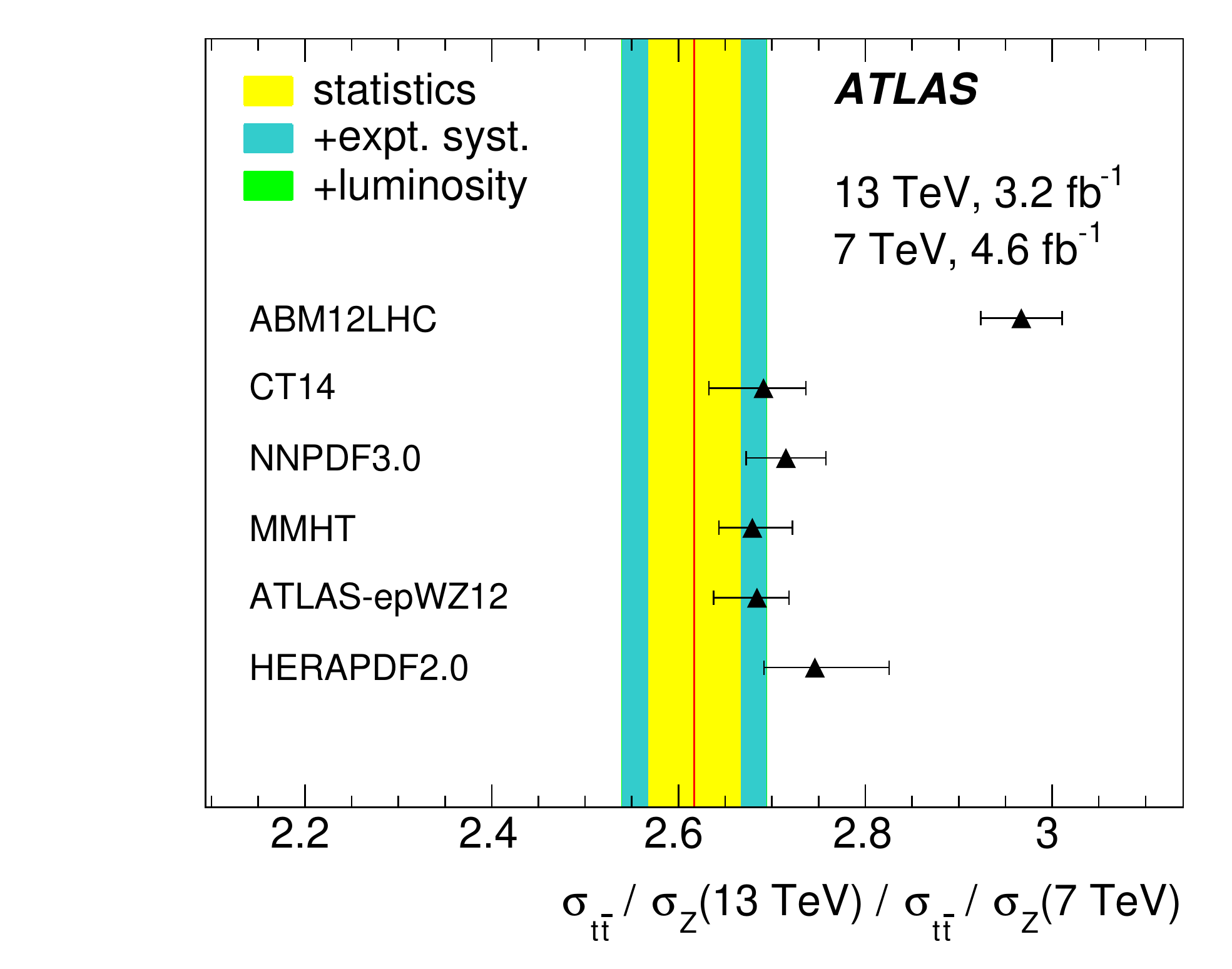}\vspace{-7mm}\center{(b)}}
\parbox{83mm}{\includegraphics[width=76mm]{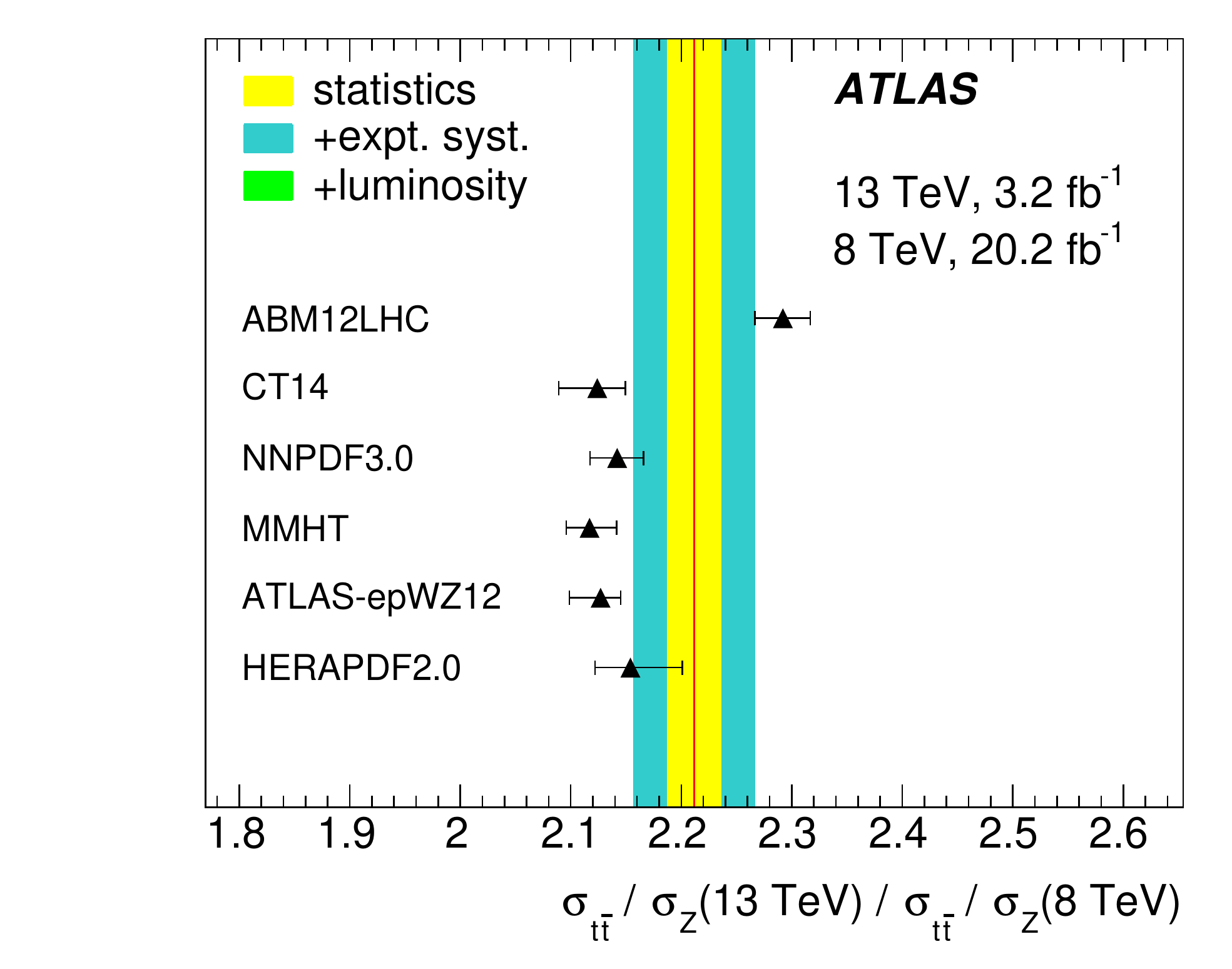}\vspace{-7mm}\center{(c)}}
\caption{\label{f:ttzratio}(a) Ratio of \ttbar\ to $Z$ production
cross-sections at \sxyt, and double ratios of \ttbar\ to $Z$ production
cross-sections at different energies: (b) \rttzyw, (c) \rttzyv.
The bands show the experimental measurements with
the statistical (inner yellow bands), statistical plus experimental and
theoretical systematic (middle cyan bands) and total including luminosity
(barely visible outer green bands)
uncertainties. The black triangles with error bars show the predictions and
uncertainties from various PDF sets. The \sxyt\ results use only the
3.2\,\ifb\ data sample recorded in 2015.}
\end{figure}

\section{Differential cross-section results}\label{s:diffres}
 
The single-lepton and dilepton absolute fiducial differential cross-section
results were obtained by solving Eqs.~(\ref{e:fidtags}) for each bin $i$ of
each distribution, using the combined 2015--16 data sample. The normalised
differential cross-sections were obtained from the absolute results using
Eqs.~(\ref{e:normx}) and~(\ref{e:binw}).
As in the inclusive cross-section analysis, the results
were found to be stable when varying the jet \pt, $|\eta|$ and $b$-tagging
requirements. The single-lepton \ptl\ and \etal\ distributions were also
measured for electrons and muons separately, instead of combining them into
lepton distributions
with two entries per event, and found to be compatible. The distributions
of bin-by-bin differences in the electron and muon differential cross-sections
have $\chi^2$ per degree of freedom of $7/10$ for \ptl\ and 13/8 for \etal,
in both cases taking statistical and uncorrelated systematic uncertainties into account.
 
\subsection{Results for measured distributions}\label{ss:diffvals}
 
The measured absolute and normalised fiducial differential cross-sections
are shown in
Table~\ref{t:insXSec1} (\ptl\ and \etal),
Table~\ref{t:insXSec2} (\ptll\ and \mll),
Table~\ref{t:insXSec3} (\rapll\ and \dphill) and
Table~\ref{t:insXSec4} (\ptsum\ and \esum)
in the Appendix. The double-differential cross-sections are shown in
Tables~\ref{t:insXSec5}--\ref{t:insXSec6} (\etalvmll),
Tables~\ref{t:insXSec7}--\ref{t:insXSec8} (\rapllvmll) and
Tables~\ref{t:insXSec9}--\ref{t:insXSec10} (\dphillvmll).
These tables show the measured cross-section values and uncertainties,
together with a breakdown of the total uncertainties into components
corresponding to data statistics (`Stat'), \ttbar\ modelling (`\ttbar\ mod.'),
lepton identification and measurement (`Lept.'), jet and $b$-tagging
uncertainties (`Jet/$b$'), background uncertainties (`Bkg.') and
luminosity/beam energy uncertainties (`$L/E_\mathrm{b}$'), matching
the categories described in Sections~\ref{ss:sytt}--\ref{ss:sylumieb}.
The rightmost columns show the cross-sections corrected using
Eq.~(\ref{e:notau}) to remove the contributions where at least one lepton
results  from a leptonic decay of a $\tau$-lepton.
As also visible in Figure~\ref{f:fracsyst}, the total uncertainties
in the normalised differential cross-sections range from 0.6\% to
around 10\%, and are typically around half those for the corresponding
distributions measured at \sxvt\ \cite{TOPQ-2015-02}.
The largest uncertainties are generally statistical, but background
uncertainties (in particular from $\ttbar/Wt$ interference) become
dominant at the high ends of the \ptl, \ptll, \mll, \ptsum\ and \esum\
distributions, and \ttbar\ modelling uncertainties from the comparison
of {\sc aMC@NLO\,+\,Pythia8} and {\sc Powheg\,+\,Pythia8}  are dominant for
most of the \dphill\ distribution. Uncertainties related to leptons and jets
generally play only a minor role; in particular those due to jet energy
measurement and $b$-tagging are suppressed due to the determination of
\epsb\ from data in Eqs.~(\ref{e:fidtags}).
The systematic uncertainties in the normalised differential cross-sections
benefit from significant cancellations between bins, and the uncertainties
in the absolute cross-sections are substantially larger. The latter also suffer
from the uncertainties in the integrated luminosity and beam energy,
which contribute 2.3--2.8\%, depending on the background level in each bin.
 
The measured normalised differential cross-sections are shown graphically
in Figures~\ref{f:distresa}--\ref{f:distresd}. The different \mll\
bins for the double-differential cross-sections are shown sequentially
on the $x$ axes, separated by vertical dotted lines. The measured
cross-sections are compared with the particle-level predictions from
the baseline {\sc Powheg\,+\,Pythia8} \ttbar\ sample, {\sc Powheg\,+\,Pythia8}
samples with more or less parton-shower radiation,
and {\sc aMC@NLO\,+\,Pythia8}.
The trends seen are similar to those visible in the reconstructed
distributions shown in Figures~\ref{f:dmcjlept} and~\ref{f:dmcdilept}, and
are discussed in the context of comparisons with a larger set of samples
in the following.

\begin{figure}[tp]
\parbox{83mm}{\includegraphics[width=76mm]{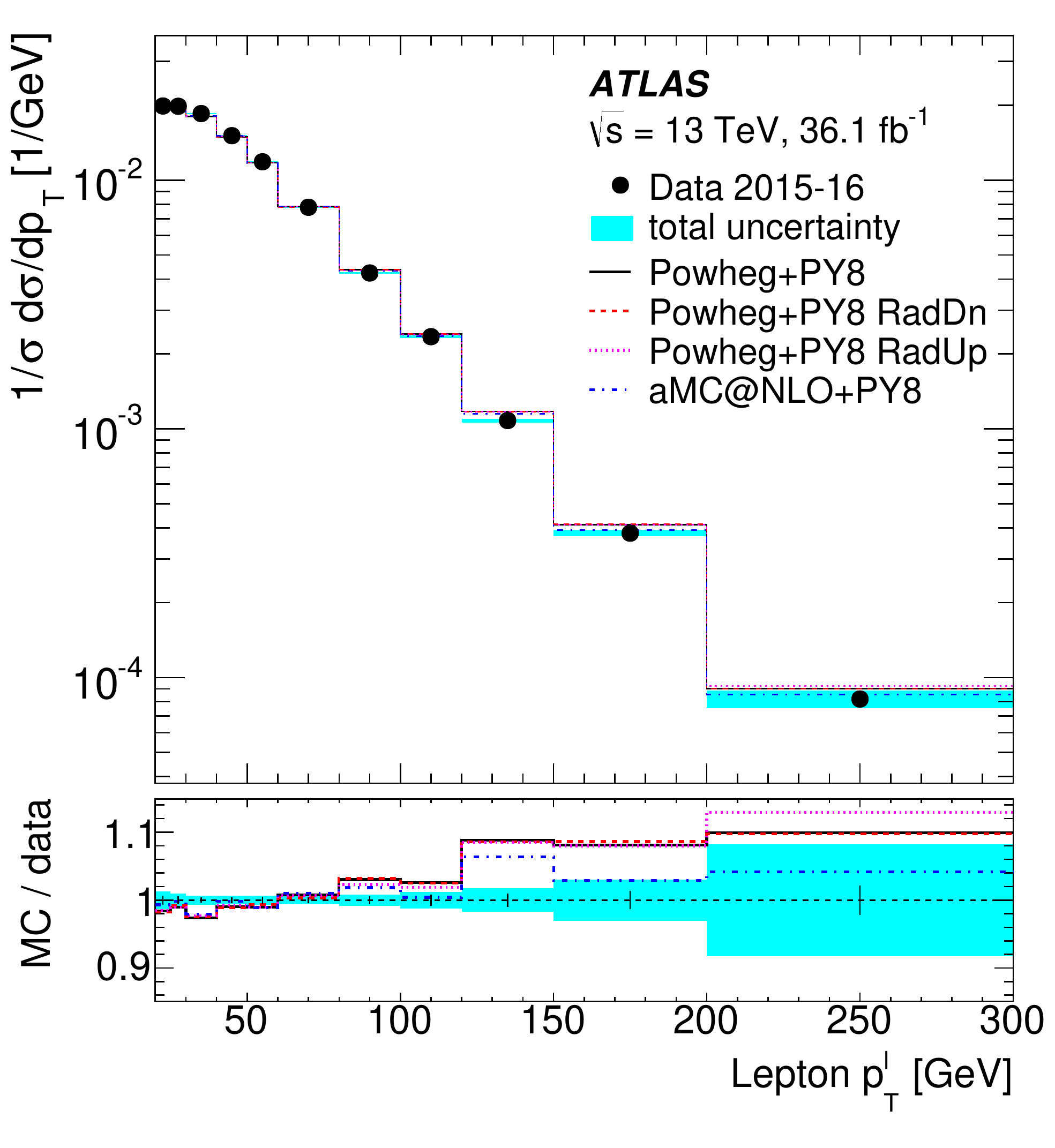}\vspace{-7mm}\center{(a)}}
\parbox{83mm}{\includegraphics[width=76mm]{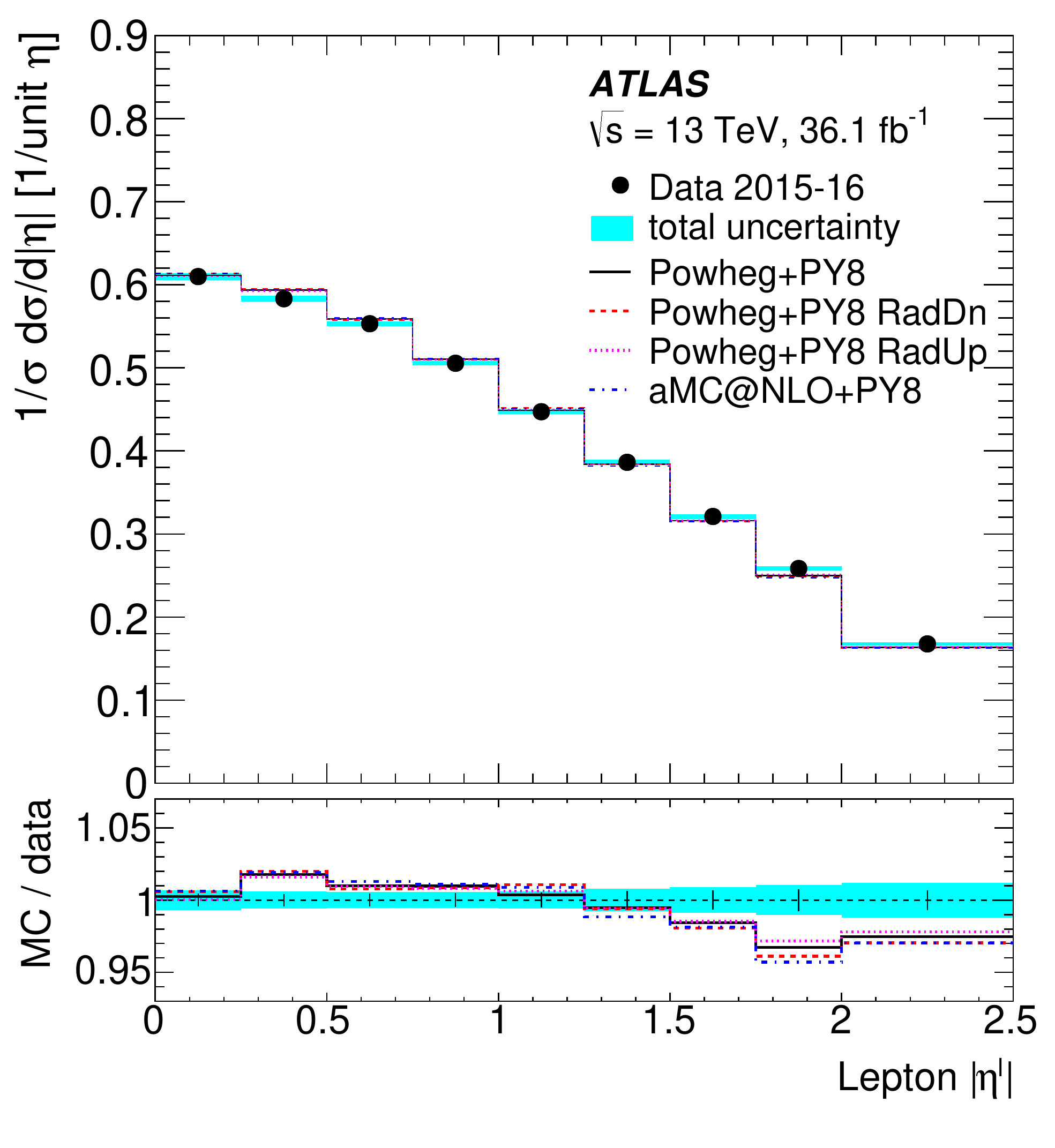}\vspace{-7mm}\center{(b)}}
\parbox{83mm}{\includegraphics[width=76mm]{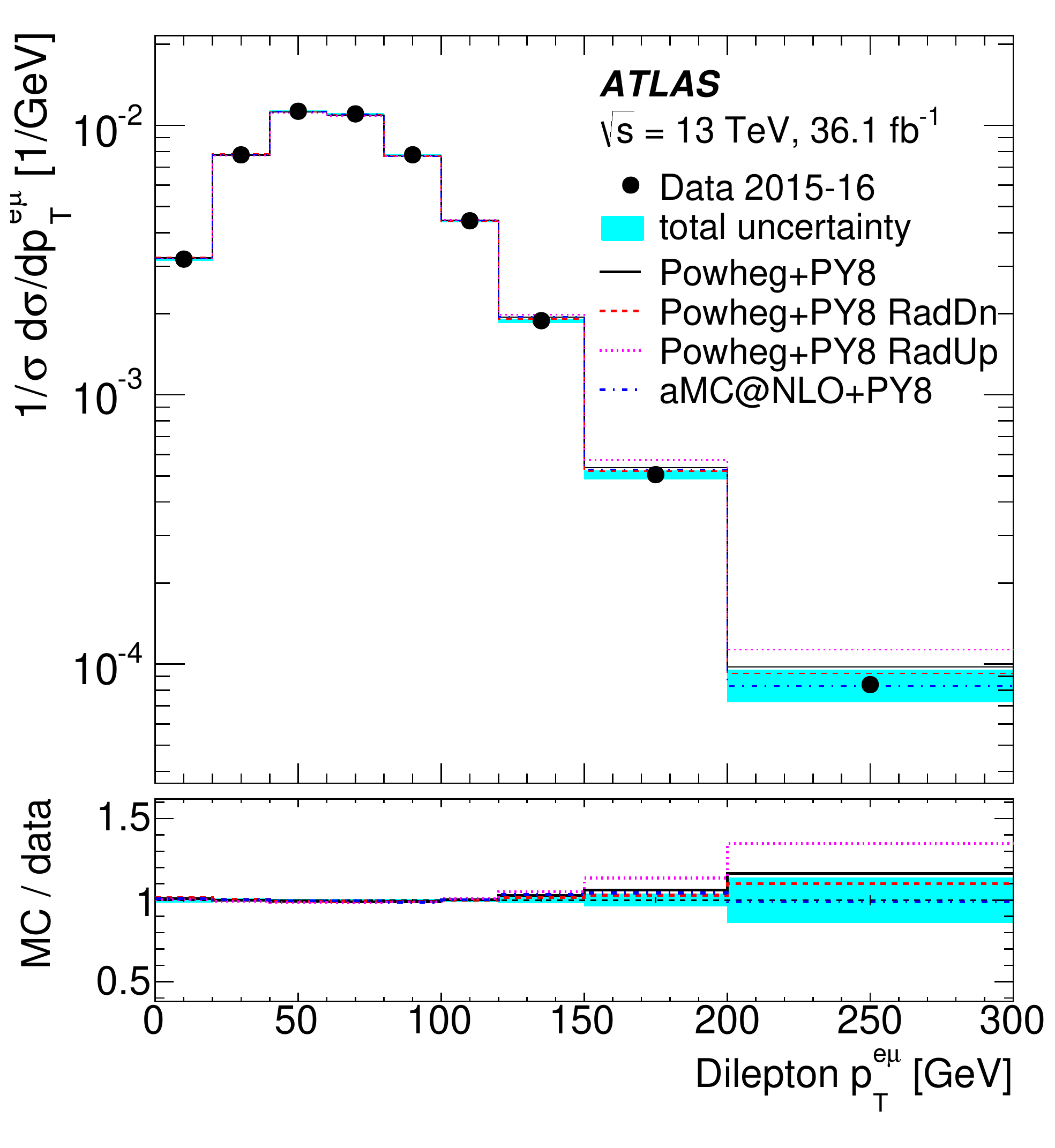}\vspace{-7mm}\center{(c)}}
\parbox{83mm}{\includegraphics[width=76mm]{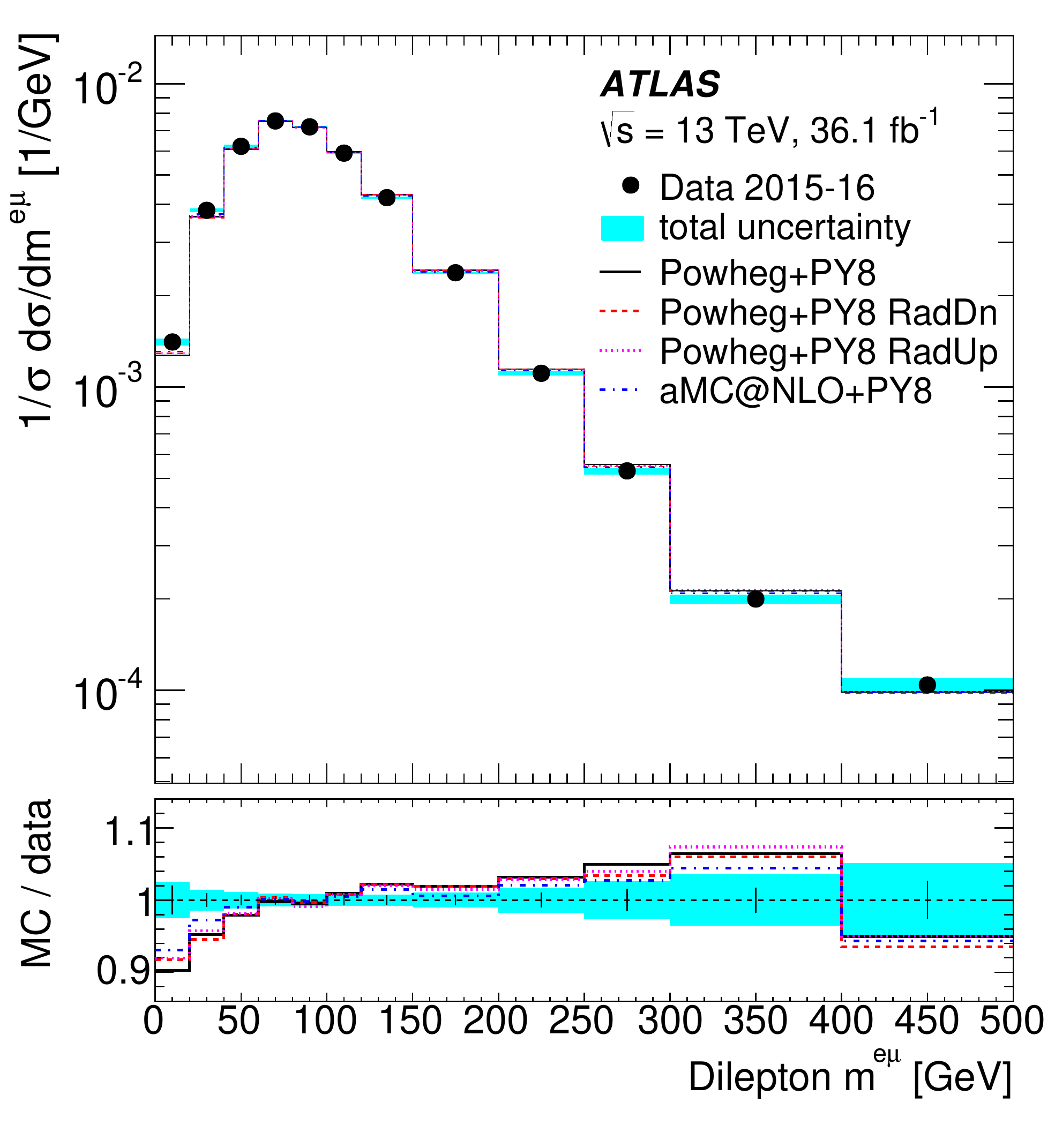}\vspace{-7mm}\center{(d)}}
\caption{\label{f:distresa}Normalised differential cross-sections as a function
of (a) \ptl, (b) \etal, (c) \ptll\ and (d) \mll. The measured values are
shown by the black points with error bars corresponding to the data statistical
uncertainties and cyan bands corresponding to the total uncertainties in each
bin, and include the contributions via $W\rightarrow\tau\rightarrow e/\mu$
decays. The data points are placed at the centre of each bin.
The results are compared with the predictions from the baseline
{\sc Powheg\,+\,Pythia8} \ttbar\ sample, {\sc Powheg\,+\,Pythia8} samples
with more or less parton-shower radiation (RadUp and RadDn), and an
{\sc aMC@NLO\,+\,Pythia8} sample. The lower plots show the ratios of
predictions to data, with the error bars indicating the data statistical
uncertainties and the
cyan bands indicating the total uncertainties in the measurements.}
\end{figure}

\begin{figure}[tp]
\parbox{83mm}{\includegraphics[width=76mm]{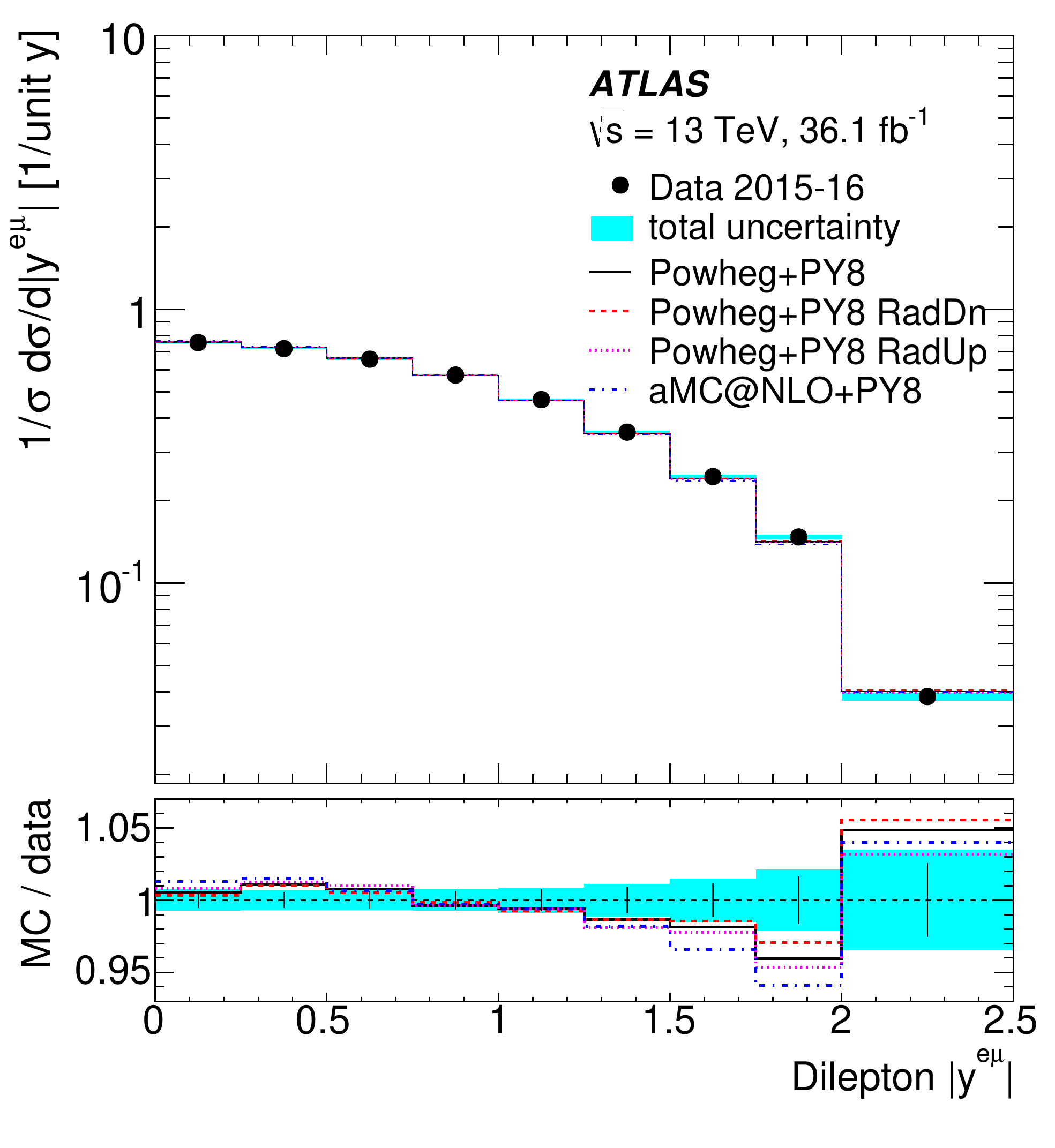}\vspace{-7mm}\center{(a)}}
\parbox{83mm}{\includegraphics[width=76mm]{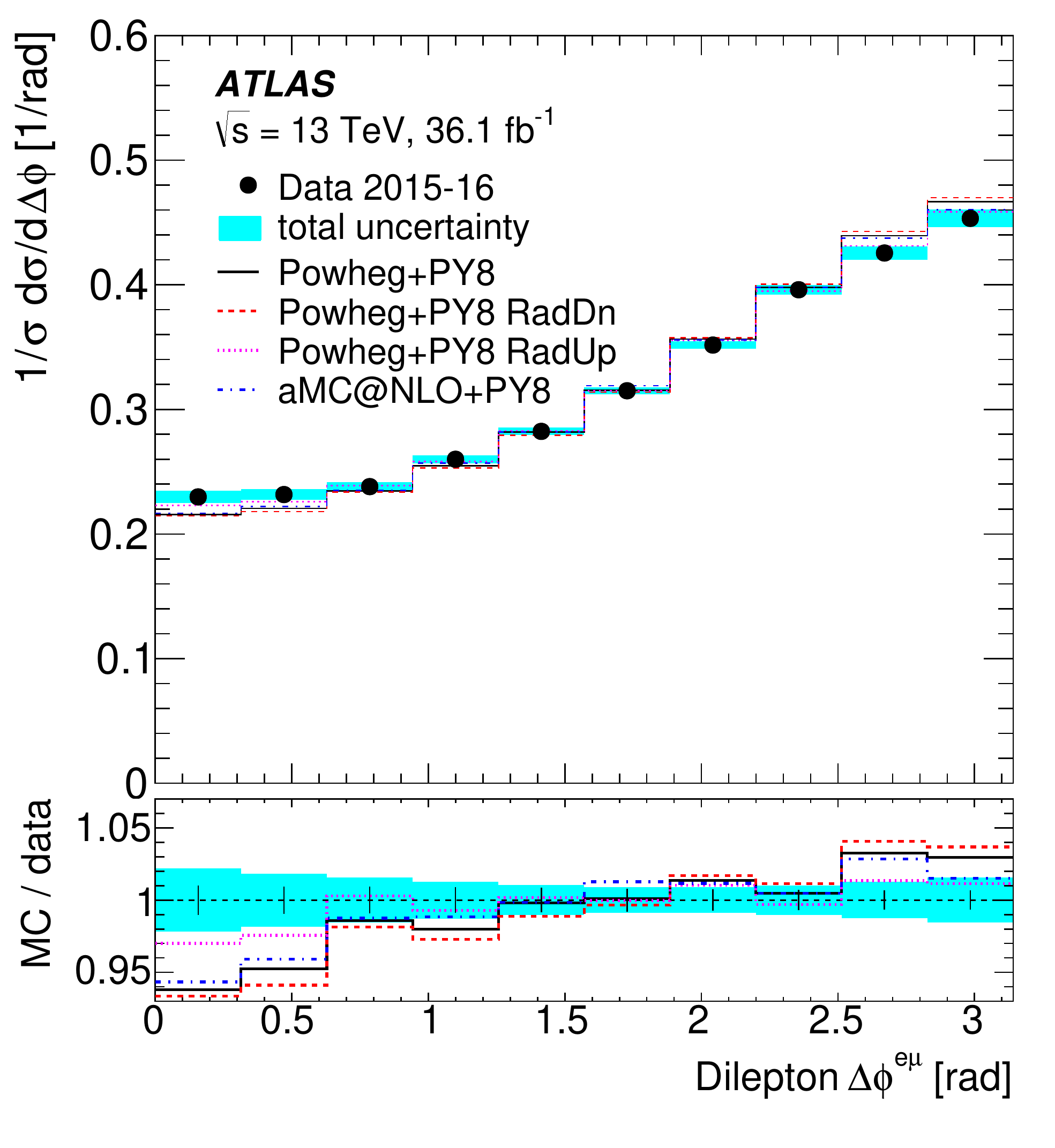}\vspace{-7mm}\center{(b)}}
\parbox{83mm}{\includegraphics[width=76mm]{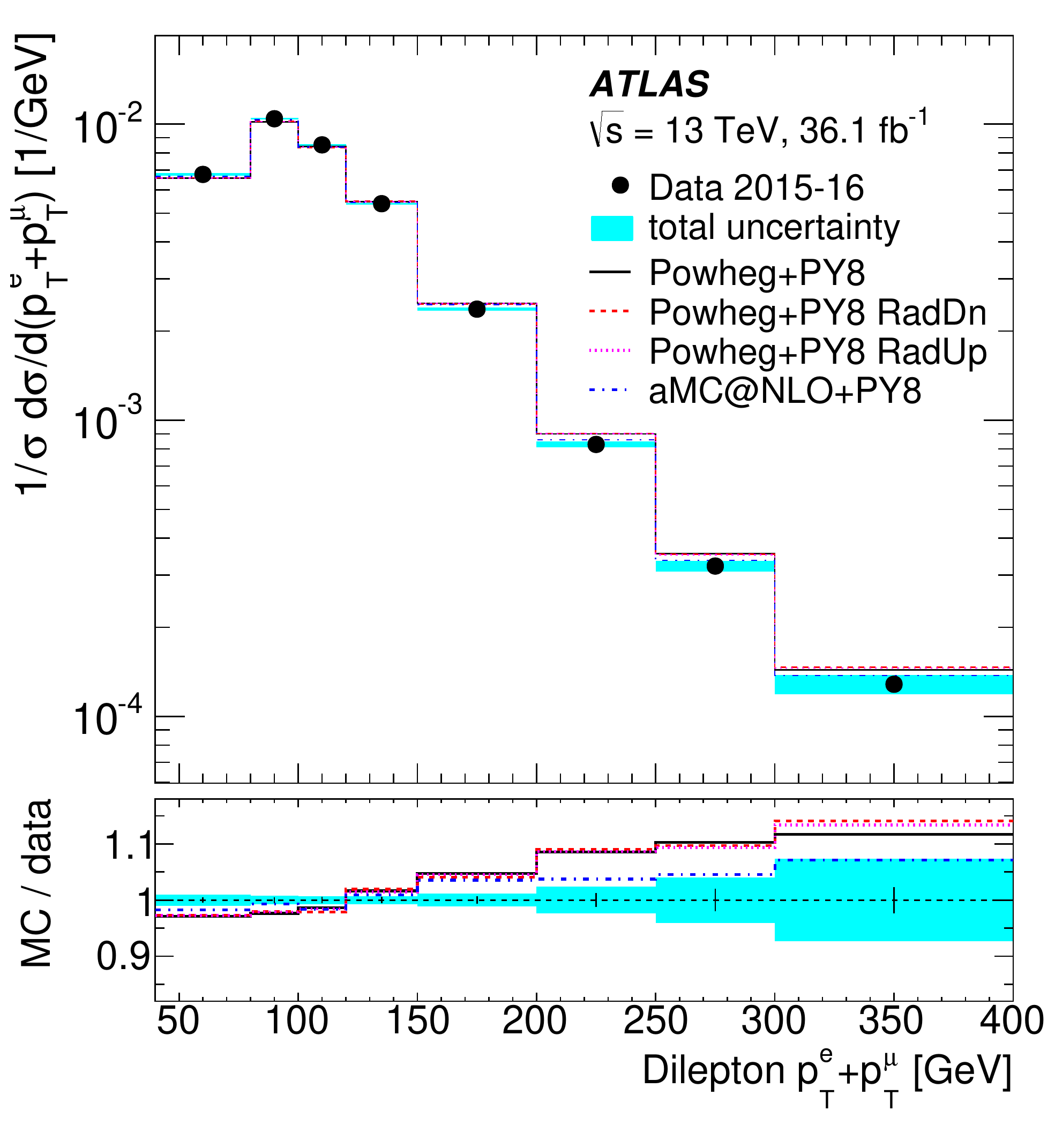}\vspace{-7mm}\center{(c)}}
\parbox{83mm}{\includegraphics[width=76mm]{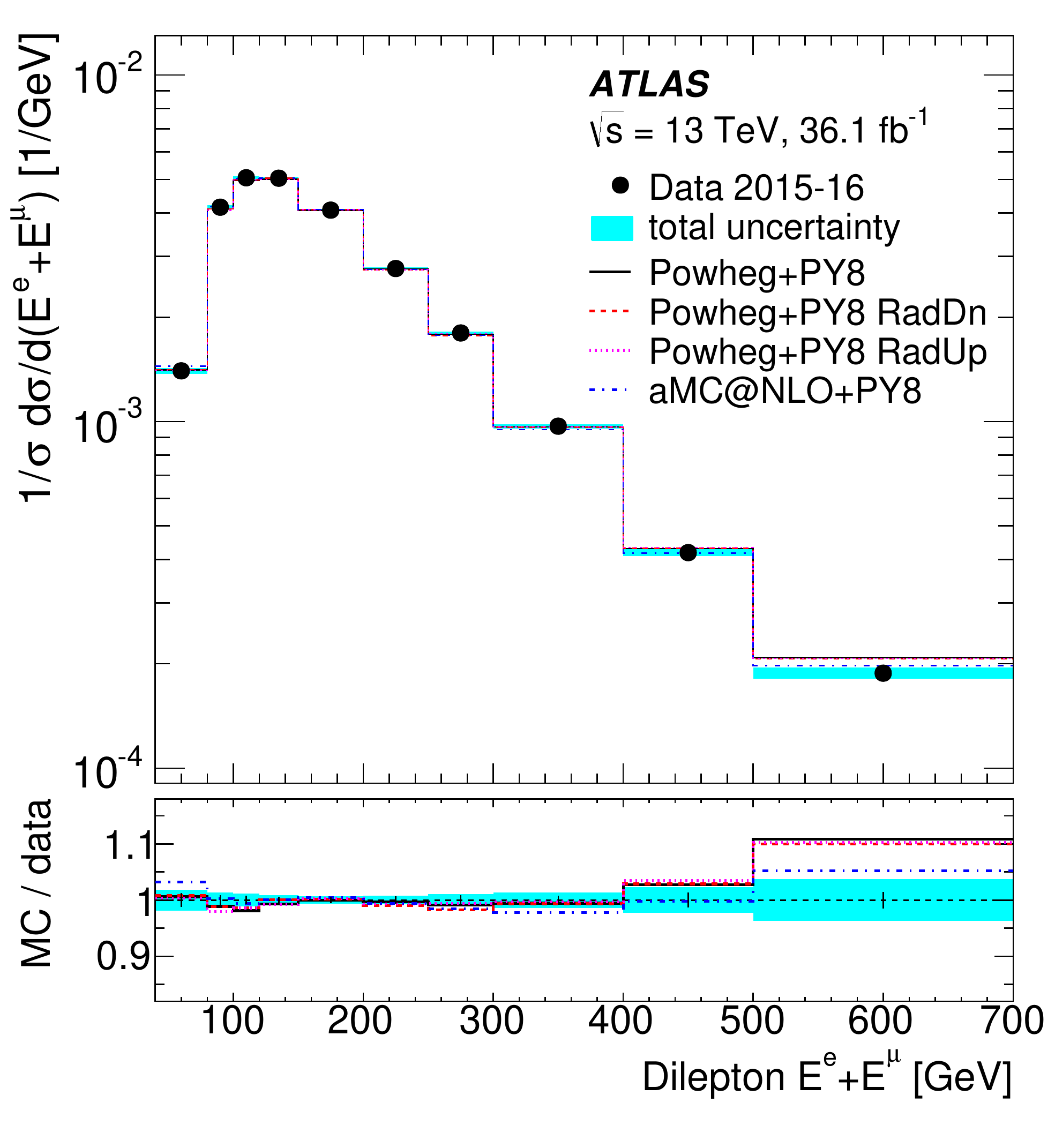}\vspace{-7mm}\center{(d)}}
\caption{\label{f:distresb}Normalised differential cross-sections as a function
of (a) \rapll, (b) \dphill, (c) \ptsum\ and (d) \esum. The measured values are
shown by the black points with error bars corresponding to the data statistical
uncertainties and cyan bands corresponding to the total uncertainties in each
bin, and include the contributions via $W\rightarrow\tau\rightarrow e/\mu$
decays. The data points are placed at the centre of each bin.
The results are compared with the predictions from the baseline
{\sc Powheg\,+\,Pythia8} \ttbar\ sample, {\sc Powheg\,+\,Pythia8} samples
with more or less parton-shower radiation (RadUp and RadDn), and an
{\sc aMC@NLO\,+\,Pythia8} sample. The lower plots show the ratios of
predictions to data, with the error bars indicating the data statistical
uncertainties and the
cyan bands indicating the total uncertainties in the measurements.}
\end{figure}

\begin{figure}[tp]
\centering
 
\includegraphics[width=160mm]{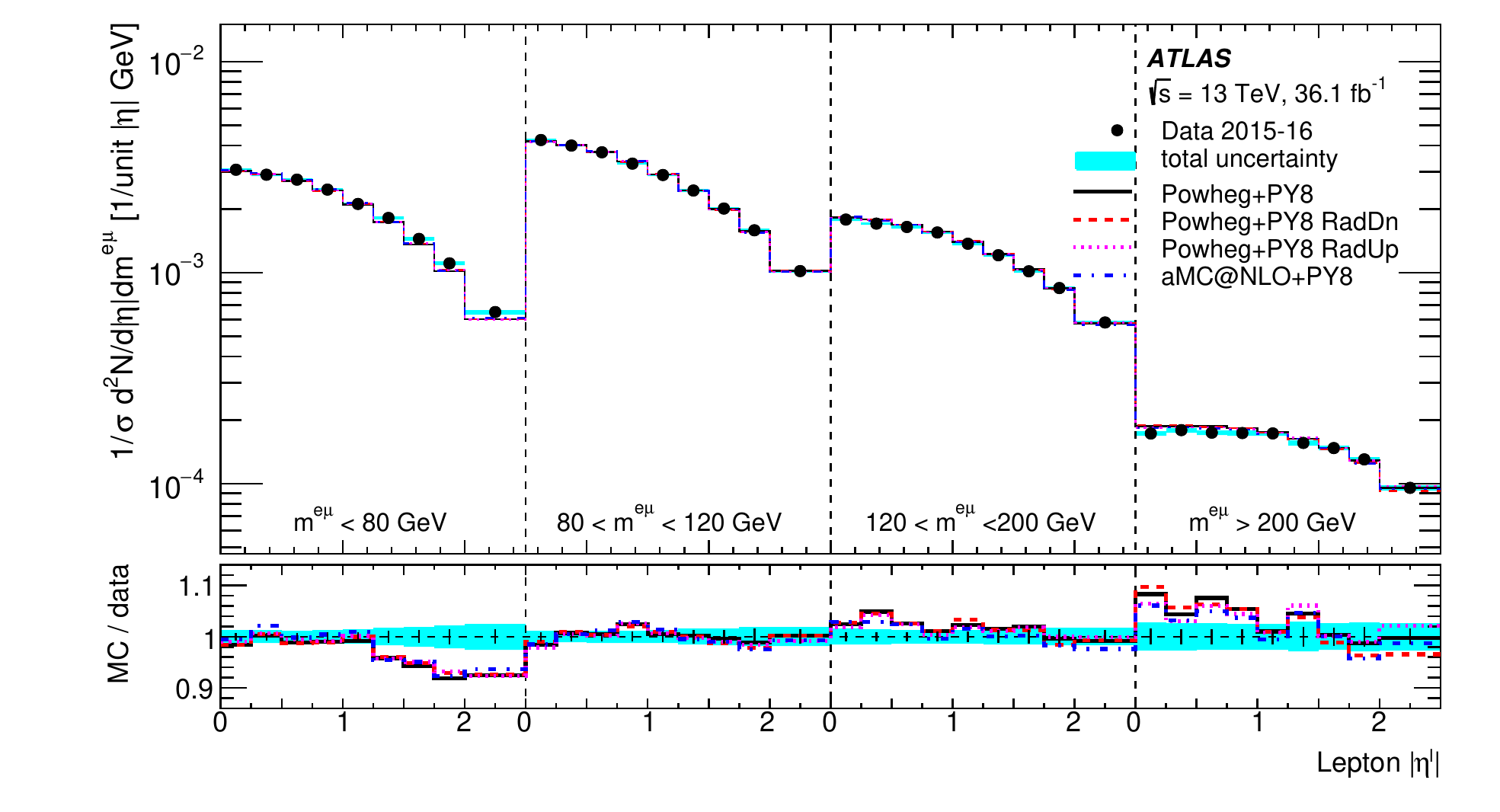}
\includegraphics[width=160mm]{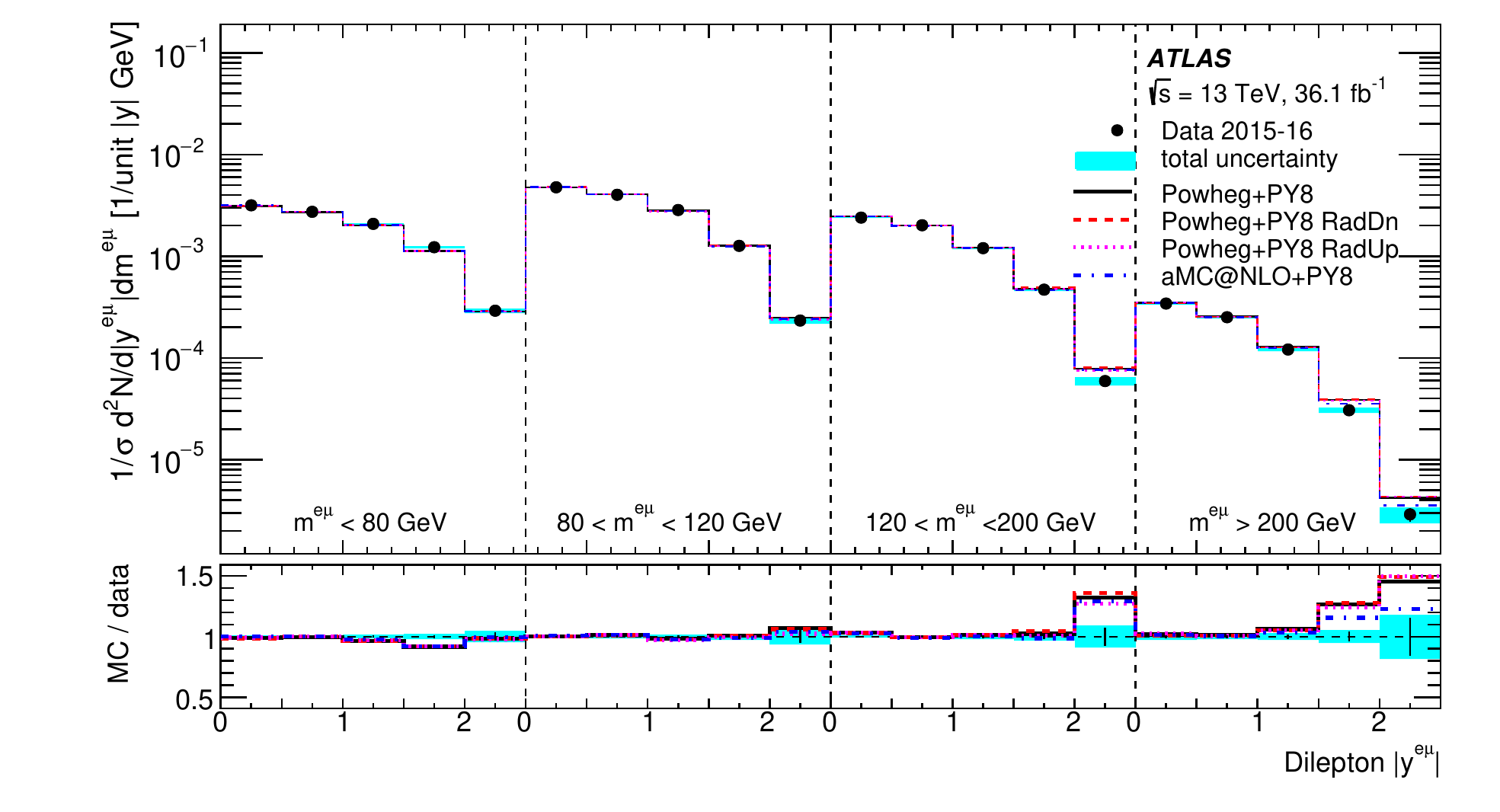}
 
\caption{\label{f:distresc}Normalised double-differential cross-sections as
functions of \etal\ and \mll\ (top), and \rapll\ and \mll\ (bottom).
The measured values are
shown by the black points with error bars corresponding to the data statistical
uncertainties and cyan bands corresponding to the total uncertainties in each
bin, and include the contributions via $W\rightarrow\tau\rightarrow e/\mu$
decays. The data points are placed at the centre of each bin.
The results are compared with the predictions from the baseline
{\sc Powheg\,+\,Pythia8} \ttbar\ sample, {\sc Powheg\,+\,Pythia8} samples
with more or less parton-shower radiation (RadUp and RadDn), and an
{\sc aMC@NLO\,+\,Pythia8} sample. The lower plots show the ratios of
predictions to data, with the error bars indicating the data statistical
uncertainties and the
cyan bands indicating the total uncertainties in the measurements.}
\end{figure}

\begin{figure}[tp]
\centering
 
\includegraphics[width=160mm]{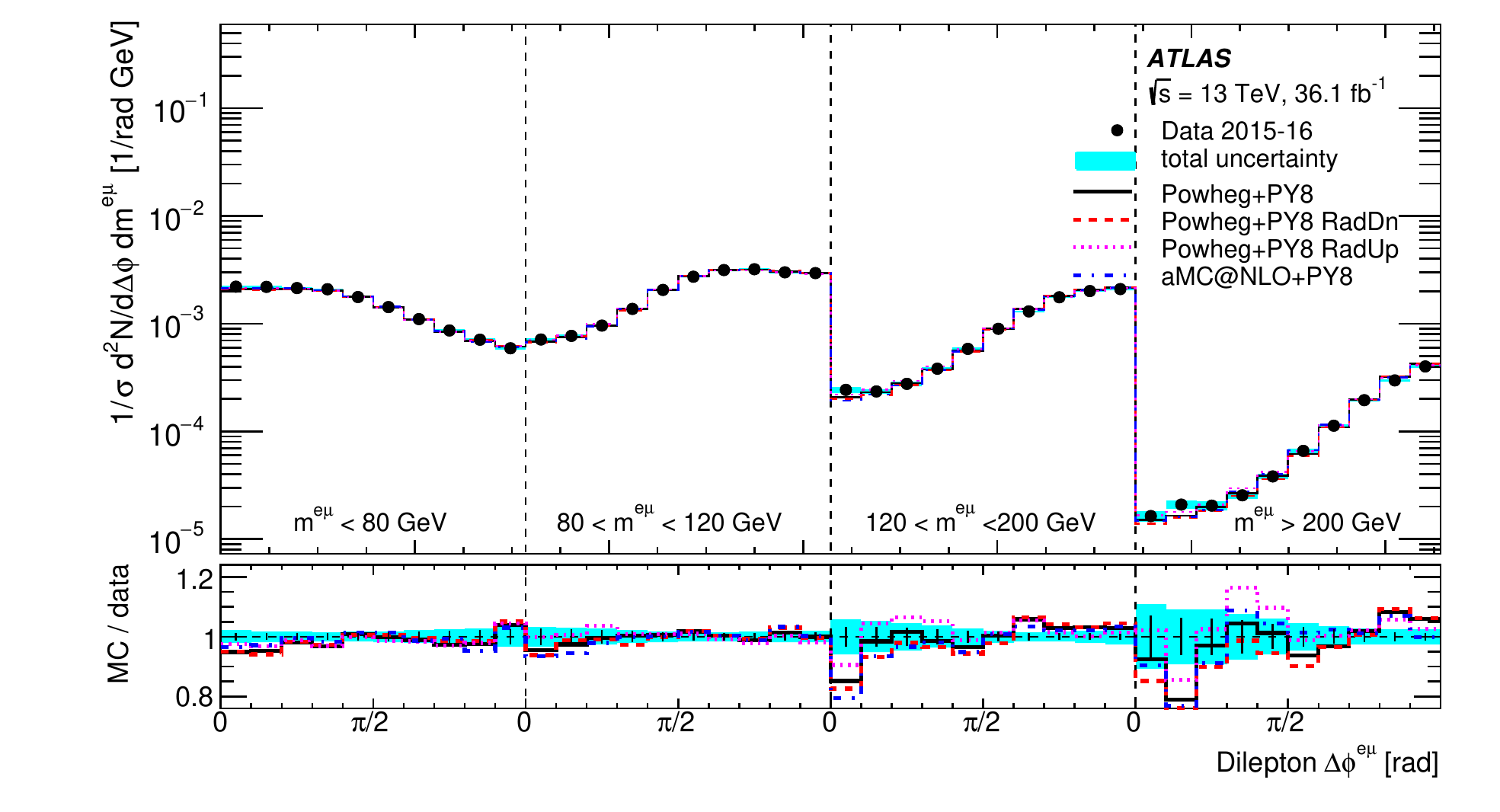}
 
\caption{\label{f:distresd}Normalised double-differential cross-sections as a
function of \dphill\ and \mll. The measured values are
shown by the black points with error bars corresponding to the data statistical
uncertainties and cyan bands corresponding to the total uncertainties in each
bin, and include the contributions via $W\rightarrow\tau\rightarrow e/\mu$
decays. The data points are placed at the centre of each bin.
The results are compared with the predictions from the baseline
{\sc Powheg\,+\,Pythia8} \ttbar\ sample, {\sc Powheg\,+\,Pythia8} samples
with more or less parton-shower radiation (RadUp and RadDn), and an
{\sc aMC@NLO\,+\,Pythia8} sample. The lower plots show the ratios of
predictions to data, with the error bars indicating the data statistical
uncertainties and the
cyan bands indicating the total uncertainties in the measurements.}
\end{figure}
 
\FloatBarrier
 
\subsection{Comparison with event generator predictions}\label{ss:gencomp}
 
The measured normalised differential cross-sections are compared to a
set of particle-level predictions from different Monte Carlo \ttbar\ event
generator configurations in Figures~\ref{f:rdistresa}--\ref{f:rdistresg}. These
figures show the ratios of each prediction to the data as functions
of the differential variables, with the comparison organised into the four
groups of samples summarised in Table~\ref{t:mcsam}. These include samples
based on {\sc Powheg} or {\sc aMC@NLO} for the NLO matrix-element
generator, interfaced to {\sc Pythia8}, {\sc Pythia6} or
{\sc Herwig7}, and using various PDF sets. As well as NNPDF3.0 \cite{nnpdf3}
used for the baseline samples, the global NLO  PDF sets
CT10 \cite{cttenpdf}, CT14 \cite{ct14}, MMHT14 \cite{mmht} and
PDF4LHC\_NLO\_30 \cite{pdf4lhc2} are shown, together with the HERAPDF 2.0
PDF set, based mainly on deep inelastic scattering data \cite{herapdf20}.
Furthermore, the {\sc Powheg\,+\,Pythia8} samples
with more (denoted `RadUp') or less (`RadDn') parton-shower radiation
described in Section~\ref{s:datmc} are included, together with samples
which differ from the baseline {\sc Powheg\,+\,Pythia8} configuration
only by changes of the factorisation and renormalisation scales
\muf\ and \mur\ up and down by factors of two.
 
\begin{table}
\centering
 
\begin{tabular}{lllll}\hline
& Matrix element & PDF & Parton shower/tune & Comments \\
\hline
1 & {\sc Powheg} & NNPDF3.0 & {\sc Pythia8} A14 & $\hdamp=\frac{3}{2}\mtop$  \\
& {\sc Powheg} & CT10 & {\sc Pythia6} P2012 & $\hdamp=\mtop$ \\
& {\sc Powheg} & NNPDF3.0 & {\sc Herwig7} H7UE & $\hdamp=\frac{3}{2}\mtop$ \\
& {\sc Powheg} & NNPDF3.0 & {\sc Pythia8} A14 & top quark \pt\ reweighted to Ref. \cite{TOPQ-2016-01} \\
\hline
2 & {\sc Powheg} & NNPDF3.0 & {\sc Pythia8} A14v3cDo & $\hdamp=\frac{3}{2}\mtop$, $2\mu_\mathrm{F,R}$ (RadDn) \\
& {\sc Powheg} & NNPDF3.0 & {\sc Pythia8} A14v3cUp & $\hdamp=3\mtop$, $\frac{1}{2}\mu_\mathrm{F,R}$ (RadUp) \\
& {\sc Powheg} & NNPDF3.0 & {\sc Pythia8} A14 & $\hdamp=\frac{3}{2}\mtop$, $2\mu_\mathrm{F,R}$ \\
& {\sc Powheg} & NNPDF3.0 & {\sc Pythia8} A14 & $\hdamp=\frac{3}{2}\mtop$, $\frac{1}{2}\mu_\mathrm{F,R}$ \\
\hline
3 & {\sc Powheg} & NNPDF3.0 & {\sc Pythia8} A14 & $\hdamp=\frac{3}{2}\mtop$ \\
& {\sc Powheg} & PDF4LHC15 & {\sc Pythia8} A14 & $\hdamp=\frac{3}{2}\mtop$ \\
& {\sc Powheg} & CT14 & {\sc Pythia8} A14 & $\hdamp=\frac{3}{2}\mtop$ \\
& {\sc Powheg} & MMHT & {\sc Pythia8} A14 & $\hdamp=\frac{3}{2}\mtop$ \\
\hline
4 & {\sc aMC@NLO} & NNPDF3.0 & {\sc Pythia8} A14 & \\
& {\sc aMC@NLO} & CT10 & {\sc Pythia8} A14 & \\
& {\sc aMC@MLO} & HERAPDF2.0 & {\sc Pythia8} A14 & \\
\hline
\end{tabular}
\caption{\label{t:mcsam}Summary of particle-level simulation samples used in the
comparison with the corrected data distributions in Section~\ref{ss:gencomp},
giving the matrix-element event generator, PDF set, parton shower and
associated tune parameter set, and other relevant settings.
The top quark mass was set to $\mtop=172.5\,\GeV$ in all samples.
The four groups shown correspond to the four panels for each measured
distribution shown in Figures~\ref{f:rdistresa}--\ref{f:rdistresg}. The baseline
{\sc Powheg\,+\,Pythia8} configuration appears in both groups~1 and~3.}
\end{table}
 
The baseline {\sc Powheg\,+\,Pythia8} configuration
is known to predict too hard a top quark \pt\ distribution compared
to data at \sxyt\ \cite{TOPQ-2016-01} and \sxvt\ \cite{TOPQ-2015-06},
and compared to NNLO QCD calculations \cite{topptnnlo}. To explore
the effect of this mismodelling on the lepton differential distributions,
the {\sc Powheg\,+\,Pythia8} \ttbar\ sample was reweighted according to
the top quark \pt\ in each event, using a linear function whose parameters
were chosen so as to reproduce the measured top quark \pt\ distribution
shown in Figure~19 of Ref. \cite{TOPQ-2016-01}; this sample is included in
the first sample group and labelled `{\sc Powheg\,+\,PY8} \pt\ rew'
in Figures~\ref{f:rdistresa}--\ref{f:rdistresg}.
 
\begin{figure}[tp]
\parbox{83mm}{\includegraphics[width=76mm]{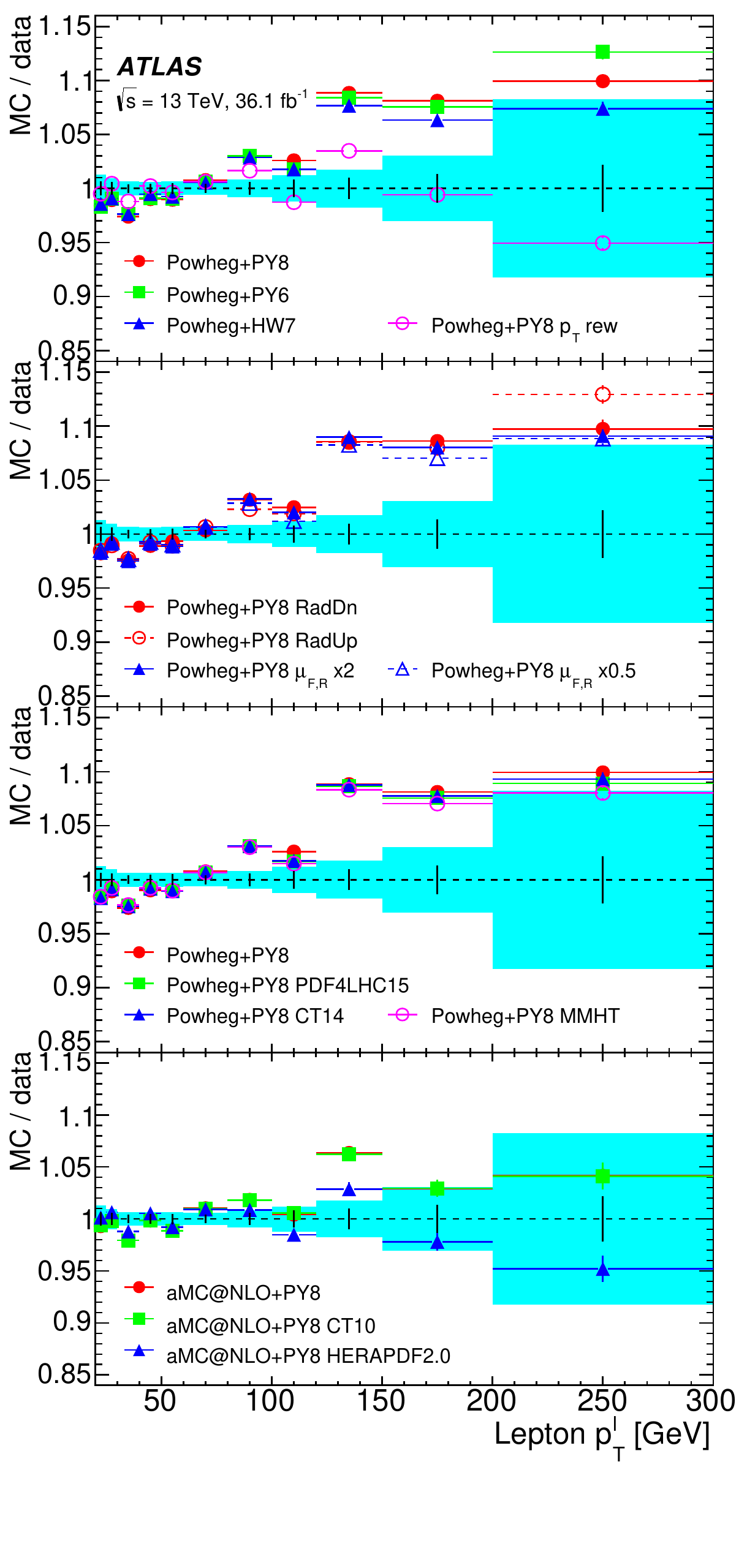}\vspace{-7mm}\center{(a)}}
\parbox{83mm}{\includegraphics[width=76mm]{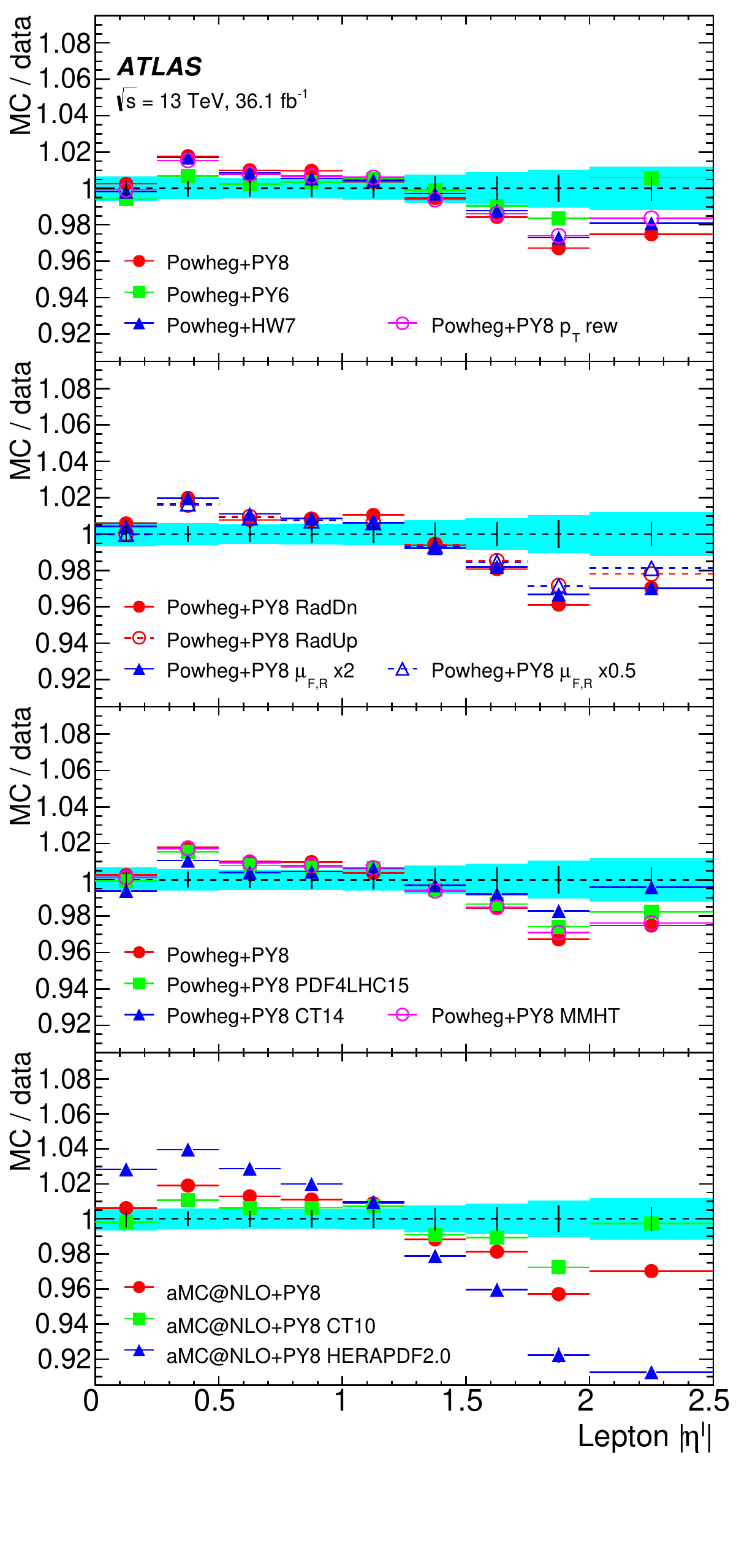}\vspace{-7mm}\center{(b)}}
\caption{\label{f:rdistresa}Ratios of predictions of normalised differential
cross-sections to data as a function of (a) \ptl\ and (b) \etal. The data
statistical uncertainties are shown by the black error bars around a ratio
of unity, and the total uncertainties are shown by the cyan bands.
Several different \ttbar\ predictions are shown in each panel, grouped
from top to bottom as shown in Table~\ref{t:mcsam},
and the error bars indicate the uncertainties due to the limited size
of the simulated samples.}
\end{figure}
 
\begin{figure}[tp]
\parbox{83mm}{\includegraphics[width=76mm]{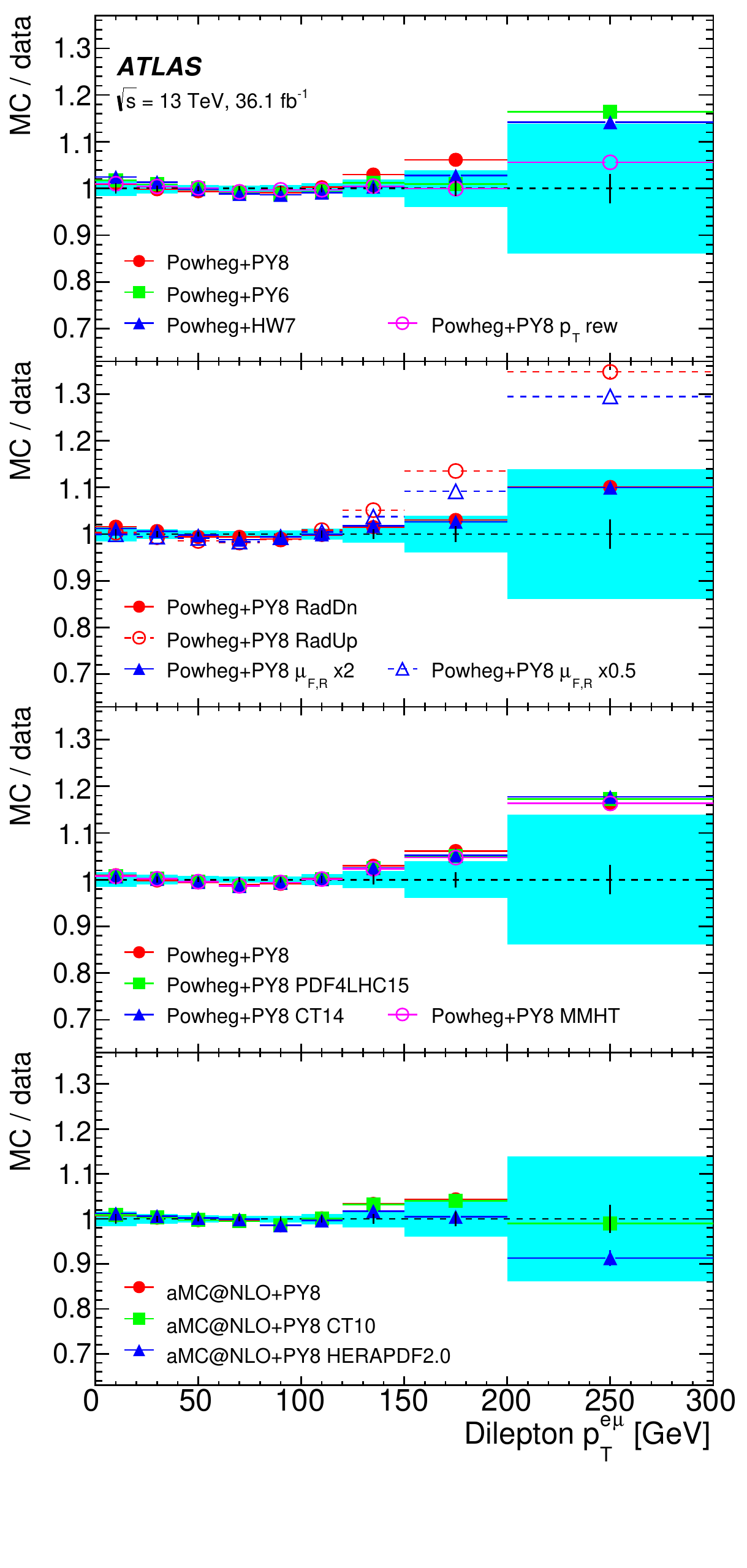}\vspace{-7mm}\center{(a)}}
\parbox{83mm}{\includegraphics[width=76mm]{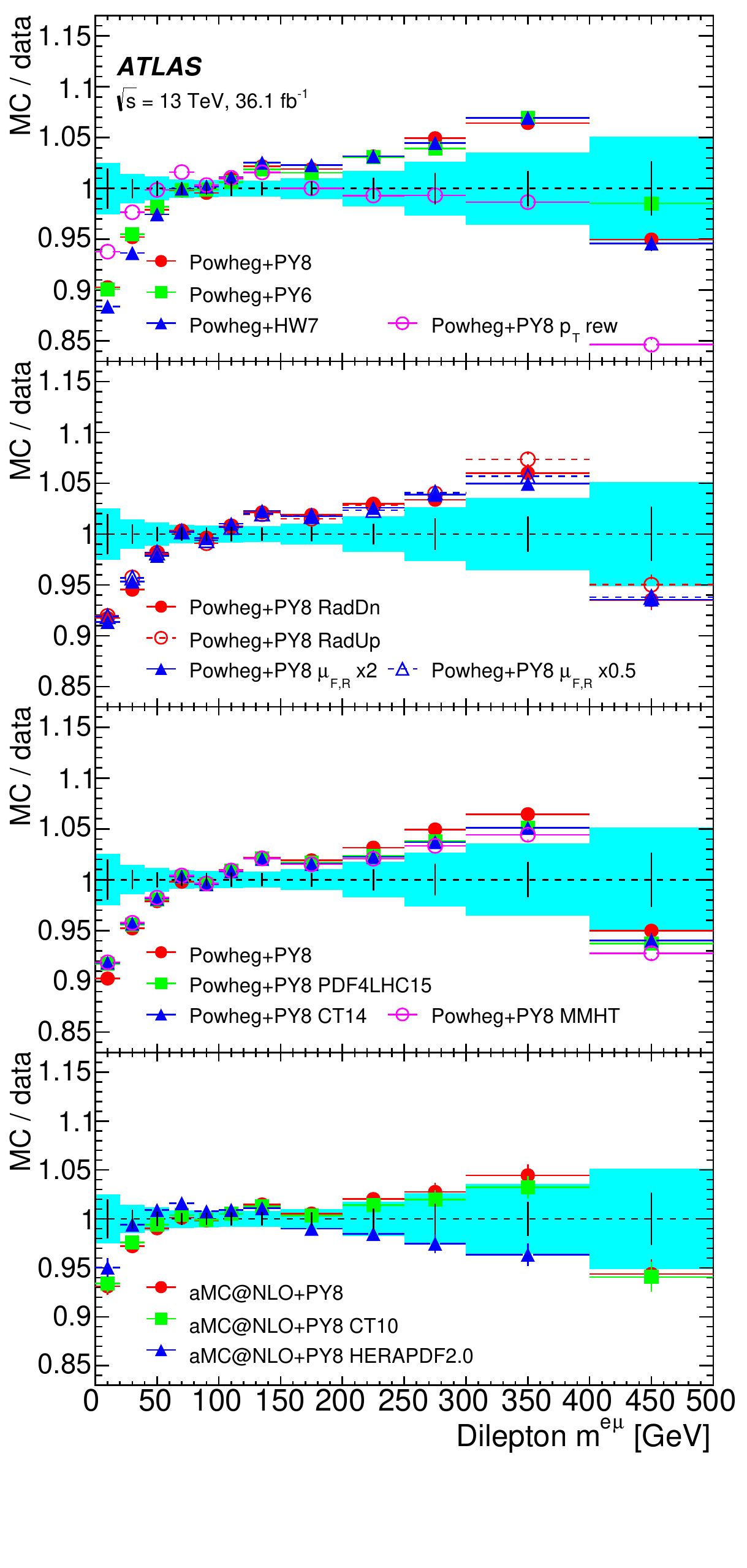}\vspace{-7mm}\center{(b)}}
\caption{\label{f:rdistresb}Ratios of predictions of normalised differential
cross-sections to data as a function of (a) \ptll\ and (b) \mll. The data
statistical uncertainties are shown by the black error bars around a ratio
of unity, and the total uncertainties are shown by the cyan bands.
Several different \ttbar\ predictions are shown in each panel, grouped
from top to bottom as shown in Table~\ref{t:mcsam},
and the error bars indicate the uncertainties due to the limited size
of the simulated samples.}
\end{figure}
 
\begin{figure}[tp]
\parbox{83mm}{\includegraphics[width=76mm]{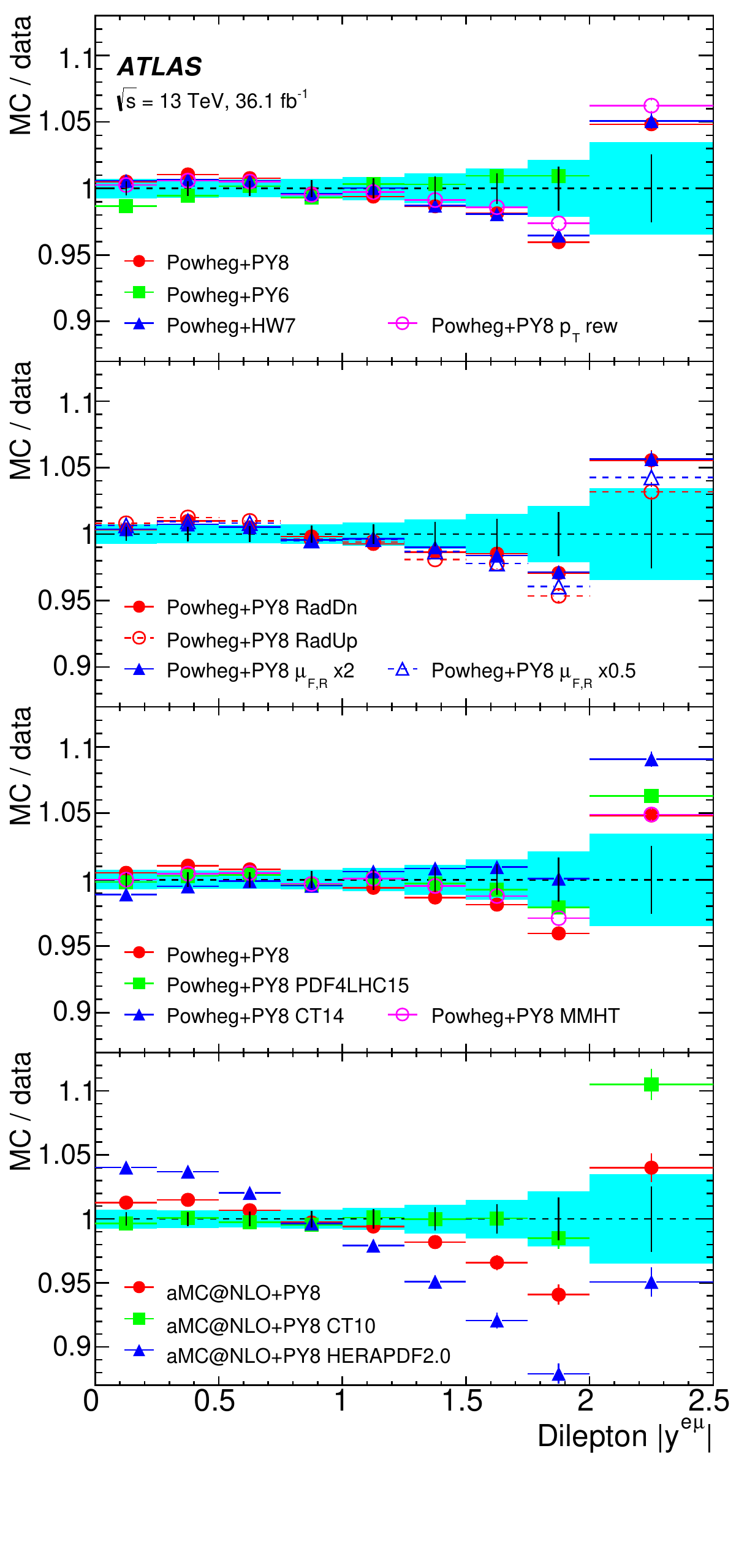}\vspace{-7mm}\center{(a)}}
\parbox{83mm}{\includegraphics[width=76mm]{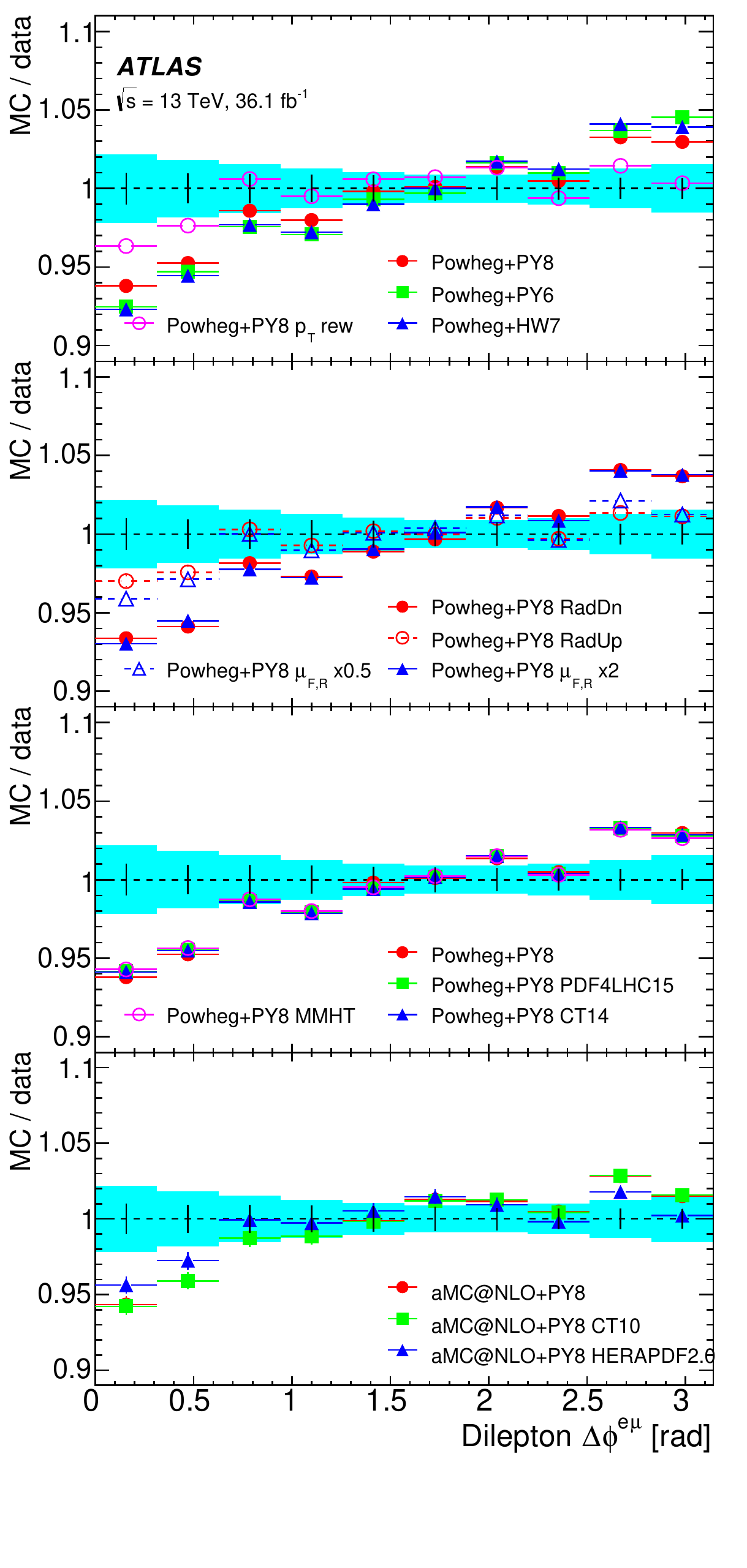}\vspace{-7mm}\center{(b)}}
\caption{\label{f:rdistresc}Ratios of predictions of normalised differential
cross-sections to data as a function of (a) \rapll\ and (b) \dphill. The data
statistical uncertainties are shown by the black error bars around a ratio
of unity, and the total uncertainties are shown by the cyan bands.
Several different \ttbar\ predictions are shown in each panel, grouped
from top to bottom as shown in Table~\ref{t:mcsam},
and the error bars indicate the uncertainties due to the limited size
of the simulated samples.}
\end{figure}
 
\begin{figure}[tp]
\parbox{83mm}{\includegraphics[width=76mm]{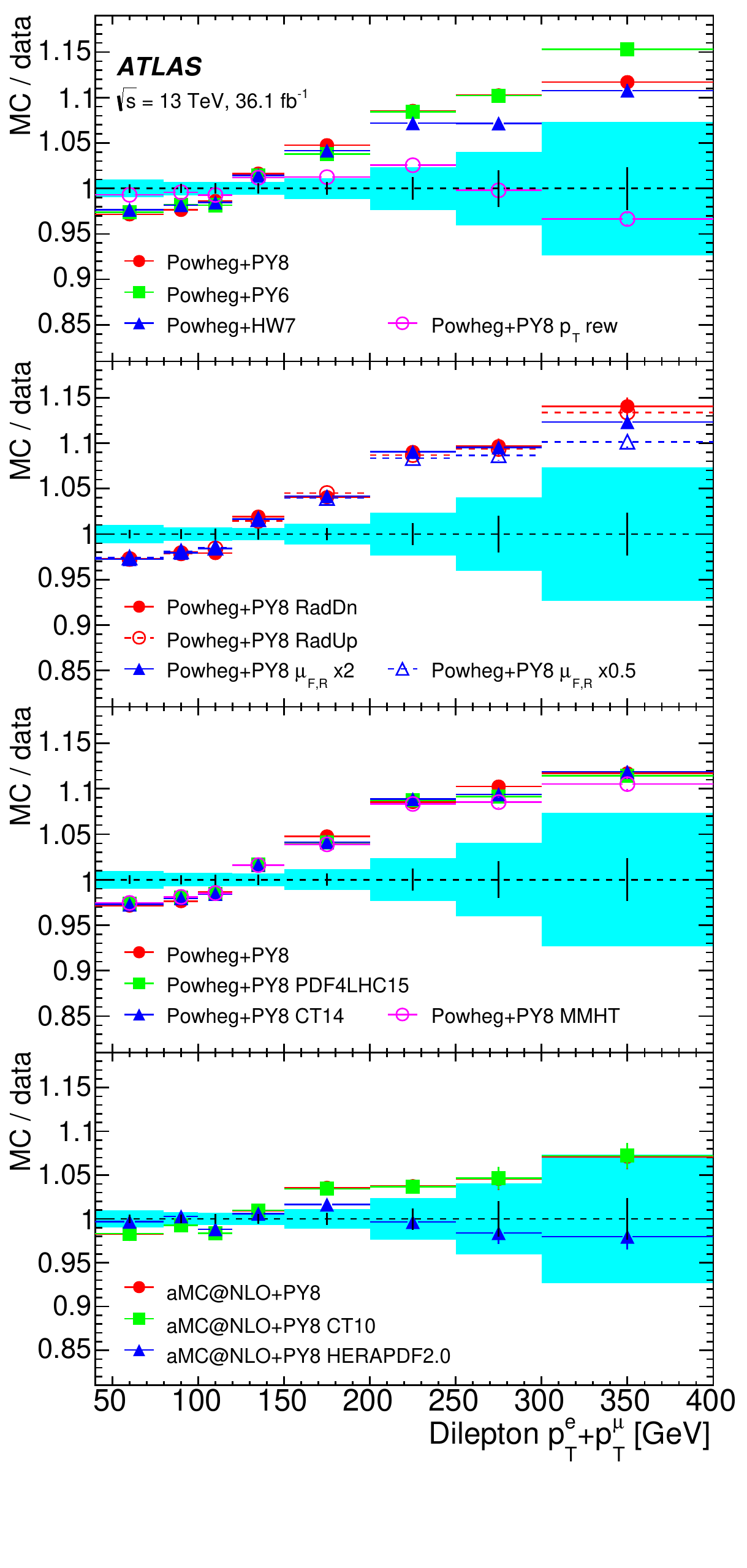}\vspace{-7mm}\center{(a)}}
\parbox{83mm}{\includegraphics[width=76mm]{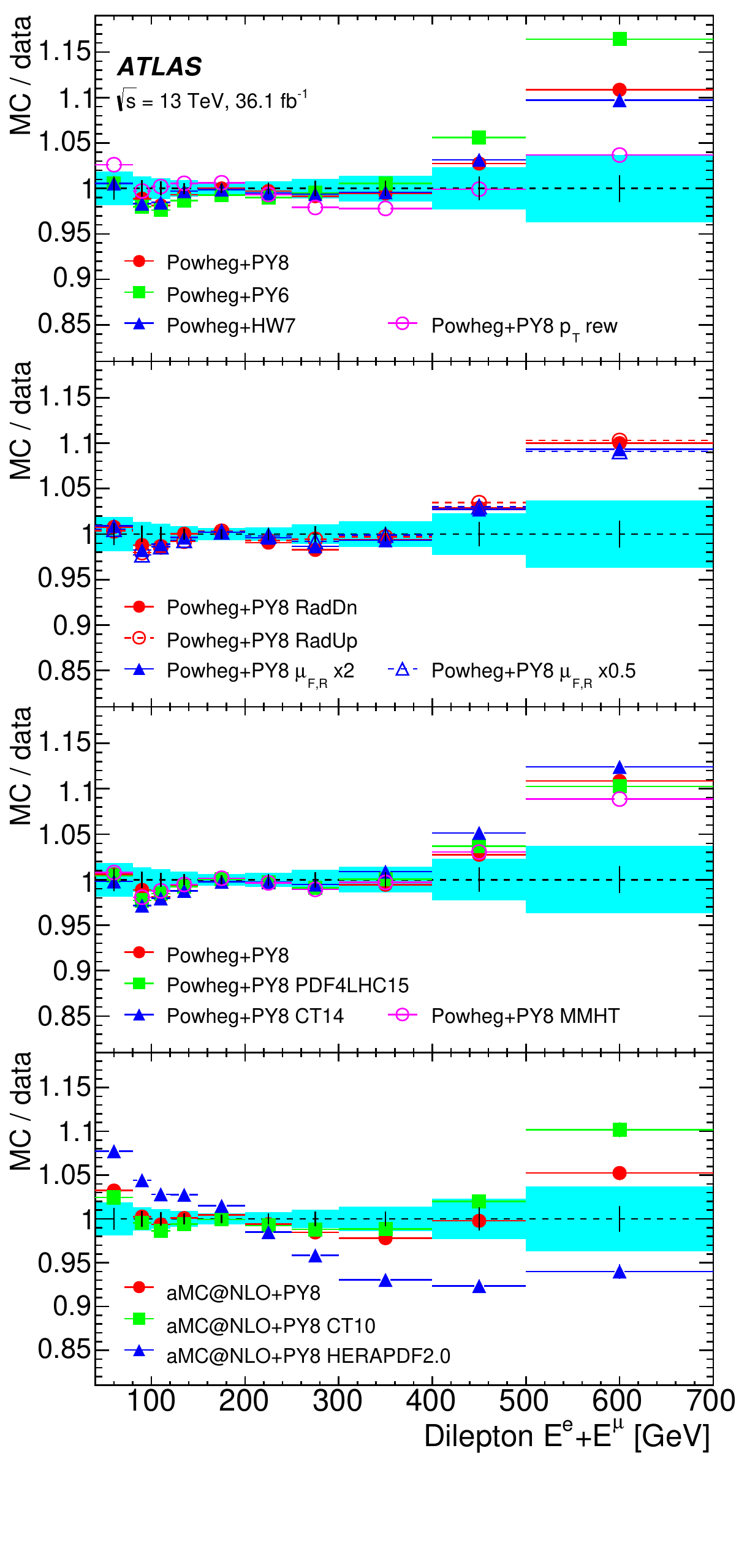}\vspace{-7mm}\center{(b)}}
\caption{\label{f:rdistresd}Ratios of predictions of normalised differential
cross-sections to data as a function of (a) \ptsum\ and (b) \esum. The data
statistical uncertainties are shown by the black error bars around a ratio
of unity, and the total uncertainties are shown by the cyan bands.
Several different \ttbar\ predictions are shown in each panel, grouped
from top to bottom as shown in Table~\ref{t:mcsam},
and the error bars indicate the uncertainties due to the limited size
of the simulated samples.}
\end{figure}
 
\begin{figure}[tp]
\includegraphics[width=160mm]{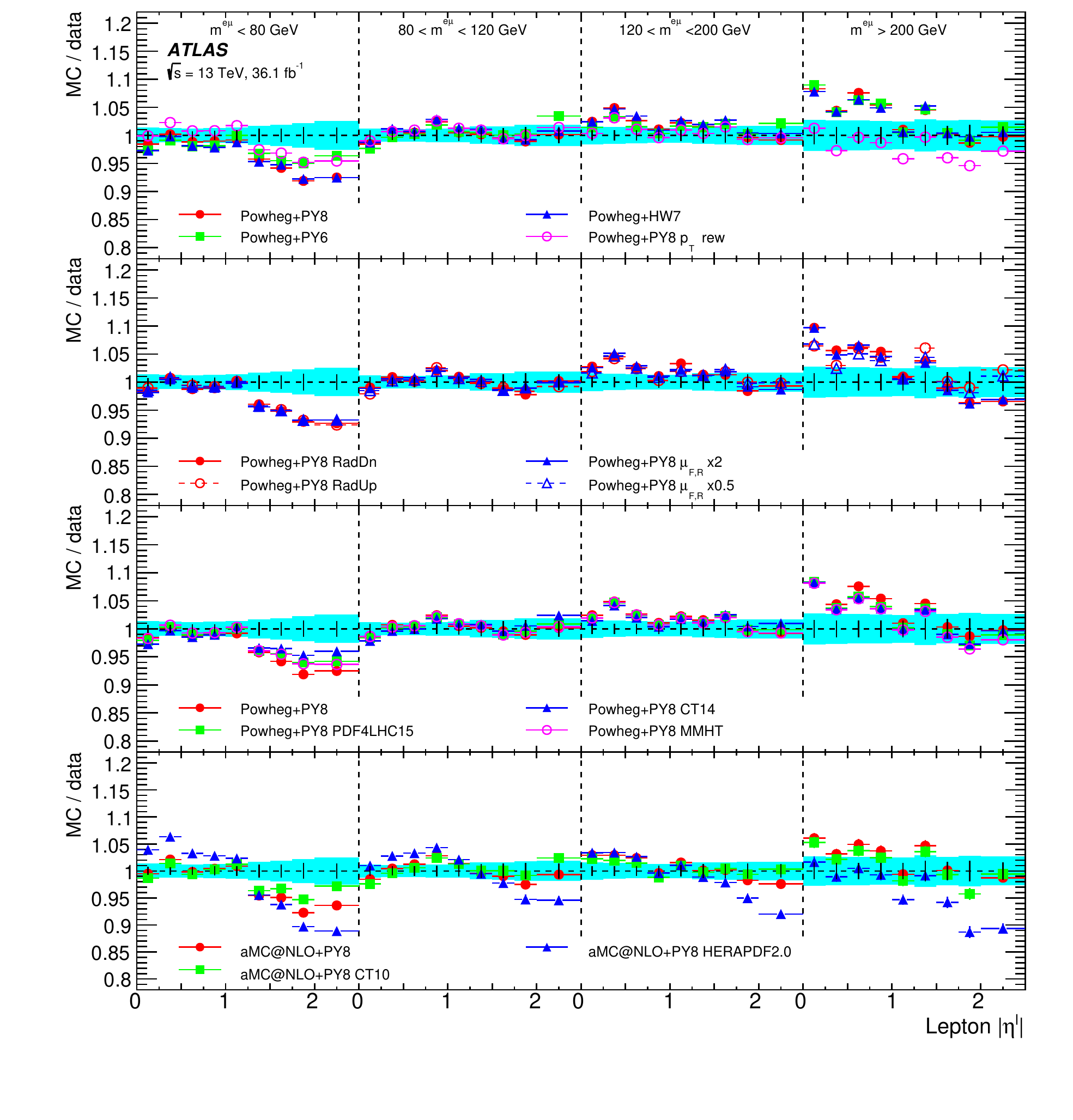}
\caption{\label{f:rdistrese}Ratios of predictions of the normalised
double-differential
cross-sections to data as a function of \etal\ and \mll. The data
statistical uncertainties are shown by the black error bars around a ratio
of unity, and the total uncertainties are shown by the cyan bands. The vertical
dotted lines indicate the four bins of \mll.
Several different \ttbar\ predictions are shown in each panel, grouped
from top to bottom as shown in Table~\ref{t:mcsam},
and the error bars indicate the uncertainties due to the limited size
of the simulated samples.}
\end{figure}
 
\begin{figure}[tp]
\includegraphics[width=160mm]{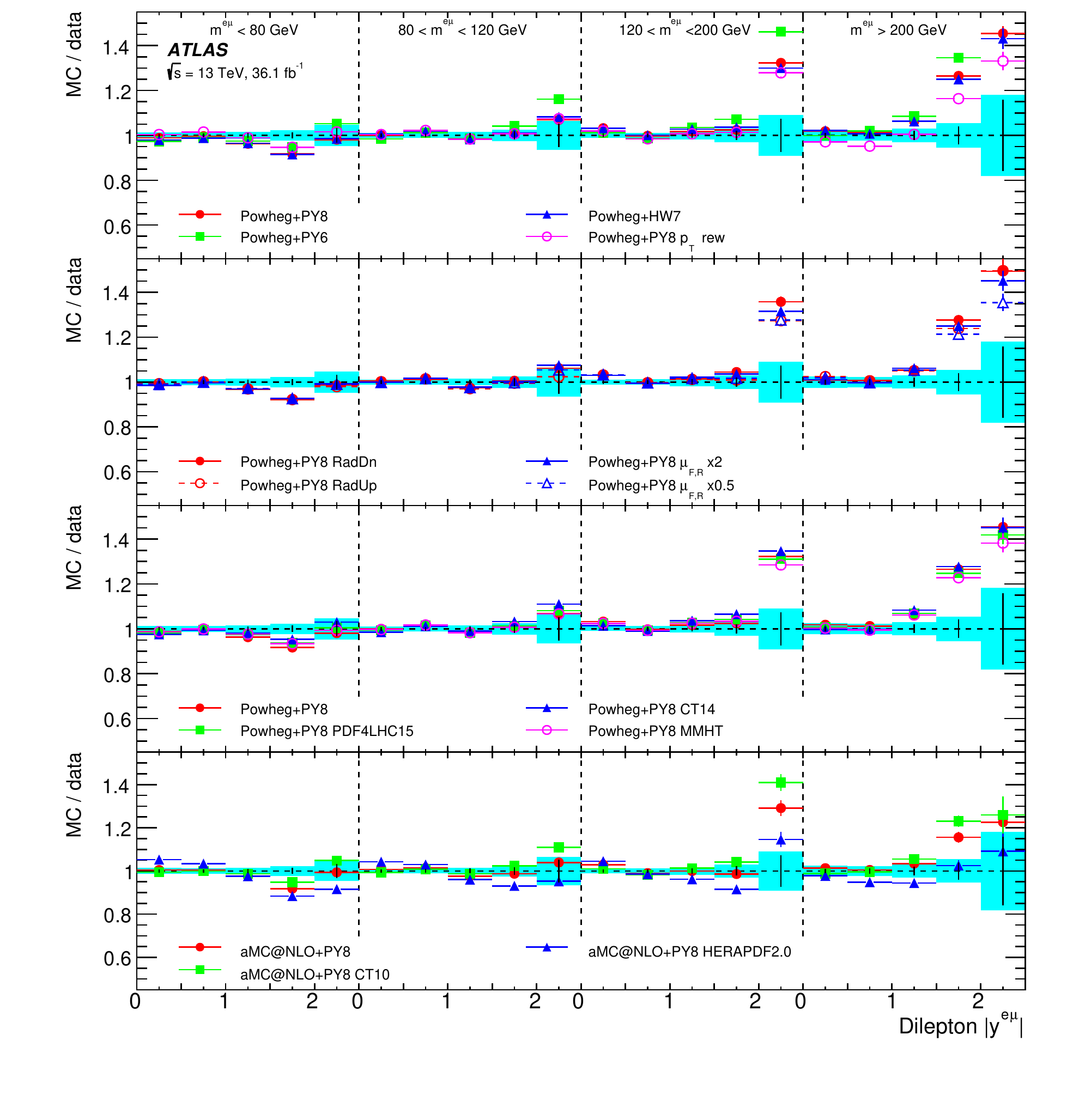}
\caption{\label{f:rdistresf}Ratios of predictions of the normalised
double-differential
cross-sections to data as a function of \rapll\ and \mll. The data
statistical uncertainties are shown by the black error bars around a ratio
of unity, and the total uncertainties are shown by the cyan bands. The vertical
dotted lines indicate the four bins of \mll.
Several different \ttbar\ predictions are shown in each panel, grouped
from top to bottom as shown in Table~\ref{t:mcsam},
and the error bars indicate the uncertainties due to the limited size
of the simulated samples.}
\end{figure}
 
\begin{figure}[tp]
\includegraphics[width=160mm]{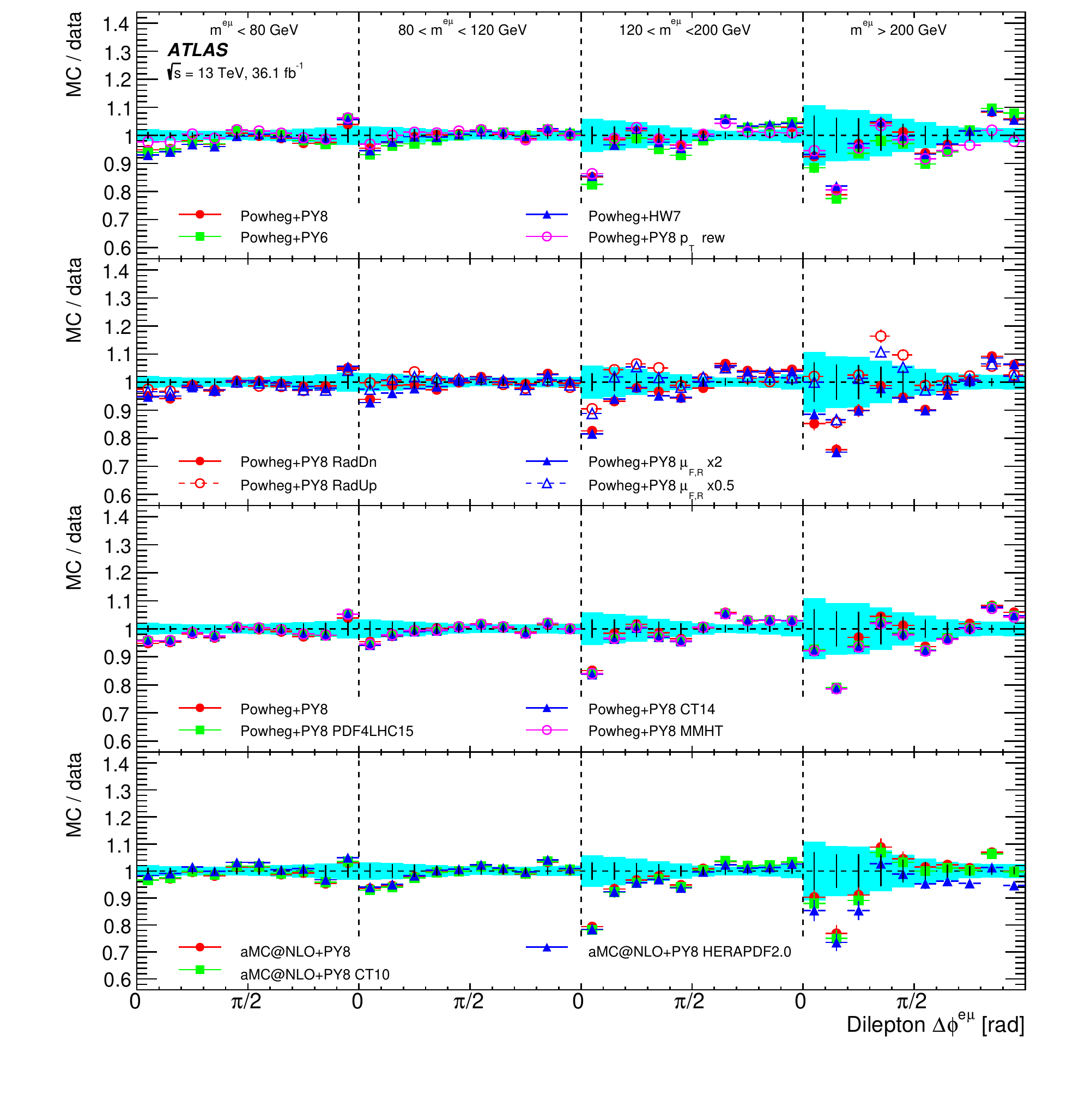}
\caption{\label{f:rdistresg}Ratios of predictions of the normalised
double-differential
cross-sections to data as a function of \dphill\ and \mll. The data
statistical uncertainties are shown by the black error bars around a ratio
of unity, and the total uncertainties are shown by the cyan bands. The vertical
dotted lines indicate the four bins of \mll.
Several different \ttbar\ predictions are shown in each panel, grouped
from top to bottom as shown in Table~\ref{t:mcsam},
and the error bars indicate the uncertainties due to the limited size
of the simulated samples.}
\end{figure}
 
The compatibility of each prediction with each measured normalised
distribution was tested using a $\chi^2$ calculated as
\begin{eqnarray}\label{e:redchi}
\chi^2 = {\mathbf\Delta}_{(n-1)}^T {\mathbf S}^{-1}_{(n-1)} {\mathbf\Delta}_{(n-1)}\,,
\end{eqnarray}
where ${\mathbf\Delta}_{(n-1)}$ is the vector of differences between the measured
and predicted normalised differential cross-section in each of the $n$
bins, excluding the last one, and ${\mathbf S}^{-1}_{(n-1)}$ is the inverse of
the corresponding
covariance matrix, including both the experimental uncertainties in the
measurement and the statistical uncertainties in the predictions.
Correlations between the measurements in different bins were incorporated
via off-diagonal terms in $\mathbf{S}$, and
the last bin of each distribution was excluded
to account for the degree of freedom lost to the normalisation condition. The
resulting $\chi^2$ values and corresponding probability $p$-values (for $n-1$
degrees of freedom) are shown for each single-differential
distribution and prediction in Table~\ref{t:NorChiXPP0Data2016Data},
and for the double-differential distributions and predictions in
Table~\ref{t:NorChiXPP8Data2016Data}.
 
\begin{table}[htp]
\small\centering
 
\begin{tabular}{l|rrrrrrrr}\hline
Generator & \ptl & \etal & \ptll & \mll & \rapll & \dphill & \ptsum & \esum \\
$N_\mathrm{dof}$ & 10 & 8 & 8 & 11 & 8 & 9 & 7 & 9 \\ \hline
{\sc Powheg\,+\,PY8}&  43.7 &  19.5 &   8.6 &  44.3 &  11.4 &  14.4 &  32.5 &  18.4 \\
{\sc Powheg\,+\,PY6} CT10&  36.1 &   7.9 &   9.3 &  33.0 &  16.2 &  16.2 &  21.9 &  30.5 \\
{\sc Powheg\,+\,HW7}&  34.8 &  15.9 &  11.5 &  62.7 &   9.4 &  17.3 &  23.0 &  14.7 \\
{\sc Powheg\,+\,PY8} \pt\ rew.&  20.2 &  14.7 &   2.3 &  38.3 &   8.4 &  12.7 &   9.4 &  14.0 \\
{\sc Powheg\,+\,PY8} RadDn&  40.0 &  24.2 &   6.1 &  44.3 &   9.2 &  16.3 &  29.0 &  20.1 \\
{\sc Powheg\,+\,PY8} RadUp&  33.0 &  16.3 &  21.9 &  35.3 &  12.3 &   6.4 &  26.7 &  16.5 \\
{\sc Powheg\,+\,PY8} $\mu_\mathrm{F,R}\times 2$&  46.5 &  21.6 &   6.2 &  42.6 &   8.5 &  16.5 &  28.9 &  17.1 \\
{\sc Powheg\,+\,PY8} $\mu_\mathrm{F,R}\times 0.5$&  39.8 &  17.3 &  11.4 &  38.0 &  10.7 &  10.9 &  27.6 &  14.2 \\
{\sc Powheg\,+\,PY8} PDF4LHC15&  43.4 &  14.6 &   7.4 &  39.0 &   6.2 &  13.5 &  28.0 &  15.9 \\
{\sc Powheg\,+\,PY8} CT14&  44.1 &   9.3 &   7.6 &  37.0 &   8.2 &  13.5 &  28.5 &  18.2 \\
{\sc Powheg\,+\,PY8} MMHT&  41.2 &  17.7 &   6.9 &  39.0 &   6.3 &  13.2 &  26.3 &  14.3 \\
{\sc aMC@NLO\,+\,PY8}&  26.2 &  25.7 &  11.4 &  19.7 &  16.7 &  13.2 &  12.5 &  14.0 \\
{\sc aMC@NLO\,+\,PY8} CT10&  24.9 &  11.7 &  10.6 &  16.9 &  10.0 &  13.4 &  12.0 &  19.0 \\
{\sc aMC@NLO\,+\,PY8} HERA2&  17.1 &  96.6 &   6.9 &  26.0 &  68.5 &  12.5 &   6.1 &  38.4 \\
\hline
{\sc Powheg\,+\,PY8}& $4\cdot 10^{-6}$ & 0.012 & 0.37 & $6\cdot 10^{-6}$ & 0.18 & 0.11 & $3\cdot 10^{-5}$ & 0.030 \\
{\sc Powheg\,+\,PY6} CT10& $8\cdot 10^{-5}$ & 0.45 & 0.32 & $5\cdot 10^{-4}$ & 0.039 & 0.062 & $3\cdot 10^{-3}$ & $4\cdot 10^{-4}$ \\
{\sc Powheg\,+\,HW7}& $1\cdot 10^{-4}$ & 0.043 & 0.18 & $3\cdot 10^{-9}$ & 0.31 & 0.045 & $2\cdot 10^{-3}$ & 0.098 \\
{\sc Powheg\,+\,PY8} \pt\ rew.& 0.028 & 0.065 & 0.97 & $7\cdot 10^{-5}$ & 0.39 & 0.18 & 0.23 & 0.12 \\
{\sc Powheg\,+\,PY8} RadDn& $2\cdot 10^{-5}$ & $2\cdot 10^{-3}$ & 0.64 & $6\cdot 10^{-6}$ & 0.32 & 0.060 & $1\cdot 10^{-4}$ & 0.017 \\
{\sc Powheg\,+\,PY8} RadUp& $3\cdot 10^{-4}$ & 0.038 & $5\cdot 10^{-3}$ & $2\cdot 10^{-4}$ & 0.14 & 0.70 & $4\cdot 10^{-4}$ & 0.057 \\
{\sc Powheg\,+\,PY8} $\mu_\mathrm{F,R}\times 2$& $1\cdot 10^{-6}$ & $6\cdot 10^{-3}$ & 0.62 & $1\cdot 10^{-5}$ & 0.39 & 0.056 & $1\cdot 10^{-4}$ & 0.048 \\
{\sc Powheg\,+\,PY8} $\mu_\mathrm{F,R}\times 0.5$& $2\cdot 10^{-5}$ & 0.027 & 0.18 & $8\cdot 10^{-5}$ & 0.22 & 0.28 & $3\cdot 10^{-4}$ & 0.12 \\
{\sc Powheg\,+\,PY8} PDF4LHC15& $4\cdot 10^{-6}$ & 0.067 & 0.49 & $5\cdot 10^{-5}$ & 0.62 & 0.14 & $2\cdot 10^{-4}$ & 0.068 \\
{\sc Powheg\,+\,PY8} CT14& $3\cdot 10^{-6}$ & 0.32 & 0.47 & $1\cdot 10^{-4}$ & 0.42 & 0.14 & $2\cdot 10^{-4}$ & 0.033 \\
{\sc Powheg\,+\,PY8} MMHT& $1\cdot 10^{-5}$ & 0.024 & 0.55 & $5\cdot 10^{-5}$ & 0.62 & 0.15 & $5\cdot 10^{-4}$ & 0.11 \\
{\sc aMC@NLO\,+\,PY8}& $3\cdot 10^{-3}$ & $1\cdot 10^{-3}$ & 0.18 & 0.049 & 0.034 & 0.15 & 0.086 & 0.12 \\
{\sc aMC@NLO\,+\,PY8} CT10& $5\cdot 10^{-3}$ & 0.16 & 0.23 & 0.11 & 0.27 & 0.15 & 0.10 & 0.025 \\
{\sc aMC@NLO\,+\,PY8} HERA2& 0.073 & 0& 0.54 & $6\cdot 10^{-3}$ & 0& 0.19 & 0.53 & $1\cdot 10^{-5}$ \\
\hline
\end{tabular}
\caption{\label{t:NorChiXPP0Data2016Data}$\chi^2$ values (top) and associated probabilities (bottom) for comparison of normalised measured single-differential
fiducial cross-sections with various \ttbar\ simulation samples.
Probabilities smaller than $10^{-10}$ are shown as zero.}
\end{table}
 
\begin{table}[htp]
\small\centering
 
\begin{tabular}{l|rrr}\hline
Generator & \etalvmll & \rapllvmll & \dphillvmll \\
$N_\mathrm{dof}$ & 35 & 19 & 39 \\ \hline
{\sc Powheg\,+\,PY8}&  53.1 &  72.3 &  65.4 \\
{\sc Powheg\,+\,PY6} CT10&  45.9 &  92.9 &  79.5 \\
{\sc Powheg\,+\,HW7}&  49.3 &  67.4 &  63.7 \\
{\sc Powheg\,+\,PY8} \pt\ rew.&  47.1 &  56.1 &  51.4 \\
{\sc Powheg\,+\,PY8} RadDn&  57.1 &  74.2 &  69.9 \\
{\sc Powheg\,+\,PY8} RadUp&  50.6 &  62.5 &  51.7 \\
{\sc Powheg\,+\,PY8} $\mu_\mathrm{F,R}\times 2$&  60.7 &  68.4 &  71.1 \\
{\sc Powheg\,+\,PY8} $\mu_\mathrm{F,R}\times 0.5$&  50.3 &  60.0 &  52.0 \\
{\sc Powheg\,+\,PY8} PDF4LHC15&  51.5 &  61.5 &  59.7 \\
{\sc Powheg\,+\,PY8} CT14&  50.6 &  67.3 &  60.0 \\
{\sc Powheg\,+\,PY8} MMHT&  53.7 &  57.9 &  58.7 \\
{\sc aMC@NLO\,+\,PY8}&  55.0 &  45.9 &  58.2 \\
{\sc aMC@NLO\,+\,PY8} CT10&  43.7 &  50.6 &  59.5 \\
{\sc aMC@NLO\,+\,PY8} HERA2& 130.3 &  97.6 &  58.0 \\
\hline
{\sc Powheg\,+\,PY8}& 0.026 & $4\cdot 10^{-8}$ & $5\cdot 10^{-3}$ \\
{\sc Powheg\,+\,PY6} CT10& 0.10 & 0& $1\cdot 10^{-4}$ \\
{\sc Powheg\,+\,HW7}& 0.055 & $2\cdot 10^{-7}$ & $8\cdot 10^{-3}$ \\
{\sc Powheg\,+\,PY8} \pt\ rew.& 0.084 & $2\cdot 10^{-5}$ & 0.088 \\
{\sc Powheg\,+\,PY8} RadDn& 0.011 & $2\cdot 10^{-8}$ & $2\cdot 10^{-3}$ \\
{\sc Powheg\,+\,PY8} RadUp& 0.042 & $2\cdot 10^{-6}$ & 0.083 \\
{\sc Powheg\,+\,PY8} $\mu_\mathrm{F,R}\times 2$& $5\cdot 10^{-3}$ & $2\cdot 10^{-7}$ & $1\cdot 10^{-3}$ \\
{\sc Powheg\,+\,PY8} $\mu_\mathrm{F,R}\times 0.5$& 0.045 & $4\cdot 10^{-6}$ & 0.079 \\
{\sc Powheg\,+\,PY8} PDF4LHC15& 0.036 & $2\cdot 10^{-6}$ & 0.018 \\
{\sc Powheg\,+\,PY8} CT14& 0.042 & $3\cdot 10^{-7}$ & 0.017 \\
{\sc Powheg\,+\,PY8} MMHT& 0.023 & $8\cdot 10^{-6}$ & 0.022 \\
{\sc aMC@NLO\,+\,PY8}& 0.017 & $5\cdot 10^{-4}$ & 0.024 \\
{\sc aMC@NLO\,+\,PY8} CT10& 0.15 & $1\cdot 10^{-4}$ & 0.019 \\
{\sc aMC@NLO\,+\,PY8} HERA2& 0& 0& 0.026 \\
\hline
\end{tabular}
\caption{\label{t:NorChiXPP8Data2016Data}$\chi^2$ values (top) and associated probabilities (bottom) for comparison of normalised measured double-differential
fiducial cross-sections with various \ttbar\ simulation samples.
Probabilities smaller than $10^{-10}$ are shown as zero.}
\end{table}

A number of observations can be made for the modelling of the individual
lepton and dilepton distributions. The single-lepton \pt\ and dilepton \ptsum\
distributions (Figures~\ref{f:rdistresa}(a) and~\ref{f:rdistresd}(a))
are softer in the data than in all the {\sc Powheg}-based
predictions, irrespective of the choice of parton shower, scale/tune settings
or PDF. The {\sc aMC@NLO\,+\,Pythia8} samples agree better with data,
especially when using the HERAPDF2.0 PDF set.
Reweighting the top quark \pt\ in the
{\sc Powheg\,+\,Pythia8} sample also gives significantly better agreement.
Similar features were seen in the comparisons of the \ptl\ and \ptsum\
distributions at \sxvt\ \cite{TOPQ-2015-02}, and in the \ptl\ distribution
measured by CMS at \sxyt\ in a different fiducial region including
requirements on jets \cite{CMS-TOP-17-014}.
 
The single-lepton \etal\ distribution (Figure~\ref{f:rdistresa}(b))
is more forward than the predictions
from either {\sc Powheg\,+\,Pythia8} or {\sc aMC@NLO\,+\,Pythia8} with the
NNPDF3.0 set, and agreement is improved by using CT10 or CT14. The MMHT
and PDF4LHC15 PDF sets lie somewhere in between, but HERAPDF2.0 predicts
much too central a distribution. The \rapll\ distribution (Figure~\ref{f:rdistresc}(a)) shows a slightly
different picture; again HERAPDF2.0 is in very poor agreement with the data,
but all the other PDFs do reasonably well. These observations differ from
those at \sxvt\ \cite{TOPQ-2015-02}, where the HERAPDF 1.5 and 2.0 PDF sets
were found to describe the data better than CT10, which was used as the default.
 
All the generators model the \ptll\ distribution well
(Figure~\ref{f:rdistresb}(a)), with the exception
of the {\sc Powheg\,+\,Pythia8} RadUp configuration, and to a lesser extent,
{\sc Powheg\,+\,Pythia8} with reduced QCD scales. This distribution shows
little sensitivity to PDFs. The \mll\ distribution
(Figure~\ref{f:rdistresb}(b)) is poorly modelled by
all {\sc Powheg}-based samples. The {\sc aMC@NLO\,+\,Pythia8} samples
do better (except when HERAPDF2.0 is used), but still fail to describe
the data at very low \mll.
 
The data have a less steep \dphill\ distribution than all the predictions
(Figure~\ref{f:rdistresc}(b)),
although the {\sc Powheg\,+\,Pythia8} RadUp and reduced QCD scale samples
come close, as does the sample with reweighted top quark \pt. The tensions
between data and predictions are smaller than in the dedicated ATLAS
\ttbar\ spin correlation analysis \cite{TOPQ-2016-10}, but the latter
analysis has a more restrictive fiducial region definition, with higher
lepton \pt\ thresholds and a requirement of at least two jets.
 
Finally, the \esum\ distribution (Figure~\ref{f:rdistresd}(b))
is reasonably described by the baseline
{\sc Powheg\,+\,Pythia8} prediction except at high \esum, where
agreement is improved by top quark \pt\ reweighting. The distribution shows some
sensitivity to PDFs, with NNPDF3.0 agreeing with data better than CT10,
and HERAPDF2.0 again agreeing very poorly with data.
 
The comparisons of normalised double-differential cross-section measurements
and predictions in Figures~\ref{f:rdistrese}--\ref{f:rdistresg} reflect
those seen in the single-differential results, although generally with
reduced significance due to the larger per-bin statistical uncertainties.
The $\chi^2$ and probabilities shown in
Table~\ref{t:NorChiXPP8Data2016Data} are all poor, driven by poor
agreement of the measured \mll\ distribution and predictions already
visible in Figure~\ref{f:rdistresb}(b).
The largest differences between the models
are seen at low \mll\ for \etal, whereas the differences become more
pronounced at high \mll\ for \rapll. Similar trends in \dphill\ are visible
in all \mll\ bins in Figure~\ref{f:rdistresg}, despite the shape of the
overall \dphill\ distribution changing significantly across the \mll\
bins, as shown in Figure~\ref{f:distresd}. This distribution is again
best described by the {\sc Powheg\,+\,Pythia8} predictions with
increased radiation (RadUp), reduced QCD scales, or reweighted top quark \pt.
 
The $\chi^2$ computation of Eq.~(\ref{e:redchi})
was extended to consider several normalised distributions simultaneously.
The statistical correlations between distributions were evaluated using
pseudo-experiments, and systematic uncertainties were assumed to be correlated
between distributions. Five sets of combined distributions were considered:
\ptl\ and \ptll; \ptll, \mll\ and \ptsum; \etal\ and \rapll; \etal, \rapll\
and \esum; and the combination of all eight single-differential
distributions.
 
\begin{table}[tp]
\small\centering
 
\begin{tabular}{l|ccccc}\hline
Generator & \ptl, \ptsum & \ptll, \mll, & \etal, \rapll & \etal, \rapll, & All 8 \\
&  & \ptsum &  & \esum & dists. \\
$N_\mathrm{dof}$ & 17 & 26 & 16 & 25 & 70 \\ \hline
{\sc Powheg\,+\,PY8}&  52.2 &  92.8 &  31.2 &  51.5 & 176.5 \\
{\sc Powheg\,+\,PY6} CT10&  42.9 &  87.9 &  31.0 &  58.0 & 176.6 \\
{\sc Powheg\,+\,HW7}&  42.5 &  97.4 &  25.7 &  41.6 & 169.8 \\
{\sc Powheg\,+\,PY8} \pt\ rew.&  27.5 &  57.4 &  25.4 &  36.5 & 137.6 \\
{\sc Powheg\,+\,PY8} RadDn&  49.7 & 110.8 &  37.8 &  58.3 & 193.9 \\
{\sc Powheg\,+\,PY8} RadUp&  42.9 &  71.8 &  25.5 &  44.2 & 151.8 \\
{\sc Powheg\,+\,PY8} $\mu_\mathrm{F,R}\times 2$&  54.5 & 111.1 &  35.6 &  54.4 & 195.0 \\
{\sc Powheg\,+\,PY8} $\mu_\mathrm{F,R}\times 0.5$&  50.5 &  71.3 &  26.3 &  42.8 & 160.4 \\
{\sc Powheg\,+\,PY8} PDF4LHC15&  52.2 &  89.7 &  26.7 &  44.1 & 167.1 \\
{\sc Powheg\,+\,PY8} CT14&  52.9 &  91.5 &  26.6 &  44.8 & 170.2 \\
{\sc Powheg\,+\,PY8} MMHT&  49.9 &  89.4 &  28.7 &  44.8 & 167.6 \\
{\sc aMC@NLO\,+\,PY8}&  33.2 &  46.3 &  37.1 &  49.6 & 131.9 \\
{\sc aMC@NLO\,+\,PY8} CT10&  31.6 &  46.7 &  26.2 &  43.0 & 122.9 \\
{\sc aMC@NLO\,+\,PY8} HERA2&  23.1 &  51.5 & 119.0 & 132.8 & 229.8 \\
\hline
{\sc Powheg\,+\,PY8}& $2\cdot 10^{-5}$ & $2\cdot 10^{-9}$ & 0.013 & $1\cdot 10^{-3}$ & 0 \\
{\sc Powheg\,+\,PY6} CT10& $5\cdot 10^{-4}$ & $1\cdot 10^{-8}$ & 0.014 & $2\cdot 10^{-4}$ & 0 \\
{\sc Powheg\,+\,HW7}& $6\cdot 10^{-4}$ & $3\cdot 10^{-10}$ & 0.058 & 0.020 & $3\cdot 10^{-10}$ \\
{\sc Powheg\,+\,PY8} \pt\ rew.& 0.052 & $4\cdot 10^{-4}$ & 0.062 & 0.064 & $3\cdot 10^{-6}$ \\
{\sc Powheg\,+\,PY8} RadDn& $5\cdot 10^{-5}$ & 0 & $2\cdot 10^{-3}$ & $2\cdot 10^{-4}$ & 0 \\
{\sc Powheg\,+\,PY8} RadUp& $5\cdot 10^{-4}$ & $4\cdot 10^{-6}$ & 0.062 & 0.010 & $6\cdot 10^{-8}$ \\
{\sc Powheg\,+\,PY8} $\mu_\mathrm{F,R}\times 2$& $8\cdot 10^{-6}$ & 0 & $3\cdot 10^{-3}$ & $6\cdot 10^{-4}$ & 0 \\
{\sc Powheg\,+\,PY8} $\mu_\mathrm{F,R}\times 0.5$& $4\cdot 10^{-5}$ & $4\cdot 10^{-6}$ & 0.049 & 0.015 & $5\cdot 10^{-9}$ \\
{\sc Powheg\,+\,PY8} PDF4LHC15& $2\cdot 10^{-5}$ & $6\cdot 10^{-9}$ & 0.045 & 0.011 & $7\cdot 10^{-10}$ \\
{\sc Powheg\,+\,PY8} CT14& $2\cdot 10^{-5}$ & $3\cdot 10^{-9}$ & 0.046 & $9\cdot 10^{-3}$ & $3\cdot 10^{-10}$ \\
{\sc Powheg\,+\,PY8} MMHT& $4\cdot 10^{-5}$ & $7\cdot 10^{-9}$ & 0.026 & $9\cdot 10^{-3}$ & $6\cdot 10^{-10}$ \\
{\sc aMC@NLO\,+\,PY8}& 0.011 & $9\cdot 10^{-3}$ & $2\cdot 10^{-3}$ & $2\cdot 10^{-3}$ & $1\cdot 10^{-5}$ \\
{\sc aMC@NLO\,+\,PY8} CT10& 0.017 & $8\cdot 10^{-3}$ & 0.051 & 0.014 & $1\cdot 10^{-4}$ \\
{\sc aMC@NLO\,+\,PY8} HERA2& 0.14 & $2\cdot 10^{-3}$ & 0 & 0 & 0 \\
\hline
\end{tabular}
\caption{\label{t:NorCombChiXPPData2016Data}$\chi^2$ values (top) and associated probabilities (bottom) for comparison of combinations of measured normalised differential fiducial cross-sections with various \ttbar\ simulation samples.
The last column gives the results for the combination of all eight measured
single-differential distributions.
Probabilities smaller than $10^{-10}$ are shown as zero.}
\end{table}
 
The resulting $\chi^2$ and $p$-values are shown for each combination and
prediction in Table~\ref{t:NorCombChiXPPData2016Data}.
The best descriptions of \pt\ and \ptsum\ are achieved by
{\sc Powheg\,+\,Pythia8} with top quark \pt\ reweighting, or by
{\sc aMC@NLO\,+\,Pythia8}, particularly with the HERAPDF2.0 PDF set.
Either NLO generator combined with  several PDF sets can describe the
\etal\ and \rapll\ distributions, although only the sample with top quark
\pt\ reweighting provides a reasonable description once \esum\ is also
included. No samples describe the combinations including \mll,
as this variable is not modelled by any of the generator configurations.
 
\FloatBarrier
\section{Conclusions}\label{s:conc}
 
The inclusive \ttbar\ production cross-section \xtt\ has been measured
in $pp$ collisions at \sxyt\ using \intlumi\,\ifb\ of data recorded
by the ATLAS experiment at the LHC in 2015--16. Using events with an
opposite-sign $e\mu$ pair and one or two $b$-tagged jets, the result is:
\begin{equation}\nonumber
\xtt = \ttxval \pm \ttxstat\,\mathrm{(stat)}\ \pm \ttxsyst\,\mathrm{(syst)}\ \pm \ttxlumi\,\mathrm{(lumi)}\ \pm \ttxebeam\,\mathrm{(beam)}\,\mathrm{pb},
\end{equation}
where the four uncertainties are due to data statistics, experimental and
theoretical systematic effects, and the knowledge of the integrated
luminosity and of the LHC beam energy. The
result is consistent with NNLO+NNLL QCD predictions. Fiducial cross-sections
corresponding to the experimental acceptance for the leptons, with and without
a correction for the contribution of leptons from leptonic $\tau$ decays,
have also been measured.
The dependence of predictions for \xtt\ on the top quark
pole mass \mtpole\ has been exploited to determine a mass value of
\begin{equation}\nonumber
\mtpole = \mtpval^{+\mtpeup}_{-\mtpedn}\,\GeV
\end{equation}
from the inclusive cross-section,
using the predictions derived with the CT14 PDF set. This result is compatible
with other top quark mass determinations using a variety of techniques.
The inclusive cross-section has also been combined with
previous  measurements at $\sqrt{s}=7$ and 8\,\TeV\ to determine ratios of
\ttbar\ cross-sections, and double ratios of \ttbar\ to $Z$ cross-sections,
at different energies, which are found to be
compatible with predictions using a range
of PDF sets.
 
The same data sample has been used to measure eight single-differential and
three double-differential cross-sections as a function of lepton and dilepton
kinematic variables, with uncertainties as small as 0.6\% for normalised
distributions in some parts of the fiducial region.
The measured distributions are generally well described by the NLO
matrix-element generators {\sc Powheg} and {\sc aMC@NLO} when interfaced to
{\sc Pythia} or {\sc Herwig} for parton shower, hadronisation and
underlying-event modelling. However, the {\sc Powheg}-based predictions give
lepton \pt\ spectra that are significantly harder than those observed in data,
and none of the predictions describe the low-mass part of the dilepton
invariant mass distribution. These differential cross-section
results have sensitivity to PDFs and can be used as the basis for a precise
determination of the top quark mass based on lepton kinematics.
 
\section*{Acknowledgements}
 
% The next lines are included from the .//acknowledgements/Acknowledgements.tex input file

We thank CERN for the very successful operation of the LHC, as well as the
support staff from our institutions without whom ATLAS could not be
operated efficiently.
 
We acknowledge the support of ANPCyT, Argentina; YerPhI, Armenia; ARC, Australia; BMWFW and FWF, Austria; ANAS, Azerbaijan; SSTC, Belarus; CNPq and FAPESP, Brazil; NSERC, NRC and CFI, Canada; CERN; CONICYT, Chile; CAS, MOST and NSFC, China; COLCIENCIAS, Colombia; MSMT CR, MPO CR and VSC CR, Czech Republic; DNRF and DNSRC, Denmark; IN2P3-CNRS and CEA-DRF/IRFU, France; SRNSFG, Georgia; BMBF, HGF and MPG, Germany; GSRT, Greece; RGC and Hong Kong SAR, China; ISF and Benoziyo Center, Israel; INFN, Italy; MEXT and JSPS, Japan; CNRST, Morocco; NWO, Netherlands; RCN, Norway; MNiSW and NCN, Poland; FCT, Portugal; MNE/IFA, Romania; MES of Russia and NRC KI, Russia Federation; JINR; MESTD, Serbia; MSSR, Slovakia; ARRS and MIZ\v{S}, Slovenia; DST/NRF, South Africa; MINECO, Spain; SRC and Wallenberg Foundation, Sweden; SERI, SNSF and Cantons of Bern and Geneva, Switzerland; MOST, Taiwan; TAEK, Turkey; STFC, United Kingdom; DOE and NSF, United States of America. In addition, individual groups and members have received support from BCKDF, CANARIE, Compute Canada and CRC, Canada; ERC, ERDF, Horizon 2020, Marie Sk{\l}odowska-Curie Actions and COST, European Union; Investissements d'Avenir Labex, Investissements d'Avenir Idex and ANR, France; DFG and AvH Foundation, Germany; Herakleitos, Thales and Aristeia programmes co-financed by EU-ESF and the Greek NSRF, Greece; BSF-NSF and GIF, Israel; CERCA Programme Generalitat de Catalunya and PROMETEO Programme Generalitat Valenciana, Spain; G\"{o}ran Gustafssons Stiftelse, Sweden; The Royal Society and Leverhulme Trust, United Kingdom.
 
The crucial computing support from all WLCG partners is acknowledged gratefully, in particular from CERN, the ATLAS Tier-1 facilities at TRIUMF (Canada), NDGF (Denmark, Norway, Sweden), CC-IN2P3 (France), KIT/GridKA (Germany), INFN-CNAF (Italy), NL-T1 (Netherlands), PIC (Spain), ASGC (Taiwan), RAL (UK) and BNL (USA), the Tier-2 facilities worldwide and large non-WLCG resource providers. Major contributors of computing resources are listed in Ref.~\cite{ATL-GEN-PUB-2016-002}.
 
% End of text imported from the .//acknowledgements/Acknowledgements.tex input file
 
\clearpage
\appendix
\part*{Appendix}
\addcontentsline{toc}{part}{Appendix}
 
The measured absolute and normalised differential cross-sections as functions
of individual lepton and dilepton variables are shown in
Tables~\ref{t:insXSec1}--\ref{t:insXSec4}.  The absolute and normalised
double-differential cross-sections as functions of \etal\ and \mll\
are shown in Tables~\ref{t:insXSec5} and~\ref{t:insXSec6}, those as a function
of \rapll\ and \mll\ in Tables~\ref{t:insXSec7} and~\ref{t:insXSec8},
and those as a function of \dphill\ and \mll\ in
Tables~\ref{t:insXSec9} and~\ref{t:insXSec10}. More details are given
in Section~\ref{s:diffres}.
 
\begin{table}
 
{\centering \small
\begin{tabular}{lrrrrrrrrr}
\hline
Absolute & $\mathrm{d}\sigma/\mathrm{d}\ptl$ & Stat. & \ttbar\ mod. & Lept. & Jet/$b$ & Bkg. & $L/E_\mathrm{b}$ & Total & $\mathrm{d}\sigma/\mathrm{d}\ptl$ (no $\tau$) \\
Bin [\GeV] & [fb/\GeV] & (\%) & (\%) & (\%) & (\%) & (\%) & (\%) & (\%) & [fb/\GeV] \\\hline
20--25 &$     564\pm      17$ & 0.7 &  0.9 &  1.4 &  0.2 &  1.0 &  2.3 &  3.1 & $     436\pm      13$\\
25--30 &$     562\pm      16$ & 0.6 &  0.9 &  1.1 &  0.2 &  0.8 &  2.3 &  2.9 & $     456\pm      13$\\
30--40 &$     525\pm      14$ & 0.4 &  0.9 &  0.7 &  0.2 &  0.7 &  2.3 &  2.7 & $     441\pm      12$\\
40--50 &$     428\pm      12$ & 0.5 &  0.9 &  0.6 &  0.2 &  0.7 &  2.3 &  2.7 & $     369\pm      10$\\
50--60 &$   336.2\pm     9.2$ & 0.5 &  0.9 &  0.6 &  0.3 &  0.7 &  2.3 &  2.7 & $   294.8\pm     7.8$\\
60--80 &$   220.6\pm     6.0$ & 0.4 &  1.0 &  0.6 &  0.2 &  0.7 &  2.3 &  2.7 & $   195.9\pm     5.2$\\
80--100 &$   120.1\pm     3.4$ & 0.6 &  1.1 &  0.7 &  0.2 &  0.8 &  2.3 &  2.9 & $   107.4\pm     3.0$\\
100--120 &$    66.6\pm     2.0$ & 0.8 &  1.1 &  0.7 &  0.2 &  1.0 &  2.3 &  3.0 & $    59.7\pm     1.7$\\
120--150 &$    30.6\pm     1.0$ & 1.0 &  1.2 &  0.8 &  0.3 &  1.5 &  2.4 &  3.3 & $    27.4\pm     0.9$\\
150--200 &$   10.80\pm    0.45$ & 1.4 &  1.2 &  1.0 &  0.2 &  2.6 &  2.5 &  4.1 & $    9.64\pm    0.39$\\
200--300+ &$    2.33\pm    0.20$ & 2.2 &  1.6 &  1.5 &  0.4 &  7.8 &  2.6 &  8.8 & $    2.07\pm    0.18$\\
\hline
Normalised & $\frac{1}{\sigma}\mathrm{d}\sigma/\mathrm{d}\ptl$ & Stat. & \ttbar\ mod. & Lept. & Jet/$b$ & Bkg. & $L/E_\mathrm{b}$ & Total & $\frac{1}{\sigma}\mathrm{d}\sigma/\mathrm{d}\ptl$ (no $\tau$) \\
Bin [\GeV] & [$10^{-2}/$\GeV] & (\%) & (\%) & (\%) & (\%) & (\%) & (\%) & (\%) & [$10^{-2}/$\GeV] \\\hline
20--25 &$   1.987\pm   0.026$ & 0.7 &  0.4 &  0.8 &  0.1 &  0.6 &  0.0 &  1.3 & $   1.796\pm   0.023$\\
25--30 &$   1.982\pm   0.019$ & 0.6 &  0.4 &  0.5 &  0.1 &  0.5 &  0.0 &  1.0 & $   1.876\pm   0.018$\\
30--40 &$   1.852\pm   0.012$ & 0.4 &  0.2 &  0.3 &  0.1 &  0.4 &  0.0 &  0.7 & $   1.817\pm   0.012$\\
40--50 &$  1.5095\pm  0.0096$ & 0.4 &  0.2 &  0.3 &  0.1 &  0.3 &  0.0 &  0.6 & $  1.5212\pm  0.0097$\\
50--60 &$  1.1855\pm  0.0079$ & 0.5 &  0.2 &  0.3 &  0.1 &  0.2 &  0.0 &  0.7 & $  1.2142\pm  0.0080$\\
60--80 &$  0.7779\pm  0.0047$ & 0.4 &  0.3 &  0.3 &  0.1 &  0.2 &  0.0 &  0.6 & $  0.8067\pm  0.0047$\\
80--100 &$  0.4234\pm  0.0035$ & 0.6 &  0.4 &  0.3 &  0.0 &  0.3 &  0.0 &  0.8 & $  0.4423\pm  0.0035$\\
100--120 &$  0.2348\pm  0.0028$ & 0.8 &  0.5 &  0.4 &  0.1 &  0.6 &  0.0 &  1.2 & $  0.2459\pm  0.0029$\\
120--150 &$  0.1080\pm  0.0019$ & 1.0 &  0.6 &  0.6 &  0.1 &  1.2 &  0.1 &  1.7 & $  0.1129\pm  0.0019$\\
150--200 &$  0.0381\pm  0.0012$ & 1.3 &  0.9 &  1.0 &  0.2 &  2.4 &  0.1 &  3.0 & $  0.0397\pm  0.0012$\\
200--300+ &$  0.0082\pm  0.0007$ & 2.2 &  1.6 &  1.6 &  0.4 &  7.6 &  0.3 &  8.2 & $  0.0085\pm  0.0007$\\
\hline
\end{tabular}
\vspace{3mm}
 
\begin{tabular}{lrrrrrrrrr}
\hline
Absolute & $\mathrm{d}\sigma/\mathrm{d}\etal$ & Stat. & \ttbar\ mod. & Lept. & Jet/$b$ & Bkg. & $L/E_\mathrm{b}$ & Total & $\mathrm{d}\sigma/\mathrm{d}\etal$ (no $\tau$) \\
Bin [unit $|\eta|$] & [fb/unit $|\eta|$] & (\%) & (\%) & (\%) & (\%) & (\%) & (\%) & (\%) & [fb/unit $|\eta|$] \\\hline
0.00--0.25 &$   17270\pm     460$ & 0.5 &  0.8 &  0.7 &  0.2 &  0.7 &  2.3 &  2.7 & $   14750\pm     390$\\
0.25--0.50 &$   16520\pm     440$ & 0.5 &  0.8 &  0.6 &  0.2 &  0.7 &  2.3 &  2.7 & $   14110\pm     370$\\
0.50--0.75 &$   15660\pm     420$ & 0.5 &  0.8 &  0.6 &  0.2 &  0.7 &  2.3 &  2.7 & $   13390\pm     350$\\
0.75--1.00 &$   14320\pm     390$ & 0.5 &  0.9 &  0.7 &  0.2 &  0.7 &  2.3 &  2.7 & $   12250\pm     330$\\
1.00--1.25 &$   12660\pm     350$ & 0.5 &  1.0 &  0.7 &  0.2 &  0.8 &  2.3 &  2.8 & $   10850\pm     290$\\
1.25--1.50 &$   10940\pm     310$ & 0.7 &  1.0 &  0.7 &  0.2 &  0.8 &  2.3 &  2.8 & $    9370\pm     260$\\
1.50--1.75 &$    9090\pm     260$ & 0.7 &  1.1 &  0.7 &  0.2 &  0.9 &  2.3 &  2.9 & $    7810\pm     220$\\
1.75--2.00 &$    7320\pm     220$ & 0.8 &  1.2 &  0.8 &  0.2 &  1.0 &  2.3 &  3.0 & $    6310\pm     180$\\
2.00--2.50 &$    4750\pm     150$ & 0.7 &  1.3 &  0.9 &  0.2 &  1.2 &  2.3 &  3.2 & $    4100\pm     130$\\
\hline
Normalised & $\frac{1}{\sigma}\mathrm{d}\sigma/\mathrm{d}\etal$ & Stat. & \ttbar\ mod. & Lept. & Jet/$b$ & Bkg. & $L/E_\mathrm{b}$ & Total & $\frac{1}{\sigma}\mathrm{d}\sigma/\mathrm{d}\etal$ (no $\tau$) \\
Bin [unit $|\eta|$] & [$10^{-1}/$unit $|\eta|$] & (\%) & (\%) & (\%) & (\%) & (\%) & (\%) & (\%) & [$10^{-1}/$unit $|\eta|$] \\\hline
0.00--0.25 &$   6.099\pm   0.041$ & 0.4 &  0.4 &  0.1 &  0.1 &  0.3 &  0.0 &  0.7 & $   6.082\pm   0.041$\\
0.25--0.50 &$   5.832\pm   0.035$ & 0.4 &  0.3 &  0.1 &  0.0 &  0.3 &  0.0 &  0.6 & $   5.816\pm   0.035$\\
0.50--0.75 &$   5.531\pm   0.031$ & 0.4 &  0.2 &  0.1 &  0.0 &  0.2 &  0.0 &  0.6 & $   5.521\pm   0.031$\\
0.75--1.00 &$   5.056\pm   0.028$ & 0.5 &  0.2 &  0.1 &  0.0 &  0.2 &  0.0 &  0.6 & $   5.049\pm   0.028$\\
1.00--1.25 &$   4.472\pm   0.027$ & 0.5 &  0.2 &  0.1 &  0.0 &  0.2 &  0.0 &  0.6 & $   4.473\pm   0.027$\\
1.25--1.50 &$   3.863\pm   0.030$ & 0.6 &  0.3 &  0.1 &  0.1 &  0.2 &  0.0 &  0.8 & $   3.863\pm   0.030$\\
1.50--1.75 &$   3.211\pm   0.028$ & 0.7 &  0.4 &  0.2 &  0.1 &  0.3 &  0.0 &  0.9 & $   3.217\pm   0.028$\\
1.75--2.00 &$   2.584\pm   0.027$ & 0.8 &  0.5 &  0.2 &  0.0 &  0.5 &  0.0 &  1.0 & $   2.599\pm   0.027$\\
2.00--2.50 &$   1.676\pm   0.020$ & 0.7 &  0.6 &  0.3 &  0.0 &  0.7 &  0.0 &  1.2 & $   1.690\pm   0.020$\\
\hline
\end{tabular}
\vspace{3mm}
 
}
\caption{\label{t:insXSec1}Absolute and normalised differential cross-sections as functions of \ptl\ (top) and \etal\ (bottom). The columns show the bin ranges, measured cross-section and total uncertainty, relative statistical uncertainty, relative systematic uncertainties in various categories (see text), total relative uncertainty, and differential cross-section corrected to remove contributions via $W\rightarrow\tau\rightarrow e/\mu$ decays. Relative uncertainties smaller than 0.05\% are indicated by `0.0'. The last bin includes overflows where indicated by the `+' sign.}
\end{table}
 
\begin{table}
\vspace{-5mm}
 
{\centering \small
\begin{tabular}{lrrrrrrrrr}
\hline
Absolute & $\mathrm{d}\sigma/\mathrm{d}\ptll$ & Stat. & \ttbar\ mod. & Lept. & Jet/$b$ & Bkg. & $L/E_\mathrm{b}$ & Total & $\mathrm{d}\sigma/\mathrm{d}\ptll$ (no $\tau$) \\
Bin [\GeV] & [fb/\GeV] & (\%) & (\%) & (\%) & (\%) & (\%) & (\%) & (\%) & [fb/\GeV] \\\hline
0--20 &$    45.1\pm     1.4$ & 1.1 &  1.2 &  0.7 &  0.3 &  0.9 &  2.3 &  3.1 & $    36.5\pm     1.1$\\
20--40 &$   110.1\pm     3.1$ & 0.7 &  1.0 &  0.7 &  0.3 &  0.8 &  2.3 &  2.9 & $    89.7\pm     2.5$\\
40--60 &$   159.8\pm     4.5$ & 0.6 &  1.0 &  0.8 &  0.3 &  0.8 &  2.3 &  2.8 & $   132.2\pm     3.6$\\
60--80 &$   156.2\pm     4.3$ & 0.6 &  1.0 &  0.7 &  0.2 &  0.7 &  2.3 &  2.8 & $   134.8\pm     3.6$\\
80--100 &$   110.1\pm     3.1$ & 0.7 &  1.1 &  0.7 &  0.2 &  0.7 &  2.3 &  2.8 & $    98.2\pm     2.7$\\
100--120 &$    62.6\pm     1.9$ & 0.8 &  1.2 &  0.7 &  0.2 &  0.9 &  2.3 &  3.0 & $    56.9\pm     1.6$\\
120--150 &$   26.67\pm    0.90$ & 1.1 &  1.4 &  0.8 &  0.3 &  1.4 &  2.4 &  3.4 & $   24.29\pm    0.80$\\
150--200 &$    7.15\pm    0.35$ & 1.7 &  2.0 &  1.0 &  0.3 &  3.2 &  2.5 &  5.0 & $    6.45\pm    0.32$\\
200--300+ &$    1.19\pm    0.17$ & 3.2 &  3.4 &  1.5 &  0.6 & 13.2 &  2.8 & 14.3 & $    1.06\pm    0.15$\\
\hline
Normalised & $\frac{1}{\sigma}\mathrm{d}\sigma/\mathrm{d}\ptll$ & Stat. & \ttbar\ mod. & Lept. & Jet/$b$ & Bkg. & $L/E_\mathrm{b}$ & Total & $\frac{1}{\sigma}\mathrm{d}\sigma/\mathrm{d}\ptll$ (no $\tau$) \\
Bin [\GeV] & [$10^{-2}/$\GeV] & (\%) & (\%) & (\%) & (\%) & (\%) & (\%) & (\%) & [$10^{-2}/$\GeV] \\\hline
0--20 &$  0.3189\pm  0.0051$ & 1.1 &  0.9 &  0.2 &  0.2 &  0.7 &  0.0 &  1.6 & $  0.3011\pm  0.0049$\\
20--40 &$  0.7776\pm  0.0082$ & 0.6 &  0.6 &  0.2 &  0.1 &  0.5 &  0.0 &  1.1 & $  0.7402\pm  0.0081$\\
40--60 &$  1.1286\pm  0.0088$ & 0.5 &  0.4 &  0.2 &  0.1 &  0.4 &  0.0 &  0.8 & $  1.0905\pm  0.0089$\\
60--80 &$  1.1035\pm  0.0070$ & 0.5 &  0.2 &  0.1 &  0.1 &  0.3 &  0.0 &  0.6 & $  1.1121\pm  0.0072$\\
80--100 &$  0.7780\pm  0.0058$ & 0.6 &  0.4 &  0.1 &  0.1 &  0.2 &  0.0 &  0.7 & $  0.8098\pm  0.0059$\\
100--120 &$  0.4425\pm  0.0051$ & 0.8 &  0.6 &  0.3 &  0.2 &  0.4 &  0.0 &  1.2 & $  0.4694\pm  0.0054$\\
120--150 &$  0.1884\pm  0.0035$ & 1.0 &  1.0 &  0.5 &  0.2 &  1.0 &  0.1 &  1.9 & $  0.2003\pm  0.0038$\\
150--200 &$  0.0505\pm  0.0020$ & 1.7 &  1.8 &  0.9 &  0.3 &  2.9 &  0.2 &  4.0 & $  0.0532\pm  0.0021$\\
200--300+ &$  0.0084\pm  0.0012$ & 3.2 &  3.4 &  1.5 &  0.5 & 13.0 &  0.5 & 13.9 & $  0.0087\pm  0.0012$\\
\hline
\end{tabular}
\vspace{3mm}
 
\begin{tabular}{lrrrrrrrrr}
\hline
Absolute & $\mathrm{d}\sigma/\mathrm{d}\mll$ & Stat. & \ttbar\ mod. & Lept. & Jet/$b$ & Bkg. & $L/E_\mathrm{b}$ & Total & $\mathrm{d}\sigma/\mathrm{d}\mll$ (no $\tau$) \\
Bin [\GeV] & [fb/\GeV] & (\%) & (\%) & (\%) & (\%) & (\%) & (\%) & (\%) & [fb/\GeV] \\\hline
0--20 &$   19.98\pm    0.74$ & 2.0 &  1.3 &  1.0 &  0.3 &  1.2 &  2.3 &  3.7 & $   16.57\pm    0.60$\\
20--40 &$    54.3\pm     1.6$ & 1.0 &  1.0 &  0.9 &  0.2 &  1.0 &  2.3 &  3.0 & $    45.6\pm     1.3$\\
40--60 &$    88.2\pm     2.6$ & 0.8 &  0.9 &  0.9 &  0.2 &  1.0 &  2.3 &  2.9 & $    73.1\pm     2.1$\\
60--80 &$   107.0\pm     3.0$ & 0.7 &  0.8 &  0.8 &  0.2 &  0.9 &  2.3 &  2.8 & $    88.9\pm     2.4$\\
80--100 &$   102.3\pm     2.9$ & 0.7 &  0.9 &  0.7 &  0.3 &  0.8 &  2.3 &  2.8 & $    86.4\pm     2.4$\\
100--120 &$    83.8\pm     2.3$ & 0.8 &  0.9 &  0.6 &  0.3 &  0.7 &  2.3 &  2.8 & $    71.7\pm     2.0$\\
120--150 &$    59.8\pm     1.7$ & 0.7 &  0.9 &  0.6 &  0.2 &  0.7 &  2.3 &  2.8 & $    51.9\pm     1.4$\\
150--200 &$   33.80\pm    0.97$ & 0.7 &  1.0 &  0.6 &  0.2 &  0.9 &  2.3 &  2.9 & $   29.96\pm    0.84$\\
200--250 &$   15.75\pm    0.51$ & 1.1 &  1.2 &  0.6 &  0.4 &  1.3 &  2.4 &  3.2 & $   14.21\pm    0.45$\\
250--300 &$    7.51\pm    0.29$ & 1.6 &  1.5 &  0.8 &  0.2 &  1.8 &  2.4 &  3.8 & $    6.85\pm    0.25$\\
300--400 &$    2.84\pm    0.13$ & 1.8 &  2.0 &  0.7 &  0.2 &  2.7 &  2.4 &  4.5 & $    2.62\pm    0.11$\\
400--500+ &$    1.48\pm    0.09$ & 2.7 &  2.7 &  0.9 &  0.3 &  3.7 &  2.4 &  5.9 & $    1.38\pm    0.08$\\
\hline
Normalised & $\frac{1}{\sigma}\mathrm{d}\sigma/\mathrm{d}\mll$ & Stat. & \ttbar\ mod. & Lept. & Jet/$b$ & Bkg. & $L/E_\mathrm{b}$ & Total & $\frac{1}{\sigma}\mathrm{d}\sigma/\mathrm{d}\mll$ (no $\tau$) \\
Bin [\GeV] & [$10^{-3}/$\GeV] & (\%) & (\%) & (\%) & (\%) & (\%) & (\%) & (\%) & [$10^{-3}/$\GeV] \\\hline
0--20 &$   1.408\pm   0.036$ & 2.0 &  1.1 &  0.4 &  0.2 &  1.0 &  0.0 &  2.5 & $   1.364\pm   0.034$\\
20--40 &$   3.826\pm   0.055$ & 1.0 &  0.7 &  0.3 &  0.2 &  0.8 &  0.0 &  1.4 & $   3.751\pm   0.053$\\
40--60 &$   6.218\pm   0.073$ & 0.7 &  0.5 &  0.3 &  0.1 &  0.7 &  0.0 &  1.2 & $   6.017\pm   0.069$\\
60--80 &$   7.541\pm   0.068$ & 0.6 &  0.4 &  0.2 &  0.1 &  0.5 &  0.0 &  0.9 & $   7.315\pm   0.066$\\
80--100 &$   7.210\pm   0.061$ & 0.6 &  0.3 &  0.1 &  0.3 &  0.3 &  0.0 &  0.8 & $   7.108\pm   0.060$\\
100--120 &$   5.907\pm   0.049$ & 0.7 &  0.3 &  0.1 &  0.1 &  0.2 &  0.0 &  0.8 & $   5.898\pm   0.049$\\
120--150 &$   4.212\pm   0.033$ & 0.7 &  0.3 &  0.2 &  0.1 &  0.2 &  0.0 &  0.8 & $   4.270\pm   0.034$\\
150--200 &$   2.382\pm   0.025$ & 0.7 &  0.5 &  0.2 &  0.1 &  0.5 &  0.0 &  1.0 & $   2.466\pm   0.024$\\
200--250 &$   1.110\pm   0.019$ & 1.1 &  0.8 &  0.3 &  0.2 &  1.1 &  0.0 &  1.8 & $   1.169\pm   0.019$\\
250--300 &$   0.529\pm   0.014$ & 1.6 &  1.2 &  0.7 &  0.1 &  1.6 &  0.1 &  2.6 & $   0.564\pm   0.014$\\
300--400 &$  0.2000\pm  0.0071$ & 1.8 &  1.6 &  0.6 &  0.1 &  2.6 &  0.1 &  3.6 & $  0.2154\pm  0.0073$\\
400--500+ &$  0.1041\pm  0.0053$ & 2.7 &  2.4 &  1.0 &  0.2 &  3.5 &  0.1 &  5.1 & $  0.1132\pm  0.0055$\\
\hline
\end{tabular}
}
\caption{\label{t:insXSec2}Absolute and normalised differential cross-sections as functions of \ptll\ (top) and \mll\ (bottom). The columns show the bin ranges, measured cross-section and total uncertainty, relative statistical uncertainty, relative systematic uncertainties in various categories (see text), total relative uncertainty, and differential cross-section corrected to remove contributions via $W\rightarrow\tau\rightarrow e/\mu$ decays. Relative uncertainties smaller than 0.05\% are indicated by `0.0'. The last bin includes overflows where indicated by the `+' sign.}
\end{table}
 
\begin{table}
 
{\centering \small
\begin{tabular}{lrrrrrrrrr}
\hline
Absolute & $\mathrm{d}\sigma/\mathrm{d}\rapll$ & Stat. & \ttbar\ mod. & Lept. & Jet/$b$ & Bkg. & $L/E_\mathrm{b}$ & Total & $\mathrm{d}\sigma/\mathrm{d}\rapll$ (no $\tau$) \\
Bin [unit $|y|$] & [fb/unit $|y|$] & (\%) & (\%) & (\%) & (\%) & (\%) & (\%) & (\%) & [fb/unit $|y|$] \\\hline
0.00--0.25 &$   10700\pm     290$ & 0.6 &  0.7 &  0.6 &  0.2 &  0.7 &  2.3 &  2.7 & $    9150\pm     240$\\
0.25--0.50 &$   10160\pm     270$ & 0.6 &  0.8 &  0.6 &  0.2 &  0.7 &  2.3 &  2.7 & $    8700\pm     230$\\
0.50--0.75 &$    9300\pm     260$ & 0.6 &  0.9 &  0.7 &  0.2 &  0.7 &  2.3 &  2.8 & $    7970\pm     210$\\
0.75--1.00 &$    8140\pm     230$ & 0.7 &  1.0 &  0.7 &  0.2 &  0.8 &  2.3 &  2.8 & $    6970\pm     190$\\
1.00--1.25 &$    6620\pm     190$ & 0.8 &  1.1 &  0.7 &  0.2 &  0.8 &  2.3 &  2.9 & $    5680\pm     160$\\
1.25--1.50 &$    5030\pm     150$ & 1.0 &  1.2 &  0.8 &  0.3 &  0.9 &  2.3 &  3.0 & $    4320\pm     130$\\
1.50--1.75 &$    3460\pm     110$ & 1.2 &  1.4 &  0.8 &  0.2 &  1.1 &  2.3 &  3.3 & $    2969\pm      96$\\
1.75--2.00 &$    2085\pm      76$ & 1.7 &  1.6 &  0.9 &  0.2 &  1.3 &  2.3 &  3.7 & $    1790\pm      65$\\
2.00--2.50 &$     545\pm      26$ & 2.6 &  2.1 &  1.1 &  0.3 &  2.2 &  2.4 &  4.8 & $     467\pm      22$\\
\hline
Normalised & $\frac{1}{\sigma}\mathrm{d}\sigma/\mathrm{d}\rapll$ & Stat. & \ttbar\ mod. & Lept. & Jet/$b$ & Bkg. & $L/E_\mathrm{b}$ & Total & $\frac{1}{\sigma}\mathrm{d}\sigma/\mathrm{d}\rapll$ (no $\tau$) \\
Bin [unit $|y|$] & [$10^{-1}/$unit $|y|$] & (\%) & (\%) & (\%) & (\%) & (\%) & (\%) & (\%) & [$10^{-1}/$unit $|y|$] \\\hline
0.00--0.25 &$   7.560\pm   0.056$ & 0.5 &  0.4 &  0.1 &  0.0 &  0.3 &  0.0 &  0.7 & $   7.550\pm   0.056$\\
0.25--0.50 &$   7.184\pm   0.050$ & 0.6 &  0.3 &  0.1 &  0.0 &  0.3 &  0.0 &  0.7 & $   7.182\pm   0.050$\\
0.50--0.75 &$   6.574\pm   0.045$ & 0.6 &  0.3 &  0.1 &  0.0 &  0.2 &  0.0 &  0.7 & $   6.574\pm   0.045$\\
0.75--1.00 &$   5.752\pm   0.042$ & 0.6 &  0.3 &  0.1 &  0.1 &  0.2 &  0.0 &  0.7 & $   5.748\pm   0.043$\\
1.00--1.25 &$   4.679\pm   0.041$ & 0.8 &  0.4 &  0.1 &  0.1 &  0.2 &  0.0 &  0.9 & $   4.688\pm   0.041$\\
1.25--1.50 &$   3.558\pm   0.040$ & 0.9 &  0.5 &  0.2 &  0.1 &  0.4 &  0.0 &  1.1 & $   3.562\pm   0.040$\\
1.50--1.75 &$   2.449\pm   0.037$ & 1.2 &  0.7 &  0.2 &  0.1 &  0.6 &  0.0 &  1.5 & $   2.449\pm   0.037$\\
1.75--2.00 &$   1.474\pm   0.031$ & 1.7 &  0.9 &  0.3 &  0.2 &  0.9 &  0.0 &  2.1 & $   1.477\pm   0.032$\\
2.00--2.50 &$   0.385\pm   0.013$ & 2.6 &  1.4 &  0.6 &  0.1 &  1.8 &  0.1 &  3.5 & $   0.386\pm   0.014$\\
\hline
\end{tabular}
\vspace{3mm}
 
\begin{tabular}{lrrrrrrrrr}
\hline
Absolute & $\mathrm{d}\sigma/\mathrm{d}\dphill$ & Stat. & \ttbar\ mod. & Lept. & Jet/$b$ & Bkg. & $L/E_\mathrm{b}$ & Total & $\mathrm{d}\sigma/\mathrm{d}\dphill$ (no $\tau$) \\
Bin [rad] & [fb/rad] & (\%) & (\%) & (\%) & (\%) & (\%) & (\%) & (\%) & [fb/rad] \\\hline
0.00--0.31 &$    3250\pm     110$ & 1.1 &  2.0 &  0.8 &  0.2 &  0.8 &  2.3 &  3.4 & $    2847\pm      90$\\
0.31--0.63 &$    3280\pm     110$ & 1.0 &  1.7 &  0.8 &  0.1 &  0.8 &  2.3 &  3.2 & $    2882\pm      87$\\
0.63--0.94 &$    3370\pm     100$ & 1.0 &  1.4 &  0.8 &  0.3 &  0.8 &  2.3 &  3.1 & $    2965\pm      88$\\
0.94--1.26 &$    3680\pm     110$ & 0.9 &  1.2 &  0.7 &  0.2 &  0.8 &  2.3 &  3.0 & $    3219\pm      93$\\
1.26--1.57 &$    4000\pm     120$ & 0.9 &  1.0 &  0.7 &  0.2 &  0.8 &  2.3 &  2.9 & $    3476\pm      98$\\
1.57--1.88 &$    4460\pm     130$ & 0.8 &  0.9 &  0.7 &  0.2 &  0.8 &  2.3 &  2.8 & $    3850\pm     110$\\
1.88--2.20 &$    4980\pm     140$ & 0.8 &  1.0 &  0.7 &  0.3 &  0.8 &  2.3 &  2.8 & $    4260\pm     120$\\
2.20--2.51 &$    5610\pm     160$ & 0.7 &  1.1 &  0.6 &  0.3 &  0.7 &  2.3 &  2.9 & $    4740\pm     130$\\
2.51--2.83 &$    6030\pm     180$ & 0.7 &  1.3 &  0.6 &  0.3 &  0.8 &  2.3 &  3.0 & $    5060\pm     150$\\
2.83--3.14 &$    6420\pm     200$ & 0.7 &  1.6 &  0.6 &  0.3 &  0.8 &  2.3 &  3.1 & $    5350\pm     170$\\
\hline
Normalised & $\frac{1}{\sigma}\mathrm{d}\sigma/\mathrm{d}\dphill$ & Stat. & \ttbar\ mod. & Lept. & Jet/$b$ & Bkg. & $L/E_\mathrm{b}$ & Total & $\frac{1}{\sigma}\mathrm{d}\sigma/\mathrm{d}\dphill$ (no $\tau$) \\
Bin [rad] & [$10^{-1}/$rad] & (\%) & (\%) & (\%) & (\%) & (\%) & (\%) & (\%) & [$10^{-1}/$rad] \\\hline
0.00--0.31 &$   2.298\pm   0.050$ & 1.0 &  1.9 &  0.1 &  0.2 &  0.4 &  0.0 &  2.2 & $   2.345\pm   0.044$\\
0.31--0.63 &$   2.316\pm   0.043$ & 0.9 &  1.5 &  0.1 &  0.2 &  0.3 &  0.0 &  1.8 & $   2.374\pm   0.040$\\
0.63--0.94 &$   2.380\pm   0.037$ & 0.9 &  1.2 &  0.1 &  0.2 &  0.3 &  0.0 &  1.6 & $   2.442\pm   0.036$\\
0.94--1.26 &$   2.600\pm   0.033$ & 0.9 &  0.9 &  0.1 &  0.1 &  0.3 &  0.0 &  1.3 & $   2.651\pm   0.034$\\
1.26--1.57 &$   2.823\pm   0.029$ & 0.8 &  0.5 &  0.1 &  0.1 &  0.2 &  0.0 &  1.0 & $   2.864\pm   0.031$\\
1.57--1.88 &$   3.149\pm   0.028$ & 0.8 &  0.3 &  0.0 &  0.1 &  0.2 &  0.0 &  0.9 & $   3.172\pm   0.029$\\
1.88--2.20 &$   3.516\pm   0.031$ & 0.7 &  0.4 &  0.0 &  0.1 &  0.2 &  0.0 &  0.9 & $   3.505\pm   0.030$\\
2.20--2.51 &$   3.960\pm   0.040$ & 0.7 &  0.7 &  0.1 &  0.1 &  0.2 &  0.0 &  1.0 & $   3.908\pm   0.038$\\
2.51--2.83 &$   4.255\pm   0.054$ & 0.7 &  1.0 &  0.1 &  0.2 &  0.3 &  0.0 &  1.3 & $   4.164\pm   0.052$\\
2.83--3.14 &$   4.533\pm   0.070$ & 0.7 &  1.4 &  0.1 &  0.1 &  0.3 &  0.0 &  1.6 & $   4.405\pm   0.071$\\
\hline
\end{tabular}
\vspace{3mm}
 
}
\caption{\label{t:insXSec3}Absolute and normalised differential cross-sections as functions of \rapll\ (top) and \dphill\ (bottom). The columns show the bin ranges, measured cross-section and total uncertainty, relative statistical uncertainty, relative systematic uncertainties in various categories (see text), total relative uncertainty, and differential cross-section corrected to remove contributions via $W\rightarrow\tau\rightarrow e/\mu$ decays. Relative uncertainties smaller than 0.05\% are indicated by `0.0'. The bin boundaries for \dphill\ correspond to exact multiples of $\pi/10$ but are quoted to two decimal places.}
\end{table}
 
\begin{table}
 
{\centering \small
\begin{tabular}{lrrrrrrrrr}
\hline
Absolute & $\mathrm{d}\sigma/\mathrm{d}(\ptsum)$ & Stat. & \ttbar\ mod. & Lept. & Jet/$b$ & Bkg. & $L/E_\mathrm{b}$ & Total & $\mathrm{d}\sigma/\mathrm{d}(\ptsum)$ (no $\tau$) \\
Bin [\GeV] & [fb/\GeV] & (\%) & (\%) & (\%) & (\%) & (\%) & (\%) & (\%) & [fb/\GeV] \\\hline
40--80 &$    96.1\pm     2.8$ & 0.6 &  0.9 &  1.0 &  0.2 &  0.9 &  2.3 &  2.9 & $    76.1\pm     2.2$\\
80--100 &$   148.2\pm     4.1$ & 0.6 &  1.0 &  0.7 &  0.2 &  0.8 &  2.3 &  2.8 & $   126.3\pm     3.4$\\
100--120 &$   120.9\pm     3.4$ & 0.6 &  1.0 &  0.7 &  0.2 &  0.7 &  2.3 &  2.8 & $   105.9\pm     2.9$\\
120--150 &$    76.5\pm     2.1$ & 0.6 &  1.0 &  0.6 &  0.2 &  0.8 &  2.3 &  2.8 & $    68.1\pm     1.8$\\
150--200 &$    33.7\pm     1.0$ & 0.7 &  1.1 &  0.7 &  0.3 &  1.1 &  2.4 &  3.0 & $    30.3\pm     0.9$\\
200--250 &$   11.77\pm    0.43$ & 1.3 &  1.2 &  0.8 &  0.3 &  2.0 &  2.4 &  3.7 & $   10.65\pm    0.38$\\
250--300 &$    4.57\pm    0.23$ & 2.0 &  1.4 &  1.0 &  0.3 &  3.3 &  2.5 &  4.9 & $    4.18\pm    0.20$\\
300--400+ &$    1.82\pm    0.14$ & 2.4 &  1.6 &  1.5 &  0.6 &  6.7 &  2.6 &  7.9 & $    1.68\pm    0.13$\\
\hline
Normalised & $\frac{1}{\sigma}\mathrm{d}\sigma/\mathrm{d}(\ptsum)$ & Stat. & \ttbar\ mod. & Lept. & Jet/$b$ & Bkg. & $L/E_\mathrm{b}$ & Total & $\frac{1}{\sigma}\mathrm{d}\sigma/\mathrm{d}(\ptsum)$ (no $\tau$) \\
Bin [\GeV] & [$10^{-2}/$\GeV] & (\%) & (\%) & (\%) & (\%) & (\%) & (\%) & (\%) & [$10^{-2}/$\GeV] \\\hline
40--80 &$  0.6764\pm  0.0066$ & 0.5 &  0.3 &  0.4 &  0.0 &  0.7 &  0.0 &  1.0 & $  0.6263\pm  0.0063$\\
80--100 &$  1.0429\pm  0.0079$ & 0.5 &  0.2 &  0.2 &  0.0 &  0.4 &  0.0 &  0.8 & $  1.0393\pm  0.0080$\\
100--120 &$  0.8512\pm  0.0062$ & 0.6 &  0.2 &  0.2 &  0.1 &  0.3 &  0.0 &  0.7 & $  0.8717\pm  0.0064$\\
120--150 &$  0.5387\pm  0.0038$ & 0.6 &  0.3 &  0.2 &  0.0 &  0.2 &  0.0 &  0.7 & $  0.5601\pm  0.0039$\\
150--200 &$  0.2373\pm  0.0027$ & 0.7 &  0.4 &  0.4 &  0.2 &  0.7 &  0.0 &  1.1 & $  0.2490\pm  0.0027$\\
200--250 &$  0.0829\pm  0.0020$ & 1.2 &  0.7 &  0.7 &  0.1 &  1.8 &  0.1 &  2.4 & $  0.0876\pm  0.0020$\\
250--300 &$  0.0322\pm  0.0013$ & 2.0 &  1.1 &  1.0 &  0.3 &  3.2 &  0.1 &  4.1 & $  0.0344\pm  0.0014$\\
300--400+ &$  0.0128\pm  0.0009$ & 2.4 &  1.5 &  1.6 &  0.5 &  6.6 &  0.3 &  7.3 & $  0.0138\pm  0.0010$\\
\hline
\end{tabular}
\vspace{3mm}
 
\begin{tabular}{lrrrrrrrrr}
\hline
Absolute & $\mathrm{d}\sigma/\mathrm{d}(\esum)$ & Stat. & \ttbar\ mod. & Lept. & Jet/$b$ & Bkg. & $L/E_\mathrm{b}$ & Total & $\mathrm{d}\sigma/\mathrm{d}(\esum)$ (no $\tau$) \\
Bin [\GeV] & [fb/\GeV] & (\%) & (\%) & (\%) & (\%) & (\%) & (\%) & (\%) & [fb/\GeV] \\\hline
40--80 &$   19.82\pm    0.63$ & 1.3 &  0.8 &  1.2 &  0.2 &  1.1 &  2.3 &  3.2 & $   14.95\pm    0.47$\\
80--100 &$    58.8\pm     1.7$ & 1.0 &  0.7 &  0.9 &  0.2 &  0.9 &  2.3 &  2.9 & $    47.3\pm     1.4$\\
100--120 &$    71.5\pm     2.0$ & 0.8 &  0.8 &  0.8 &  0.2 &  0.8 &  2.3 &  2.8 & $    59.4\pm     1.6$\\
120--150 &$    71.3\pm     2.0$ & 0.7 &  0.8 &  0.7 &  0.2 &  0.8 &  2.3 &  2.8 & $    60.5\pm     1.6$\\
150--200 &$    57.7\pm     1.6$ & 0.6 &  0.9 &  0.7 &  0.2 &  0.7 &  2.3 &  2.7 & $    49.8\pm     1.3$\\
200--250 &$    39.1\pm     1.1$ & 0.7 &  1.0 &  0.7 &  0.3 &  0.7 &  2.3 &  2.8 & $    34.2\pm     0.9$\\
250--300 &$   25.49\pm    0.76$ & 0.9 &  1.2 &  0.6 &  0.2 &  0.9 &  2.3 &  3.0 & $   22.57\pm    0.65$\\
300--400 &$   13.72\pm    0.44$ & 0.9 &  1.3 &  0.6 &  0.2 &  1.3 &  2.4 &  3.2 & $   12.24\pm    0.38$\\
400--500 &$    5.92\pm    0.22$ & 1.4 &  1.6 &  0.8 &  0.2 &  1.9 &  2.4 &  3.8 & $    5.33\pm    0.19$\\
500--700+ &$    2.66\pm    0.13$ & 1.5 &  1.7 &  1.0 &  0.3 &  3.4 &  2.4 &  4.9 & $    2.42\pm    0.12$\\
\hline
Normalised & $\frac{1}{\sigma}\mathrm{d}\sigma/\mathrm{d}(\esum)$ & Stat. & \ttbar\ mod. & Lept. & Jet/$b$ & Bkg. & $L/E_\mathrm{b}$ & Total & $\frac{1}{\sigma}\mathrm{d}\sigma/\mathrm{d}(\esum)$ (no $\tau$) \\
Bin [\GeV] & [$10^{-3}/$\GeV] & (\%) & (\%) & (\%) & (\%) & (\%) & (\%) & (\%) & [$10^{-3}/$\GeV] \\\hline
40--80 &$   1.401\pm   0.026$ & 1.2 &  0.8 &  0.6 &  0.1 &  1.0 &  0.1 &  1.9 & $   1.234\pm   0.023$\\
80--100 &$   4.157\pm   0.056$ & 0.9 &  0.6 &  0.3 &  0.1 &  0.7 &  0.0 &  1.3 & $   3.905\pm   0.053$\\
100--120 &$   5.054\pm   0.057$ & 0.8 &  0.5 &  0.2 &  0.1 &  0.6 &  0.0 &  1.1 & $   4.900\pm   0.056$\\
120--150 &$   5.039\pm   0.045$ & 0.6 &  0.3 &  0.1 &  0.1 &  0.5 &  0.0 &  0.9 & $   4.995\pm   0.045$\\
150--200 &$   4.076\pm   0.027$ & 0.5 &  0.2 &  0.1 &  0.0 &  0.3 &  0.0 &  0.7 & $   4.107\pm   0.027$\\
200--250 &$   2.765\pm   0.021$ & 0.7 &  0.3 &  0.1 &  0.1 &  0.2 &  0.0 &  0.8 & $   2.826\pm   0.022$\\
250--300 &$   1.802\pm   0.019$ & 0.8 &  0.5 &  0.2 &  0.1 &  0.4 &  0.0 &  1.1 & $   1.863\pm   0.020$\\
300--400 &$   0.970\pm   0.014$ & 0.8 &  0.6 &  0.3 &  0.2 &  0.9 &  0.0 &  1.4 & $   1.010\pm   0.014$\\
400--500 &$  0.4187\pm  0.0097$ & 1.3 &  0.9 &  0.6 &  0.1 &  1.5 &  0.1 &  2.3 & $  0.4397\pm  0.0099$\\
500--700+ &$  0.1879\pm  0.0070$ & 1.5 &  1.2 &  0.9 &  0.1 &  3.1 &  0.1 &  3.7 & $  0.1998\pm  0.0074$\\
\hline
\end{tabular}
\vspace{3mm}
 
}
\caption{\label{t:insXSec4}Absolute and normalised differential cross-sections as functions of \ptsum\ (top) and \esum\ (bottom). The columns show the bin ranges, measured cross-section and total uncertainty, relative statistical uncertainty, relative systematic uncertainties in various categories (see text), total relative uncertainty, and differential cross-section corrected to remove contributions via $W\rightarrow\tau\rightarrow e/\mu$ decays. Relative uncertainties smaller than 0.05\% are indicated by `0.0'. The last bin includes overflows where indicated by the `+' sign.}
\end{table}
 
\begin{table}
 
{\centering \small
\begin{tabular}{lrrrrrrrrr}
\hline
Absolute & $\mathrm{d^2}\sigma/\mathrm{d}\etal\mathrm{d}\mll$ & Stat. & \ttbar\ mod. & Lept. & Jet/$b$ & Bkg. & $L/E_\mathrm{b}$ & Total & $\mathrm{d^2}\sigma/\mathrm{d}\etal\mathrm{d}\mll$ \\
& & & & & & & & & (no $\tau$) \\
Bin [unit $|\eta|$ \GeV] & [fb/unit $|\eta|$ \GeV] & (\%) & (\%) & (\%) & (\%) & (\%) & (\%) & (\%) & [fb/unit $|\eta|$ \GeV] \\\hline
\multicolumn{10}{l}{$0 < \mll < 80$\,\GeV} \\
0.00--0.25 &$    87.2\pm     2.5$ & 0.8 &  0.7 &  0.8 &  0.1 &  1.0 &  2.3 &  2.8 & $    72.3\pm     2.0$\\
0.25--0.50 &$    82.6\pm     2.3$ & 0.8 &  0.8 &  0.8 &  0.1 &  0.9 &  2.3 &  2.8 & $    68.5\pm     1.9$\\
0.50--0.75 &$    78.2\pm     2.2$ & 0.8 &  0.9 &  0.8 &  0.2 &  0.8 &  2.3 &  2.9 & $    65.0\pm     1.8$\\
0.75--1.00 &$    70.2\pm     2.1$ & 0.9 &  1.1 &  0.9 &  0.2 &  0.9 &  2.3 &  3.0 & $    58.3\pm     1.7$\\
1.00--1.25 &$    60.0\pm     1.9$ & 0.9 &  1.3 &  0.9 &  0.2 &  0.9 &  2.3 &  3.1 & $    50.0\pm     1.5$\\
1.25--1.50 &$    51.5\pm     1.7$ & 1.2 &  1.5 &  0.9 &  0.1 &  1.0 &  2.3 &  3.3 & $    42.9\pm     1.3$\\
1.50--1.75 &$    41.0\pm     1.4$ & 1.3 &  1.7 &  1.0 &  0.2 &  1.0 &  2.3 &  3.4 & $    34.2\pm     1.1$\\
1.75--2.00 &$    31.4\pm     1.2$ & 1.5 &  1.8 &  1.0 &  0.2 &  1.2 &  2.3 &  3.7 & $    26.2\pm     0.9$\\
2.00--2.50 &$   18.45\pm    0.73$ & 1.5 &  2.2 &  1.1 &  0.3 &  1.5 &  2.3 &  4.0 & $   15.42\pm    0.57$\\
\hline
\multicolumn{10}{l}{$80 < \mll < 120$\,\GeV} \\
0.00--0.25 &$   120.6\pm     3.4$ & 0.9 &  0.9 &  0.7 &  0.3 &  0.8 &  2.3 &  2.9 & $   102.2\pm     3.0$\\
0.25--0.50 &$   113.5\pm     3.2$ & 0.9 &  0.9 &  0.6 &  0.3 &  0.7 &  2.3 &  2.8 & $    96.3\pm     2.7$\\
0.50--0.75 &$   105.5\pm     3.0$ & 0.9 &  0.8 &  0.6 &  0.3 &  0.7 &  2.3 &  2.8 & $    89.5\pm     2.5$\\
0.75--1.00 &$    93.2\pm     2.7$ & 1.0 &  0.9 &  0.7 &  0.3 &  0.7 &  2.3 &  2.9 & $    79.1\pm     2.3$\\
1.00--1.25 &$    82.3\pm     2.4$ & 1.1 &  0.9 &  0.7 &  0.3 &  0.7 &  2.3 &  2.9 & $    69.9\pm     2.0$\\
1.25--1.50 &$    69.5\pm     2.1$ & 1.3 &  1.0 &  0.7 &  0.3 &  0.8 &  2.3 &  3.0 & $    58.9\pm     1.8$\\
1.50--1.75 &$    57.1\pm     1.8$ & 1.4 &  1.0 &  0.7 &  0.3 &  0.9 &  2.3 &  3.1 & $    48.4\pm     1.5$\\
1.75--2.00 &$    45.0\pm     1.5$ & 1.6 &  1.0 &  0.8 &  0.3 &  1.0 &  2.3 &  3.2 & $    38.4\pm     1.2$\\
2.00--2.50 &$   28.85\pm    0.97$ & 1.5 &  1.1 &  0.9 &  0.3 &  1.3 &  2.3 &  3.4 & $   24.53\pm    0.83$\\
\hline
\multicolumn{10}{l}{$120 < \mll < 200$\,\GeV} \\
0.00--0.25 &$    50.6\pm     1.6$ & 0.9 &  1.7 &  0.6 &  0.3 &  0.8 &  2.3 &  3.2 & $    44.5\pm     1.3$\\
0.25--0.50 &$    48.4\pm     1.5$ & 0.9 &  1.6 &  0.5 &  0.2 &  0.8 &  2.3 &  3.1 & $    42.5\pm     1.3$\\
0.50--0.75 &$    46.7\pm     1.4$ & 0.9 &  1.5 &  0.6 &  0.2 &  0.7 &  2.3 &  3.1 & $    40.9\pm     1.2$\\
0.75--1.00 &$    44.0\pm     1.3$ & 1.0 &  1.4 &  0.6 &  0.2 &  0.7 &  2.3 &  3.0 & $    38.6\pm     1.1$\\
1.00--1.25 &$    38.9\pm     1.2$ & 1.1 &  1.3 &  0.6 &  0.3 &  0.8 &  2.3 &  3.1 & $    34.1\pm     1.0$\\
1.25--1.50 &$    34.3\pm     1.1$ & 1.3 &  1.3 &  0.6 &  0.2 &  0.8 &  2.3 &  3.2 & $    30.0\pm     0.9$\\
1.50--1.75 &$   28.79\pm    0.92$ & 1.3 &  1.3 &  0.7 &  0.3 &  1.0 &  2.3 &  3.2 & $   25.20\pm    0.79$\\
1.75--2.00 &$   23.97\pm    0.78$ & 1.4 &  1.3 &  0.7 &  0.2 &  1.0 &  2.3 &  3.2 & $   20.98\pm    0.67$\\
2.00--2.50 &$   16.46\pm    0.54$ & 1.3 &  1.2 &  0.8 &  0.2 &  1.3 &  2.3 &  3.3 & $   14.43\pm    0.47$\\
\hline
\multicolumn{10}{l}{$200 < \mll < 500+$\,\GeV} \\
0.00--0.25 &$    4.90\pm    0.18$ & 1.6 &  1.3 &  0.6 &  0.3 &  1.8 &  2.4 &  3.7 & $    4.47\pm    0.17$\\
0.25--0.50 &$    5.09\pm    0.19$ & 1.5 &  1.2 &  0.6 &  0.3 &  1.9 &  2.4 &  3.7 & $    4.65\pm    0.17$\\
0.50--0.75 &$    4.95\pm    0.18$ & 1.5 &  1.2 &  0.6 &  0.3 &  2.0 &  2.4 &  3.7 & $    4.52\pm    0.17$\\
0.75--1.00 &$    4.93\pm    0.18$ & 1.5 &  1.2 &  0.6 &  0.4 &  1.9 &  2.4 &  3.7 & $    4.50\pm    0.16$\\
1.00--1.25 &$    4.91\pm    0.18$ & 1.5 &  1.3 &  0.6 &  0.2 &  1.9 &  2.4 &  3.7 & $    4.46\pm    0.16$\\
1.25--1.50 &$    4.42\pm    0.18$ & 1.9 &  1.5 &  0.7 &  0.6 &  2.0 &  2.4 &  4.0 & $    4.01\pm    0.16$\\
1.50--1.75 &$    4.18\pm    0.16$ & 1.7 &  1.5 &  0.7 &  0.3 &  1.8 &  2.3 &  3.9 & $    3.80\pm    0.14$\\
1.75--2.00 &$    3.70\pm    0.15$ & 1.9 &  1.8 &  0.7 &  0.3 &  1.9 &  2.3 &  4.0 & $    3.36\pm    0.13$\\
2.00--2.50 &$    2.71\pm    0.11$ & 1.6 &  2.0 &  0.8 &  0.3 &  1.9 &  2.3 &  4.0 & $    2.46\pm    0.10$\\
\hline
\end{tabular}
\vspace{3mm}
 
}
\caption{\label{t:insXSec5}Absolute differential cross-sections as a function of \etalvmll. The columns show the bin ranges, measured cross-section and total uncertainty, relative statistical uncertainty, relative systematic uncertainties in various categories (see text), total relative uncertainty, and differential cross-section corrected to remove contributions via $W\rightarrow\tau\rightarrow e/\mu$ decays. Relative uncertainties smaller than 0.05\% are indicated by `0.0'. The last bin includes overflows where indicated by the `+' sign.}
\end{table}
 
\begin{table}
 
{\centering \small
\begin{tabular}{lrrrrrrrrr}
\hline
Normalised & $\frac{1}{\sigma}\mathrm{d^2}\sigma/\mathrm{d}\etal\mathrm{d}\mll$ & Stat. & \ttbar\ mod. & Lept. & Jet/$b$ & Bkg. & $L/E_\mathrm{b}$ & Total & $\frac{1}{\sigma}\mathrm{d^2}\sigma/\mathrm{d}\etal\mathrm{d}\mll$ \\
& & & & & & & & & (no $\tau$) \\
Bin [unit $|\eta|$ \GeV] & [$10^{-3}/$unit $|\eta|$ \GeV] & (\%) & (\%) & (\%) & (\%) & (\%) & (\%) & (\%) & [$10^{-3}/$unit $|\eta|$ \GeV] \\\hline
\multicolumn{10}{l}{$0 < \mll < 80$\,\GeV} \\
0.00--0.25 &$   3.072\pm   0.042$ & 0.8 &  0.8 &  0.2 &  0.1 &  0.8 &  0.0 &  1.4 & $   2.977\pm   0.039$\\
0.25--0.50 &$   2.909\pm   0.037$ & 0.7 &  0.7 &  0.2 &  0.1 &  0.7 &  0.0 &  1.3 & $   2.819\pm   0.034$\\
0.50--0.75 &$   2.757\pm   0.034$ & 0.8 &  0.7 &  0.2 &  0.1 &  0.6 &  0.0 &  1.2 & $   2.675\pm   0.031$\\
0.75--1.00 &$   2.473\pm   0.032$ & 0.8 &  0.8 &  0.2 &  0.1 &  0.6 &  0.0 &  1.3 & $   2.398\pm   0.030$\\
1.00--1.25 &$   2.115\pm   0.030$ & 0.9 &  0.9 &  0.3 &  0.1 &  0.6 &  0.0 &  1.4 & $   2.057\pm   0.028$\\
1.25--1.50 &$   1.815\pm   0.031$ & 1.2 &  1.0 &  0.3 &  0.1 &  0.6 &  0.0 &  1.7 & $   1.765\pm   0.029$\\
1.50--1.75 &$   1.444\pm   0.028$ & 1.2 &  1.2 &  0.4 &  0.1 &  0.7 &  0.0 &  1.9 & $   1.406\pm   0.025$\\
1.75--2.00 &$   1.106\pm   0.025$ & 1.5 &  1.4 &  0.4 &  0.1 &  0.9 &  0.0 &  2.2 & $   1.079\pm   0.023$\\
2.00--2.50 &$   0.650\pm   0.016$ & 1.4 &  1.6 &  0.5 &  0.2 &  1.2 &  0.0 &  2.5 & $   0.635\pm   0.015$\\
\hline
\multicolumn{10}{l}{$80 < \mll < 120$\,\GeV} \\
0.00--0.25 &$   4.248\pm   0.051$ & 0.9 &  0.7 &  0.2 &  0.1 &  0.4 &  0.0 &  1.2 & $   4.208\pm   0.054$\\
0.25--0.50 &$   3.999\pm   0.045$ & 0.9 &  0.6 &  0.2 &  0.1 &  0.4 &  0.0 &  1.1 & $   3.963\pm   0.046$\\
0.50--0.75 &$   3.716\pm   0.043$ & 0.9 &  0.6 &  0.1 &  0.1 &  0.4 &  0.0 &  1.1 & $   3.685\pm   0.042$\\
0.75--1.00 &$   3.283\pm   0.040$ & 1.0 &  0.6 &  0.1 &  0.2 &  0.4 &  0.0 &  1.2 & $   3.255\pm   0.039$\\
1.00--1.25 &$   2.900\pm   0.036$ & 1.0 &  0.6 &  0.1 &  0.1 &  0.3 &  0.0 &  1.3 & $   2.878\pm   0.036$\\
1.25--1.50 &$   2.448\pm   0.038$ & 1.3 &  0.7 &  0.1 &  0.2 &  0.4 &  0.0 &  1.6 & $   2.425\pm   0.038$\\
1.50--1.75 &$   2.011\pm   0.032$ & 1.4 &  0.7 &  0.2 &  0.1 &  0.4 &  0.0 &  1.6 & $   1.992\pm   0.032$\\
1.75--2.00 &$   1.586\pm   0.029$ & 1.5 &  0.8 &  0.2 &  0.2 &  0.6 &  0.0 &  1.9 & $   1.581\pm   0.030$\\
2.00--2.50 &$   1.017\pm   0.019$ & 1.4 &  0.8 &  0.4 &  0.1 &  0.8 &  0.0 &  1.9 & $   1.010\pm   0.019$\\
\hline
\multicolumn{10}{l}{$120 < \mll < 200$\,\GeV} \\
0.00--0.25 &$   1.783\pm   0.028$ & 0.9 &  1.2 &  0.2 &  0.1 &  0.5 &  0.0 &  1.6 & $   1.832\pm   0.025$\\
0.25--0.50 &$   1.706\pm   0.025$ & 0.9 &  1.0 &  0.2 &  0.1 &  0.4 &  0.0 &  1.5 & $   1.751\pm   0.023$\\
0.50--0.75 &$   1.645\pm   0.022$ & 0.9 &  0.9 &  0.2 &  0.1 &  0.4 &  0.0 &  1.4 & $   1.685\pm   0.022$\\
0.75--1.00 &$   1.549\pm   0.021$ & 1.0 &  0.8 &  0.2 &  0.1 &  0.3 &  0.0 &  1.3 & $   1.587\pm   0.021$\\
1.00--1.25 &$   1.370\pm   0.018$ & 1.0 &  0.7 &  0.2 &  0.1 &  0.4 &  0.0 &  1.3 & $   1.402\pm   0.019$\\
1.25--1.50 &$   1.208\pm   0.019$ & 1.3 &  0.7 &  0.2 &  0.1 &  0.4 &  0.0 &  1.6 & $   1.235\pm   0.019$\\
1.50--1.75 &$   1.014\pm   0.016$ & 1.3 &  0.7 &  0.3 &  0.1 &  0.5 &  0.0 &  1.6 & $   1.037\pm   0.016$\\
1.75--2.00 &$   0.845\pm   0.015$ & 1.4 &  0.7 &  0.2 &  0.1 &  0.6 &  0.0 &  1.7 & $   0.863\pm   0.015$\\
2.00--2.50 &$  0.5801\pm  0.0098$ & 1.3 &  0.6 &  0.3 &  0.1 &  0.9 &  0.0 &  1.7 & $  0.5938\pm  0.0100$\\
\hline
\multicolumn{10}{l}{$200 < \mll < 500+$\,\GeV} \\
0.00--0.25 &$  0.1728\pm  0.0047$ & 1.6 &  1.5 &  0.5 &  0.2 &  1.6 &  0.1 &  2.7 & $  0.1840\pm  0.0050$\\
0.25--0.50 &$  0.1793\pm  0.0047$ & 1.5 &  1.3 &  0.6 &  0.2 &  1.7 &  0.1 &  2.6 & $  0.1914\pm  0.0050$\\
0.50--0.75 &$  0.1744\pm  0.0046$ & 1.5 &  1.1 &  0.5 &  0.1 &  1.8 &  0.1 &  2.6 & $  0.1859\pm  0.0048$\\
0.75--1.00 &$  0.1736\pm  0.0044$ & 1.5 &  1.0 &  0.5 &  0.2 &  1.7 &  0.0 &  2.6 & $  0.1853\pm  0.0047$\\
1.00--1.25 &$  0.1729\pm  0.0043$ & 1.5 &  0.9 &  0.5 &  0.2 &  1.7 &  0.0 &  2.5 & $  0.1837\pm  0.0046$\\
1.25--1.50 &$  0.1559\pm  0.0045$ & 1.9 &  1.0 &  0.6 &  0.6 &  1.8 &  0.0 &  2.9 & $  0.1652\pm  0.0048$\\
1.50--1.75 &$  0.1474\pm  0.0039$ & 1.7 &  1.0 &  0.5 &  0.2 &  1.6 &  0.0 &  2.6 & $  0.1563\pm  0.0041$\\
1.75--2.00 &$  0.1303\pm  0.0037$ & 1.9 &  1.1 &  0.5 &  0.2 &  1.6 &  0.0 &  2.8 & $  0.1382\pm  0.0039$\\
2.00--2.50 &$  0.0955\pm  0.0026$ & 1.6 &  1.3 &  0.4 &  0.3 &  1.7 &  0.1 &  2.7 & $  0.1012\pm  0.0027$\\
\hline
\end{tabular}
\vspace{3mm}
 
}
\caption{\label{t:insXSec6}Normalised differential cross-sections as a function of \etalvmll. The columns show the bin ranges, measured cross-section and total uncertainty, relative statistical uncertainty, relative systematic uncertainties in various categories (see text), total relative uncertainty, and differential cross-section corrected to remove contributions via $W\rightarrow\tau\rightarrow e/\mu$ decays. Relative uncertainties smaller than 0.05\% are indicated by `0.0'. The last bin includes overflows where indicated by the `+' sign.}
\end{table}
 
\begin{table}
 
{\centering \small
\begin{tabular}{lrrrrrrrrr}
\hline
Absolute & $\mathrm{d^2}\sigma/\mathrm{d}\rapll\mathrm{d}\mll$ & Stat. & \ttbar\ mod. & Lept. & Jet/$b$ & Bkg. & $L/E_\mathrm{b}$ & Total & $\mathrm{d^2}\sigma/\mathrm{d}\rapll\mathrm{d}\mll$ \\
& & & & & & & & & (no $\tau$) \\
Bin [unit $|y|$ \GeV] & [fb/unit $|y|$ \GeV] & (\%) & (\%) & (\%) & (\%) & (\%) & (\%) & (\%) & [fb/unit $|y|$ \GeV] \\\hline
\multicolumn{10}{l}{$0 < \mll < 80$\,\GeV} \\
0.00--0.50 &$    44.9\pm     1.3$ & 0.7 &  0.8 &  0.8 &  0.2 &  0.9 &  2.3 &  2.8 & $    37.1\pm     1.0$\\
0.50--1.00 &$    38.7\pm     1.1$ & 0.8 &  1.0 &  0.8 &  0.2 &  0.9 &  2.3 &  2.9 & $    32.1\pm     0.9$\\
1.00--1.50 &$   29.48\pm    0.95$ & 1.0 &  1.4 &  0.9 &  0.2 &  1.0 &  2.3 &  3.2 & $   24.56\pm    0.74$\\
1.50--2.00 &$   17.38\pm    0.65$ & 1.5 &  1.9 &  1.0 &  0.4 &  1.3 &  2.3 &  3.8 & $   14.58\pm    0.52$\\
2.00--2.50 &$    4.13\pm    0.24$ & 3.4 &  2.9 &  1.2 &  0.5 &  2.5 &  2.4 &  5.8 & $    3.49\pm    0.20$\\
\hline
\multicolumn{10}{l}{$80 < \mll < 120$\,\GeV} \\
0.00--0.50 &$    67.4\pm     1.9$ & 0.8 &  0.7 &  0.7 &  0.3 &  0.8 &  2.3 &  2.7 & $    57.0\pm     1.6$\\
0.50--1.00 &$    56.9\pm     1.6$ & 0.9 &  0.8 &  0.7 &  0.3 &  0.7 &  2.3 &  2.8 & $    48.2\pm     1.4$\\
1.00--1.50 &$    40.3\pm     1.2$ & 1.2 &  1.2 &  0.7 &  0.3 &  0.7 &  2.3 &  3.0 & $    34.5\pm     1.1$\\
1.50--2.00 &$   17.92\pm    0.68$ & 1.9 &  1.7 &  0.8 &  0.3 &  1.4 &  2.3 &  3.8 & $   15.33\pm    0.60$\\
2.00--2.50 &$    3.30\pm    0.24$ & 5.3 &  2.9 &  1.2 &  1.1 &  2.8 &  2.4 &  7.3 & $    2.86\pm    0.21$\\
\hline
\multicolumn{10}{l}{$120 < \mll < 200$\,\GeV} \\
0.00--0.50 &$   33.95\pm    0.98$ & 0.8 &  1.2 &  0.6 &  0.2 &  0.7 &  2.3 &  2.9 & $   29.66\pm    0.84$\\
0.50--1.00 &$   28.57\pm    0.85$ & 0.9 &  1.3 &  0.6 &  0.3 &  0.8 &  2.3 &  3.0 & $   25.02\pm    0.72$\\
1.00--1.50 &$   17.01\pm    0.56$ & 1.2 &  1.5 &  0.6 &  0.2 &  1.1 &  2.3 &  3.3 & $   14.98\pm    0.48$\\
1.50--2.00 &$    6.65\pm    0.28$ & 2.1 &  2.0 &  0.8 &  0.3 &  1.6 &  2.3 &  4.2 & $    5.89\pm    0.25$\\
2.00--2.50 &$    0.84\pm    0.08$ & 7.4 &  3.6 &  1.6 &  1.2 &  4.2 &  2.5 &  9.8 & $    0.74\pm    0.07$\\
\hline
\multicolumn{10}{l}{$200 < \mll < 500+$\,\GeV} \\
0.00--0.50 &$    4.85\pm    0.18$ & 1.1 &  1.7 &  0.6 &  0.3 &  1.8 &  2.4 &  3.6 & $    4.41\pm    0.16$\\
0.50--1.00 &$    3.56\pm    0.13$ & 1.3 &  1.2 &  0.6 &  0.3 &  1.9 &  2.4 &  3.6 & $    3.25\pm    0.11$\\
1.00--1.50 &$    1.72\pm    0.07$ & 2.0 &  1.5 &  0.9 &  0.3 &  2.1 &  2.4 &  4.1 & $    1.57\pm    0.06$\\
1.50--2.00 &$    0.43\pm    0.03$ & 4.1 &  2.6 &  1.2 &  0.3 &  2.9 &  2.4 &  6.2 & $    0.40\pm    0.02$\\
2.00--2.50 &$    0.04\pm    0.01$ &15.9 &  7.1 &  2.4 &  1.5 &  5.2 &  2.7 & 18.6 & $    0.04\pm    0.01$\\
\hline
\end{tabular}
\vspace{3mm}
 
}
\caption{\label{t:insXSec7}Absolute differential cross-sections as a function of \rapllvmll. The columns show the bin ranges, measured cross-section and total uncertainty, relative statistical uncertainty, relative systematic uncertainties in various categories (see text), total relative uncertainty, and differential cross-section corrected to remove contributions via $W\rightarrow\tau\rightarrow e/\mu$ decays. Relative uncertainties smaller than 0.05\% are indicated by `0.0'. The last bin includes overflows where indicated by the `+' sign.}
\end{table}
 
\begin{table}
 
{\centering \small
\begin{tabular}{lrrrrrrrrr}
\hline
Normalised & $\frac{1}{\sigma}\mathrm{d^2}\sigma/\mathrm{d}\rapll\mathrm{d}\mll$ & Stat. & \ttbar\ mod. & Lept. & Jet/$b$ & Bkg. & $L/E_\mathrm{b}$ & Total & $\frac{1}{\sigma}\mathrm{d^2}\sigma/\mathrm{d}\rapll\mathrm{d}\mll$ \\
& & & & & & & & & (no $\tau$) \\
Bin [unit $|y|$ \GeV] & [$10^{-3}/$unit $|y|$ \GeV] & (\%) & (\%) & (\%) & (\%) & (\%) & (\%) & (\%) & [$10^{-3}/$unit $|y|$ \GeV] \\\hline
\multicolumn{10}{l}{$0 < \mll < 80$\,\GeV} \\
0.00--0.50 &$   3.165\pm   0.044$ & 0.7 &  1.0 &  0.2 &  0.1 &  0.7 &  0.0 &  1.4 & $   3.061\pm   0.040$\\
0.50--1.00 &$   2.733\pm   0.035$ & 0.8 &  0.8 &  0.2 &  0.1 &  0.6 &  0.0 &  1.3 & $   2.649\pm   0.032$\\
1.00--1.50 &$   2.080\pm   0.032$ & 1.0 &  0.9 &  0.3 &  0.1 &  0.6 &  0.0 &  1.5 & $   2.024\pm   0.029$\\
1.50--2.00 &$   1.226\pm   0.027$ & 1.4 &  1.3 &  0.4 &  0.4 &  0.9 &  0.0 &  2.2 & $   1.202\pm   0.026$\\
2.00--2.50 &$   0.291\pm   0.014$ & 3.4 &  2.3 &  0.7 &  0.4 &  2.2 &  0.1 &  4.7 & $   0.287\pm   0.013$\\
\hline
\multicolumn{10}{l}{$80 < \mll < 120$\,\GeV} \\
0.00--0.50 &$   4.758\pm   0.049$ & 0.8 &  0.5 &  0.2 &  0.1 &  0.4 &  0.0 &  1.0 & $   4.696\pm   0.048$\\
0.50--1.00 &$   4.017\pm   0.044$ & 0.9 &  0.5 &  0.1 &  0.1 &  0.4 &  0.0 &  1.1 & $   3.974\pm   0.045$\\
1.00--1.50 &$   2.846\pm   0.041$ & 1.1 &  0.8 &  0.2 &  0.2 &  0.4 &  0.0 &  1.5 & $   2.840\pm   0.043$\\
1.50--2.00 &$   1.264\pm   0.031$ & 1.8 &  1.3 &  0.2 &  0.1 &  1.0 &  0.0 &  2.5 & $   1.263\pm   0.033$\\
2.00--2.50 &$   0.233\pm   0.015$ & 5.3 &  2.6 &  0.7 &  1.1 &  2.5 &  0.1 &  6.5 & $   0.236\pm   0.016$\\
\hline
\multicolumn{10}{l}{$120 < \mll < 200$\,\GeV} \\
0.00--0.50 &$   2.395\pm   0.024$ & 0.8 &  0.5 &  0.2 &  0.1 &  0.3 &  0.0 &  1.0 & $   2.444\pm   0.025$\\
0.50--1.00 &$   2.016\pm   0.024$ & 0.8 &  0.7 &  0.2 &  0.1 &  0.4 &  0.0 &  1.2 & $   2.062\pm   0.023$\\
1.00--1.50 &$   1.200\pm   0.021$ & 1.2 &  1.0 &  0.2 &  0.1 &  0.7 &  0.0 &  1.7 & $   1.234\pm   0.021$\\
1.50--2.00 &$   0.469\pm   0.014$ & 2.1 &  1.6 &  0.4 &  0.3 &  1.2 &  0.1 &  3.0 & $   0.485\pm   0.015$\\
2.00--2.50 &$  0.0592\pm  0.0054$ & 7.4 &  3.4 &  1.3 &  1.2 &  3.9 &  0.2 &  9.2 & $  0.0611\pm  0.0056$\\
\hline
\multicolumn{10}{l}{$200 < \mll < 500+$\,\GeV} \\
0.00--0.50 &$  0.3425\pm  0.0084$ & 1.1 &  1.5 &  0.4 &  0.1 &  1.6 &  0.0 &  2.5 & $  0.3634\pm  0.0086$\\
0.50--1.00 &$  0.2513\pm  0.0058$ & 1.3 &  0.7 &  0.5 &  0.3 &  1.6 &  0.0 &  2.3 & $  0.2676\pm  0.0061$\\
1.00--1.50 &$  0.1211\pm  0.0036$ & 2.0 &  1.0 &  0.7 &  0.1 &  1.8 &  0.1 &  2.9 & $  0.1292\pm  0.0038$\\
1.50--2.00 &$  0.0306\pm  0.0017$ & 4.1 &  2.3 &  1.0 &  0.3 &  2.5 &  0.1 &  5.4 & $  0.0327\pm  0.0018$\\
2.00--2.50 &$  0.0029\pm  0.0005$ &15.9 &  6.9 &  2.1 &  1.5 &  4.9 &  0.4 & 18.2 & $  0.0031\pm  0.0006$\\
\hline
\end{tabular}
\vspace{3mm}
 
}
\caption{\label{t:insXSec8}Normalised differential cross-sections as a function of \rapllvmll. The columns show the bin ranges, measured cross-section and total uncertainty, relative statistical uncertainty, relative systematic uncertainties in various categories (see text), total relative uncertainty, and differential cross-section corrected to remove contributions via $W\rightarrow\tau\rightarrow e/\mu$ decays. Relative uncertainties smaller than 0.05\% are indicated by `0.0'. The last bin includes overflows where indicated by the `+' sign.}
\end{table}
 
\begin{table}
 
{\centering \small
\begin{tabular}{lrrrrrrrrr}
\hline
Absolute & $\mathrm{d^2}\sigma/\mathrm{d}\dphill\mathrm{d}\mll$ & Stat. & \ttbar\ mod. & Lept. & Jet/$b$ & Bkg. & $L/E_\mathrm{b}$ & Total & $\mathrm{d^2}\sigma/\mathrm{d}\dphill\mathrm{d}\mll$ \\
& & & & & & & & & (no $\tau$) \\
Bin [rad \GeV] & [fb/rad \GeV] & (\%) & (\%) & (\%) & (\%) & (\%) & (\%) & (\%) & [fb/rad \GeV] \\\hline
\multicolumn{10}{l}{$0 < \mll < 80$\,\GeV} \\
0.00--0.31 &$    31.2\pm     1.0$ & 1.2 &  1.6 &  0.8 &  0.1 &  1.0 &  2.3 &  3.4 & $    27.0\pm     0.9$\\
0.31--0.63 &$   31.13\pm    0.99$ & 1.1 &  1.4 &  0.8 &  0.1 &  0.9 &  2.3 &  3.2 & $   27.04\pm    0.82$\\
0.63--0.94 &$   30.34\pm    0.95$ & 1.2 &  1.3 &  0.8 &  0.2 &  0.9 &  2.3 &  3.1 & $   26.31\pm    0.79$\\
0.94--1.26 &$   29.60\pm    0.92$ & 1.2 &  1.1 &  0.8 &  0.2 &  0.9 &  2.3 &  3.1 & $   25.31\pm    0.76$\\
1.26--1.57 &$   25.05\pm    0.80$ & 1.3 &  1.1 &  0.9 &  0.2 &  1.0 &  2.3 &  3.2 & $   20.89\pm    0.64$\\
1.57--1.88 &$   20.21\pm    0.66$ & 1.5 &  1.1 &  0.9 &  0.3 &  1.0 &  2.3 &  3.3 & $   16.31\pm    0.52$\\
1.88--2.20 &$   15.63\pm    0.55$ & 1.8 &  1.3 &  1.0 &  0.2 &  1.1 &  2.3 &  3.5 & $   12.16\pm    0.42$\\
2.20--2.51 &$   12.27\pm    0.46$ & 2.1 &  1.4 &  1.0 &  0.3 &  1.1 &  2.3 &  3.7 & $    9.27\pm    0.34$\\
2.51--2.83 &$   10.07\pm    0.40$ & 2.3 &  1.6 &  1.1 &  0.2 &  1.3 &  2.3 &  4.0 & $    7.53\pm    0.29$\\
2.83--3.14 &$    8.38\pm    0.37$ & 2.6 &  1.8 &  1.1 &  0.6 &  1.7 &  2.3 &  4.4 & $    6.20\pm    0.27$\\
\hline
\multicolumn{10}{l}{$80 < \mll < 120$\,\GeV} \\
0.00--0.31 &$   10.11\pm    0.44$ & 2.8 &  1.9 &  0.8 &  0.4 &  1.3 &  2.3 &  4.3 & $    9.13\pm    0.40$\\
0.31--0.63 &$   10.92\pm    0.45$ & 2.6 &  1.7 &  0.8 &  0.4 &  1.0 &  2.3 &  4.1 & $    9.84\pm    0.40$\\
0.63--0.94 &$   13.61\pm    0.52$ & 2.3 &  1.5 &  0.7 &  0.4 &  1.0 &  2.3 &  3.8 & $   12.28\pm    0.46$\\
0.94--1.26 &$   19.44\pm    0.68$ & 1.9 &  1.3 &  0.7 &  0.2 &  0.9 &  2.3 &  3.5 & $   17.50\pm    0.60$\\
1.26--1.57 &$   29.17\pm    0.94$ & 1.6 &  1.1 &  0.6 &  0.4 &  0.9 &  2.3 &  3.2 & $   26.19\pm    0.83$\\
1.57--1.88 &$    38.7\pm     1.2$ & 1.4 &  1.0 &  0.7 &  0.2 &  0.8 &  2.3 &  3.1 & $    34.1\pm     1.0$\\
1.88--2.20 &$    44.6\pm     1.4$ & 1.3 &  0.9 &  0.7 &  0.4 &  0.8 &  2.3 &  3.1 & $    37.9\pm     1.1$\\
2.20--2.51 &$    45.4\pm     1.4$ & 1.3 &  0.9 &  0.7 &  0.3 &  0.8 &  2.3 &  3.0 & $    37.3\pm     1.1$\\
2.51--2.83 &$    42.5\pm     1.3$ & 1.4 &  0.9 &  0.7 &  0.4 &  0.8 &  2.3 &  3.1 & $    34.1\pm     1.1$\\
2.83--3.14 &$    41.8\pm     1.3$ & 1.4 &  1.1 &  0.7 &  0.4 &  0.8 &  2.3 &  3.1 & $    33.1\pm     1.1$\\
\hline
\multicolumn{10}{l}{$120 < \mll < 200$\,\GeV} \\
0.00--0.31 &$    3.44\pm    0.22$ & 3.3 &  4.8 &  0.8 &  0.3 &  1.3 &  2.3 &  6.5 & $    3.17\pm    0.21$\\
0.31--0.63 &$    3.33\pm    0.20$ & 3.3 &  4.1 &  0.8 &  0.2 &  1.3 &  2.3 &  6.0 & $    3.05\pm    0.19$\\
0.63--0.94 &$    3.92\pm    0.21$ & 3.1 &  3.4 &  0.8 &  0.5 &  1.5 &  2.3 &  5.5 & $    3.59\pm    0.20$\\
0.94--1.26 &$    5.41\pm    0.25$ & 2.6 &  2.7 &  0.7 &  0.2 &  1.2 &  2.3 &  4.6 & $    4.94\pm    0.24$\\
1.26--1.57 &$    8.28\pm    0.32$ & 2.1 &  2.0 &  0.7 &  0.4 &  1.0 &  2.3 &  3.9 & $    7.55\pm    0.31$\\
1.57--1.88 &$   12.70\pm    0.44$ & 1.7 &  1.5 &  0.6 &  0.2 &  1.0 &  2.4 &  3.5 & $   11.52\pm    0.40$\\
1.88--2.20 &$   18.44\pm    0.59$ & 1.4 &  1.3 &  0.6 &  0.2 &  0.9 &  2.3 &  3.2 & $   16.55\pm    0.52$\\
2.20--2.51 &$   24.92\pm    0.80$ & 1.2 &  1.5 &  0.6 &  0.3 &  0.8 &  2.3 &  3.2 & $   21.86\pm    0.68$\\
2.51--2.83 &$   28.48\pm    0.99$ & 1.2 &  2.0 &  0.6 &  0.4 &  0.8 &  2.3 &  3.5 & $   24.35\pm    0.85$\\
2.83--3.14 &$    29.7\pm     1.1$ & 1.1 &  2.7 &  0.6 &  0.2 &  0.8 &  2.3 &  3.8 & $    25.0\pm     1.0$\\
\hline
\multicolumn{10}{l}{$200 < \mll < 500+$\,\GeV} \\
0.00--0.31 &$    0.23\pm    0.03$ & 7.1 &  6.3 &  0.9 &  1.0 &  5.6 &  2.3 & 11.4 & $    0.22\pm    0.02$\\
0.31--0.63 &$    0.30\pm    0.03$ & 6.3 &  5.6 &  0.8 &  0.7 &  4.3 &  2.3 &  9.8 & $    0.28\pm    0.03$\\
0.63--0.94 &$    0.29\pm    0.03$ & 6.2 &  4.8 &  0.8 &  0.8 &  5.0 &  2.4 &  9.7 & $    0.27\pm    0.03$\\
0.94--1.26 &$    0.36\pm    0.03$ & 5.6 &  3.9 &  0.7 &  0.8 &  3.7 &  2.3 &  8.2 & $    0.34\pm    0.03$\\
1.26--1.57 &$    0.54\pm    0.04$ & 4.3 &  3.1 &  0.8 &  0.5 &  3.2 &  2.4 &  6.7 & $    0.51\pm    0.03$\\
1.57--1.88 &$    0.93\pm    0.05$ & 3.3 &  2.4 &  0.8 &  0.4 &  2.7 &  2.3 &  5.5 & $    0.87\pm    0.05$\\
1.88--2.20 &$    1.60\pm    0.07$ & 2.4 &  1.7 &  0.7 &  0.5 &  2.1 &  2.3 &  4.4 & $    1.48\pm    0.07$\\
2.20--2.51 &$    2.76\pm    0.11$ & 1.9 &  1.4 &  0.7 &  0.4 &  1.9 &  2.4 &  3.9 & $    2.54\pm    0.10$\\
2.51--2.83 &$    4.22\pm    0.16$ & 1.5 &  1.5 &  0.6 &  0.3 &  1.7 &  2.4 &  3.7 & $    3.83\pm    0.14$\\
2.83--3.14 &$    5.68\pm    0.21$ & 1.3 &  2.0 &  0.6 &  0.3 &  1.3 &  2.4 &  3.7 & $    5.08\pm    0.18$\\
\hline
\end{tabular}
\vspace{3mm}
 
}
\caption{\label{t:insXSec9}Absolute differential cross-sections as a function of \dphillvmll. The columns show the bin ranges, measured cross-section and total uncertainty, relative statistical uncertainty, relative systematic uncertainties in various categories (see text), total relative uncertainty, and differential cross-section corrected to remove contributions via $W\rightarrow\tau\rightarrow e/\mu$ decays. Relative uncertainties smaller than 0.05\% are indicated by `0.0'. The bin boundaries for \dphill\ correspond to exact multiples of $\pi/10$ but are quoted to two decimal places.}
\end{table}
 
\begin{table}
 
{\centering \small
\begin{tabular}{lrrrrrrrrr}
\hline
Normalised & $\frac{1}{\sigma}\mathrm{d^2}\sigma/\mathrm{d}\dphill\mathrm{d}\mll$ & Stat. & \ttbar\ mod. & Lept. & Jet/$b$ & Bkg. & $L/E_\mathrm{b}$ & Total & $\frac{1}{\sigma}\mathrm{d^2}\sigma/\mathrm{d}\dphill\mathrm{d}\mll$ \\
& & & & & & & & & (no $\tau$) \\
Bin [rad \GeV] & [$10^{-3}/$rad \GeV] & (\%) & (\%) & (\%) & (\%) & (\%) & (\%) & (\%) & [$10^{-3}/$rad \GeV] \\\hline
\multicolumn{10}{l}{$0 < \mll < 80$\,\GeV} \\
0.00--0.31 &$   2.204\pm   0.047$ & 1.2 &  1.6 &  0.2 &  0.2 &  0.7 &  0.0 &  2.1 & $   2.224\pm   0.045$\\
0.31--0.63 &$   2.196\pm   0.041$ & 1.1 &  1.3 &  0.1 &  0.2 &  0.6 &  0.0 &  1.8 & $   2.227\pm   0.038$\\
0.63--0.94 &$   2.140\pm   0.036$ & 1.1 &  1.1 &  0.1 &  0.1 &  0.6 &  0.0 &  1.7 & $   2.167\pm   0.034$\\
0.94--1.26 &$   2.088\pm   0.033$ & 1.2 &  0.9 &  0.2 &  0.1 &  0.6 &  0.0 &  1.6 & $   2.084\pm   0.032$\\
1.26--1.57 &$   1.767\pm   0.029$ & 1.3 &  0.8 &  0.2 &  0.2 &  0.7 &  0.0 &  1.7 & $   1.720\pm   0.028$\\
1.57--1.88 &$   1.426\pm   0.026$ & 1.5 &  0.7 &  0.3 &  0.1 &  0.7 &  0.0 &  1.8 & $   1.343\pm   0.024$\\
1.88--2.20 &$   1.103\pm   0.023$ & 1.7 &  0.9 &  0.3 &  0.1 &  0.8 &  0.0 &  2.1 & $   1.001\pm   0.022$\\
2.20--2.51 &$   0.865\pm   0.021$ & 2.0 &  1.0 &  0.4 &  0.1 &  0.8 &  0.0 &  2.5 & $   0.763\pm   0.019$\\
2.51--2.83 &$   0.710\pm   0.020$ & 2.3 &  1.2 &  0.5 &  0.2 &  1.1 &  0.1 &  2.8 & $   0.620\pm   0.018$\\
2.83--3.14 &$   0.591\pm   0.020$ & 2.5 &  1.4 &  0.5 &  0.5 &  1.5 &  0.0 &  3.3 & $   0.511\pm   0.017$\\
\hline
\multicolumn{10}{l}{$80 < \mll < 120$\,\GeV} \\
0.00--0.31 &$   0.713\pm   0.023$ & 2.7 &  1.4 &  0.2 &  0.3 &  1.0 &  0.0 &  3.3 & $   0.751\pm   0.026$\\
0.31--0.63 &$   0.770\pm   0.023$ & 2.6 &  1.3 &  0.2 &  0.4 &  0.7 &  0.0 &  3.0 & $   0.810\pm   0.025$\\
0.63--0.94 &$   0.960\pm   0.026$ & 2.3 &  1.1 &  0.1 &  0.4 &  0.6 &  0.0 &  2.7 & $   1.011\pm   0.027$\\
0.94--1.26 &$   1.371\pm   0.030$ & 1.9 &  0.9 &  0.2 &  0.2 &  0.4 &  0.0 &  2.2 & $   1.441\pm   0.031$\\
1.26--1.57 &$   2.057\pm   0.037$ & 1.6 &  0.7 &  0.1 &  0.3 &  0.4 &  0.0 &  1.8 & $   2.156\pm   0.038$\\
1.57--1.88 &$   2.729\pm   0.042$ & 1.4 &  0.6 &  0.1 &  0.1 &  0.4 &  0.0 &  1.5 & $   2.809\pm   0.043$\\
1.88--2.20 &$   3.145\pm   0.049$ & 1.3 &  0.6 &  0.2 &  0.3 &  0.5 &  0.0 &  1.6 & $   3.124\pm   0.048$\\
2.20--2.51 &$   3.201\pm   0.052$ & 1.3 &  0.8 &  0.2 &  0.2 &  0.5 &  0.0 &  1.6 & $   3.075\pm   0.050$\\
2.51--2.83 &$   2.998\pm   0.053$ & 1.4 &  0.9 &  0.2 &  0.3 &  0.5 &  0.0 &  1.8 & $   2.811\pm   0.051$\\
2.83--3.14 &$   2.948\pm   0.056$ & 1.4 &  1.1 &  0.2 &  0.3 &  0.5 &  0.0 &  1.9 & $   2.722\pm   0.058$\\
\hline
\multicolumn{10}{l}{$120 < \mll < 200$\,\GeV} \\
0.00--0.31 &$   0.243\pm   0.014$ & 3.3 &  4.7 &  0.3 &  0.4 &  1.2 &  0.1 &  5.9 & $   0.261\pm   0.016$\\
0.31--0.63 &$   0.235\pm   0.012$ & 3.3 &  3.9 &  0.3 &  0.3 &  1.2 &  0.0 &  5.3 & $   0.251\pm   0.014$\\
0.63--0.94 &$   0.276\pm   0.013$ & 3.1 &  3.2 &  0.3 &  0.4 &  1.3 &  0.0 &  4.7 & $   0.296\pm   0.015$\\
0.94--1.26 &$   0.382\pm   0.014$ & 2.6 &  2.4 &  0.2 &  0.2 &  0.9 &  0.0 &  3.7 & $   0.407\pm   0.017$\\
1.26--1.57 &$   0.584\pm   0.016$ & 2.1 &  1.7 &  0.3 &  0.4 &  0.7 &  0.0 &  2.8 & $   0.621\pm   0.020$\\
1.57--1.88 &$   0.896\pm   0.018$ & 1.7 &  1.0 &  0.2 &  0.1 &  0.6 &  0.0 &  2.0 & $   0.948\pm   0.022$\\
1.88--2.20 &$   1.300\pm   0.021$ & 1.3 &  0.7 &  0.2 &  0.1 &  0.5 &  0.0 &  1.6 & $   1.363\pm   0.022$\\
2.20--2.51 &$   1.758\pm   0.029$ & 1.2 &  1.1 &  0.2 &  0.2 &  0.3 &  0.0 &  1.6 & $   1.800\pm   0.028$\\
2.51--2.83 &$   2.009\pm   0.042$ & 1.1 &  1.7 &  0.2 &  0.2 &  0.3 &  0.0 &  2.1 & $   2.005\pm   0.043$\\
2.83--3.14 &$   2.095\pm   0.057$ & 1.1 &  2.5 &  0.2 &  0.1 &  0.3 &  0.0 &  2.7 & $   2.062\pm   0.063$\\
\hline
\multicolumn{10}{l}{$200 < \mll < 500+$\,\GeV} \\
0.00--0.31 &$  0.0164\pm  0.0018$ & 7.1 &  6.0 &  0.7 &  1.0 &  5.5 &  0.0 & 10.9 & $  0.0179\pm  0.0020$\\
0.31--0.63 &$  0.0209\pm  0.0019$ & 6.3 &  5.3 &  0.4 &  0.7 &  4.1 &  0.0 &  9.2 & $  0.0227\pm  0.0021$\\
0.63--0.94 &$  0.0205\pm  0.0019$ & 6.1 &  4.5 &  0.5 &  0.8 &  4.9 &  0.1 &  9.1 & $  0.0224\pm  0.0021$\\
0.94--1.26 &$  0.0255\pm  0.0019$ & 5.6 &  3.6 &  0.3 &  0.8 &  3.5 &  0.0 &  7.6 & $  0.0279\pm  0.0022$\\
1.26--1.57 &$  0.0382\pm  0.0023$ & 4.2 &  2.8 &  0.6 &  0.6 &  3.1 &  0.0 &  6.0 & $  0.0416\pm  0.0026$\\
1.57--1.88 &$  0.0659\pm  0.0031$ & 3.3 &  2.0 &  0.6 &  0.2 &  2.5 &  0.0 &  4.7 & $  0.0717\pm  0.0034$\\
1.88--2.20 &$  0.1130\pm  0.0039$ & 2.4 &  1.3 &  0.5 &  0.4 &  2.0 &  0.0 &  3.4 & $  0.1221\pm  0.0043$\\
2.20--2.51 &$  0.1949\pm  0.0053$ & 1.8 &  1.0 &  0.6 &  0.2 &  1.7 &  0.0 &  2.7 & $  0.2094\pm  0.0056$\\
2.51--2.83 &$  0.2974\pm  0.0075$ & 1.5 &  1.3 &  0.5 &  0.3 &  1.5 &  0.1 &  2.5 & $  0.3154\pm  0.0075$\\
2.83--3.14 &$   0.401\pm   0.011$ & 1.3 &  2.0 &  0.4 &  0.1 &  1.0 &  0.1 &  2.6 & $   0.418\pm   0.010$\\
\hline
\end{tabular}
\vspace{3mm}
 
}
\caption{\label{t:insXSec10}Normalised differential cross-sections as a function of \dphillvmll. The columns show the bin ranges, measured cross-section and total uncertainty, relative statistical uncertainty, relative systematic uncertainties in various categories (see text), total relative uncertainty, and differential cross-section corrected to remove contributions via $W\rightarrow\tau\rightarrow e/\mu$ decays. Relative uncertainties smaller than 0.05\% are indicated by `0.0'. The bin boundaries for \dphill\ correspond to exact multiples of $\pi/10$ but are quoted to two decimal places.}
\end{table}

\clearpage
 
\printbibliography

\clearpage
% ATLAS Collaboration author list
% Reference date of TOPQ-2018-17 is 2019-07-15
% Author list last updated on date 02-MAR-20
% Data extracted on 02-Mar-2020 for paper reference TOPQ-2018-17
% at 12:08pm
 
\begin{flushleft}
{\Large The ATLAS Collaboration}

\bigskip

G.~Aad$^\textrm{\scriptsize 102}$,    
B.~Abbott$^\textrm{\scriptsize 129}$,    
D.C.~Abbott$^\textrm{\scriptsize 103}$,    
A.~Abed~Abud$^\textrm{\scriptsize 71a,71b}$,    
K.~Abeling$^\textrm{\scriptsize 53}$,    
D.K.~Abhayasinghe$^\textrm{\scriptsize 94}$,    
S.H.~Abidi$^\textrm{\scriptsize 167}$,    
O.S.~AbouZeid$^\textrm{\scriptsize 40}$,    
N.L.~Abraham$^\textrm{\scriptsize 156}$,    
H.~Abramowicz$^\textrm{\scriptsize 161}$
% ...
\bigskip
\\

% ...
$^{40}$Niels Bohr Institute, University of Copenhagen, Copenhagen; Denmark.\\
% ...
$^{53}$II. Physikalisches Institut, Georg-August-Universit\"{a}t G\"ottingen, G\"ottingen; Germany.\\
% ...
$^{71}$$^{(a)}$INFN Sezione di Pavia;$^{(b)}$Dipartimento di Fisica, Universit\`a di Pavia, Pavia; Italy.\\
% ...
$^{94}$Department of Physics, Royal Holloway University of London, Egham; United Kingdom.\\
% ...
$^{102}$CPPM, Aix-Marseille Universit\'e, CNRS/IN2P3, Marseille; France.\\
$^{103}$Department of Physics, University of Massachusetts, Amherst MA; United States of America.\\
% ...
$^{129}$Homer L. Dodge Department of Physics and Astronomy, University of Oklahoma, Norman OK; United States of America.\\
% ...
$^{156}$Department of Physics and Astronomy, University of Sussex, Brighton; United Kingdom.\\
% ...
$^{161}$Raymond and Beverly Sackler School of Physics and Astronomy, Tel Aviv University, Tel Aviv; Israel.\\
% ...
$^{167}$Department of Physics, University of Toronto, Toronto ON; Canada.\\
% ...
$^{a}$ Also at Borough of Manhattan Community College, City University of New York, New York NY; United States of America.\\
$^{b}$ Also at CERN, Geneva; Switzerland.\\
% ...

\end{flushleft}

% Created with Glance <Atlas.Glance@cern.ch>
 
\end{document}